\documentclass[11pt]{sar-dissertation}

\usepackage{epsfig,amssymb,amsfonts,citesort,amsmath,amsbsy}
\allowdisplaybreaks[2]

\newcommand{\veck}{\text{$\vec{k}$}}

\newcommand{\dens}{\text{${\boldsymbol \rho}$}}
\newcommand{\dvphi}{\text{$\dot{\varphi}$}}
\newcommand{\pih}{\text{$\hat{\pi}$}}
\newcommand{\psip}{\text{$\psi_{+}$}}
\newcommand{\psim}{\text{$\psi_{-}$}}
\newcommand{\psibp}{\text{$\bar{\psi}_{+}$}}
\newcommand{\psibm}{\text{$\bar{\psi}_{-}$}}
\newcommand{\yc}{\text{$f$}}
\newcommand{\psib}{\text{$\bar{\psi}$}}
\newcommand{\vphiH}{\text{$\varphi_{\text{{\tiny H}}}$}}
\newcommand{\PsibH}{\text{$\bar{\Psi}_{\text{{\tiny H}}}$}}
\newcommand{\PsiH}{\text{$\Psi_{\text{{\tiny H}}}$}}
\newcommand{\PhiH}{\text{$\Phi_{\text{{\tiny H}}}$}}

\newcommand{\vk}{\text{$\vec{k}$}}

\newcommand{\xp}{\text{$x^{\prime}$}}

\newcommand{\Mpl}{\text{$M_{\text{{\tiny P}}}$}}

\newcommand{\phip}{\text{$\phi_{+}$}}
\newcommand{\phim}{\text{$\phi_{-}$}}

\newcommand{\phipm}{\text{$\phi_{\pm}$}}

\newcommand{\Jpm}{\text{$J_{\pm}$}}
\newcommand{\phih}{\text{$\hat{\phi}$}}
\newcommand{\phihp}{\text{$\hat{\phi}_{+}$}}
\newcommand{\phihm}{\text{$\hat{\phi}_{-}$}}
\newcommand{\phihpm}{\text{$\hat{\phi}_{\pm}$}}
\newcommand{\vphi}{\text{$\varphi$}}
\newcommand{\vecphi}{\text{$\vec{\phi}$}}
\newcommand{\vphip}{\text{$\varphi_{+}$}}
\newcommand{\vphim}{\text{$\varphi_{-}$}}

\newcommand{\sqmg}{\text{$\sqrt{-g}$}}
\newcommand{\sqmgp}{\text{$\sqrt{-g'}$}}

\title{NONEQUILIBRIUM DYNAMICS OF QUANTUM \\
FIELDS IN INFLATIONARY COSMOLOGY}

\author{Stephen Allen Ramsey}

\department{Department of Physics}

\advisor{Professor Bei-Lok B. Hu}

\committee{
Professor Thomas D. Cohen\\
Professor Theodore A. Jacobson\\
Professor Charles W. Misner\\
Professor John C. Wang}

\dedication{\centering In memory of Te,\\who never stopped learning.}

\acknowledgements{
This dissertation describes research carried out in collaboration with,
and under the skillful direction of, Prof.\ Bei-Lok Hu.  Without his
encouragement, guidance, and vision, this work would not have been possible.  
I am privileged to have had the opportunity to work with him and with
the Maryland Gravitation Theory Group.

The research described in this dissertation was 
supported and facilitated by many institutions and individuals.  
The Surface Physics Group and the Solid State Theory Group at the University 
of Maryland, and D.~Benton of the National Scalable Computing Project 
at the University of Pennsylvania generously permitted the use of their 
respective groups' computing resources, making possible the numerical 
work described in Chapter~\ref{chap-preheat}.  A grant from the National
Science Foundation (PHY94-21849) provided partial support for this work during 
the past two years.  The hospitality of the Hong Kong University of 
Science and Technology, the University of Buenos Aires, and the Los Alamos 
Center for Nonlinear Studies, where some of the research described in 
Chapters~\ref{chap-oncst}, \ref{chap-preheat}, and 
\ref{chap-fermion} was carried out, is gratefully acknowledged.  
Helpful advice and suggestions on various aspects of this research
were provided by Daniel Boyanovsky, Esteban Calzetta, Salman Habib, 
Diego Mazzitelli, Emil Mottola, and Alpan Raval.  

I wish to thank Nicholas Phillips, Kazutomu Shiokawa, and Philip Johnson
for their helpful advice and encouragement.

I would like to give a special word of thanks to
several colleagues who played important roles in making this dissertation 
research possible.  First, I would like to thank Prof.~Ted 
Jacobson and Dr.~Jonathan Simon, whom I frequently consulted for sage 
advice on quantum field theory in curved spacetime.
Their kindness and extraordinary patience is sincerely 
appreciated.  Second, I would like to thank Prof.~Sankar Das Sarma and 
Prof.~Ellen Williams, who took a chance in hiring me three years ago, and 
for whom I have been employed during the past four years.  I am especially 
grateful for their patience and understanding during the past few months.  
Finally, I would like to thank Greg Stephens, who 
has been an unfailing source of moral support, helpful advice, and great 
physics discussions.

I would also like to thank two individuals who played an important 
part in determining my academic worldline.  For first inspiring me to study 
physics, I thank Dr.~David Workman of the Illinois Mathematics and Science 
Academy; for introducing me to general relativity and for inspiring me to 
study cosmology, I thank Prof.~Robert Brandenberger of Brown University.

Most of all, I thank my parents and Elain.
}

\comment{This dissertation was prepared by the author on a
Macintosh Quadra 650 computer running NetBSD/mac68k, 
using the \LaTeX2e\ document preparation system and a modified 
version of the {\tt dissertation} \LaTeX2e\ document 
class by Pablo A. Straub of the University of Maryland, College Park.}

\abstract{The nonequilibrium dynamics of quantum fields is studied in
inflationary cosmology, with particular emphasis on applications to
the problem of post-inflation reheating.  
The Schwinger-Keldysh closed-time-path (CTP) formalism is utilized 
along with the two-particle-irreducible (2PI) effective action in order 
to obtain coupled, nonperturbative equations for the mean field and variance 
in a general curved background spacetime, both as a closed system in the
case of a self-interacting inflaton field, and as an open system in the
case of coarse-grained dynamics of the inflaton field interacting with
fermions.  For a model consisting of a quartically self-interacting 
O$(N)$ field theory (with unbroken symmetry) in spatially flat FRW spacetime,
the dynamics of the mean field is studied numerically, at leading order
in the large-$N$ expansion, with initial conditions appropriate to the 
end state of slow roll in chaotic inflation scenarios.  The time evolution 
of the scale factor is determined self-consistently using the semiclassical 
Einstein equation.  It is found that cosmic expansion
can dramatically affect the efficiency of parametric resonance-induced
particle production.  The production of fermions due to the oscillating
inflaton mean field is studied for the case of a scalar inflaton
coupled to a fermion field via a Yukawa coupling $f$.  The dissipation and
noise kernels appearing at $O(f^2)$ in the one-loop CTP effective action are 
shown to satisfy a zero-temperature fluctuation-dissipation relation (FDR).  
The normal-threshold $O(f^4)$ parts of the one-loop CTP effective action are 
also shown to satisfy a FDR.  The effective stochastic equation obeyed by the
inflaton zero mode at $O(f^4)$ contains multiplicative noise.  It is 
shown that stochasticity becomes important to the dynamics of the inflaton 
zero mode before the end of reheating.  The thermalization problem
is discussed, and a strategy is presented for obtaining time-local equations 
for equal-time correlation functions which goes beyond the Hartree-Fock 
approximation.  For the $\lambda \Phi^4$ field theory, the correlation 
entropy associated with a particular coarse graining scheme consisting of 
slaving the three-point function to the mean field and two-point function 
is computed, and found not to be conserved.
}

\begin{document}
\makefrontmatter

\chapter{Introduction}
\label{chap-intro}

\section{Background}
The inflationary Universe 
\cite{guth:1981a,sato:1981a,albrecht:1982b,linde:1982a,linde:1985a,starobinsky:1986a,bardeen:1987a,brandenberger:1985a,abbott:1986a,kolb:1990a,linde:1990a,linde:1996a} 
has for over a decade been the new
paradigm for addressing many basic issues in cosmology such as the 
spatial flatness-oldness problem, the large-scale homogeneity (horizon)
problem, and the small-scale inhomogeneity (structure formation) problem. 
The linkage between observations, especially
those from the recent Cosmic Background Explorer (COBE) 
data, and theory, based on grand unified theories
(GUT's) and Friedmann-Robertson-Walker-- (FRW--)de~Sitter 
models, has been pursued in earnest, but most
theoretical discussions to date are largely phenomenological and somewhat
utilitarian in nature 
\cite{steinhardt:1984a,lidsey:1997a}. This lack of rigor and precision
is understandable for at least two reasons: the precise physical
conditions between the Planck and GUT scales
(when the most cosmologically significant inflationary evolutions
are believed to have taken place) have not been clearly understood, and the
theoretical framework for the treatment of processes affecting the inception
and completion of inflation were not well developed. As stressed earlier 
by \cite{hu:1986d,hu:1986a}, the important physical
processes which can determine whether inflation can occur,
sustain, and finish with the necessary features are affected by at least
three aspects: the geometry, topology, and dynamics of the spacetime
\cite{birrell:1982a},
the quantum field theory aspects pertaining to the analysis of infrared
behavior, and the statistical mechanical aspects pertaining to
nonequilibrium processes.
These quantum and statistical processes include phase transition, particle
creation, entropy generation, fluctuation or stochastic dynamics, and structure
formation  \cite{hu:1993b,hu:1995b}.
 Most of these invoke the quantum field and statistical mechanical
aspects, and for processes occurring at the Planck scale (which are 
instrumental in starting certain models of inflation, such as proposed in
\cite{linde:1985a,starobinsky:1980a,hu:1986b}), 
also the geometry and topology of spacetime.
Two important problems involving field theory in curved spacetime
\cite{birrell:1982a}, namely, the back reaction of cosmological particle
creation 
\cite{parker:1969a,sexl:1969a,zeldovich:1970a,zeldovich:1971a,hu:1974a} 
on the structure and dynamics of spacetimes
\cite{starobinsky:1980a,zeldovich:1971a,zeldovich:1977a,hu:1977a,hu:1978a,fischetti:1979a,hartle:1979a,hartle:1980a,hu:1981a,anderson:1984a},
and the effects of geometry and topology of spacetime on cosmological
phase transitions
\cite{hu:1986a,hu:1986b,oconnor:1983a,shen:1985a,hu:1987b,berkin:1992a},
were investigated systematically and comprehensively in the 1970s and 1980s.
The statistical mechanical aspect has not been considered with equal mastery.

The statistical mechanical aspect enters into all three stages of inflationary
cosmology:
(i) At the inception: What conditions would be most conducive to starting
inflation? Do there exist metastable states for the Higgs boson field which can
generate inflation \cite{hu:1986c}? 
 Can thermal or quantum
fluctuations assist the inflaton in hopping or tunneling out of the
potential barrier in the spinodal or nucleation pictures? Most depictions so
far have been based on the finite temperature effective potential, 
which assumes an unrealistic equilibrium condition and a constant background 
field.  However, when asking such questions in critical dynamics one
should be using a Langevin or Fokker-Planck equation (a generalized
time-dependent Landau-Ginzberg equation \cite{calzetta:1997a})
incorporating dynamic dissipation and intrinsic noise consistently.
(ii) During inflation, the dynamics of the inflaton field can be more
easily understood
in terms of a Kadanoff-Migdal exponential scaling transform \cite{hu:1993d}.
The reason why the inflaton evolves as a classical stochastic field 
\cite{guth:1985a,cornwall:1988a,buryak:1996a} at late times involves the
process of decoherence,
caused by noise and fluctuations from environmental fields \cite{hu:1996a};
this necessitates statistical mechanical considerations. 
The evolution of the classical density contrast (containing the seedings
of structures) from quantum fluctuations of the inflaton also requires
both quantum and stochastic field theory considerations
\cite{guth:1982a,starobinsky:1982a,hawking:1982a,bardeen:1983a,brandenberger:1983a,mukhanov:1992a,deruelle:1992a,hu:1993a,calzetta:1995a,matacz:1996a}.
(iii) In the reheating epoch, particle creation induces
dissipation of the inflaton field, and the interaction of quantum fields is 
the source for reheating the Universe. This last epoch is the focus 
of this dissertation, as we shall detail below.

The construction of a viable theoretical framework for treating quantum
statistical processes in the  early Universe has been underway
for the past decade (for a review, see \cite{hu:1994a}).
This framework has now been successfully established,
and its application to the problems mentioned above has just begun.
The cornerstones are the Schwinger-Keldysh closed-time-path (CTP)
\cite{schwinger:1961a,bakshi:1963a,keldysh:1964a,niemi:1981a,niemi:1984a,landsman:1987a,chou:1985a,su:1988a,dewitt:1986a,jordan:1986a,calzetta:1987a,calzetta:1988b,calzetta:1989a}
effective action and the Feynman-Vernon influence functional 
\cite{feynman:1963a,feynman:1965a,caldeira:1983a,grabert:1988a,hu:1992a,hu:1993c,calzetta:1994a}
formalisms.
They are useful for treating particle creation back reaction 
\cite{calzetta:1987a,calzetta:1989a},
fluctuation or noise, and dissipation or entropy problems 
\cite{hu:1993a,calzetta:1995a,calzetta:1994a}.
Other essential ingredients include the Wigner function 
\cite{wigner:1932a,hillery:1984a}, the $n$-particle-irreducible 
($n$PI) effective action
\cite{cornwall:1974a,hu:1987b,calzetta:1988b,ramsey:1997a},
and the correlation hierarchy
\cite{calzetta:1993a,calzetta:1995b}
for treating kinetic theory processes
\cite{calzetta:1988b,calzetta:1988a} and phase transition problems
\cite{calzetta:1989b,calzetta:1995a}.
In this dissertation we apply these techniques to the
problems of inflaton damping due to back reaction from 
parametric particle creation (Chapter~\ref{chap-preheat})
and dissipation due to particle creation (Chapter~\ref{chap-fermion}),
which are relevant in in the third epoch depicted above. In parallel,
these newly developed methods in statistical field theory are now being 
applied to derive the classical  stochastic dynamics of the inflaton 
(in the second epoch) \cite{hu:1996a}, and the statistical field theory of 
spinodal decomposition (in the first epoch) \cite{calzetta:1997a}.

\section{Issues}
\label{sec-backissu}
Most all inflationary cosmologies share the feature of a period of
cosmic expansion driven by a nearly constant vacuum energy density $\rho$
(a ``vacuum-dominated'' era with effective equation of state, $p = -\rho$):
In a Friedmann-Robertson-Walker (FRW) spacetime,
the scale factor expands exponentially in cosmic time, resulting in 
extreme redshifting of the energy density of all other
forms of matter and fields. As long as the interaction time scale of any 
physical process involving given fields is longer than the cosmic
expansion time $H^{-1}$, the fields will remain in disequilibrium.
This condition can prevail in all three stages of inflation,
and one should use a fully nonequilibrium, nonperturbative treatment of
the dynamics of the inflation field.
The physics of the reheating epoch is important because it directly
determines several important cosmological parameters which are relevant
to later evolution of the Universe, and in principle verifiable by
observational data. For example, the reheating temperature is a vital link
between the inflationary Universe scenario and GUT scale baryogenesis
\cite{kolb:1996a}, and may provide a mechanism to explain the origin
of dark matter \cite{kofman:1994a,allahverdi:1997a}.

It is generally believed that at the end of inflation, the state of the
inflaton field can be approximately described by a condensate of 
zero-momentum particles undergoing coherent quasioscillations about the true 
minimum of the effective potential 
\cite{kolb:1990a,linde:1990a,traschen:1990a}.  
The reheating problem involves describing the processes  by which the many 
light fields coupled to the inflaton become populated with quanta, and
eventually thermalize.  It is commonly 
believed that if the fields interact sufficiently rapidly and strongly,  
the Universe thermalizes and turns into the radiation-dominated condition 
described by the standard Friedmann solution, but this has not been proven
satisfactorily.

There has been a great deal of work over the past 15 years on the reheating
problem, and in attempting to understand reheating, a wealth of interesting
physics has been revealed (see, e.g., \cite{kofman:1996a}).
To date, the work on
particle production during reheating largely follows two distinct approaches,
each pursued in two stages.

In the first stage of work on the reheating problem
(group 1A, \cite{abbott:1982a,albrecht:1982a,dolgov:1982a,dolgov:1990a}),
time-dependent perturbation theory was used to compute the rate of particle
production into light fields (usually fermions) coupled to the
inflaton.  Particle production rates were computed in flat space assuming
an eternally sinusoidally oscillating inflaton field.  The inflaton evolution
in FRW spacetime was modeled with a phenomenological c-number 
equation involving the Hubble parameter $H$ and the classical inflaton
amplitude $\phi$,
\begin{equation} 
\ddot{\phi} + m^2 \phi + (\Gamma_{\phi} + 3H) \dot{\phi} = 0,
\label{eq-phenom}
\end{equation}
where $\Gamma_{\phi}$, given by the imaginary part of the self-energy of 
$\phi$, is the total perturbative decay rate, and $\dot{\phi} = d\phi/dt$. 
Bose enhancement of particle production
into the spatial Fourier modes of the inflaton fluctuation field $\vphi$
(and light Bose fields coupled to the inflaton) was not taken into account.  

In the second stage of this first approach to the reheating problem
(group 1B, 
\cite{kofman:1994a,shtanov:1995a,dolgov:1994a,allahverdi:1997a})
Eq.\ (\ref{eq-phenom}) was still utilized to model the mean-field dynamics, 
but with $\Gamma_{\phi}$ computed  beyond first-order in
perturbation theory.  In the work of Shtanov, Traschen, and 
Brandenberger \cite{shtanov:1995a} and Kofman, Linde, and Starobinsky (KLS)
\cite{kofman:1994a}, $\Gamma_{\phi}$ was computed for a real self-interacting 
scalar inflaton field $\phi$ which was both Yukawa-coupled to
a spinor field $\psi$, and bi-quadratically coupled to a scalar field $\chi$
(KLS studied both the $\phi \rightarrow -\phi$ symmetry-breaking and
unbroken symmetry cases).  From the one-loop equations for the quantum
modes of the $\chi$, $\vphi$, and $\psi$ fields (in which the mean field
$\phih$ appears quadratically as an effective mass), approximate expressions
for the growth
rate of occupation numbers were derived, assuming a quasi-oscillatory
mean field $\phih$. For bosonic decay-product fields, it was found that
first-order time-dependent perturbation theory drastically underestimates
the particle production rate for modes which are in an instability band for
parametric resonance.\footnote{The first study to point out that
parametric resonance effects can dramatically effect particle production
in an out-of-equilibrium phase transition was \cite{traschen:1990a}.}
Parametric amplification of quantum fluctuations\footnote{Parametric
amplification of quantum fluctuations refers to the increase in expectation
values of occupation numbers for parametric oscillators, due to a 
time-dependent perturbing frequency.} in Bose decay-product fields
can result in rapid out-of-equilibrium transfer of energy from the inflaton
mean field to the (spatially) inhomogeneous inflaton modes and light Bose
fields coupled to the inflaton.  This phenomenon was called 
{\em preheating\/} by
KLS.  It has been suggested that exponential growth of 
quantum fluctuations can in some cases lead to out-of-equilibrium 
(nonthermal) symmetry restoration in the ``new'' inflation models with a 
spontaneously broken symmetry \cite{kofman:1996c,tkachev:1996a}.
(See, however, the work of Boyanovsky
{\em et al.,\/} which reached a different conclusion on the possibility of 
nonequilibrium symmetry restoration \cite{boyanovsky:1996b}.)
This may have interesting implications for baryogenesis,
defect formation, and generation of primordial density perturbations 
\cite{kofman:1996a,kofman:1994a,tkachev:1996a}.

In both stages of this first approach, the back reaction of the variance
of the inflaton on the mean-field dynamics, and of the variance on the
quantum mode functions, were not treated self-consistently.
The effect of spacetime dynamics was either excluded
entirely, or not included self-consistently using the semiclassical Einstein
equation.  
Due to the potentially large initial inflaton amplitude at the onset of 
reheating, particularly in the case of chaotic inflation \cite{linde:1990a},
the effect of cosmic expansion on quantum particle production
needs to be included.  Since the mean field and variance (mean-squared
fluctuations) are coupled,
the back reaction of particle production on the mean-field dynamics
must be accounted for in a self-consistent manner. 

In the decade before the advent of inflationary cosmology, there was active
research on quantum processes in curved spacetimes. An important class of
problems is vacuum particle creation
\cite{parker:1969a,sexl:1969a,zeldovich:1970a,zeldovich:1971a,hu:1974a}
and its effect on the dynamics and structure of the early Universe 
\cite{zeldovich:1971a,zeldovich:1977a,hu:1977a,hu:1978a,fischetti:1979a,hartle:1979a,hartle:1980a,hu:1981a,anderson:1984a} at
the Planck time. The effect of spacetime dynamics and the importance of
parametric amplification on cosmological particle creation
were realized very early \cite{parker:1969a,zeldovich:1970a,hu:1974a}.
Most of the effort in the latter part of the 1970s
was focused on obtaining both a regularized energy-momentum tensor and a 
viable formalism for the treatment of back reaction effects. The wisdom 
gained from work in that period before the inflationary cosmology program was
initiated is particularly relevant to the
reheating problem. Simply put, for obtaining a finite energy-momentum 
tensor for a quantum field in a cosmological spacetime, the adiabatic
\cite{hu:1974a,parker:1974a,fulling:1974a,fulling:1974b} and dimensional
\cite{brown:1977a} regularization methods are the most useful. 
For studying the back reaction of particle creation,
the Schwinger-Keldysh (CTP, ``in-in'') effective action formalism
\cite{schwinger:1961a,keldysh:1964a,chou:1985a,su:1988a,dewitt:1986a,jordan:1986a,calzetta:1987a,calzetta:1989a} is more appropriate than the usual 
Schwinger-DeWitt (``in-out'') method \cite{schwinger:1951a,dewitt:1964a}.

The second approach to the post-inflationary reheating problem is built upon
the body of earlier work on cosmological particle creation.
Following the application of closed-time-path techniques to nonequilibrium
relativistic field theory problems 
\cite{calzetta:1987a,calzetta:1988b},
several authors (which we call group 2A) derived perturbative
mean-field equations for a scalar inflaton with cubic \cite{calzetta:1989a}
and quartic \cite{paz:1990a} self-couplings, as well as for
a scalar inflaton Yukawa-coupled to fermions \cite{stylianopoulos:1991a}.
The closed-time-path method yields a real and causal mean-field 
equation with back reaction from quantum particle
creation taken into account.  For the case
of Bose particle production, perturbation theory in the coupling constant is
known to break down for sufficiently large occupation numbers, which occurs
on the time scale $\tau_1$ for parametric resonance effects to become
important \cite{boyanovsky:1995a,son:1996a}.  It is, therefore, necessary
to employ nonperturbative techniques in order to study reheating in most 
inflationary models.

The second stage of work in this second approach to the reheating problem
used the closed-time-path
method to derive self-consistent mean-field equations for
an inflaton coupled to lighter quantum fields (group 2B, 
\cite{boyanovsky:1995b,boyanovsky:1995c,boyanovsky:1995d,baacke:1997a,kaiser:1996a,mazzitelli:1989a,khlebnikov:1997a}).
In the first of these studies
\cite{boyanovsky:1995b,boyanovsky:1995c,boyanovsky:1995d,baacke:1997a}, 
the coupled one-loop mean-field and mode-function 
equations were solved numerically in Minkowski space, implicitly carrying out
an {\em ad hoc\/} nonperturbative resummation in $\hbar$.  In the 
one-loop equations, the variances
for the inflaton $\langle\varphi^2\rangle$ and light Bose fields
$\langle\chi^2\rangle$
do not back-react on the mode functions directly.  However, mean-field
equations were derived for an O$(N)$-invariant linear $\sigma$ model (with a
$\lambda \Phi^4$ self-interaction) at leading order in the large-$N$ 
approximation by Boyanovsky {\em et al.\/} \cite{boyanovsky:1995a}.
In this approximation, the variance 
does back-react on the quantum mode functions.  At leading order in 
the $1/N$ expansion, the unbroken symmetry dynamical equations for the
quartic O$(N)$
model are formally similar to the dynamical equations for a single
$\lambda \Phi^4$ field theory in the time-dependent Hartree-Fock 
approximation 
\cite{cornwall:1974a}.  The nonequilibrium dynamics of the quartically 
self-interacting O$(N)$ field theory in Minkowski space has been numerically
studied at leading order in the $1/N$ expansion in both the unbroken symmetry 
\cite{boyanovsky:1996b,boyanovsky:1995a,cooper:1994a}
and symmetry-broken 
\cite{boyanovsky:1996b,boyanovsky:1995a,cooper:1997b} cases.
Some analytic work has been done on the self-consistent Hartree-Fock 
mean-field equations for a quartic scalar field in Minkowski space
\cite{son:1996a}.
In addition, the Hartree-Fock equations for a $\lambda \Phi^4$ field in the 
slow-roll regime have been studied numerically in Minkowski space 
\cite{boyanovsky:1993a} and in FRW spacetime \cite{boyanovsky:1994a}.
However, the effect of spacetime dynamics on reheating in the O$(N)$ field
theory has not (to our knowledge) been studied using the coupled,
self-consistent semiclassical Einstein equation and matter-field dynamical
equations, though some simple analytic work has been done on curvature
effects in reheating \cite{dolgov:1994a,kaiser:1996a}.  The semiclassical
equations for one-loop reheating in FRW spacetime were derived in 
\cite{mazzitelli:1989a}.  The $\phi^2 \chi^2$ theory has been studied
in FRW spacetime by \cite{hu:1993a,khlebnikov:1997a,zhang:1991a}.
In addition, numerical work has been done on
symmetry-breaking phase transitions in both
a $\lambda \Phi^4$ scalar field in de~Sitter spacetime \cite{boyanovsky:1997e},
and an  O$(N)$ theory in FRW spacetime \cite{mazenko:1986a,boyanovsky:1996c}.

\section{Organization}
This dissertation is organized as follows.  In Chapter~\ref{chap-oncst},
which describes work published in Ref.~\cite{ramsey:1997a},
we construct the two-particle-irreducible (2PI), closed-time-path (CTP)
effective action for the O$(N)$ field theory in a general curved spacetime.
From this we derive a set of coupled equations for the mean field and the
variance, which are useful for studying the nonperturbative, nonequilibrium
dynamics of a quantum field when full back reactions of the quantum field
on the curved spacetime, as well as the fluctuations on the mean field,
are required.  Renormalization of the effective action at leading order
in the $1/N$ expansion is then discussed.

In Chapter~\ref{chap-preheat}, which describes work published in 
Ref~\cite{ramsey:1997b}, we study the nonperturbative, nonequilibrium
dynamics of a quantum field in the preheating phase of inflationary cosmology,
including full back reactions of the quantum field on the curved spacetime,
as well as the fluctuations on the mean field.  We use the O$(N)$ field theory
with unbroken symmetry in a spatially flat FRW spacetime to study the
dynamics of the inflaton in the post-inflation, preheating stage.  Oscillations
of the inflaton's zero mode induce parametric amplification of quantum 
fluctuations, resulting in a rapid transfer of energy to the inhomogeneous
modes of the inflaton field.  The large-amplitude oscillations of the mean
field, as well as stimulated emission effects require a nonperturbative
formulation of the quantum dynamics, while the nonequilibrium evolution 
requires a statistical field theory treatment.  We adopt the coupled 
nonperturbative equations for the mean field and variance derived in
Chapter~\ref{chap-oncst} while specialized to a dynamical FRW background,
up to leading order in the $1/N$ expansion.  Adiabatic regularization
is employed.  The renormalized dynamical equations are evolved numerically
from initial data which are generic to the end state of slow roll in
many inflationary cosmological scenarios.  We find that for sufficiently large
initial mean-field amplitudes $\gtrsim \Mpl/300$ (where $\Mpl$ is the 
Planck mass) in this model, the parametric resonance effect alone (in a
collisionless approximation) is not an efficient mechanism of energy
transfer from the mean field to the inhomogeneous modes of the
quantum field.  For small initial mean-field amplitude, damping of the
mean field due to particle creation is seen to occur, and in this case
can be adequately described by prior analytic studies with approximations
based on field theory in Minkowski spacetime.  

In Chapter~\ref{chap-fermion}, which describes work to be published
\cite{ramsey:1997c}, we present a detailed and systematic 
analysis of the coarse-grained, nonequilibrium dynamics of a scalar inflaton
field coupled to a fermion field in the late stages (dominated by fermion
particle production) of the reheating period of inflationary cosmology with
unbroken symmetry.  We derive coupled nonperturbative
equations for the inflaton mean field and variance at two loops in a general
curved spacetime, and show that the equations of motion are real and causal,
and that the gap equation for the two-point function is dissipative due to
fermion particle production.  We then specialize to the case of Minkowski
space and small-amplitude inflaton oscillations, and derive the perturbative
one-loop dissipation and noise kernels to fourth order in the Yukawa coupling
constant; the normal-threshold dissipation and noise kernels are shown to
satisfy a zero-temperature fluctuation-dissipation relation.  We derive a
Langevin equation for the dynamics of the inflaton zero mode.  We then show
that the variance of the inflaton zero moe can be  non-negligible during 
reheating, which is the primary physical result of the chapter.

In Chapter~\ref{chap-entropy}, which describes work to be published
\cite{ramsey:1997d}, we set the stage for a study of the 
thermalization process in reheating by investigating how entropy can be defined
for an interacting quantum field.  We discuss various definitions of entropy
but focus our attention on the {\em correlation entropy.\/} 
We discuss how an effectively open system
arises when hierarchy of correlation functions is truncated and one of
the higher correlation functions is slaved to the lower correlation functions.
We show how the dynamics of a nonperturbative truncation of the Schwinger-Dyson
equations can be reduced to coupled equations for the equal-time correlation
functions.  We then compute the correlation entropy for the case of the
$\lambda \Phi^4$ truncated at third order in the correlation hierarchy,
where the three-point function is slaved to the mean field and two-point
function, for the case of a (spatially) translationally invariant Gaussian 
density matrix.  We then discuss the possible benefits of this approach
to the thermalization problem.

\section{Notation}
Throughout this dissertation we use units in which 
$c = k_{\text{{\tiny B}}} = 1$. Planck's constant $\hbar$ is shown explicitly 
(i.e., not set equal to 1) except in Chapter~\ref{chap-entropy}.
In relativistic units where $c=1$, Newton's constant is 
$G = \hbar \Mpl^{-2}$, where $\Mpl$ is the Planck mass.
We work with a four-dimensional spacetime manifold, and follow
the sign conventions\footnote{In the classification
scheme of Misner, Thorne and Wheeler \cite{misner:1973a}, the 
sign convention of Birrell and Davies \cite{birrell:1982a} is 
classified as $(+,+,+)$.}  
of Birrell and Davies \cite{birrell:1982a}
for the metric tensor $g_{\mu\nu}$, 
the Riemann curvature tensor $R_{\mu\nu\sigma\rho}$, and
the Einstein tensor $G_{\mu\nu}$.  We use greek letters to 
denote spacetime indices. The beginning latin letters 
$a,b,c,d,e,f$ are used as time branch indices (see Sec.~\ref{sec-ctpform}),
and in Chapters~\ref{chap-oncst} and \ref{chap-preheat}, 
the middle latin letters $i,j,k,l,m,n$ are used as indices in the
O$(N)$ space (see Sec.~\ref{sec-ontheory}).  In
Section~\ref{sec-detcf} the middle latin letters are used
as indices to indicate spatial coordinate. 
The Einstein summation convention over repeated indices is employed.
Covariant differentiation is denoted with a nabla $\nabla_{\mu}$ or
a semicolon. 

\chapter{O$(N)$ Quantum fields in curved spacetime}
\label{chap-oncst}

\section{Introduction}
One major direction of research on quantum field theory in curved spacetime
\cite{dewitt:1975a,birrell:1982a,fulling:1989a} since the 1980s has been the 
application of interacting 
quantum fields to the consideration of symmetry breaking and phase transitions
in the early Universe, from the Planck to the grand unified energy scales
\cite{toms:1980a,ford:1982a,shore:1980a,vilenkin:1982a,vilenkin:1983a,allen:1983a,anderson:1985c,ratra:1985a,mazenko:1986a}.
In a series of work, Hu, O'Connor, Shen, Sinha, and Stylianopoulos
\cite{hu:1983b,oconnor:1983a,shen:1985a,hu:1986a,hu:1986c,hu:1986b,hu:1987b,sinha:1988a,stylianopoulos:1989a} 
systematically investigated the effect of spacetime curvature, dynamics, and 
finite temperature in causing a symmetry restoration of interacting quantum 
fields in curved spacetime. In general one wants to see how quantum 
fluctuations $\varphi$ around a mean field $\hat \phi$ change as a function of 
these parameters. For this purpose, the two-particle-irreducible (2PI) 
effective action was constructed for an $N$-component scalar O$(N)$ model 
with quartic interaction \cite{oconnor:1983a,anderson:1985c,hu:1987b}. Hu 
and O'Connor \cite{hu:1987b} found 
that the spectrum of the small-fluctuation operator contains interesting 
information concerning how infrared behavior of the system depends on the 
geometry and topology. The equation for $\hat \phi$ containing contributions 
from the variance of the fluctuation field $\langle\varphi^2\rangle$ depicts 
how the mean field evolves in time. This program  explored two of the three
essential elements of an investigation of 
a phase transition \cite{hu:1986a},
the geometry and topology and the field theory and infrared behavior aspects,
but not the nonequilibrium statistical-mechanical aspect.

For this and other reasons, 
Calzetta and Hu \cite{calzetta:1987a} started exploring the
closed-time-path (CTP) or Schwinger-Keldysh formalism 
\cite{schwinger:1961a,bakshi:1963a,keldysh:1964a,chou:1985a}, 
which is formulated with an ``in-in'' 
boundary condition. Because the CTP effective action produces a real
and causal equation of motion \cite{dewitt:1986a,jordan:1986a}, 
it is well suited for 
particle production back-reaction problems 
\cite{hartle:1979a,hartle:1980a,calzetta:1989a}.  Use of the
CTP formalism in conjunction with the 2PI effective action 
\cite{cornwall:1974a} and the Wigner function \cite{wigner:1932a} enabled
Calzetta and Hu to 
construct a quantum kinetic field theory (in flat spacetime), deriving
the Boltzmann field equation from first principles \cite{calzetta:1988b}. 
The necessary ingredients
were then in place for an analysis of nonequilibrium phase transitions
\cite{calzetta:1989b}.
In recent years these tools (CTP, 2PI) have indeed been applied to the
problems of heavy-ion collisions, pair production in strong electric fields
\cite{cooper:1994a}, disoriented chiral condensates  
\cite{boyanovsky:1994b,boyanovsky:1997a}, and reheating in inflationary 
cosmology \cite{boyanovsky:1995e}.  However, none of these recent works has 
included curved spacetime effects in a self-consistent manner, where the 
spacetime governs the evolution of a quantum field and is, in turn, governed 
by the quantum field dynamics.  This is especially important for Planck scale 
processes involving quantum fluctuations with back reaction, such as
particle creation \cite{calzetta:1994a}, galaxy formation 
\cite{calzetta:1995a}, preheating, and thermalization in chaotic inflation 
\cite{boyanovsky:1997e,boyanovsky:1996c}.

In this Chapter we return to the problems begun by Calzetta, Hu and 
O'Connor
a decade ago.  We wish to derive the coupled equations for the evolution
of the mean field and the variance for the O$(N)$ model in curved spacetime,
which should provide a solid and versatile platform for studies of phase 
transitions in the
early Universe.  The first order of business is to construct
the CTP-2PI effective action in a general curved spacetime.  The evolution
equations are derived from it. We must also deal with the divergences
arising in it.  From the vantage point of the correlation hierarchy (and the
associated master effective action) as applied to a nonequilibrium quantum
field \cite{calzetta:1995b}, there is {\em a priori\/} no reason why one 
should stop at the 2PI effective action.  Indeed, the 2PI effective action
corresponds to a further approximation from the two-loop truncation
of the master effective action constructed from the full Schwinger-Dyson
hierarchy \cite{calzetta:1993a,calzetta:1995b}. For problems where the mean 
field and the two-point function give an adequate description (which
is not the case near the critical point, where higher-order correlation
functions become important \cite{ma:1976a}), 
the CTP-2PI effective action is sufficient. In particular, the 2PI 
effective action contains the commonly used leading-order 
large-$N$, time-dependent 
Hartree-Fock, and one-loop approximations.

The O$(N)$ model has been usefully applied to a great variety of problems in
field theory and statistical mechanics \cite{eyal:1996a}.  At leading
order in the large-$N$ expansion, the O$(N)$ field theory yields 
nonperturbative,\footnote{By this, we mean that the solution to the coupled
equations for the mean field and inhomogeneous modes represents a 
nonperturbative resummation of an infinite subclass of diagrams in the
ordinary 1PI effective action, which is a functional of the mean field only.}
local evolution equations for the mean field and the modes
of the fluctuation field, which are valid in the regime of strong mean
field \cite{cornwall:1974a,cooper:1994a}.  This approximation has recently 
been applied to problems of nonequilibrium phase transitions 
\cite{boyanovsky:1996b,boyanovsky:1995a,cooper:1997b}.
In the ``preheating'' problem studied in Chapter~\ref{chap-preheat}, 
we shall see that it is particularly useful for describing the 
nonperturbative dynamics of the inflaton field in chaotic inflation 
scenarios \cite{linde:1985a},
where the inflaton mean-field amplitude can be on the order of the Planck 
mass at the end of the slow roll period \cite{linde:1990a,linde:1994a}. 
The $1/N$ expansion has many attractive features, as it is known to preserve
the Ward identities for the O$(N)$ theory \cite{coleman:1974a}
and to yield a covariantly conserved energy-momentum tensor 
\cite{hartle:1981a}.
Furthermore, in the limit of large $N$, the quantum effective action for the
matter fields can be interpreted as a leading-order term in the expansion of 
the full (matter plus gravity) quantum effective action \cite{hartle:1981a}.

Mazzitelli and Paz \cite{mazzitelli:1989b} have studied the $\lambda \Phi^4$ 
and O$(N)$ field theories in a general curved spacetime in the Gaussian
and large-$N$ approximations, respectively.  Their approach differs from ours
in that it is based on a Gaussian factorization which does not permit 
systematic improvement either in the loop expansion or in the $1/N$ 
approximation.  In contrast, our approach is based on a closed-time-path
formulation of the correlation dynamics.  The evolution equation we obtain
for the two-point function contains a two-loop dissipative contribution
(due to multiparticle production) which is not present in the large-$N$ 
approximation.  At leading order in the large-$N$ approximation, our results 
agree with theirs, so that their renormalization counterterms can be directly 
applied to the leading-order-large-$N$ limit of the 
mean field and gap equations derived here.

This chapter is organized as follows.  In Secs.~\ref{sec-ctpform} and
\ref{sec-cjtform} we present self-contained summaries of the two essential 
theoretical methodologies employed in this study, the closed-time-path 
formalism and the two-particle-irreducible effective action.  The adaptation 
of these tools to the quantum dynamics of a $\lambda \Phi^4$ field theory 
in curved spacetime is presented in Sec.~\ref{sec-lpfcst}.  The O$(N)$
scalar field theory is treated in Sec.~\ref{sec-ontheory}, where we study
the two-loop truncation of the 2PI effective action.

\section{Schwinger-Keldysh formalism}
\label{sec-ctpform}
The Schwinger-Keldysh or ``closed-time-path'' (CTP) formalism is a powerful 
method for deriving real and causal evolution equations for expectation values 
of operators for quantum fields in disequilibrium.  A quantum
field may be defined to be out of equilibrium whenever its density 
matrix ${\boldsymbol \rho}$ and Hamiltonian ${\boldsymbol H}$ 
fail to commute, i.e.,
$[{\boldsymbol H},{\boldsymbol \rho}] \neq 0$.  
Such conditions can occur, for example,  
in a field theory quantized on a dynamical background spacetime, 
and also in an interacting field theory with nonequilibrium initial 
conditions.  Although in Chapters~\ref{chap-oncst} and 
\ref{chap-preheat} we shall be concerned with closed-system,
unitary dynamics of a single self-interacting quantum field,
the methods discussed here are also well suited to studying the 
dynamics of an open quantum system, as shown in 
Chapter~\ref{chap-fermion} below.
Excellent reviews of the Schwinger-Keldysh method are Chou {\em et al.} 
\cite{chou:1985a} as applied to nonequilibrium quantum field theory and 
Calzetta and Hu \cite{calzetta:1987a} as applied to the back reaction 
problem in semiclassical gravity.  

Let us briefly review the Schwinger-Keldysh method as applied to the 
effective mean-field dynamics of a self-interacting scalar field theory 
in Minkowski space.  
The classical action for a scalar $\lambda \Phi^4$ theory in Minkowski space 
${\mathbb M}^4$ is
\begin{equation}
S^{\text{{\tiny F}}}[\phi] = -\frac{1}{2} \int \! d^{\, 4} x \left[ 
\phi \eta^{\mu\nu} \partial_{\mu} \partial_{\nu} \phi + m^2 \phi^2  
+ \frac{\lambda}{12} \phi^4 \right],
\label{eq-lpfca}
\end{equation}
where $\lambda$ is a coupling constant with dimensions of $1/\hbar$ (inverse
mass times inverse length), 
$m$ is the ``mass'' with dimensions of inverse length, and 
$\eta^{\mu\nu} \equiv \text{diag}(+,-,-,-)$ is the Minkowski space metric
tensor.  The Euler-Lagrange equations are obtained
by variation of the action, where it is understood that the variations 
of $\phi$ must 
satisfy boundary conditions in order that surface terms can be discarded.

Let us denote the Heisenberg field operator for the canonically quantized
theory with classical action (\ref{eq-lpfca}) by $\Phi_{\text{{\tiny H}}}(x)$.
By ``effective mean-field dynamics'' we mean that we seek a dynamical equation
for the mean field $\phih$, which is the expectation value of 
$\Phi_{\text{{\tiny H}}}$, 
\begin{equation}
\phih(x) \equiv \langle \Omega,\text{in}|\Phi_{\text{{\tiny H}}}(x)|
\Omega,\text{in}\rangle,
\label{eq-defmf3}
\end{equation}
in a quantum state $|\Omega,\text{in}\rangle$ 
for which $\phih$ is initially displaced from zero.
In what follows, we shall assume that the quantum state initially
corresponds to the vacuum state for the {\em fluctuation field,\/}
defined as the difference between the Heisenberg
field operator $\Phi_{\text{{\tiny H}}}$ and the mean field,
\begin{equation}
\varphi_{\text{{\tiny H}}}(x) \equiv \Phi_{\text{{\tiny H}}}(x) - \phih(x).
\label{eq-deffluc1pi}
\end{equation}
It is important to note that even if the quantum state 
$|\Omega,\text{in}\rangle$ initially corresponds to the vacuum state for
the fluctuation field, in a nonequilibrium setting (e.g., time-dependent
background field $\phih$), it will not remain so.  At later times,
$|\Omega,\text{in}\rangle$ will not correspond to the no-particle state for 
the fluctuation field.  In what follows, we will simply refer to 
$|\Omega,\text{in}\rangle$ as the initial ``vacuum state,'' though it should 
be understood as the
vacuum state for modes of the {\em fluctuation field,\/} and not of the
field operator $\Phi_{\text{{\tiny H}}}$.

An initial quantum state $|\Omega,\text{in}\rangle$ with nonvanishing mean
field such as described above may arise in the following way.
Let us suppose that coupled to the scalar field $\phi$ there is an external 
source\footnote{We assume the source is sufficiently weak that we do not need 
to take into account nonperturbative Schwinger-type particle production 
effects.} which is nonvanishing for $t < t_0$, and which is removed at $t_0$,
\begin{equation}
F(t) = F\theta(t_0-t).
\end{equation}
Let us denote the quantum state for $t < t_0$ by $|\Omega,\text{in}\rangle$, 
and suppose that in this state, for $t < t_0$, the expectation value of 
$\Phi_{\text{{\tiny H}}}$ is given by a constant $\phih_0$.  We will assume
that the constant $F$ satisfies
\begin{equation}
F = -\frac{\delta V_{\text{eff}}}{\delta \phi}[\phih_0],
\end{equation}
where $V_{\text{eff}}$ is the vacuum effective potential \cite{ryder:1985a} 
for the theory with classical action (\ref{eq-lpfca}), and that for
$t < t_0$, $|\Omega,\text{in}\rangle$ corresponds to the vacuum state for the
fluctuation field $\Phi_{\text{{\tiny H}}} - \phih_0$.  Then the expectation
state $\phih$ is equal to the constant stable equilibrium configuration
$\phih_0$ for $t < t_0$.  The expectation value $\phih$ is spatially 
homogeneous for all times due to the spatial translation invariance
of the fluctuation-field vacuum state $|\Omega,\text{in}\rangle$ and
the action (\ref{eq-lpfca}). Because of 
the instantaneous change in the external source at $t=t_0$, we may use the
sudden approximation, in which $|\Omega,\text{in}\rangle$ is taken to be the 
initial quantum state for the $t > t_0$ evolution.  The physical picture
here is that the expectation value of the scalar field operator is like
a classical field initially held fixed at $\phih = \phih_0$ for $t < t_0$
and which is suddenly released at $t=t_0$.  Let us ask whether there
is an action $\Gamma[\phih]$ whose variation gives the dynamical equation
governing the subsequent evolution of the mean field $\phih$, including
quantum corrections.  As stressed above, because of the time-dependence
of background mean field $\phih$, the condition $[{\boldsymbol H},{\boldsymbol
\rho}] \neq 0$ for $t > t_0$, and in this case, the conventional 
``in-out'' generating functional will not yield the correct dynamics of
the mean field $\phih$.  Nevertheless, it is instructive to see why this is 
so.

In the conventional Schwinger-DeWitt or ``in-out'' approach 
\cite{schwinger:1951a,dewitt:1975a}, one couples an arbitrary $c$-number 
source $J$ (which is a function on ${\mathbb M}^4$)
to the field $\phi$ and computes the
vacuum persistence amplitude in the presence of the source $J$.  This
amplitude has a path integral representation
\begin{equation}
Z[J] = \exp \left( \frac{i}{\hbar} W[J] \right) =
\int D\phi \exp \left[ \frac{i}{\hbar} 
\left( S^{\text{{\tiny F}}}[\phi] + 
\int \! d^{\, 4} x J(x) \phi(x) \right) \right],
\label{eq-lpfgf}
\end{equation}
where the functional integral is a sum over classical histories of
the $\phi$ field for which $\phi-\phih$ is pure negative 
frequency [i.e., all spatial Fourier modes of $\phi-\phih$ have a time 
dependence like $\exp(i \omega t)$, $\omega > 0$]
for $t \leq t_0$ and pure positive frequency [$\sim \exp(-i \omega t)$]
in the asymptotic future.\footnote{These
boundary conditions on the functional integral are equivalent (up to
an overall normalization) to adding a small imaginary term
$-i\epsilon \phi^2$ to the classical action, where $\epsilon > 0$.}
The generating functional of normalized amplitudes is given by
\begin{equation}
W[J] = -i \hbar \text{ln} Z[J],
\end{equation}
and it is well known that $W[J]$ 
is the sum of all connected diagrams of the
field theory in the presence of the source $J$ \cite{jackiw:1974a}.
The ``classical'' field $\phih_{\text{{\tiny $J$}}}
= \langle \Phi \rangle_J$ is a function of $J$ defined by
\begin{equation}
\phih_{\text{{\tiny $J$}}} = \frac{\delta W[J]}{\delta J}.
\label{eq-defcfio}
\end{equation}
Assuming the functional relation (\ref{eq-defcfio}) 
can be inverted to yield $J$ in terms of $\phih$,
one can define an effective action [whose variation gives the inverse
of Eq.~(\ref{eq-defcfio})] as the Legendre transform of $W$,
\begin{equation}
\Gamma[\phih] = W[J] - \int \! d^{\, 4} x  J(x) \phih(x),
\label{eq-ltw}
\end{equation}
where we have dropped the $J$ subscript on $\phih$ for simplicity of
notation.  By differentiating Eq.~(\ref{eq-ltw}), we find that 
\begin{equation}
J_{\text{{\tiny $\phih$}}} = -\frac{\delta \Gamma}{\delta \phih},
\label{eq-defjio}
\end{equation}
where $J$ is understood to be a functional of $\phih$.
The functional relation
$\phih = \delta W / \delta J$ can indeed be inverted if $\Gamma$
is well defined.  
The field equation satisfied by $\phih$ in the 
physical case of interest, $J=F(t)$, is given by\footnote{
While it is correct to include $F(t)$ in the equation of motion as
we have shown, one may simply set $F(t)=0$ in the equation of motion, 
since $F(t) = 0$ for $t > t_0$, and just incorporate $F(t_0)=F$ into the
initial data for $\phih$.  Henceforth, this is what we shall do.}
\begin{equation}
F(t) = -\frac{\delta \Gamma[\phih]}{\delta \phih(x)}.
\label{eq-wrong}
\end{equation}
It should be emphasized that the $\Gamma[\phih]$ appearing in 
Eq.~(\ref{eq-wrong}) follows directly, by way of the Legendre transform,
from the ``in-out'' generating functional defined in Eq.~(\ref{eq-lpfgf}).
The problem with Eq.~(\ref{eq-wrong}) is that its solution, $\phih$,
is {\em not\/} the time-dependent expectation value of the Heisenberg
field operator for the quantum field.  Let us see why this is so, and what
the interpretation of $\phih$, as defined by Eq.~\ref{eq-defcfio}, should be.

At one loop (in Minkowski space), the vacuum state is determined by an 
expansion of the fluctuation field operator in terms of spatial eigenmodes of 
the  Klein-Gordon operator with time-dependent frequency 
\begin{equation}
\omega_{\vec{k}}^2 = \vec{k}^2 + m^2 + \frac{\lambda}{2} \phih^2.
\end{equation}
In a nonequilibrium setting, such as in a curved or dynamical spacetime 
(where the scale factor is an additional time-dependent parameter in the
effective frequency), or when $\hat{\phi}$ is time-dependent, the notion of 
positive frequency in the asymptotic past is in general different
from that in the asymptotic future, in the present case because 
$\phih(t=\infty) \neq \phih_0$.  Hence, the ``in'' vacuum state
for the fluctuation field, $|\Omega,\text{in}\rangle$,
and the ``out'' vacuum state for the fluctuation field,
$|\Omega,\text{out}\rangle$ [defined as the vacuum state for
$\Phi_{\text{{\tiny H}}} - \phih(t=\infty)$], are {\em not\/} equivalent.
It is useful at this point to go over to an ``interaction picture,'' where
the ``interaction'' is the coupling $J \phi$ to the external source. 
In this representation, the evolution of the field operator 
$\Phi_{\text{{\tiny H}}}$ is just the Heisenberg evolution for the
theory without the external source $J$ (hence we retain the $H$ subscript),
and the evolution of a state vector is due to the coupling 
$J\Phi_{\text{{\tiny H}}}$.  In particular, the quantum state
$|\Omega,\text{in}\rangle$ defined for $t\leq t_0$ will evolve to
a state $|\Omega,\text{in}\rangle_{\text{{\tiny J}}}(t)$ at time $t$,
given by
\begin{equation}
|\Omega,\text{in}\rangle_{\text{{\tiny J}}}(t) \equiv T \exp \left(
\frac{i}{\hbar} \int_{x^0 \leq t} 
d^{\,4}x J(x) \Phi_{\text{{\tiny H}}}(x) \right)
| \Omega,\text{in}\rangle,
\end{equation}
where $T$ denotes temporal ordering.  For convenience we will denote the
$t=\infty$ limit of this state by
\begin{equation}
|\Omega,\text{in}\rangle_{\text{{\tiny J}}}(+\infty) =
|\Omega,\text{in}\rangle_{\text{{\tiny J}}}.
\label{eq-devs}
\end{equation}
In this interaction picture, the
``in-out'' generating functional takes a particularly simple form,
\begin{equation}
Z[J] = \langle \Omega,\text{out}|\Omega,\text{in}\rangle_{\text{{\tiny $J$}}}
(+\infty)
= \langle \Omega,\text{out}| T \exp \left(\frac{i}{\hbar}
\int \! d^{\, 4} x J(x) \Phi_{\text{{\tiny H}}} (x) \right) |\Omega,\text{in}
\rangle,
\label{eq-zjio}
\end{equation}
where $T$ denotes temporal ordering. This amplitude is in general complex, even
in the $J=0$ limit.  The imaginary part of $W[0]$ gives the integrated 
probability $P$ to produce a particle pair \cite{hartle:1979a,hartle:1980a}
over the entire time range integrated in the classical action $S$, 
\begin{equation}
P = 2 \; \text{Im} W[0].
\end{equation}
It follows that the ``classical field'' $\phih_{\text{{\tiny $J$}}}$ 
defined as the functional derivative
of $W[J]$ with respect to $J$,
\begin{equation}
\langle \Omega,\text{out}| \Phi_{\text{{\tiny H}}}(x) | \Omega,\text{in}\rangle
= \frac{\delta W[J]}{\delta J(x)}
\end{equation}
is a {\em matrix element\/} which will in general be complex.
In addition, the dependence of 
$\langle \Omega,\text{out}|\Phi_{\text{{\tiny H}}}(x)|\Omega,\text{in}\rangle$
on $J$ will not, in general, be causal \cite{jordan:1986a,dewitt:1986a}.
In curved spacetime, the quantum expectation value of the
energy-momentum tensor, $\langle T_{\mu\nu} \rangle$,
is obtained by functional differentiation of $W$ with respect to
$g^{\mu\nu}$, which yields a complex matrix element of 
$T_{\mu\nu}(\Phi_{\text{{\tiny H}}})$ between the ``in'' and ``out'' 
vacua, where $T_{\mu\nu}(\phi)$ is the classical energy-momentum tensor
for the field \cite{dewitt:1975a,birrell:1982a,calzetta:1987a}.  
This is problematic because the quantum-corrected energy-momentum tensor, 
suitably regularized, constitutes the right-hand side of the semiclassical 
Einstein equation.

If instead of the above ``in-out'' approach one uses the closed-time-path 
formalism, one can obtain a real and causal evolution equation for the
mean field $\phih$ (the expectation value of $\Phi_{\text{{\tiny H}}}$), 
as well as for the expectation value of the 
energy-momentum tensor.  Here we briefly illustrate the procedure for the 
case of the $\lambda \Phi^4$ theory in Minkowski
space; the generalizations to curved spacetime and to higher correlation
functions will be discussed in Secs.~\ref{sec-lpfcst} and \ref{sec-cjtform},
respectively.  Let $x^0 = x^0_{\star}$ be far to the future of any 
dynamics we wish to study.  
It is not necessary to assume that $\lambda = 0$ or
that the Hamiltonian is time independent at $x^0 = x^0_{\star}$.  
Here, we will specify initial conditions
at $x^0 = -\infty$, for simplicity.
As in the previous ``in-out'' approach, suppose we wish to compute
the quantum-corrected equation governing the classical field $\phih$.
Let $M = \{ (x^0,\vec{x}) | -\infty \leq x^0 \leq x^0_{\star} \}$ be
the portion of Minkowski space to the past of time $x^0_{\star}$.
We start by defining a new manifold as a quotient space
\begin{equation}
{\mathcal M} = (M \times \{+,-\} )/\sim,
\label{eq-lpfctpmn}
\end{equation}
where $\sim$ is an equivalence relation defined by 
\begin{align}
& (x,+) \sim (x',+) \qquad \text{iff $x = x'$}, \nonumber \\
& (x,-) \sim (x',-) \qquad \text{iff $x = x'$},  \label{eq-lpfctper} \\
& (x,+) \sim (x',-) \qquad \text{iff $x = x'$ and $x^0 = x^0_{\star}$.}
\nonumber
\end{align}
It is straightforward to define an orientation on ${\mathcal M}$, provided
we reverse the sign of the volume form between the $+$ and $-$
pieces of the manifold.  Choosing an overall sign, we take the volume
form on the $+$ branch to be 
\begin{equation}
{\boldsymbol \epsilon} = dx^0 \wedge dx^1 \wedge dx^2 \wedge dx^3,
\end{equation}
and the volume form on the $-$ branch to be $-{\boldsymbol \epsilon}$.  
The next step is to generalize the usual effective
action construction of Eqs.~(\ref{eq-lpfgf})--(\ref{eq-ltw}) 
to the new manifold ${\mathcal M}$.  With the volume form on ${\mathcal M}$
thus defined, we can generalize the classical action $S^{\text{{\tiny F}}}$ 
(which is a functional on $M$) to a functional 
${\mathcal S}^{\text{{\tiny F}}}$ on ${\mathcal M}$ as follows,
\begin{equation}
{\mathcal S}^{\text{{\tiny F}}}[\phi_{+},\phi_{-}] =
S^{\text{{\tiny F}}}[\phi_{+}] - S^{\text{{\tiny F}}}[\phi_{-}],
\label{eq-lpfctpca}
\end{equation}
where $\phi_{+}$ and $\phi_{-}$ denote the $\phi$  
field on the $+$ and $-$ branches of ${\mathcal M}$,
respectively.  
For a function $\phi$ on ${\mathcal M}$, let us define the
restrictions of $\phi$ to $M \times \{+\}$ and $M \times \{-\}$ by
$\phi_{+}$ and $\phi_{-}$, respectively.
In order for $\phi$ to be a function on 
${\mathcal M}$, the restrictions must satisfy
\begin{equation}
\left.\phi_{+}(x)\right|_{x^0_{\star}} = 
\left.\phi_{-}(x)\right|_{x^0_{\star}}.
\end{equation}
The generating functional of $n$-point functions
(i.e., expectation values in the $|\Omega,\text{in}\rangle$ quantum state)
for this theory is then defined by
\begin{equation}
Z[J_{+},J_{-}] = \int_{\text{{\tiny ctp}}} D\phi_{+} D\phi_{-}
\exp \left[ \frac{i}{\hbar} \left( 
{\mathcal S}^{\text{{\tiny F}}}[\phi_{+},\phi_{-}]
+ \int_M \! d^{\, 4} x (J_{+} \phi_{+} - J_{-} \phi_{-}) \right) \right],
\label{eq-gflpf}
\end{equation}
where $J_{+}$ and $J_{-}$ are $c$-number sources on the $+$ and $-$ 
branches of ${\mathcal M}$, respectively. The designation ``ctp'' on the
functional integral in Eq.~(\ref{eq-gflpf}) indicates that one integrates 
over all field
configurations $(\phi_{+},\phi_{-})$ such that (i) $\phi_{+} = \phi_{-}$
at the $x^0 = x^0_{\star}$ hypersurface and (ii) $\phi_{+}$ ($\phi_{-}$)
consists of pure negative (positive) frequency modes at $x^0 = -\infty$.
It is not necessary for the normal derivatives of $\phi_{+}$ and
$\phi_{-}$ to be equal at $x^0 = x^0_{\star}$ \cite{calzetta:1987a}.
Because the theory is free in the asymptotic past,\footnote{The vacuum
boundary conditions for a theory which is not free in the
asymptotic past are more complicated, but can be treated by 
methods discussed in Sec.~\ref{sec-ontheory}.}
a positive frequency mode\footnote{Here, the choice of
vacuum boundary conditions corresponds to adding
a small imaginary part $i\epsilon (\phip^2 - \phim^2)$ to the classical
action ${\mathcal S}^{\text{{\tiny F}}}$.  Alternatively, the
boundary conditions correspond to the usual prescription
$m^2 \rightarrow m^2 - i\epsilon$ in $S^{\text{{\tiny F}}}[\phi]$,
but with ${\mathcal S}^{\text{{\tiny F}}}$ now redefined as
${\mathcal S}^{\text{{\tiny F}}}[\phip,\phim] = S^{\text{{\tiny F}}}[\phip]
- S^{\text{{\tiny F}}}[\phim]^{\star}$, where $\star$ denotes complex
conjugation \cite{calzetta:1987a}.}
is a solution to the spatial-Fourier transformed
Euler-Lagrange equation for $\phi$ whose asymptotic behavior at
$x^0 = -\infty$ is $\exp(-i\omega x^0)$, for $\omega > 0$.

\begin{figure}[htb]
\begin{center}
\epsfig{file=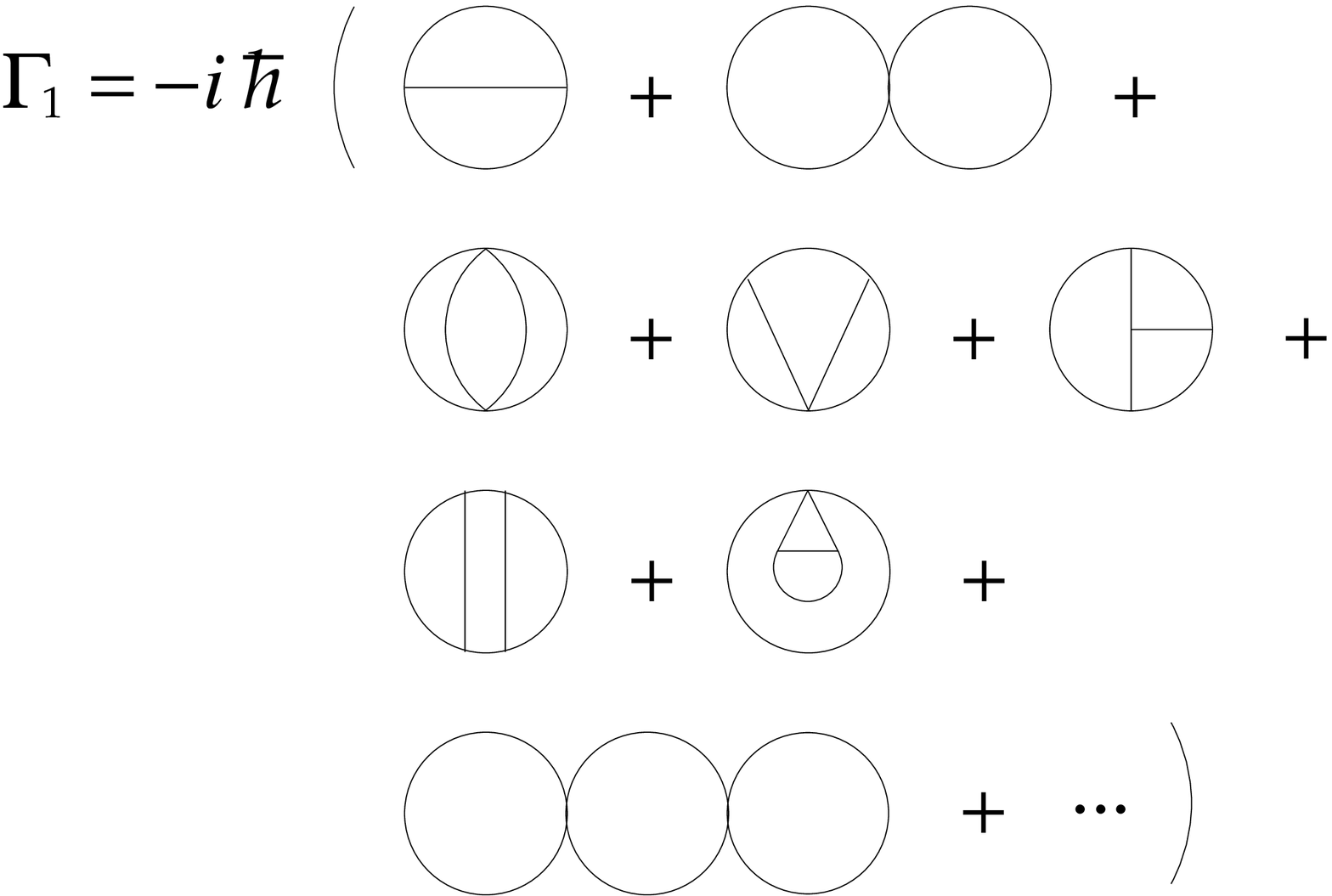,width=4.0in} 
\end{center}
\caption{Diagrammatic expansion for the $\Gamma_1$ part of the 1PI-CTP 
effective action}
\label{fig-opiea}
\end{figure}

In the interaction picture where the sources $J_{+}$ and $J_{-}$ govern
the state vector evolution, the expression $Z[J_{+},J_{-}]$ in 
(\ref{eq-gflpf}) is seen to be the amplitude for the quantum state to 
evolve forward in time under the source $J_{+}$
from $|\Omega,\text{in}\rangle$ at $x^0 = -\infty$, to some arbitrary state
$|\psi\rangle$ at $x^0 = x^0_{\star}$, times the amplitude for the state
$|\psi\rangle$ at $x^0 = x^0_{\star}$ to evolve backwards
in time under the source $J_{-}$ to the state $|\Omega,\text{in}\rangle$ 
at $x^0 = -\infty$.  The state $|\psi\rangle$ must be summed over a 
complete, orthonormal basis of the Hilbert space of the $\phi$ field.  In
this picture, the CTP generating functional takes the form
\begin{equation}
\begin{split}
Z[J_{+},J_{-}] = \sum_{\psi} \Biggl[ &
\langle\Omega,\text{in}|\tilde{T} \exp \left( -\frac{i}{\hbar}
\int^{x^0_{\star}}_{-\infty} \! dx^0 \int \! d^{\, 3}x
J_{-} \Phi_{\text{{\tiny H}}}(x) \right)|\psi\rangle   \\ &
 \times  \langle\psi|T \exp \left( \frac{i}{\hbar}
\int^{x^0_{\star}}_{-\infty} \! dx^0 \int \! d^{\, 3}x
J_{+} \Phi_{\text{{\tiny H}}}(x) \right) |\Omega,\text{in}\rangle\Biggr],
\end{split}
\end{equation}
where $T$ and $\tilde{T}$ denote temporal and anti-temporal ordering,
respectively.  
The generating functional for connected diagrams is then defined by
\begin{equation}
W[J_{+},J_{-}] = -i \hbar \text{ln} Z[J_{+},J_{-}].
\end{equation}
Classical fields on both ${+}$ and ${-}$ branches are then defined as
\begin{equation}
\phih_a(x)_{\text{{\tiny $J_{\pm}$}}} = c^{ab} 
\frac{\delta W[J_{+},J_{-}]}{\delta J_b (x)},
\label{eq-lpfph}
\end{equation}
where $a,b$ are time branch indices with index set $\{+,-\}$.
The matrix $c^{ab}$ is the $n=2$ case of the $n$-index symbol
$c^{a_1 a_2 \cdots a_n}$ defined by
\begin{equation}
c^{a_1 a_2 \cdots a_n} = 
\left\{
\begin{array}{cc}
1 & \text{if $a_1 = a_2 = \cdots = a_n = +$}, \\
-1 & \text{if $a_1 = a_2 = \cdots = a_n = -$}, \\
0 & \text{otherwise}.
\end{array}
\right.
\label{eq-cabcdef}
\end{equation}
The functional differentiation in 
Eq.~(\ref{eq-lpfph}) is carried out with variations $\delta J_{+}$ and 
$\delta J_{-}$ which satisfy the constraint that $\delta J_{+} =
\delta J_{-}$ on the $x^0 = x^0_{\star}$ hypersurface.  The $\Jpm$ subscript
in Eq.~(\ref{eq-lpfph}) indicates the functional dependence on $\Jpm$, which 
has been shown to be causal \cite{jordan:1986a,dewitt:1986a}.
In the limit $J_{+} = J_{-} \equiv J$, the classical fields on the $+$ and
$-$ time branches become equal, 
\begin{equation}
\left.
\left(\phih_{+}(x)_{\text{{\tiny $J_{\pm}$}}}\right)
\right|_{J_{+} = J_{-} \equiv J}
 = 
\left.
\left(\phih_{-}(x)_{\text{{\tiny $J_{\pm}$}}}\right)
\right|_{J_{+} = J_{-} \equiv J} 
\equiv \phih(x)_{\text{{\tiny $J$}}} = \;
 _{\text{{\tiny $J$}} }\langle \Omega,\text{in}|
\Phi_{\text{{\tiny H}}}(x)|\Omega,\text{in}\rangle_{\text{{\tiny $J$}}},
\label{eq-ident}
\end{equation}
where $|\Omega,\text{in}\rangle_{\text{{\tiny J}}}$, defined above
in Eq.~(\ref{eq-devs}),
is the state which has evolved from the vacuum at $x^0 = -\infty$
under the interaction $\Phi_{\text{{\tiny H}}} J$,
and Eq.~(\ref{eq-ident}) 
becomes the expectation value $\langle\Phi_{\text{{\tiny H}}}\rangle$ in the 
limit $J = 0$.  The effective action is again defined
via a Legendre transform, with $c^{ab}$ now acting as a ``metric''
on the internal $2 \times 2$ ``CTP'' space $\{+,-\}$,
\begin{equation}
\Gamma[\phihp,\phihm] = W[J_{+},J_{-}] - c^{ab} \int_M \! d^{\, 4} x J_a(x) \phih_b(x),
\label{eq-lpfead}
\end{equation}
where the $J$ subscripts on $\phih_{\pm}$ are suppressed for
simplicity of notation.  The functional
dependence of $J_{\pm}$ on $\phih$ via inversion of Eq.~(\ref{eq-lpfph}) is 
understood.   By direct computation, the inverse of Eq.~(\ref{eq-lpfph})
is found to be
\begin{equation}
J_a (x)_{\phihpm} = -c_{ab} \frac{\delta \Gamma[\phihp,\phihm]}{\delta 
\phih_b (x)},
\label{eq-lpfjd}
\end{equation}
where we have indicated the explicit functional dependence of $\Jpm$ 
on $\phihpm$ with a subscript, and $c_{ab}$ is the inverse of the matrix
$c^{ab}$ defined above.  In the limit $\phihp = \phihm \equiv \phih$, 
Eq.~(\ref{eq-lpfjd}) yields the evolution
equation for the expectation value $_{\text{{\tiny $J$}}}\langle
\Phi_{\text{{\tiny H}}}\rangle_{\text{{\tiny $J$}}}
\equiv \phih_{\text{{\tiny $J$}}}$ in the state which has evolved 
from $|\Omega,\text{in}\rangle$ under the source interaction
$J \Phi_{\text{{\tiny H}}}$.  The evolution equation for $\phih$, the 
vacuum expectation value $\langle \Omega,\text{in}|\Phi_{\text{{\tiny H}}}
|\Omega, \text{in}\rangle$, is therefore
\begin{equation}
\left.
\frac{\delta \Gamma[\phihp,\phihm]}{\delta \phihp}\right|_{
\phihp = \phihm \equiv
\phih} = 
\left.-\frac{\delta \Gamma[\phihp,\phihm]}{\delta \phihm}\right|_{\phihp 
= \phihm \equiv \phih} = 0.
\label{eq-lpfctpeom}
\end{equation}
Using Eqs.~(\ref{eq-lpfjd}) and (\ref{eq-lpfead}), 
an integro-differential equation for $\Gamma$ can
be derived \cite{jordan:1986a},
\begin{equation}
\Gamma[\phihp,\phihm] = -i \hbar \text{ln} \left\{ 
\int_{\text{{\tiny ctp}}} D\phi_{+} D\phi_{-} e^{ \frac{i}{\hbar} 
\bigl({\mathcal S}^{\text{{\tiny F}}}[\phip,\phim] -
\int_M d^{\,4}x 
\frac{\delta\Gamma[\phihp,\phihm]}{\delta\phih_a}(\phi_a - \phih_a)
\bigr)}\right\},
\end{equation}
in which the functional differentiations of $\Gamma$ 
with respect to $\phihpm$ are carried out with the constraint that the 
variations of $\phihpm$ satisfy $\delta \phihp = \delta \phihm$
when $x^0 = x^0_{\star}$.  The difference $\phi_a - \phih_a$ is naturally
interpreted as the fluctuations of a particular history $\phi_a$ about
the ``classical'' field configuration 
$\phih_a$.  Let us, therefore, define the {\em fluctuation field \/}
\begin{equation}
\vphi_a \equiv \phi_a - \phih_a
\label{eq-dctpfluc}
\end{equation}
in analogy with Eq.~(\ref{eq-deffluc1pi}.
Performing the change of variables $\phi_a \rightarrow \vphi_a$ in the
functional integrand, we obtain
\begin{equation}
\Gamma[\phihp,\phihm] = -i \hbar \text{ln} \left\{ 
\int_{\text{{\tiny ctp}}} D\vphi_{+} D\vphi_{-} e^{\frac{i}{\hbar} 
\bigl(
{\mathcal S}^{\text{{\tiny F}}}[\phihp + \vphip, \phihm + \vphim] -
\int_M d^{\,4}x \frac{\delta\Gamma[\phihp,\phihm]}{\delta\phih_a}\vphi_a
\bigr)}\right\}.
\label{eq-eafie}
\end{equation}
This functional integro-differential equation has a formal solution
\cite{jackiw:1974a}
\begin{equation}
\Gamma[\phihp,\phihm] = {\mathcal S}^{\text{{\tiny F}}}[\phihp,\phihm]
- \frac{i \hbar}{2} \text{ln}\,\text{det}({\mathcal A}^{-1}_{ab}) + 
\Gamma_{1}[\phihp,\phihm],
\label{eq-opieasol}
\end{equation}
where ${\mathcal A}^{ab}(x,\xp)$ is the second functional derivative
of the classical action with respect to the field $\phi_{\pm}$,
\begin{equation}
i{\mathcal A}^{ab}(x,\xp) = 
\frac{\delta^2 {\mathcal S}^{\text{{\tiny F}}}}{\delta
\phi_a(x)\phi_b(\xp)}[\phihp,\phihm].
\label{eq-dolp}
\end{equation}
The inverse of ${\mathcal A}^{ab}$ is the one-loop propagator for the
fluctuation field $\phi$.  The functional $\Gamma_{1}$ in 
Eq.~(\ref{eq-eafie}) is defined as $-i\hbar$ times the
sum of all one-particle-irreducible vacuum-to-vacuum graphs with
propagator given by ${\mathcal A}^{-1}_{ab}(x,\xp)$ and 
vertices given by a shifted action
${\mathcal S}^{\text{{\tiny F}}}_{\text{{\tiny int}}}$,
defined by
\begin{multline}
{\mathcal S}^{\text{{\tiny F}}}_{\text{{\tiny int}}}[\vphip,\vphim]
= {\mathcal S}^{\text{{\tiny F}}}[\vphip + \phihp, \vphim +  \phihm] -
{\mathcal S}^{\text{{\tiny F}}}[\phihp,\phihm] - \int_M \! d^{\, 4} x \left(
\frac{\delta {\mathcal S}^{\text{{\tiny F}}}}{\delta \phi_a}[\phihpm]\right)
\vphi_a 
\\  - \frac{1}{2}\int_M \! d^{\, 4} x \int_M \! d^{\, 4} x' \left( 
\frac{\delta^2 
{\mathcal S}^{\text{{\tiny F}}}}{\delta \phi_a(x)\delta 
\phi_b(\xp)}[\phihpm]\right) \vphi_a(x)
\vphi_b(\xp).
\label{eq-defshfa}
\end{multline}
For simplicity of notation, we do not explicitly indicate the functional 
dependence of ${\mathcal S}^{\text{{\tiny F}}}_{\text{{\tiny int}}}$ on 
$\phihpm$\@.  
Figure~\ref{fig-opiea} shows the diagrammatic expansion for $\Gamma_{1}$,  
where lines represent the propagator ${\mathcal A}^{-1}_{ab}(x,\xp)$,
and vertices terminating three lines are proportional to $\phih$
\cite{jackiw:1974a}.
Each vertex carries a spacetime label in $M$ and a time branch label in 
$\{+,-\}$.  The lowest-order contribution to $\Gamma_1$ is of order $\hbar^2$,
i.e., a two loop graph.  The one-loop propagator ${\mathcal A}^{-1}$ does 
not depend on $\hbar$.  The $\text{ln}(\det {\mathcal A})$ term in Eq.\
(\ref{eq-opieasol}) is the one-loop [order $\hbar$] term in the
CTP effective action.  The CTP effective action, as a functional of $\phihpm$, 
can be computed to any desired order in the loop expansion using 
Eq.~(\ref{eq-opieasol}).  As with the ordinary in-out effective action
\cite{peskin:1995a}, the CTP effective action contains divergences at each 
order in the loop expansion, which need to be regularized.  The 
renormalizability of the field theory in the ``in-out'' representation is 
sufficient to guarantee renormalizability of the ``in-in'' equations of motion 
for expectation values \cite{jordan:1986a,calzetta:1987a,calzetta:1988b}\@.

Functionally differentiating $\Gamma[\phihp,\phihm]$ with respect to
either $\phihp$ or $\phihm$ and making the identification 
$\phihp=\phihm=\phih$ [as shown in Eq.~(\ref{eq-lpfctpeom})]
yields a real and causal dynamical equation for the mean field $\phih$.  
Thus the 1PI effective action $\Gamma[\phihpm]$ 
yields {\em mean-field\/} dynamics for the theory, which is a lowest-order 
truncation of the correlation hierarchy 
\cite{calzetta:1993a,calzetta:1995b}.
However, for a detailed study of nonperturbative growth of quantum 
fluctuations relevant to nonequilibrium mean-field dynamics
(or a symmetry-breaking phase transition), 
it is also necessary to obtain dynamical equations for
the {\em variance\/} of the quantum field,
\begin{equation}
\langle\Phi_{\text{{\tiny H}}}^2\rangle-\langle\Phi_{\text{{\tiny H}}}
\rangle^2 = 
\langle\Phi_{\text{{\tiny H}}}^2\rangle-\phih^2 =
\langle \vphi_{\text{{\tiny H}}}^2 \rangle 
\equiv \hbar G_{++}(x,x),
\label{eq-lpf2pitpf}
\end{equation}
where $\hbar G_{++}(x,\xp)$ is the
time-ordered Green function for the fluctuation field 
$\vphi_{\text{{\tiny H}}}$, $\langle T (\vphi(x)_{\text{{\tiny H}}} 
\vphi(\xp)_{\text{{\tiny H}}}) \rangle$.
A higher-order truncation of the correlation hierarchy is 
needed in order to explicitly follow the growth of quantum fluctuations; the
two-particle-irreducible (2PI) effective action formalism, to which we now 
turn, serves this purpose.

\section{Two-Particle-Irreducible formalism}
\label{sec-cjtform}
In a nonperturbative study of nonequilibrium field dynamics in the
regime where quantum fluctuations are significant, the 1PI
effective action is inadequate because it does not permit a 
derivation of the evolution equations for the mean field 
$\langle \Phi_{\text{{\tiny H}}} \rangle$ {\em and\/} 
variance $\langle \vphi_{\text{{\tiny H}}}^2 \rangle$, at a
{\em consistent\/} order in a nonperturbative expansion scheme.
In addition, the initial data for the mean field $\phih$ do not contain
any information about the quantum state for fluctuations $\vphi$
around the mean field. 
The two-particle-irreducible (2PI) formalism can be used 
to obtain nonperturbative dynamical equations for both 
the mean field $\phih(x)$ and two-point function $G(x,y)$,
which contains the variance, as shown in Eq.~(\ref{eq-lpf2pitpf}).
The 2PI method generalizes the 1PI effective action 
to an action $\Gamma[\phih,G]$ which 
is a functional of possible histories for both $\phih$ and $G$.  
Alternatively, the 2PI effective action can be viewed as a truncation of the
master effective action to second order in the correlation hierarchy
\cite{calzetta:1995b}. In this section we briefly review how the 
2PI method works; more thorough presentations can be
found in \cite{cornwall:1974a,calzetta:1988b}.

Generalizing the 1PI method where the mean field is fixed to
be $\phih$, the 2PI method fixes the mean field to be $\phih$ and
the sum of all self-energy diagrams to be $G$. This leads to a drastic
compression of the diagrammatic expansion containing the full information
of the field theory \cite{calzetta:1993a}.  Coupled dynamical
equations for $\phih$ and $G$ are obtained by 
separately varying $\Gamma[\phih,G]$ with respect to $G$ and $\phih$.  
Imposing
$\delta\Gamma /\delta \phih = 0$ yields an equation for the mean
field $\phih$,
and setting $\delta \Gamma/\delta G = 0$ yields an equation
for $G$, the ``gap'' equation.  The variance 
$\langle\vphi_{\text{{\tiny H}}}^2\rangle$
is the coincidence limit of the 
two-point function $\hbar G$, as seen from  Eq.~(\ref{eq-lpf2pitpf}).
In a nonequilibrium setting, the closed-time-path method should be
used in conjunction with the 2PI formalism in order to 
obtain real and causal dynamical equations for $\phih$ and $G$ 
\cite{calzetta:1988b,calzetta:1989b,calzetta:1993a}.  

Let us apply the 2PI method to a scalar $\lambda \Phi^4$ theory
in Minkowski space, with vacuum initial conditions.
In a direct generalization of Sec.~\ref{sec-ctpform}, we now couple
both a local source $J_a(x)$ and nonlocal source $K_{ab}(x,\xp)$ 
(which are $c$-number functions on ${\mathcal M}$) to the scalar field 
via interactions of the form $\hbar c^{ab} J_a \phi_a$ and $\hbar
c^{ab} c^{cd} K_{ac}(x,\xp) \phi_b(x) \phi_d(\xp)$\@.  Following
Eq.~(\ref{eq-gflpf}), the CTP generating functional is defined as a 
vacuum persistence amplitude in the presence of the sources $J$ and $K$,
which has the path integral representation
\begin{equation}
\begin{split}
Z[J,K] = \int_{\text{{\tiny ctp}}} D\phi_{+} D\phi_{-}
\exp \Biggl[ \frac{i}{\hbar} \biggl( 
& {\mathcal S}^{\text{{\tiny F}}}
[\phi_{+},\phi_{-}] + \int_M \! d^{\, 4} x c^{ab} J_a \phi_b  \\ &
+ \frac{1}{2} \int_M \! d^{\, 4} x \int_M \! d^{\, 4} x' c^{ab} c^{cd} K_{ac}
(x,\xp)
\phi_b(x) \phi_d(\xp) \biggr) \Biggr].
\label{eq-lpf2pigf}
\end{split}
\end{equation}
Here, ${\mathcal S}^{\text{{\tiny F}}}$ is as defined in 
Eq.~(\ref{eq-lpfctpca}), and we are using $Z[J,K]$ as a shorthand for 
$Z[J_{+},J_{-};K_{++},K_{--},K_{+-},
K_{-+}]$.
The generating functional for normalized $n$-point functions
(connected diagrams) is defined by 
\begin{equation}
W[J,K] = -i\hbar \text{ln} Z[J,K].
\label{eq-wgf2pi}
\end{equation}
The ``classical'' field $\phih_a(x)_{\text{{\tiny $JK$}}}$ and two-point 
function 
$G_{ab}(x,\xp)_{\text{{\tiny $JK$}}}$ are then given by
\begin{align}
\phih_a(x)_{\text{{\tiny $JK$}}} &= c_{ab} \frac{\delta W[J,K]}{
\delta J_b(x)},
\label{eq-lpf2piphidef} 
\\
\hbar G_{ab}(x,\xp)_{\text{{\tiny $JK$}}} &= 2 c_{ac} c_{bd} \frac{\delta 
W[J,K]}{
\delta K_{cd}(x,\xp)} - \phih_a(x)_{\text{{\tiny $JK$}}}\phih_b(\xp)_{
\text{{\tiny $JK$}}}, 
\label{eq-lpf2piphidefb}
\end{align}
where we use the subscript $JK$ to indicate that $\phih_a$ and
$G_{ab}$ are functions of the sources $J$ and $K$.  

\begin{figure}[htb]
\begin{center}
\epsfig{file=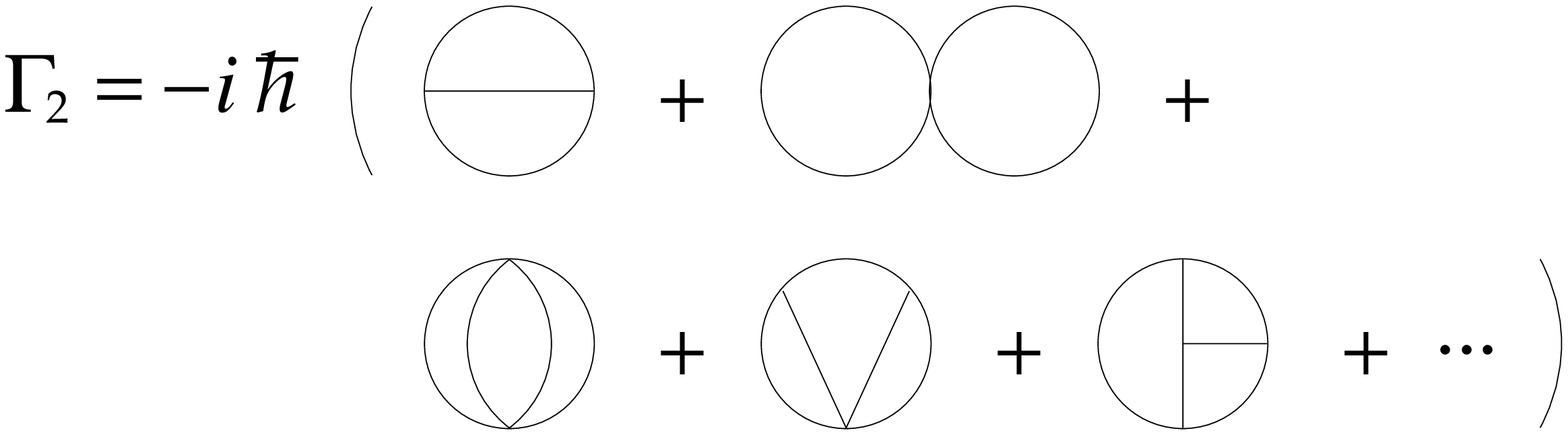,width=4.0in} 
\end{center}
\caption{Diagrammatic expansion for $\Gamma_2$ part of the 2PI-CTP effective
action}
\label{fig-gamma2}
\end{figure}

In the limit $K = J = 0$, the classical field $\phih_a$ satisfies\footnote{
For simplicity, the ``in'' vacuum for the fluctuation field is henceforth
denoted by $|\Omega\rangle$ instead of $|\Omega,\text{in}\rangle$, since
we will not have any need to refer to the ``out'' vacuum in what follows.}
\begin{equation}
(\phihp)_{J=K=0} = 
(\phihm)_{J=K=0} = \langle \Omega | \Phi_{\text{{\tiny H}}} | \Omega \rangle
\equiv \phih;
\label{eq-ljkzmf}
\end{equation}
i.e., $\phih_a$ becomes the expectation value of the Heisenberg field operator 
$\Phi_{\text{{\tiny H}}}$ in the quantum state 
$|\Omega\rangle$, i.e., the mean field.
In the same limit, the two-point function $G_{ab}$ is the 
CTP propagator for the fluctuation field defined by
Eq.~(\ref{eq-deffluc1pi}).
The four components of the CTP propagator are, for $J = K = 0$,
\begin{align}
\hbar G_{++}(x,x')_{|J=K=0} &= \langle \Omega |
T\left(\vphi_{\text{{\tiny H}}}(x)
\vphi_{\text{{\tiny H}}}(x')\right) | \Omega \rangle,
\label{eq-ljkzgfa}\\
\hbar G_{--}(x,x')_{|J=K=0} &= \langle \Omega |
\tilde{T}\left(\vphi_{\text{{
\tiny H}}}(x) \vphi_{\text{{\tiny H}}}(x')\right) | \Omega \rangle,\\
\hbar G_{+-}(x,x')_{|J=K=0} &= \langle \Omega | \vphi_{\text{{\tiny H}}}(x')
\vphi_{\text{{\tiny H}}}(x) | \Omega \rangle,\\
\hbar G_{-+}(x,x')_{|J=K=0} &= \langle \Omega | \vphi_{\text{{\tiny H}}}(x) 
\vphi_{\text{{\tiny H}}}(x') | \Omega \rangle,
\label{eq-ljkzgfd}
\end{align}
in the Heisenberg picture.  In the coincidence limit $x' = x$, all
four components above are equivalent to the variance
$\langle \vphi_{\text{{\tiny H}}}^2 \rangle$ defined in
Eq.~(\ref{eq-lpf2pitpf}),
\begin{equation}
\langle \varphi_{\text{{\tiny H}}}^2(x) \rangle = \hbar G_{ab}(x,x).
\label{eq-dvrc}
\end{equation}
Provided we can invert Eqs.~(\ref{eq-lpf2piphidef}) and 
(\ref{eq-lpf2piphidefb})
to obtain $J$ and $K$ in terms of $\phih$ and $G$, the 2PI effective
action can be defined as the double Legendre transform (in both $J$ and
$K$) of $W[J,K]$,
\begin{multline}
\Gamma[\phih,G] = W[J,K] -  \int_M \! d^{\, 4} x c^{ab} J_a(x) \phih_b(x) 
 \\ 
- \frac{1}{2} \int_M \! d^{\, 4} x \int_M \! d^{\, 4} x' c^{ab} c^{cd} 
K_{ac}(x,\xp) [ \hbar
G_{bd}(x,\xp) + \phih_b(x) \phih_d(\xp) ].
\label{eq-lpf2pilegtrn}
\end{multline}
As with $W[J,K]$ above,
$\Gamma[\phih,G]$ denotes
$\Gamma[\phihp,\phihm;G_{++},G_{--},G_{+-},G_{-+}]$.  The $JK$ 
subscripting of $\phih$ and $G$ is suppressed, but the functional
dependence of $\phih$ and $G$ on $J$ and $K$ through inversion of
Eqs.~(\ref{eq-lpf2piphidef}) and (\ref{eq-lpf2piphidefb}) is understood.
By direct functional differentiation of Eq.~(\ref{eq-lpf2pilegtrn}), 
the inverses of Eqs.\~(\ref{eq-lpf2piphidef}) and 
(\ref{eq-lpf2piphidefb}) are found to be
\begin{align}
&\frac{\delta \Gamma[\phih,G]}{\delta \phih_a(x)}
=  -c^{ab} J_b(x)_{\phih G} - \frac{1}{2} c^{ab} c^{cd} \int_M \! d^{\, 4} x'
( K_{bd}(x,\xp)_{\phih G} + K_{db}(\xp,x)_{\phih G})\phih_c(\xp),
\label{eq-lpf2pidgam} \\
& \frac{\delta \Gamma[\phih,G]}{\delta G_{ab}(x,\xp)} =
-\frac{\hbar}{2}c^{ac}c^{bd}K_{cd}(x,\xp)_{\phih G},
\label{eq-lpf2pidgamb} 
\end{align}
where the subscript ``$\phih G$'' indicates that $K$ and $J$ are functionals
of $\phih$ and $G$.  Once $\Gamma[\phih,G]$ has been calculated,
the evolution equations for $\phih$ and $G$ are given by
\begin{align}
\left.
\frac{\delta \Gamma[\phih,G]}{\delta \phih_a(x)}\right|_{
\phihp = \phihm \equiv 
\phih} &= 0,
\label{eq-lpf2pieom} \\
\left.\frac{\delta \Gamma[\phih,G]}{\delta G_{ab}(x,y)}\right|_{ 
\phihp = \phihm \equiv \phih} &= 0.
\label{eq-lpf2pieomb} 
\end{align}
Of course, the two equations contained in Eq.~(\ref{eq-lpf2pieom}) 
(corresponding to $a=+$ and $a=-$, respectively) are not independent,
just as in Eq.~(\ref{eq-lpfctpeom}).  In addition, only two of
equations (\ref{eq-lpf2pieomb}) are independent, one
on the diagonal and one off diagonal in the ``internal'' CTP space.
Using both Eq.~(\ref{eq-lpf2pilegtrn}) and 
Eq.~(\ref{eq-lpf2pigf}), an equation for $\Gamma[\phih,G]$ in
terms of the sources $K$ and $J$ can be derived,
\begin{equation}
\begin{split}
\Gamma[&\phih,G] = -i \hbar \text{ln} 
\Biggl\{ \int_{\text{{\tiny ctp}}} D\phip D\phim \exp
\biggl[ \frac{i}{\hbar} \Bigl( {\mathcal S}^{\text{{\tiny F}}}[\phip,\phim]
+ c^{ab} \int_M \! d^{\, 4} x J_a(x) [\phi_b(x) - \phih_b(x)]  
\\ & +
\frac{1}{2} c^{ac} c^{bd} \int_M \! d^{\, 4} x \int_M \! d^{\, 4} x' 
K_{ab}(x,\xp) [ \phi_c(x)\phi_d(\xp)
- \phih_c(x)\phih_d(\xp) - \hbar G_{cd}(x,\xp)]\Bigr)\biggr]\Biggr\}.
\label{eq-lpf2pigam1}
\end{split}
\end{equation}
The sources $K$ and $J$ in the right-hand side of Eq.~(\ref{eq-lpf2pigam1})
are functionals of $\phih$, through Eqs.~(\ref{eq-lpf2pidgam}) and
(\ref{eq-lpf2pidgamb}).
Expressing this functional dependence, we 
obtain a functional integro-differential equation for $\Gamma$,
\begin{equation}
\begin{split}
\Gamma[\phih,G] = & \int_M \! d^{\, 4} x \int_M \! d^{\, 4} x' \frac{\delta 
\Gamma[\phih,G] }{
\delta G_{ba}(\xp,x)} G_{ab}(x,\xp)  \\ & 
- i \hbar \text{ln} \Biggl\{ \int_{\text{{\tiny ctp}}} D\phip D\phim \exp
\biggl[ \frac{i}{\hbar} \Bigl( {\mathcal S}^{\text{{\tiny F}}}[\phi_{+},
\phi_{-}] 
- \int_M \! d^{\, 4} x \frac{\delta \Gamma[\phih,G]}{\delta \phih_a} 
(\phi_a - \phih_a)  \\ &
- \frac{1}{\hbar}\int_M \! d^{\, 4} x\int_M \! d^{\, 4} x' \frac{\delta 
\Gamma[\phih,G]}{\delta
G_{ba}(\xp,x)} [\phi_a(x) - \phih_a(x)][\phi_b(\xp) - \phih_b(\xp)]\Bigr)
\biggr] \Biggr\}.
\end{split}
\end{equation}
We have dropped the $JK$ subscripting because the functional 
derivatives in the equation are only with respect to $\phih$ and $G$.
As in Sec.~\ref{sec-ctpform}, a change of variables
$D\phi_{\pm} \rightarrow D\vphi_{\pm}$ is carried out in the 
functional integral, with the fluctuation field 
$\vphi_{a}$ defined by Eq.~(\ref{eq-dctpfluc}).  The resulting equation
\begin{equation}
\begin{split}
\Gamma[\phih,G] = \int_M \! & d^{\, 4} x \int_M \! d^{\, 4} x' \frac{\delta 
\Gamma[\phih,G] }{
\delta G_{ba}(\xp,x)} G_{ab}(x,\xp)  \\ &
- i \hbar \text{ln} \Biggl\{ \int_{\text{{\tiny ctp}}} D\vphip D\vphim \exp
\biggl[ \frac{i}{\hbar} \Bigl( {\mathcal S}^{\text{{\tiny F}}}[\vphip +
\phihp, \vphim + \phihm]  \\ & - 
\int_M \! d^{\, 4} x \frac{\delta \Gamma[\phih,G]}{\delta \phih_a} \vphi_a
- \frac{1}{\hbar}\int_M \! d^{\, 4} x\int_M \! d^{\, 4} x' \frac{\delta 
\Gamma[\phih,G]}{\delta
G_{ba}(\xp,x)} \vphi_a(x) \vphi_b(\xp) \Bigr)
\biggr] \Biggr\}
\label{eq-lpf2pifideg}
\end{split}
\end{equation}
has the formal solution \cite{cornwall:1974a}
\begin{multline}
\Gamma[\phih,G] = {\mathcal S}^{\text{{\tiny F}}}[\phih] -
\frac{i \hbar}{2} \text{ln}\,\text{det} 
(G_{ab})  \\ 
+ \frac{i \hbar}{2} \int_M \! d^{\, 4} x \int_M \! d^{\, 4} x' 
{\mathcal A}^{ab}(\xp,x) G_{ab}(x,\xp) 
+ \Gamma_2 [\phih,G],
\end{multline}
where ${\mathcal A}^{ab}$ is the second functional derivative of the
classical action ${\mathcal S}^{\text{{\tiny F}}}$,
evaluated at $\phih_a$ [as defined in Eq.~(\ref{eq-dolp})].  The functional 
$\Gamma_2$ is $-i  \hbar$ times the sum of all two-particle-irreducible
vacuum-to-vacuum diagrams with lines given by $G_{ab}$ and vertices
given by a shifted action 
${\mathcal S}_{\text{{\tiny int}}}^{\text{{\tiny F}}}$ defined by
Eq.~(\ref{eq-defshfa}).
The shifted action for the $\lambda \Phi^4$ scalar field theory is
\begin{equation}
{\mathcal S}_{\text{{\tiny int}}}^{\text{{\tiny F}}}[\vphi] =
S^{\text{{\tiny F}}}_{\text{{\tiny int}}}[\vphip] - 
S^{\text{{\tiny F}}}_{\text{{\tiny int}}}[\vphim],
\end{equation}
in terms of
\begin{equation}
S^{\text{{\tiny F}}}_{\text{{\tiny int}}}[\vphi] =
-\frac{\lambda}{6} \int_M \! d^{\, 4} x
\left( \frac{1}{4} \vphi^4  + \phih \vphi^3 \right),
\label{eq-lpf2pisa} 
\end{equation}
where the functional dependence of ${\mathcal S}^{\text{{\tiny F}}}_{\text{{
\tiny int}}}$ on $\phihpm$ is not shown explicitly.
Two types of vertices appear
in Eq.~(\ref{eq-lpf2pisa}): a vertex
which terminates four lines and a vertex terminating three lines which
is proportional to the mean field $\phih$.
The expansion for $\Gamma_2$ in terms of $G$ and $\phih$ is depicted 
graphically up to three-loop order in Fig.~\ref{fig-gamma2},
where lines represent the propagator $G$ and vertices are given by 
${\mathcal S}^{\text{{\tiny F}}}_{\text{{\tiny int}}}$ \cite{cornwall:1974a}. 
The vertices terminating three lines are proportional to $\phih$\@.
Each vertex carries a spacetime label in $M$ and a time branch label 
in $\{ +,- \}$.
In general, the 2PI effective action contains divergences at each order
in the loop expansion.  It has been shown that if a field theory is 
renormalizable in the ``in-out'' formulation, then the ``in-in'' equations
of motion derived from the 2PI effective action are renormalizable 
\cite{calzetta:1988b}.  We will discuss the renormalization 
of the 2PI effective action below in Sec.~\ref{sec-onrenorm}.

Various approximations to the full quantum dynamics can be obtained
by truncating the diagrammatic expansion for $\Gamma_2$.  Throwing
away $\Gamma_2$ in its entirety yields the one-loop approximation.
In Fig.~\ref{fig-gamma2}, there are two two-loop diagrams, the 
``setting sun'' and the ``double-bubble.'' Retaining just the double-bubble
diagram yields the time-dependent Hartree-Fock approximation 
\cite{cornwall:1974a}.  Retaining both diagrams gives the
two-loop truncation of the theory.\footnote{A different approximation,
the $1/N$ expansion, is used in Sec.~\ref{sec-ontheory} to study
the nonequilibrium dynamics of the O$(N)$ field theory.} 
  This approximation will yield a
non-time-reversal-invariant mean-field equation above threshold,
due to the setting sun diagram \cite{calzetta:1995b}. 

The time-reversal noninvariance of the mean-field and gap equations 
generated by the 2PI effective action is a consequence of the fact that
the 2PI effective action really corresponds to a further approximation 
from the two-loop truncation (in the sense of topology of vacuum graphs) of
the {\em master\/} effective action \cite{calzetta:1995b}.  
The two-loop truncation of the master 
effective action is a functional $\Gamma_{l=2}[\phih,G,C_3]$
which depends on the mean field $\phih$, the two-point function $G$, and
the three-point function $C_3$.  The four-point function
$C_4$ also appears, but is not dynamical
due to a constraint.   The full set of equations,
\begin{align}
\frac{\delta\Gamma_{l=2}[\phih,G,C_3]}{\delta \phih_a} &= 0, \\
\frac{\delta\Gamma_{l=2}[\phih,G,C_3]}{\delta G_{ab}} &= 0, \\
\frac{\delta\Gamma_{l=2}[\phih,G,C_3]}{\delta (C_3)_{abc}} &= 0,
\label{eq-lpf2pimtrn} 
\end{align}
is time-reversal invariant.  However, the 2PI effective action is obtained
by solving Eq.~(\ref{eq-lpf2pimtrn}) with {\em a given choice of causal
boundary conditions\/} and substituting the resulting $C_3$ into 
$\Gamma_{l=2}$, to obtain $\Gamma_{\text{2}}[\phih,G]$.  This ``slaving''
of $C_3$ to $\phih$ and $G$ with a particular choice of boundary conditions
is what breaks the time-reversal invariance of the theory 
\cite{calzetta:1995b}.  This is shown explicitly below
in Chapter~\ref{chap-entropy}.  In Chapter~\ref{chap-preheat} where we 
discuss the preheating dynamics of the inflaton field, we work with further
approximations which discard the setting sun diagram, and thus regain
time-reversal invariance of the dynamical equations.

\section{$\lambda \Phi^4$ field theory in curved spacetime}
\label{sec-lpfcst}

In this section the quantum dynamics of a scalar $\lambda \Phi^4$ field theory 
is formulated in semiclassical gravity, which means that the matter fields are
quantized on a curved classical background spacetime.\footnote{
The semiclassical approximation 
is consistent with a truncation of the quantum effective action for matter 
fields and gravity perturbations at one loop 
[i.e., at order $O(\hbar)$] \cite{hartle:1979a}
or [in the case of the O$(N)$ field theory studied here] at leading order 
in the $1/N$ expansion \cite{hartle:1981a,hu:1987b}.}
The two-particle-irreducible effective action is used in conjunction with the 
CTP formalism to obtain manifestly covariant, coupled evolution equations
for the mean field $\langle \Phi_{\text{{\tiny H}}} \rangle$ and
variance $\langle \Phi_{\text{{\tiny H}}}^2 \rangle - \langle
\Phi_{\text{{\tiny H}}} \rangle^2$ in the $\lambda \Phi^4$ model.

Let us consider the $\lambda \Phi^4$ scalar field theory
in a globally hyperbolic, curved background spacetime with
metric tensor $g_{\mu\nu}$.  The diffeomorphism-invariant classical action
for this system is
\begin{equation}
S[\phi,g^{\mu\nu}] = S^{\text{{\tiny G}}}[g^{\mu\nu}] + S^{\text{{\tiny F}}}[
\phi,g^{\mu\nu}],
\end{equation}
where $g^{\mu\nu}$ is the contravariant metric tensor, and 
$S^{\text{{\tiny G}}}$ and $S^{\text{{\tiny F}}}$ are the classical actions
of the gravity and scalar field sectors of the theory, respectively.
For the scalar field action, we have
\begin{equation}
S^{\text{{\tiny F}}}[\phi,g^{\mu\nu}] = -\frac{1}{2} \int \! d^{\, 4} x 
\sqmg \left[
\phi (\square + m^2 + \xi R) \phi + \frac{\lambda}{12}\phi^4\right],
\label{eq-lpfcstca}
\end{equation}
where $\xi$ is the (dimensionless) coupling to gravity
(necessary in order for the field theory to be renormalizable
\cite{toms:1982a}), $\square$ is the Laplace-Beltrami operator defined 
in terms of the covariant derivative
$\nabla_{\mu}$, and $R$ is the scalar curvature.  The constant $m$ has
units of inverse length, and the $\phi$ self-coupling $\lambda$ has units
of $1/\hbar$.
Following standard procedure in semiclassical gravity \cite{birrell:1982a},
we define the semiclassical action for gravity to be
\begin{equation}
S^{\text{{\tiny G}}}[g^{\mu\nu}] = \frac{1}{16 \pi G} \int \! d^{\, 4} x 
\sqmg \left[
R - 2 \Lambda + c R^2 + b R^{\alpha\beta} R_{\alpha\beta} +
a R^{\alpha\beta\gamma\delta} R_{\alpha\beta\gamma\delta} \right],
\label{eq-lpfcstga}
\end{equation}
where $a$, $b$, and $c$ are constants with dimensions of length squared,
$R_{\alpha\beta\gamma\delta}$ is the Riemann tensor, $R_{\alpha\beta}$ is
the Ricci tensor, $\Lambda$ is the ``cosmological constant'' (with
units of inverse length-squared), $\sqmg$ is the square root of the
negative of the determinant of $g_{\mu\nu}$,
and $G$ (with units of length divided by mass) is Newton's 
constant.  As a result of the generalized Gauss-Bonnet theorem
\cite{chern:1962a}, the constants $a$, $b$, and $c$ are not all independent
in four spacetime dimensions; let us, therefore, set $a=0$.  Classical Einstein
gravity is obtained by setting $b = 0$ and $c = 0$.
Minimal and conformal coupling (for the $\phi$ field
to gravity) correspond to setting $\xi = 0$ and $\xi = 1/6$,
respectively.  

The motivation for including the arbitrary coupling
$\xi$ and the higher-order curvature terms $R^2$ and 
$R^{\alpha\beta}R_{\alpha\beta}$ in the classical action $S$ is that
we wish to study the semiclassical dynamics of the theory.  
In the semiclassical gravity field equation and matter field equations,
divergences arise which require a renormalization of
$b$, $c$, $G$, $\Lambda$, $m$, $\xi$, and $\lambda$
\cite{birrell:1982a}.  These quantities, as they appear in
Eqs.~\ref{eq-lpfcsta}--\ref{eq-lpfcstga}, are understood to be bare;
their observable  counterparts are renormalized.

The classical Euler-Lagrange equation
for $\phi$ is obtained by functionally differentiating
$S^{\text{{\tiny F}}}[\phi,g^{\mu\nu}]$ with
respect to $\phi$, and setting $\delta S^{\text{{\tiny F}}}/\delta \phi = 0$,
\begin{equation}
\left(\square + m^2 + \xi R + \frac{\lambda}{6} \phi^2\right) \phi = 0.
\end{equation}
The Euler-Lagrange equation for the metric $g_{\mu\nu}$
is obtained by functional differentiation  of $S$ with respect to $g^{\mu\nu}$
(it is assumed that
the variations $\delta\phi$ and $\delta g^{\mu\nu}$ are restricted 
so that no boundary terms arise),
\begin{equation}
 G_{\mu\nu} + \Lambda g_{\mu\nu} + c \, ^{(1)}H_{\mu\nu} + 
b \, ^{(2)}H_{\mu\nu} = -8 \pi G T_{\mu\nu},
\label{eq-cee}
\end{equation}
where the tensors $G_{\mu\nu}$, $^{(1)}H_{\mu\nu}$, and $^{(2)}H_{\mu\nu}$
are defined by \cite{fulling:1974a,fulling:1974b}
\begin{align}
^{(1)}H_{\mu\nu} &= 2R_{;\mu\nu} +  2g_{\mu\nu}\square R - 
\frac{1}{2}g_{\mu\nu}R^2 + 2 R R_{\mu\nu}, \label{eq-csteetns} \\
^{(2)}H_{\mu\nu} &= 2R_{\mu\;\;\nu\alpha}^{\;\;\alpha} - \square R_{\mu\nu}
- \frac{1}{2}g_{\mu\nu}\square R + 2 R_{\;\;\mu}^{\alpha}R_{\alpha\nu}
-\frac{1}{2}g_{\mu\nu}R^{\alpha\beta} R_{\alpha\beta}, \\
G_{\mu\nu} &= R_{\mu\nu} - \frac{1}{2}R g_{\mu\nu}.
\end{align}
In Eq.~\ref{eq-cee}, $T_{\mu\nu}$ is the classical energy-momentum tensor,
\begin{equation}
\begin{split}
 T_{\mu\nu} = (1 - 2\xi) \phi_{;\mu} \phi_{;\nu} + &
\left( 2 \xi - \frac{1}{2} \right) g_{\mu\nu} g^{\rho\sigma}
\phi_{;\rho}\phi_{;\sigma} - 2\xi \phi_{;\mu\nu} \phi
+ 2 \xi g_{\mu\nu} \phi \square \phi 
 \\ & - \xi G_{\mu\nu} \phi^2 +
\frac{1}{2} g_{\mu\nu} \left( m^2+ \frac{\lambda}{12}\phi^2\right)\phi^2.
\end{split}
\end{equation}
We are interested in the dynamics of the expectation value of the scalar field
operator and its higher moments,
which in nonequilibrium field theory does {\em not\/} follow directly
from functional differentiation of the usual Schwinger-DeWitt or ``in-out'' 
effective action.  Instead, as discussed above in Sec.~\ref{sec-ctpform}, the
Schwinger-Keldysh formalism should be used.  Here we discuss the 
implementation of the Schwinger-Keldysh method in curved spacetime.    

The first step is to generalize the closed-time-path
(CTP) manifold ${\mathcal M}$, defined in Eq.~(\ref{eq-lpfctpmn}),
to curved spacetime.  Let $\Sigma_{\star}$ be a
Cauchy hypersurface chosen so that its past domain of dependence
\cite{wald:1984a},
$D_{-}(\Sigma_{\star})$, contains all of the dynamics we wish to study.
Let us then define the manifold (with boundary)
\begin{equation}
M \equiv D_{-}(\Sigma_{\star}).
\label{eq-cstmdef}
\end{equation}
The CTP manifold ${\mathcal M}$ is defined 
following Eq.~(\ref{eq-lpfctpmn}) as a quotient space constructed by
identification on the hypersurface $\Sigma_{\star} \subset \partial M$ as in
Eq.~(\ref{eq-lpfctpmn})
\begin{equation}
{\mathcal M} \equiv (M \times \{ +, - \})/\sim,
\label{eq-dctpm2}
\end{equation}
where the equivalence relation is the same as Eq.~(\ref{eq-lpfctper}) 
except that
the matching of $+$ and $-$ time branches is now done on $\Sigma_{\star}$.
$M \times \{+,-\}$ defined by
\begin{align}
(x,+) & \sim (x',+)  \;\;\;\;\;\;\;\; \text{iff $x$} = x' \nonumber \\
(x,-) & \sim (x',-)  \;\;\;\;\;\;\;\; \text{iff $x$} = x' \\
(x,+) & \sim (x',-)  \;\;\;\;\;\;\;\; \text{iff $x$} = \text{$x'$ and
$x \in \Sigma_{\star}$}. \nonumber
\end{align}
We construct an orientation on ${\mathcal M}$ using the canonical volume
form from $M$, ${\boldsymbol \epsilon}_M$,
\begin{equation}
{\boldsymbol \epsilon}_M = 
\sqmg \; dx^0 \wedge dx^1 \wedge dx^2 \wedge dx^3,
\label{eq-lpfcstvf}
\end{equation}
and define the volume form on ${\mathcal M}$ to be
\begin{equation}
{\boldsymbol \epsilon}_{{\mathcal M}} = 
\left\{
\begin{array}{cc}
{\boldsymbol \epsilon}_M & \text{on $M\times \{+\}$}, \\
-{\boldsymbol \epsilon}_M & \text{on $M\times \{-\}$}.
\end{array}
\right.
\end{equation}
Finally, we let $\phi$ and $g^{\mu\nu}$ be independent
on the $+$ and $-$ branches of ${\mathcal M}$, provided that
$g^{\mu\nu}_{+} = g^{\mu\nu}_{-}$ and $\phi_{+} = \phi_{-}$ on
$\Sigma_{\star}$.  In other words,
$\phi$ and $g^{\mu\nu}$ must be a scalar and
a tensor, respectively, on ${\mathcal M}$.
In terms of the volume form ${\boldsymbol \epsilon}_M$, 
we can write a scalar field action on ${\mathcal M}$, 
\begin{equation}
{\mathcal S}^{\text{{\tiny F}}}[\phipm,g^{\mu\nu}_{\pm}] = 
S^{\text{{\tiny F}}}[\phip,g^{\mu\nu}_{+}] - 
S^{\text{{\tiny F}}}[\phim,g^{\mu\nu}_{-}], 
\label{eq-sfamcm}
\end{equation}
where $S^{\text{{\tiny F}}}[\phi]$ is given by Eq.~(\ref{eq-lpfcstca}), and
$g^{\mu\nu}_{\pm}$ is the metric tensor on the $+$ and $-$ branches
of ${\mathcal M}$.
Using Eq.~(\ref{eq-lpfcstga}) we can similarly define the gravity action 
${\mathcal S}^{\text{{\tiny G}}}$ on ${\mathcal M}$,
\begin{equation}
{\mathcal S}^{\text{{\tiny G}}}[g^{\mu\nu}_{+},g^{\mu\nu}_{-}] = 
S^{\text{{\tiny G}}}[g^{\mu\nu}_{+}] - S^{\text{{\tiny G}}}[g^{\mu\nu}_{-}],
\end{equation}
where it is understood that only configurations of $g^{\mu\nu}_{\pm}$
satisfying the constraint $g^{\mu\nu}_{+} = g^{\mu\nu}_{-}$
on $\Sigma_{\star}$ are permitted.

In semiclassical gravity the scalar field theory (with action 
$S^{\text{{\tiny F}}}$)
is quantized on a classical background spacetime, with metric
$g_{\mu\nu}$, whose dynamics is determined self-consistently 
by the semiclassical geometrodynamical field equation.  
Let us denote the Heisenberg-picture field operator for the
canonically quantized $\phi$ field by $\Phi_{\text{{\tiny H}}}$.
We wish to compute the quantum effective action
$\Gamma$ for this scalar field theory, using
the two-particle-irreducible (2PI) 
method described in Sec.~\ref{sec-cjtform}.
In terms of ${\mathcal S}^{\text{{\tiny F}}}$ (now defined on the curved 
manifold ${\mathcal M}$), we define the
2PI, CTP generating functional $Z[J,K,g^{\mu\nu}]$ as follows:
\begin{equation}
\begin{split}
Z[J,K,g^{\mu\nu}] = & \int_{\text{{\tiny ctp}}} D\phip D\phim
\exp \Biggl[ \frac{i}{\hbar} \biggl( {\mathcal S}^{\text{{\tiny F}}}[\phipm,
g^{\mu\nu}_{\pm}] + \int_M \! d^{\, 4} x \sqrt{-g_c} c^{abc} J_a \phi_b 
 \\ & +
\frac{1}{2} \int_M \! d^{\, 4} x \sqrt{-g_{a'}} \int_M \! d^{\, 4} x' 
\sqrt{-g'_{c'}} c^{aba'}
c^{cdc'} K_{ac}(x,\xp) \phi_b(x) \phi_d(\xp) \biggr) \Biggr],
\label{eq-lpfcstzf}
\end{split}
\end{equation}
where we have written $Z[J,K,g^{\mu\nu}]$ as a shorthand for
$Z[\Jpm,K_{\pm\pm},g^{\mu\nu}_{\pm}]$. The three-index symbol $c^{abc}$ is
the $n=3$ case of the $n$-index symbol $c^{a_1 a_2 \cdots a_n}$ defined by
Eq.~(\ref{eq-cabcdef}).
The boundary conditions on the functional integral define the initial quantum
state (assumed here to be pure).  In this and in
Chapters~\ref{chap-preheat}--\ref{chap-fermion},
we are interested in the case of a quantum
state corresponding to a nonzero mean field $\phih$, with vacuum
fluctuations around the mean field.  This entails a definition of the vacuum
state for the {\em fluctuation field,\/} defined in Eq.~(\ref{eq-deffluc1pi}).
In a general curved spacetime, there does not exist a unique
Poincar\'{e}-invariant vacuum state for a quantum field
\cite{dewitt:1975a,fulling:1989a}.  For an asymptotically
free field theory, a choice of ``in'' vacuum state corresponds to a choice 
of a particular orthonormal basis of solutions of the covariant Klein-Gordon
equation with which to canonically quantize the field.

From Eq.~(\ref{eq-lpfcstzf}), we can derive the two-particle-irreducible
(2PI) effective action
$\Gamma[\phih,G,g^{\mu\nu}]$ following the method of
Sec.~\ref{sec-cjtform}, with the understanding that
$\Gamma$ now depends functionally
on the metric $g^{\mu\nu}_{\pm}$ on the $+$ and $-$ time branches.
The functional differentiations should be carried out using a covariant
generalization of the Dirac $\delta$ function to the manifold $M$ 
\cite{birrell:1982a}.
\begin{equation}
\int_M \! d^{\, 4} x \sqrt{-g(x)} \delta^4(x-x') \frac{1}{\sqrt{-g(x')}} 
\equiv 1.
\label{eq-cdfd}
\end{equation}
The functional integro-differential equation (\ref{eq-lpf2pifideg}) for the
CTP-2PI effective action can then be generalized to the curved spacetime $M$ 
in a straightforward fashion, modulo the curved-spacetime ambiguities in the
boundary conditions of the functional integral (\ref{eq-lpfcstzf}).

The (bare) semiclassical field equations for the variance, mean field,
and metric can then be obtained by functional differentiation of
${\mathcal S}^{\text{{\tiny G}}}[g^{\mu\nu}] + \Gamma[\phih,G,g^{\mu\nu}]$
with respect to $G_{\pm\pm}$, $\phi_{\pm}$, and $g^{\mu\nu}_{\pm}$,
respectively,
followed by metric and mean-field identifications between the $+$ and $-$
time branches,
\begin{align}
\left.\frac{\delta ({\mathcal S}^{\text{{\tiny G}}}[g^{\mu\nu}] + 
\Gamma[\phih,G,g^{\mu\nu}])}{\delta g^{\mu\nu}_a}\right|_{ 
\phihp = \phihm = \phih ;\;\;\;
g^{\mu\nu}_{+} = g^{\mu\nu}_{-} = g^{\mu\nu}} &= 0,
\label{eq-lpfseea} \\
\left.\frac{\delta \Gamma[\phih,G,g^{\mu\nu}]}{\delta \phih_a}\right|_{
\phihp = \phihm = \phih ;\;\;\;
g^{\mu\nu}_{+} = g^{\mu\nu}_{-} = g^{\mu\nu}} &= 0, 
\label{eq-lpfseeb} \\
\left.\frac{\delta \Gamma[\phih,G,g^{\mu\nu}]}{\delta G_{ab}}\right|_{ 
\phihp = \phihm = \phih ;\;\;\; 
g^{\mu\nu}_{+} = g^{\mu\nu}_{-} = g^{\mu\nu}} &= 0.
\label{eq-lpfseec}
\end{align}
As above, CTP indices are suppressed inside the functional arguments.
Eqs.~(\ref{eq-lpfseea}), (\ref{eq-lpfseeb}), and (\ref{eq-lpfseec}) constitute
the semiclassical approximation to the full quantum dynamics for the
system described by the classical action
$S^{\text{{\tiny G}}}[g^{\mu\nu}] + S^{\text{{\tiny F}}}[\phi,g^{\mu\nu}]$.
The semiclassical field equation (with bare parameters) for $g^{\mu\nu}$ 
is obtained directly from 
Eq.~(\ref{eq-lpfseea}),
\begin{equation}
G_{\mu\nu} + \Lambda g_{\mu\nu} + 
c\; ^{(1)\!}H_{\mu\nu} + b\; ^{(2)\!}H_{\mu\nu} = -8 \pi G
\langle T_{\mu\nu} \rangle ,
\label{eq-lpfcstsee2}
\end{equation}
where $\langle T_{\mu\nu} \rangle$ is the (unrenormalized)  
energy-momentum tensor defined by
\begin{equation}
\langle T_{\mu\nu} \rangle = \left.\frac{2}{\sqmg} \left( \frac{\delta
\Gamma[\phih,G,g^{\mu\nu}]}{\delta g^{\mu\nu}_{+}} \right)\right|_{
\phihp = \phihm = \phih; \;\;\; g^{\mu\nu}_{+} = g^{\mu\nu}_{-} =
g^{\mu\nu}}.
\label{eq-lpfcstemt}
\end{equation}
Equation (\ref{eq-lpfcstsee2}) gives the spacetime dynamics; the
dynamics of $\phih$ and $G$ are given by the mean-field 
(\ref{eq-lpfseec}) and gap (\ref{eq-lpfseeb}) equations.
In Eq.~(\ref{eq-lpfcstemt}), the angle brackets denote an expectation 
value of the energy-momentum tensor (with the Heisenberg field operator 
$\Phi_{\text{{\tiny H}}}$ substituted for $\phi$ in the classical theory) 
with respect to a quantum state $|\Omega\rangle$ defined by the boundary
conditions on the functional integral in Eq.~(\ref{eq-lpfcstzf}).
In four spacetime dimensions the unrenormalized $\langle T_{\mu\nu} \rangle$
has divergences which can be absorbed by the renormalization of $G$, $\Lambda$,
$b$, and $c$ \cite{birrell:1982a}.  It is often useful to carry out the
renormalization in the evolution equations for expectation values rather 
than in the CTP effective action \cite{jordan:1986a}.

The energy-momentum tensor $\langle T_{\mu\nu} \rangle$
is obtained by variation of the 2PI effective action $\Gamma$, which is a
functional of the metric $g^{\mu\nu}_{\pm}$ on both the $+$ and $-$
time branches.  From Eq.~(\ref{eq-lpfcstzf}), it is 
possible to derive $\Gamma[\phih,G,g^{\mu\nu}]$ as an arbitrary functional of 
$g^{\mu\nu}_{+}$ and $g^{\mu\nu}_{-}$.  However, in practice it is often
easier to fix the metric to be the
same on both the $+$ and $-$ time branches, i.e.,
\begin{equation}
g^{\mu\nu}_{+} = g^{\mu\nu}_{-} \equiv g^{\mu\nu},
\end{equation}
in the computation of $\Gamma[\phih,G,g^{\mu\nu}]$.
Once $\Gamma[\phih,G,g^{\mu\nu}]$ (or some consistent truncation of
the full quantum effective action for ${\mathcal S}^{\text{{\tiny F}}}$) 
has been computed, it is then straightforward to 
determine how $\Gamma[\phih,G,g^{\mu\nu}]$
should be generalized to the case of an arbitrary metric on ${\mathcal M}$,
for which $g^{\mu\nu}_{+}$ and $g^{\mu\nu}_{-}$ are independent.
The bare energy-momentum tensor $\langle T_{\mu\nu}\rangle$
can then be computed using Eq.~(\ref{eq-lpfcstemt}).  
Accordingly, in Sec.~\ref{sec-ontheory} below, we
fix $g^{\mu\nu}_{+} = g^{\mu\nu}_{-} \equiv g^{\mu\nu}$ 
in the calculation of $\Gamma[\phih,G,g^{\mu\nu}]$.  

The semiclassical Einstein equation is a subcase of the general
geometrodynamical field equation (\ref{eq-lpfcstsee2}), obtained (after
renormalization) by setting the renormalized $b = c = \Lambda = 0$ 
\cite{birrell:1982a}:
\begin{equation}
G_{\mu\nu} = -8 \pi G \langle T_{\mu\nu} \rangle,
\label{eq-seesimp}
\end{equation}
where we assume that the cosmological constant vanishes.
Having shown how to derive coupled evolution equations for the mean field,
variance, and metric tensor in semiclassical gravity, we now turn our 
attention to the scalar O$(N)$ model in curved spacetime. 

\section{O$(N)$ field theory in curved spacetime}
\label{sec-ontheory}
In this section we derive coupled nonperturbative dynamical equations for 
the mean field $\phih$ and variance 
$\langle \vphi_{\text{{\tiny H}}}^2 \rangle$ 
for the minimally coupled O$(N)$ scalar field theory with
quartic self-interaction and unbroken symmetry.  The background spacetime 
dynamics is given by the semiclassical Einstein equation.  These equations 
self-consistently account for the back reaction of quantum particle production 
on the mean field, and quantum fields on the dynamical spacetime. 
In our model the Heisenberg-picture quantum
state $|\phi\rangle$  is a coherent state for the field 
$\Phi_{\text{{\tiny H}}}$ at the initial time $\eta_0$, 
in which the expectation value $\langle \Phi_{\text{{\tiny H}}} \rangle$    
is spatially homogeneous.  The coherent state is defined with respect
to the adiabatic vacuum constructed via matching of WKB and exact mode
functions for the fluctuation field in some asymptotic region of spacetime.

The O$(N)$ field theory has the property that a systematic expansion in powers
of $1/N$ yields a nonperturbative reorganization of the diagrammatic
hierarchy which preserves
the Ward identities order by order \cite{coleman:1974a}.  
Unlike perturbation theory in the coupling
constant, which is an expansion of the theory around the vacuum configuration,
the $1/N$ expansion entails an enhancement of the mean field by $\sqrt{N}$;
this corresponds to the limit of strong mean field.  
(This is precisely the situation which can arise in chaotic inflation
at the end of the slow-roll period, where the inflaton mean field amplitude
can be as large as $\Mpl/3$ \cite{ramsey:1997b}.)
As discussed in Secs.~\ref{sec-ctpform} and \ref{sec-cjtform}, the
nonequilibrium initial conditions for the mean field as well as the
nonperturbative aspect of the dynamics requires use of both closed-time-path
and two-particle-irreducible methods.  The $1/N$ expansion can be 
achieved as a further approximation from the two-loop, two-particle-irreducible
truncation of the Schwinger-Dyson equations.

Although in this study we assume a pure state, the 2PI formalism 
is also useful for an open system calculation,
in which the mean field is defined as the trace of the product of the
reduced density matrix ${\boldsymbol \rho}$ and the Heisenberg field operator 
$\Phi_{\text{{\tiny H}}}$,
$\text{Tr}({\boldsymbol \rho} \Phi_{\text{{\tiny H}}})$.
When the position-basis matrix element
$\langle \phi_1 | {\boldsymbol \rho}(\eta_0) | \phi_2 \rangle$
can be expressed as a Gaussian functional of $\phi_1$ and $\phi_2$, the
nonlocal source $K$ can encompass the initial conditions coming from
${\boldsymbol \rho}(t_0)$ in a natural way \cite{calzetta:1988b}.  
In order to incorporate a density matrix whose initial condition is beyond 
Gaussian order in the position basis, one should work with a higher-order 
truncation of the master effective action \cite{calzetta:1995b}.  This
we will do for the $\lambda \Phi^4$ theory in Chapter~\ref{chap-entropy}. The 
leading-order $1/N$ approximation is equivalent to assuming a Gaussian initial 
density matrix; therefore, the 2PI effective action is adequate for our 
purposes.

\subsection{Classical action for the O$(N)$ theory}
\label{sec-onclass}
The O$(N)$ field theory consists of $N$ spinless fields $\vec{\phi} =
\{ \phi^i \}$, $i = 1, \ldots, N$, with an action $S^{\text{{\tiny F}}}$ 
which is invariant under the $N$-dimensional real orthogonal group.  
The generally covariant classical action for the O$(N)$ theory (with
quartic self-interaction) plus gravity is given by
\begin{equation}
S[ \phi^i, g^{\mu\nu} ] = S^{\text{{\tiny G}}}[g^{\mu\nu}] +
S^{\text{{\tiny F}}}[ \phi^i, g^{\mu\nu} ],
\end{equation}
where $S^{\text{{\tiny G}}}[g^{\mu\nu}]$ is defined in Eq.~(\ref{eq-lpfcstga})
for the spacetime manifold $M$ with metric $g_{\mu\nu}$,
and the matter field action 
$S^{\text{{\tiny F}}}[\phi^i,g^{\mu\nu}]$ is given by
\begin{equation}
S^{\text{{\tiny F}}}[\phi^i, g_{\mu\nu}] = -\frac{1}{2} \int_M \! d^{\, 4} x \sqmg
\left[ \vecphi \cdot (\square + m^2 + \xi R ) \vecphi + \frac{\lambda}{4 N} 
( \vecphi \cdot \vecphi )^2 \right]. \label{eq-onsm} 
\end{equation}
The O$(N)$ inner product is defined by\footnote{
In our index notation, the latin letters
$i,j,k,l,m,n$  are used to designate O$(N)$ indices 
(with index set $\{1,\ldots, N\}$), while the latin letters
$a,b,c,d,e,f$ are used below to designate CTP indices 
(with index set $\{+,-\}$).}
\begin{equation}
\vecphi \cdot \vecphi = \phi^i \phi^j \delta_{ij}.
\end{equation}
In Eq.~(\ref{eq-onsm}), 
$\lambda$ is a (bare) coupling constant with dimensions 
of $1/\hbar$, and $\xi$ is the (bare) dimensionless coupling to gravity.
The classical Euler-Lagrange
equations are obtained by variation of the action $S$ 
separately with respect to the metric tensor $g_{\mu\nu}$ and the matter 
fields $\phi^i$.
In the classical action (\ref{eq-onsm}), the O$(N)$
symmetry is unbroken.  However, the O$(N)$ symmetry can be spontaneously
broken, for example, by changing $m^2$ to $-m^2$ in $S^{\text{{\tiny F}}}$.
In the symmetry-breaking case with tachyonic mass, the stable, static 
equilibrium configuration is found to be
\begin{equation}
\vecphi \cdot \vecphi = \frac{2 N m^2}{\lambda} \equiv v^2,
\end{equation}
which is a constant.
If we wish to study the action for small oscillations about the 
symmetry-broken equilibrium
configuration, the O$(N)$ invariance of Eq.~(\ref{eq-onsm}) implies that
we can choose the minimum to be in any direction; we choose it to be
in the first, i.e., $(\phi^1)^2  = v^2$.
In terms of the shifted field $\sigma = \phi^1 - v$ and the unshifted 
fields (the ``pions'')
$\pi^i = \phi^i$, $i = 1, \ldots, N-1$, the action becomes
\begin{multline}
S^{\text{{\tiny F}}}[\sigma,\vec{\pi},g^{\mu\nu}] = -\frac{1}{2} \int_M \! 
  d^{\, 4} x \sqmg
\Biggl[ \sigma (\square + m^2 + \xi R)\sigma + \vec{\pi}\cdot(\square + m^2 
 + \xi R)\vec{\pi} \\
+ 2 (m^2 + \xi R) \sigma^2 
+ 2 \sqrt{\frac{\lambda}{2}} M( \sigma^3  + 
 \vec{\pi}\cdot \vec{\pi} \sigma) 
+ \frac{\lambda}{4}\sigma^4 - \frac{\lambda}{2}\vec{\pi}\cdot\vec{\pi}
\sigma^2 + \frac{\lambda}{4}(\vec{\pi}\cdot\vec{\pi})^2 \Biggr].
\end{multline}
One can show that the effective mass of each of the ``pions'' $\vec{\pi}$
(defined as the second derivative of the potential) is zero, due to 
Goldstone's theorem.  The theorem holds for the quantum-corrected effective
potential as well \cite{peskin:1995a}.
In this dissertation we study the unbroken symmetry case, in order to 
focus on the parametric amplification of quantum fluctuations; this
avoids the additional complications which arise in spontaneous symmetry
breaking, e.g., infrared divergences
\cite{hu:1987b,calzetta:1997a,calzetta:1989b}.

\subsection{Quantum generating functional}
In this section we derive the mean-field and gap equations at two-loop
order.  The 2PI generating functional for the O$(N)$ theory in
curved spacetime is defined using the closed-time-path method in terms of 
$c$-number sources $J^{i}_{a}$ and nonlocal $c$-number
sources $K^{ij}_{ab}$ on the CTP manifold ${\mathcal M}$,
\begin{equation}
\begin{split}
Z[J^i_a, K^{ij}_{ab}, g_{\mu\nu}] = 
& \prod_{i,a} \int_{\text{{\tiny ctp}}} D \phi^i_a
\exp \Bigg[ \frac{i}{\hbar} \biggl( 
{\mathcal S}^{\text{{\tiny F}}}\left[ \phi, g^{\mu\nu} \right] 
+ \int_M \! d^{\, 4} x \sqmg c^{ab} \vec{J}_a \cdot \vec{\phi}_b 
 \\ &
+ \frac{1}{2} \int_M \! d^{\, 4} x \sqmg \int_M \! d^{\, 4} x' \sqmgp c^{ab} 
c^{cd} K_{ac}^{ij}(x,\xp)
\phi^k_b(x) \phi^l_d(\xp) \delta_{ik} \delta_{jl} 
\biggr) \Biggr],
\end{split}
\end{equation}
where we have (as discussed above) suppressed time branch indices on the 
metric tensor, and the CTP classical action is defined as in 
Eq.~(\ref{eq-sfamcm}), 
\begin{equation}
{\mathcal S}^{\text{{\tiny F}}}[\phi^i_a,g_{\mu\nu}] = S^{\text{{\tiny F}}}[
\phi^i_{+},g_{\mu\nu}] - S^{\text{{\tiny F}}}[\phi^i_{-},
g_{\mu\nu}].
\end{equation}
The sources $J^i_a$ are coupled to the field by the O$(N)$ 
vector inner product
\begin{equation}
\vec{J}_a \cdot \vec{\phi}_b = J^i_a \phi^j_b \delta_{ij}.
\end{equation}
For simplicity, we shall suppress all indices inside
functional arguments.  In addition, the boundary conditions on the
asymptotic past field configurations for $\phi^i_{\pm}$ in the functional
integral correspond to a choice of ``in'' quantum state 
$|\phi\rangle$ for the system. 
The generating functional for normalized expectation
values is given by
\begin{equation}
W[J, K, g_{\mu\nu}] = 
-i \hbar \text{ln} Z[J, K, g_{\mu\nu}],
\end{equation}
with the additional functional dependence of both $W$ and $Z$ on $g^{\mu\nu}$ 
understood.  In terms of this functional, we can define the ``classical''
field $\phih$ and two-point function $G$ by functional differentiation,
\begin{equation}
\phih^i_a (x) = \frac{c_{ab}}{\sqmg}\frac{\delta W}{\delta J^j_b(x)}
\delta^{ij},
\label{eq-pkdef1} 
\end{equation}
\begin{equation}
\phih^i_a (x) \phih^j_b (\xp) + \hbar G^{ij}_{ab}(x,\xp) =
2\frac{c_{ac}}{\sqmg}\frac{c_{bd}}{\sqmgp} \frac{\delta W}{\delta 
K^{lm}_{cd}(x,\xp)} \delta^{ik} \delta^{jl}. 
\label{eq-pkdef2}
\end{equation}
In the zero-source limit $K^{ij}_{ab} = J^i_a = 0$,
the classical field $\phih_a^i$ satisfies
\begin{equation}
(\phih^i_{+})_{J=K=0} = (\phih^i_{-})_{J=K=0}
= \langle \phi |\Phi_{\text{{\tiny H}}}^i | \phi \rangle
\equiv \phih^i
\end{equation}
as an expectation value of the Heisenberg field operator 
$\Phi_{\text{{\tiny H}}}^i$ in the quantum state $|\Omega\rangle$. 
The fluctuation field is defined [following Eq.~(\ref{eq-deffluc1pi})]
in terms of the Heisenberg field operator $\Phi_{\text{{\tiny H}}}$
and the mean field $\phih$ (times the identity operator),
\begin{equation}
\vphi_{\text{{\tiny H}}}^i = \Phi_{\text{{\tiny H}}}^i - \phih^i.
\end{equation}
In the same limit $J = K = 0$, 
the two-point function $G_{ab}^{ij}$ becomes the CTP propagator for 
the fluctuation field.
The four components of the CTP propagator are (for $J^i_a = K^{ij}_{ab} = 0$)
\begin{align}
\hbar G^{ij}_{++}(x,x')_{|J=K=0} &=
\langle \Omega | T(\vphi_{\text{{\tiny H}}}^i(x)
\vphi_{\text{{\tiny H}}}^j(x')) | \Omega \rangle,
\label{eq-ontpfa}  \\
\hbar G^{ij}_{--}(x,x')_{|J=K=0} &=
\langle \Omega | \tilde{T}(\vphi_{\text{{
\tiny H}}}^i(x) \vphi_{\text{{\tiny H}}}^j(x')) | \Omega \rangle,  \\
\hbar G^{ij}_{+-}(x,x')_{J=K=0} &=
\langle \Omega | \vphi_{\text{{\tiny H}}}^j(x')
\vphi_{\text{{\tiny H}}}^i(x) | \Omega \rangle, \\
\hbar G^{ij}_{-+}(x,x')_{J=K=0} &=
\langle \Omega | \vphi_{\text{{\tiny H}}}^i(x)
\vphi_{\text{{\tiny H}}}^j(x') | \Omega \rangle.
\label{eq-ontpfd}
\end{align}
In the coincidence limit $x' = x$, all four components 
(\ref{eq-ontpfa})--(\ref{eq-ontpfd}) are equivalent to the mean-squared 
fluctuations (variance) about the mean field $\phih^i$,
\begin{equation}
\hbar G^{ii}_{++}(x,x)_{|J=K=0} =
\langle \Omega | (\vphi_{\text{{\tiny H}}}^i)^2 | \Omega \rangle 
= \langle (\vphi_{\text{{\tiny H}}}^i)^2 \rangle.
\label{eq-onfldef}
\end{equation}
Provided that Eqs.~(\ref{eq-pkdef1})--(\ref{eq-pkdef2}) 
can be inverted to give 
$J^i_a$ and $K^{ij}_{ab}$ in terms of $\phih^i_a$ and $G^{ij}_{ab}$, 
we can define the 2PI effective action as a double
Legendre transform of $W$,
\begin{equation}
\begin{split}
\Gamma[&\phih, G, g^{\mu\nu}] = W[J, K, g^{\mu\nu}] -
\int_M \! d^{\, 4} x \sqmg c^{ab} J^i_a \phih^j_b \delta_{ij} 
 \\ & -
\frac{1}{2} \int_M \! d^{\, 4} x \sqmg \int_M \! d^{\, 4} x' \sqmgp c^{ab} 
c^{cd} K^{ij}_{ac}(x,\xp) 
\Bigl[
\hbar G^{kl}_{bd}(x,\xp) + \phih^k_b(x) \phih^l_d (\xp) \Bigr] 
\delta_{ik} \delta_{jl}, \label{eq-ea2pidef}
\end{split}
\end{equation}
where $J^i_a$ and $K^{ij}_{ab}$ above denote the inverses of
Eqs.~(\ref{eq-pkdef1}) and (\ref{eq-pkdef2}).
From this equation, it is clear that the inverses of Eqs.~
(\ref{eq-pkdef1}) and (\ref{eq-pkdef2})
can be obtained by straightforward functional differentiation of 
$\Gamma$,
\begin{equation}
\frac{1}{\sqmg}\frac{\delta \Gamma}{
\delta \phih^i_a (x)} = c^{ab} \delta_{ij} \Bigl( -J^j_b (x) 
- \frac{1}{2} c^{cd} \int_M \! d^{\, 4} x' \sqmgp \bigl[ K^{jk}_{bd}(x,\xp) +
K^{jk}_{db} (\xp,x) \bigr] \phih^l_d \delta_{kl}\Bigr)
\label{eq-onkdef1} 
\end{equation}
\begin{equation}
\frac{1}{\sqmg} \frac{\delta \Gamma}{\delta
G^{ij}_{ab}(x,\xp)}\frac{1}{\sqmgp} =
-\frac{\hbar}{2}c^{ac}c^{bd} K^{jk}_{bd}(\xp,x).
\label{eq-onkdef2}
\end{equation}
Performing the usual field shifting involved in the background field 
approach \cite{jackiw:1974a}, it can be shown that the 2PI effective action
which satisfies Eqs.~(\ref{eq-ea2pidef}), 
(\ref{eq-onkdef1}), and (\ref{eq-onkdef2}) can be written
\begin{multline}
\Gamma[\phih,G,g^{\mu\nu}] = 
{\mathcal S}^{\text{{\tiny F}}}[\phih,g^{\mu\nu}] 
- \frac{i \hbar}{2} \text{ln}\, 
\text{det} \left[ G^{ij}_{ab} \right] 
 \\ 
+ \frac{i \hbar}{2} \int_M \! d^{\, 4} x \sqmg \int_M \! d^{\, 4} x' 
\sqmgp {\mathcal A}_{ij}^{ab}(\xp,x)
G^{ij}_{ab}(x,\xp) + \Gamma_2[\phih,G,g^{\mu\nu}],\label{eq-ea2piai}
\end{multline}
where the kernel ${\mathcal A}$ is the second functional derivative
of the classical action with respect to the field $\phi$, 
\begin{equation}
i {\mathcal A}_{ij}^{ab}(x,\xp) = \frac{1}{\sqmg} 
\Biggl(\frac{\delta^2 {\mathcal S}^{\text{{\tiny F}}}}{
\delta \phi^i_a (x) \phi^j_b (\xp)} [\phih,g^{\mu\nu}]\Biggr) \frac{1}{\sqmgp},
\end{equation}
and $\Gamma_2$ is a functional to be defined below.
Evaluating ${\mathcal A}^{ab}_{ij}$
by differentiation of Eq.~(\ref{eq-onsm}), we find
\begin{multline}
i {\mathcal A}_{ij}^{ab}(x,x') 
= -\biggl\{\delta_{ij}c^{ab}[\square_x + m^2 + 
\xi R(x)] 
 \\ 
+ \frac{\lambda}{2N} c^{abcd} \left[ [\phih_c^k(x) \phih_d^l(x)]\delta_{ij}
\delta_{kl} + 2 \phih^k_c(x)\phih^l_d(x)\delta_{ik}\delta_{jl}\right]\biggr\}
\delta^4(x-x')\frac{1}{\sqmgp},
\end{multline}
where the four-index symbol $c^{abcd}$ is defined by Eq.~(\ref{eq-cabcdef}).
In Eq.~(\ref{eq-ea2piai}), the symbol $\Gamma_2$ is defined as 
$-i\hbar$ times the sum of all two-particle-irreducible
vacuum-to-vacuum graphs with propagator $G$ and vertices given by the 
shifted action ${\mathcal S}^{\text{{\tiny F}}}_{\text{{\tiny int}}}$, 
which takes the form
\begin{equation}
\begin{split}
{\mathcal S}^{\text{{\tiny F}}}_{\text{{\tiny int}}}[\vphi,g^{\mu\nu}&] = 
{\mathcal S}^{\text{{\tiny F}}}[\vphi + \phih, g^{\mu\nu}] - 
{\mathcal S}^{\text{{\tiny F}}}[\phih, g^{\mu\nu}]
-\int_M \! d^{\, 4} x \biggl( \frac{\delta {\mathcal S}^{\text{{\tiny F}}}}{
\delta \phi^i_a} 
[\phih,g^{\mu\nu}]\biggr) \vphi^i_a  \\ &
- \frac{1}{2} \int_M \! d^{\, 4} x \int_M \! d^{\, 4} x' \biggl( 
\frac{\delta^2 {\mathcal S}^{
\text{{\tiny F}}}
}{\delta \phi^i_a(x) \phi^j_b (\xp)}[\phih, g^{\mu\nu}]\biggr)
\vphi^i_a(x)\vphi^j_b(\xp). \label{eq-sintctp}
\end{split}
\end{equation}
The expansion of $\Gamma_2$ in terms of $G$ and $\phih$ is depicted 
graphically in Fig.~\ref{fig-gamma2} for the $\lambda \Phi^4$ theory.  In the
present case of the O$(N)$ field theory, each vertex now also carries an
O$(N)$ label.  From Eqs.~(\ref{eq-sintctp}) and (\ref{eq-onsm}),
${\mathcal S}^{\text{{\tiny F}}}_{\text{{\tiny int}}}$
is easily evaluated, and we find
\begin{align}
& {\mathcal S}^{\text{{\tiny F}}}_{\text{{\tiny int}}}
[\vphi,g^{\mu\nu}] =
S^{\text{{\tiny F}}}_{\text{{\tiny int}}}[\vphip,g^{\mu\nu}] - 
S^{\text{{\tiny F}}}_{\text{{\tiny int}}}[\vphim,g^{\mu\nu}],
\label{eq-sinteval}\\
& S^{\text{{\tiny F}}}_{\text{{\tiny int}}}[\vphi,g^{\mu\nu}] =
-\frac{\lambda}{2 N} \int_M \! d^{\, 4} x \sqmg \left[
\frac{1}{4} (\vec{\vphi}\cdot\vec{\vphi})^2 +
(\vec{ \hat{\phi} }\cdot\vec{\vphi})(\vec{\vphi}\cdot\vec{\vphi})\right].
\label{eq-sintevals}
\end{align}
The two types of vertices in Fig.~\ref{fig-gamma2} are readily
apparent in Eq.~(\ref{eq-sintevals}).  The first term corresponds to the
vertex which terminates four lines; the second term corresponds
to the vertex which terminates three lines and is proportional to
$\phih$.

The action $\Gamma$ including the full diagrammatic series for $\Gamma_2$
gives the full dynamics for $\phih$ and $G$ in the O$(N)$ theory.
It is of course not feasible to obtain an exact, closed-form expression
for $\Gamma_2$ in this model.  Instead, various approximations to the full 
2PI effective action can be obtained by
truncating the diagrammatic expansion for $\Gamma_2$.  Which approximation
is most appropriate depends on the physical problem under consideration.
\begin{enumerate}
\item
Retaining both the ``setting-sun'' and the ``double-bubble'' diagrams of
Fig.~\ref{fig-gamma2} corresponds to the two-loop, two-particle-irreducible
approximation \cite{calzetta:1995b}.  This approximation contains
two-particle scattering through the setting-sun diagram.
\item
A truncation of $\Gamma_2$ retaining only the ``double-bubble'' diagram
of Fig.~\ref{fig-gamma2} yields equations for $\phih$ and $G$ which
correspond to the time-dependent Hartree-Fock approximation to the full
quantum dynamics \cite{cornwall:1974a,cooper:1994a}.  This approximation does 
not preserve Goldstone's theorem in the symmetry-breaking case, but it is
energy conserving (i.e., leads to a covariantly conserved
energy-momentum tensor) \cite{cooper:1994a}.  
\item
Retaining only the $(\text{Tr}G^{ij}_{ab})^2$ piece of the double-bubble
diagram corresponds to taking the leading order $1/N$ approximation,
as will be shown below in Sec.~\ref{sec-onlna}.
\item
A much simpler approximation consists of 
discarding $\Gamma_2$ altogether.  This yields 
the one-loop approximation.  The limitations of the one-loop approximation
for nonequilibrium quantum field dynamics 
have been extensively documented in the literature
\cite{calzetta:1988b,calzetta:1993a,calzetta:1995b,calzetta:1989b}. 
\end{enumerate}

Let us first evaluate the 2PI effective action at two loops
\cite{hu:1987b,calzetta:1995a}.  This is the most general
of the various approximations described above, i.e.,
both two-loop diagrams in
Fig.~\ref{fig-gamma2} are retained.  The 2PI effective action is given
by Eq.~(\ref{eq-ea2piai}), and in this approximation, $\Gamma_2$ is given by
%
%
\begin{equation}
\begin{split}
\Gamma_2 & [\phih,G,g^{\mu\nu}] = 
\frac{\lambda \hbar^2}{4 N} \Biggl[ -\frac{1}{2}c^{abcd} \int_M \! d^{\, 4} 
x \sqmg
[ G^{ij}_{ab}(x,x)G^{kl}_{cd}(x,x) + 2 G^{ik}_{ab}(x,x)G^{jl}_{cd}(x,x)
] \delta_{ij} \delta_{kl}  \\ &
+ \frac{i \lambda}{N} c^{abcd} c^{a'b'c'd'} \int_M \! d^{\, 4} x \sqmg 
\int_M \! d^{\, 4} x' \sqmgp 
\phih^i_a(x) \phih^{i'}_{a'}(x') [ G^{i i'}_{bb'}(x,x') 
G^{jj'}_{cc'}(x,x')G^{kk'}_{dd'}(x,x')  \\  &
\qquad\qquad\qquad\qquad\qquad\qquad\qquad + 2 G^{ij'}_{bd'}(x,x') 
G^{jk'}_{cc'}(x,x')G^{ki'}_{db'}(x,x')]\delta_{jk}\delta_{j'k'} \Biggr].
\end{split}
\end{equation}
Functional differentiation of $\Gamma[\phih,G,g^{\mu\nu}]$ 
with respect to $\phih$ and $G$ leads
to the mean-field and gap equations, respectively.  The two-loop gap
equation is given by
\begin{gather}
(G^{-1})^{ab}_{ij}  (x,x') = {\mathcal A}^{ab}_{ij}(x,x') + 
\frac{i\lambda\hbar}{2N}c^{abcd} \delta^4(x-x') 
\left[ \delta^{ij}\delta_{kl} G^{kl}_{cd}(x,x)
 + 2 G^{ij}_{cd}(x,x) \right]  \notag 
\\[6pt]
\begin{split}
\qquad\qquad\qquad + \frac{\hbar\lambda^2}{2 N^2}  
c^{acde}c^{bc'd'e'} \delta_{kk'} \delta_{ll'} \Bigl[&
\phih^i_c(x)\phih^j_{c'}(x') G^{kl}_{dd'}(x,x')G^{k'l'}_{ee'}(x,x') 
  \\ & +
2\phih^k_c(x)\phih^l_{c'}(x') G^{k'l'}_{dd'}(x,x')G^{ij}_{ee'}(x,x') 
  \\ & + 
2\phih^i_c(x)\phih^k_{c'}(x') G^{lj}_{dd'}(x,x')G^{l'k'}_{ee'}(x,x') 
  \\ & + 
2\phih^k_c(x)\phih^l_{c'}(x') G^{k'j}_{dd'}(x,x')G^{il'}_{ee'}(x,x') 
  \\ & + 
2\phih^k_c(x)\phih^j_{c'}(x') G^{k'l}_{dd'}(x,x')G^{jl'}_{ee'}(x,x') 
\Bigr]. \label{eq-on2lge}
\end{split}
\end{gather}
The mean-field equation is found to be
\begin{multline}
\Bigl( c^{cb} (\square + m^2 + \xi R) + 
 c^{abcd} \frac{\lambda}{2N} \phih^i_a \phih^j_d \delta_{ij} \Bigr) \phih^m_b 
- \frac{i \hbar^2 \lambda^2}{4 N^2} \int_M d^{\,4}x' \sqmgp \Sigma^{cm}(x,x')
 \\ 
 + \frac{\hbar \lambda c^{abcd}}{2 N} \biggl\{
\delta_{ij} \phih^m_d G^{ij}_{ab}(x,x) 
+ \delta_{jl} \delta^m_i \phih^l_d [ G^{ij}_{ab}(x,x) + G^{ji}_{ab}(x,x)]
\biggr\} = 0,
\label{eq-on2lmfe}
\end{multline}
where the nonlocal function $\Sigma^{cm}(x,x')$ is defined by
\begin{equation}
\begin{split}
\Sigma^{em}(x,x') = c^{ebcd} c^{a'b'c'd'} \phih^i_{a'}(x') \Bigl[ & 
  G^{mi'}_{bb'}(x,x') G^{jj'}_{cc'}(x,x') G^{kk'}_{dd'}(x,x') 
 \\ & +
2 G^{mj'}_{bd'}(x,x') G^{jk'}_{cc'}(x,x') G^{ki'}_{b'd}(x,x') 
 \\ & +
G^{i'm}_{b'b}(x',x) G^{jj'}_{c'c}(x',x) G^{kk'}_{d'd}(x',x) 
 \\ & +
2 G^{i'j'}_{b'd}(x',x) G^{kk'}_{c'c}(x',x) G^{jm}_{d'b}(x',x) \Bigr] 
\delta_{jk} \delta_{j'k'} \delta_{ii'}.
\end{split}
\end{equation}
Taking the limit $\phih^i_{+} = \phih^i_{-} = \phih^i$ in 
Eqs.~(\ref{eq-on2lge}) and (\ref{eq-on2lmfe}) 
yields coupled equations for the mean field $\phih^i$
and the CTP propagators $G^{ij}_{ab}$, on the fixed background spacetime 
$g^{\mu\nu}$.  The equations, as well as the semiclassical Einstein  
equation obtained by differentiating $\Gamma[\phih,G,g^{\mu\nu}]$ with 
respect to $g^{\mu\nu}$, are real and causal, and correspond
to expectation values in the limits $\phihp = \phihm = \phih$.
The $O(\lambda^2)$ parts of the above equations 
are nonlocal and dissipative.  The nonlocal aspect
makes numerical solution difficult;
the dissipative aspect will be addressed below in Chapter~\ref{chap-fermion}.
One can regain the perturbative (amplitude) expansion
for the CTP effective action at two loops 
by expanding the one-loop CTP propagators in Eq.~(\ref{eq-on2lmfe})
in a functional power series in the mean field $\phih$.

\subsection{Large-$N$ approximation}
\label{sec-onlna}
We now carry out the $1/N$ expansion to obtain local, covariant,
nonperturbative  mean-field and gap equations for the O$(N)$ field theory
in a general curved spacetime.  The $1/N$ expansion is a controlled
nonperturbative approximation scheme which can be used to study 
nonequilibrium quantum 
field dynamics in the regime of strong quasiclassical field amplitude 
\cite{boyanovsky:1996b,boyanovsky:1995a,cooper:1994a,cooper:1997b}.   
In the large-$N$ approach, the large-amplitude quasiclassical field is 
modeled by $N$ fields, and the quantum-field-theoretic generating functional 
is expanded in powers of $1/N$.  In this sense the method is a controlled 
expansion in a small parameter.  Unlike perturbation theory in the coupling 
constant $\lambda$, which corresponds to an expansion of the theory around the
vacuum, the large-$N$ approximation corresponds to an expansion of the field 
theory about a strong quasiclassical field configuration \cite{cooper:1994a}.
At a particular order in the $1/N$ expansion, the approximation yields
truncated Schwinger-Dyson equations which are O$(N)$-
and renormalization-group--invariant, 
unitary, and (in Minkowski space) energy conserving 
\cite{cooper:1994a}.  In contrast, the Hartree-Fock approximation cannot
be systematically improved beyond leading order, and in the case of
spontaneous symmetry breaking, it violates Goldstone's theorem and 
incorrectly predicts the order of the phase transition \cite{cooper:1997b}.

Let us implement the leading order large-$N$ approximation in
the two-loop, 2PI mean-field and gap equations (\ref{eq-on2lmfe}) 
and (\ref{eq-on2lge}) derived above.  This amounts to computing the
leading-order part of $\Gamma$ in the limit of large $N$, which is 
$O(N)$.  In the unbroken symmetry case, this is easily carried out
by scaling $\phih$ by $\sqrt{N}$ and leaving $G$ unscaled 
\cite{cornwall:1974a},
\begin{align}
\phih^i_a(x) & \rightarrow  \sqrt{N} \phih_a(x), \\
G^{ij}_{ab}(x,x') & \rightarrow  G_{ab}(x,x')\delta^{ij},  \\
{\mathcal A}_{ij}^{ab}(x,x') & \rightarrow  {\mathcal A}^{ab}(x,x')
\delta_{ij}, \\
\vphi^i_a(x) & \rightarrow  \vphi_a(x), 
\label{eq-onscl} 
\end{align}
for all $i,j$.
The Heisenberg field operator $\vphi_{\text{{\tiny H}}}^i$ 
scales like $\vphi^i_a$ in Eq.~(\ref{eq-onscl}).
In the above equations, the connection between the large-$N$ limit and
the strong mean-field limit is clear.

The truncation of the $1/N$ expansion should be carried out in the 2PI
effective action, where it can be shown that the three-loop and higher-order
diagrams do not contribute (at leading order in the $1/N$ expansion).  
Let us now also allow the metric $g_{\mu\nu}$ to be specified 
independently on the $+$ and $-$ time branches.  We find, for the classical 
action,
\begin{equation}
{\mathcal S}^{\text{{\tiny F}}}[\phi,g^{\mu\nu}] =
S^{\text{{\tiny F}}}[\phi_{+},g^{\mu\nu}_{+}] - 
S^{\text{{\tiny F}}}[\phi_{-},g^{\mu\nu}_{-}], 
\end{equation}
where
\begin{equation}
S^{\text{{\tiny F}}}[\phi,g^{\mu\nu}] =
-\frac{N}{2} \int_M \! d^{\, 4} x \sqmg \left[ \phi ( \square + m^2 + 
\xi R ) \phi +
\frac{\lambda}{2} \phi^4 \right].
\end{equation} 
The inverse of the one-loop propagator is\footnote{
Note that the index $b$ is not to be summed in the right-hand side of
Eq.~(\ref{eq-onloolp}), and the $c$ subscript on $\square$ and $R$ is a CTP
index.}
\begin{equation}
i{\mathcal A}^{ab}(x,x') = -\left[ c^{abc}(\square^x_c + m^2 + \xi R_c(x))
+ \frac{\lambda}{2} c^{abcd} \phih_c(x)\phih_d(x) \right] \delta^4(x-x')
\frac{1}{\sqmgp_b}. \label{eq-onloolp}
\end{equation}
Finally, for the CTP-2PI effective action
at leading order in the $1/N$ expansion, we obtain
\begin{equation}
\begin{split}
\Gamma[\phih,G,g^{\mu\nu}] = & {\mathcal S}^{\text{{\tiny F}}}
[\phih,g^{\mu\nu}] - \frac{i \hbar N}{2} \text{ln}\,\text{det}
\left[ G_{ab} \right] 
 \\ & + \frac{i \hbar N}{2}
\int_M \! d^{\, 4} x \sqmg_a \int_M \! d^{\, 4} x' \sqmgp_b 
{\mathcal A}^{ab}(x',x) G_{ab}(x,x')   \\ & -
\frac{\lambda \hbar^2 N}{8} c^{abcde} \int_M \! d^{\, 4} x 
\sqmg_e G_{ab}(x,x) G_{cd}(x,x)
+ O(1).
\label{eq-lolnea}
\end{split}
\end{equation}
Applying Eq.~(\ref{eq-lpfseeb}) and taking the limits $\phihp = \phihm =\phih$
and $g^{\mu\nu}_{+} = g^{\mu\nu}_{-} = g^{\mu\nu}$, we obtain the
gap equation for $G_{ab}$ at leading order in the $1/N$ expansion,
\begin{equation}
(G^{-1})^{ab}(x,x') = \hat{{\mathcal A}}^{ab}(x,x') 
+ \frac{i \hbar \lambda}{2}
c^{abcd} G_{cd}(x,x) \delta^4(x-x')\frac{1}{\sqmgp} + O\left(
\frac{1}{N}\right),
\label{eq-ongelm} 
\end{equation}
where
\begin{equation}
i\hat{{\mathcal A}}^{ab}(x,x') \equiv -\left[ c^{ab} [ \square + m^2 + 
\xi R(x) ]
+ \frac{\lambda}{2}c^{abcd} \phih_c(x) \phih_d(x) \right] \delta^4(x-x')
\frac{1}{\sqmgp}.
\end{equation}
Similarly, we obtain the mean-field equation for $\phih$ at leading order
in the $1/N$ expansion,
\begin{equation}
\left(\square + m^2 + \xi R + \frac{\lambda}{2} \phih^2 + \frac{\hbar 
\lambda}{2}
G(x,x) \right) \phih(x) + O\left(\frac{1}{N}\right) = 0,
\label{eq-onmflm}
\end{equation}
where $G(x,x) \equiv G_{ab}(x,x)$; this definition is the same for all 
$a,b \in \{+,-\}$, which can be seen from Eq.~(\ref{eq-ongelm}).
To get a consistent set of dynamical equations, we need only consider the 
$++$ component of Eq.~(\ref{eq-ongelm}).  It should also be noted that 
$G_{ab}(x,x)$ is 
formally divergent.  Regularization of the coincidence limit of the
two-point function and the energy-momentum tensor is necessary, and will be
carried out in Chapter~\ref{chap-preheat} below.
Multiplying Eq.~(\ref{eq-ongelm}) by $G$ and
integrating over spacetime, we obtain a differential equation for
the $++$ CTP Green function,
\begin{equation}
\left( \square_x + m^2 + \xi R(x) + \frac{\lambda}{2}\phih^2(x) + 
\frac{\hbar \lambda}{2} G(x,x) 
\right) G(x,x') + O\left(\frac{1}{N}\right) = 
-\frac{i \delta^4 (x-x')}{\sqmgp},
\label{eq-onige}
\end{equation}
where appropriate boundary conditions must be specified for $G_{++}$ to
obtain the time-ordered propagator as the solution to Eq.~(\ref{eq-onige}). 

Equations (\ref{eq-onmflm}) and (\ref{eq-onige}) are the covariant
evolution equations for the mean field $\phih$ and the two-point function
$G_{++}$ at leading order in the $1/N$ expansion.  
Following Eq.~(\ref{eq-onfldef}), we denote the coincidence limit 
$\hbar G(x,x)$ by $\langle \vphi_{\text{{\tiny H}}}^2 \rangle$.
With the inclusion of the semiclassical gravity field equation
(\ref{eq-lpfcstsee2}),
these equations form a consistent, closed set of dynamical equations
for the mean field $\phih$, the time-ordered fluctuation-field
Green function $G_{++}$, and the metric $g_{\mu\nu}$.

The one-loop equations for $\phih$ and $G$ can be obtained from 
the leading-order equations by dropping the $\hbar G(x,x)$ term from 
Eq.~(\ref{eq-onige}),
while leaving the mean-field equation (\ref{eq-onmflm}) unchanged.  In
the Hartree approximation, the gap equation is 
unchanged from Eq.~(\ref{eq-ongelm}), and the mean-field equation is obtained
from Eq.~(\ref{eq-onmflm}) by changing $\hbar \rightarrow 3\hbar$
\cite{boyanovsky:1995a}.  The principal limitation of the leading-order
large-$N$ approximation is that it neglects the setting-sun diagram which
is the lowest-order contribution to collisional thermalization of the
system \cite{calzetta:1988b}.  The system, therefore, does not thermalize
at leading order in the $1/N$ expansion, and the approximation breaks down on a
time scale $\tau_2$ which is on the order of the mean free time for binary
scattering \cite{cooper:1997b}.

Let us now use Eq.~(\ref{eq-lpfseea}) 
to derive the bare semiclassical Einstein
equation for the O$(N)$ theory at leading order in $1/N$.  This equation
contains two parts $\delta {\mathcal S}^{\text{{\tiny G}}}/\delta 
g^{\mu\nu}_{+}$ and $\delta \Gamma /\delta g^{\mu\nu}_{+}$.  The latter
part is related to the bare energy-momentum tensor $ \langle T_{\mu\nu}
\rangle$ by Eq.~(\ref{eq-lpfcstemt}).  At leading order in $1/N$, 
$\langle T_{\mu\nu}\rangle$ is given by a sum of ``classical'' and 
``quantum'' parts (distinguished by the latter having an overall factor
of $\hbar$),
\begin{equation}
\langle T_{\mu\nu} \rangle = T^{\text{{\tiny C}}}_{\mu\nu} +
T^{\text{{\tiny Q}}}_{\mu\nu} - \frac{\lambda N}{8} \langle \vphi_{\text{{\tiny
H}}}^2 \rangle^2 g_{\mu\nu},
\end{equation}
where we define the classical part of $\langle T_{\mu\nu}
\rangle$ by
\begin{multline}
T^{\text{{\tiny C}}}_{\mu\nu} = N \Biggl[ (1-2\xi) \phih_{;\mu}
\phih_{;\nu} + \left(2\xi - \frac{1}{2}\right) g_{\mu\nu}
g^{\rho\sigma} \phih_{;\rho} \phih_{;\sigma} -
2 \xi \phih_{;\mu\nu} \phih  \\  +
2 \xi g_{\mu\nu} \phih \square \phih - \xi G_{\mu\nu} \phih^2
+ \frac{1}{2} g_{\mu\nu} \left( m^2 + \frac{\lambda}{4}\phih^2\right)
\phih^2 \Biggr]
\label{eq-onemtcl} 
\end{multline}
and the quantum part of $\langle T_{\mu\nu} \rangle$ by
\begin{equation}
\begin{split}
T^{\text{{\tiny Q}}}_{\mu\nu} = N \hbar \lim_{x' \rightarrow x} \Biggl\{
\Biggl[ &
(1 - 2\xi) \nabla_{\mu} \nabla_{\nu}' + \left( 2 \xi - \frac{1}{2} \right)
g_{\mu\nu} g^{\rho\sigma} \nabla_{\rho} \nabla_{\sigma}' 
-2\xi \nabla_{\mu} \nabla_{\nu} + 2 \xi g_{\mu\nu}g^{\rho\sigma}
\nabla_{\rho} \nabla_{\sigma}   
\\ & - \xi G_{\mu\nu} + 
\frac{1}{2} g_{\mu\nu} \left( m^2 + \frac{\lambda}{2} \phih^2 +
\frac{\hbar\lambda}{4} G_{++}(x,x') \right) \Biggr] G_{++}(x,x') 
\Biggr\} + O(1).
\label{eq-onemtqm}
\end{split}
\end{equation}
The above expression for $T^{\text{{\tiny Q}}}_{\mu\nu}$ is divergent
in four spacetime dimensions, and needs to be regularized or renormalized.
The energy-momentum
tensor in the one-loop approximation is obtained by neglecting the
$O(\hbar^2)$ terms in Eq.~(\ref{eq-onemtqm}).  It can be shown
using (\ref{eq-onige}) that the energy-momentum tensor 
at leading order in the $1/N$ expansion is covariantly conserved, up to
terms of order $O(1)$ (next-to-leading-order).  The bare
semiclassical Einstein equation is then given (in terms of $\langle
T_{\mu\nu}\rangle$ shown above) by Eq.~(\ref{eq-lpfcstsee2}).

At this point we formally set $N = 1$ since we are not including 
next-to-leading-order diagrams in the $1/N$ expansion.  This can be 
envisioned as a simple rescaling of the Planck mass by $\sqrt{N}$, since
the matter field effective action $\Gamma$ is entirely of order $O(N)$.
We now turn to the issue of renormalization.

\subsection{Renormalization}
\label{sec-onrenorm}
To renormalize the leading-order, large-$N$, CTP effective action in a general 
curved spacetime, one can use dimensional regularization \cite{thooft:1972a}, 
which requires formulating effective action in $n$ spacetime 
dimensions.  This necessitates the introduction of a length parameter 
$\mu^{-1}$ into the classical action, $\lambda \rightarrow \lambda \mu^{4-n}$,
in order for the classical action to have consistent units.  
As before, we maintain the restriction 
$g^{\mu\nu}_{+} = g^{\mu\nu}_{-} = g^{\mu\nu}$, and we suppress indices inside
functional arguments.

Making a substitution of the gap equation (\ref{eq-onige}) into the 
leading-order-large-$N$, 2PI effective action (\ref{eq-lolnea}), we obtain
\begin{multline}
\Gamma[\phih,g^{\mu\nu}] = {\mathcal S}^{\text{{\tiny F}}}[\phih,
g^{\mu\nu}] + \frac{i\hbar N}{2} \text{Tr}\,\text{ln} \left[
\left( G^{-1} \right)^{ab} \right]  \\  +
\frac{\hbar^2 N \lambda \mu^{4-n}}{8}
\int_M d^{\,n}x \sqrt{-g} c^{abcd} \left[ G_{ab}(x,x) G_{cd}(x,x) \right],
\label{eq-eaphi}
\end{multline}
in terms of the CTP propagator $G_{ab}(x,x')$.  For convenience we shall 
rewrite the gap equation as
\begin{equation}
\left( G^{-1} \right)^{ab} = i \left( \square_x c^{ab} + \chi^{ab}(x) \right)
\delta(x-x')\frac{1}{\sqrt{-g'}},
\end{equation}
in terms of a four-component ``effective mass''
\begin{equation}
\chi^{ab}(x) = \left( m^2 + \xi R \right) c^{ab} +
\frac{\lambda \mu^{4-n}}{2} c^{abcd} [ \phih_c \phih_d + \hbar
G_{cd}(x,x) ].
\end{equation}
The divergences in the effective action can now be made explicit with the
use of a CTP generalization of the heat kernel
\cite{dewitt:1964a,thooft:1972a,dowker:1976a}. Let us define a function
$K^a_{\;\;b}(x,y,s)$ which satisfies the equation
\begin{equation}
\frac{\partial K^a_{\;\;b}(x,y;s)}{\partial s} + \int_M d^{\,n}z
\sqrt{-g_z} c_{cd} (G^{-1})^{ac}(x,z) K^d_{\;\;b}(z,y;s) = 0,
\label{eq-hk}
\end{equation}
(where $s$ is a real parameter) with boundary conditions
\begin{equation}
K^a_{\;\;b}(x,y;0) = \delta^a_{\;\;b} \delta(x-y)\frac{1}{\sqrt{-g_y}}
\end{equation}
at $s = 0$ \cite{toms:1982a}.  From Eqs.~(\ref{eq-hk}) and 
(\ref{eq-eaphi}) it follows that
$K^{+}_{\;\;-} = K^{-}_{\;\;+} = 0$ for all $x,$ $y,$ and $s,$ and that
$K^{+}_{\;\;+}$ ($K^{-}_{\;\;-}$) is a functional of $\phih_{+}$ ($\phih_{-}$)
only.  The CTP effective action can then be expressed as 
\begin{equation}
\Gamma[\phih,g^{\mu\nu}] = \Gamma^{+}_{\text{{\tiny SD}}}[\phih_{+},g^{\mu\nu}]
- \Gamma^{-}_{\text{{\tiny SD}}}[\phih_{-},g^{\mu\nu}], 
\end{equation}
in terms of a functional $\Gamma_{\text{{\tiny SD}}}$ on $M$ defined by
\begin{multline}
\Gamma^{+}_{\text{{\tiny SD}}}[\phih_{+},g^{\mu\nu}] = S^{\text{{\tiny F}}}[
\phih_{+},g^{\mu\nu}] -  \frac{i \hbar N}{2} \int_M d^{\,n}x \sqrt{-g}
\int_0^{\infty} \frac{ds}{s} K^{+}_{\;\;+}(x,x;s)  \\ 
+ \frac{\hbar^2 N \lambda \mu^{4-n}}{8} \int_M d^{\,n}x \sqrt{-g} 
\left[ \int_0^{\infty} ds 
K^{+}_{\;\;+}(x,x;s)\right]^2,
\end{multline}
and similarly for $\Gamma^{-}_{\text{{\tiny SD}}}$.  It follows from
Eq.~(\ref{eq-hk}) that $K^{+}_{\;\;+}(x,x;s)[\phih_{+}]$ is exactly the same
functional of $\phih_{+}$ as $K^{-}_{\;\;-}(x,x;s)[\phih_{-}]$ is of 
$\phih_{-}$; we denote it by $K(x,x;s)[\phih]$, where $\phih$ is a function
on $M$.

The divergences in the effective action $\Gamma_{\text{{\tiny SD}}}$
arise in the small-$s$ part of the integrations, so that only the first term
on the right-hand side of the equation
\begin{equation}
\int_0^{\infty} \frac{ds}{s} K(x,x;s)= \int_0^{s_0} \frac{ds}{s} K(x,x;s) + 
\int_{s_0}^{\infty} \frac{ds}{s} K(x,x;s)
\end{equation}
is divergent.  Using the
$s \rightarrow 0^{+}$ asymptotic expansion for $K(x,x;s)$ \cite{toms:1982a},
one has (for a scalar field, such as the unbroken symmetry, large-$N$ limit
of the O$(N)$ model)
\begin{equation}
K(x,x;s) \sim (4\pi s)^{-\frac{n}{2}} \sum_{m=0}^{\infty} s^m a_m(x),
\end{equation}
where the $a_n(x)$ are the well-known ``HaMiDeW coefficients'' 
made up of scalar invariants of the spacetime curvature
\cite{dewitt:1964a,dewitt:1975a,fulling:1989a}.
The divergences then show up as poles in $1/(n-4)$ after the $s$ integrations
are performed.  They have been evaluated for the $\lambda \Phi^4$ theory in 
a general spacetime by many authors (see, e.g., 
\cite{toms:1982a,hu:1984a,paz:1988a}) and in the large-$N$ limit of
the O$(N)$ model \cite{mazzitelli:1989b}.
At leading order in the $1/N$ expansion, the renormalization of 
$\lambda$, $\xi$, $m$, $G$, $\Lambda$, $b$, and $c$ is required,
but no field amplitude renormalization is required 
\cite{mazzitelli:1989b,cooper:1994a}.  In Chapter~\ref{chap-preheat} below
we carry out an explicit renormalization of the large-$N$ dynamics in
spatially flat FRW spacetime.

\section{Summary}
In this chapter, we have presented a method for deriving causal, coupled 
equations of motion for the mean field and two-point function for an
interacting quantum field in an arbitrary, classical background spacetime.
We derived the equations at two loops at leading order in the $1/N$ 
expansion, and in the latter case showed that renormalization counterterms
for the ``in-out'' formulation of the theory are all that is necessary
to renormalize the effective action.  In Chapter~\ref{chap-preheat}, we
apply these equations to the study of inflaton dynamics during the 
reheating period of inflationary cosmology.  The mean field and gap
equations derived here are also useful, by changing $m^2 \rightarrow -m^2$,
for describing the dynamics of symmetry-breaking phase transition
\cite{boyanovsky:1996b,cooper:1997b}.

\chapter{Parametric particle creation and curved spacetime effects}
\label{chap-preheat}

\section{Introduction}
\label{sec-intro}
In this Chapter we study the nonperturbative, out-of-equilibrium 
dynamics of a minimally coupled scalar O$(N)$ field theory, 
with quartic self-interaction, in a spatially flat FRW spacetime whose
dynamics is given self-consistently by the semiclassical Einstein equation.
The purpose of this study is to understand the preheating period in 
inflationary cosmology, with particular emphasis on the effect of spacetime 
dynamics on the phenomenon of particle production via parametric 
amplification of quantum fluctuations.  Of primary interest is obtaining the 
dynamics of the inflaton (including back reaction from created particles) 
using rigorous methods of nonequilibrium field theory in curved spacetime 
\cite{calzetta:1988b,calzetta:1988a}.  We have chosen to focus in this 
Chapter on parametric amplification of quantum fluctuations 
because this phenomenon can be the dominant effect in the 
preheating stage of unbroken symmetry inflationary scenarios, among which
the chaotic inflation scenarios most directly necessitate (through
initial conditions) considerations of Planck-scale physics.
``New'' inflationary scenarios which involve a spontaneously broken symmetry
often contain additional subtleties (e.g., infrared divergences, spinodal 
instabilities), and are the subject of ongoing investigation 
\cite{calzetta:1997a}.
The results of our work are, therefore, particularly relevant to chaotic
inflation scenarios \cite{linde:1983a}.  
The additional interactions which should be included to treat the 
broken-symmetry case are discussed in \cite{ramsey:1997a}.

In Chapter~\ref{chap-oncst} we derived the evolution equations for the 
mean field $\langle \Phi_{\text{{\tiny H}}} \rangle$ (subscript H denotes
the Heisenberg field operator) and mean-squared fluctuations
(variance) $\langle \Phi_{\text{{\tiny H}}}^2 
\rangle - \langle \Phi_{\text{{\tiny H}}} \rangle^2$  using the
closed-time-path (CTP), two-particle-irreducible (2PI) effective
action \cite{cornwall:1974a} in a fully covariant form.
Here we use these results for the case of spatially flat FRW spacetime.
The quantum state for the field theory (in the case of FRW spacetime) 
consists of a coherent state for the spatially homogeneous field mode, and the
adiabatic vacuum state for the spatially inhomogeneous modes.
At conformal past infinity, the spacetime is assumed to be asymptotically
de~Sitter, and the mean field is chosen to be asymptotically constant.

In this Chapter we study the O$(N)$ field theory using the
$1/N$ expansion, which yields nonperturbative dynamics in the regime of
strong mean field.  This is particularly important for chaotic inflation
scenarios \cite{linde:1985a},
in which the inflaton mean-field amplitude can be as large as $\Mpl/3$ at the
end of the slow-roll period \cite{linde:1990a,linde:1994a}.  Treatments of
the reheating problem which rely on time-dependent perturbation theory do not
apply to such cases where the inflaton undergoes large-amplitude oscillations,
in contrast with nonperturbative methods such as large $N$.

We now summarize the principal distinctions between our work and previous
treatments of  preheating in inflationary cosmology, which were summarized
in Section~\ref{sec-backissu}.  Our work
improves on the group 1A methodology by including parametric resonance
effects.  As it is based on first-order, time-dependent perturbation theory,
the group 1A approach cannot correctly describe the inflaton dynamics
with large initial mean-field amplitude.
In addition, our work improves on both the group 1A and group 1B
studies by including the effect of back reaction from quantum particle
creation on both the mean field and the inhomogeneous modes.
We are treating the inflaton dynamics from first principles, without
assuming a phenomenological equation (with a damping term
$\Gamma \dot{\phi}$ put in by hand) for the
mean field.  In our approach the damping of the mean field is due
to back reaction from quantum particle production in the self-consistent
equations for the mean field and its variance.  In contrast, the analytic 
results of
the group 1A and 1B work are based on the assumption of either large-amplitude
mean-field oscillations ($\lambda \phih^2/2 \gg m^2$) or harmonic oscillations
($m^2 \gg \phih^2/2$), and, therefore, cannot describe the interesting case
of inflaton dynamics in which neither term dominates the tree-level
effective mass, i.e., $m^2 \sim \lambda \phih^2$. Furthermore, our work
improves on group 1A, group 1B, and group 2A in that the closed-time-path
effective action is computed in {\em curved spacetime\/} without assuming that 
$H^{-1} \gg \tau_0$ (where $\tau_0$ is the period of mean-field oscillations).
In our work, the dynamics of the two-point function
(which reflects quantum particle production) is formulated in curved 
spacetime assuming only that semiclassical gravity is valid,  
i.e., $\Mpl \gg H$.

Most significantly, our work improves on all the previous treatments 
in that it includes curved spacetime effects
systematically using the coupled, {\em self-consistent\/}
semiclassical Einstein equation and matter field equations.
Among the group 2B studies of preheating dynamics, inflaton
dynamics has been studied primarily in {\em fixed\/} background spacetimes:
Minkowski space \cite{boyanovsky:1996b,boyanovsky:1995a}, de~Sitter space
\cite{boyanovsky:1997e,boyanovsky:1996c,boyanovsky:1997c}, and in 
radiation-dominated, spatially flat FRW spacetime \cite{boyanovsky:1996c}.
In the present work, the spacetime is {\em dynamical,\/} with the renormalized
trace of the semiclassical Einstein equation governing the dynamics of the 
scale factor $a$.  This permits quantitative study of the transition of the 
spacetime from the (slow-roll) period of vacuum-dominated expansion to the 
radiation-dominated (``standard'') FRW cosmology.  In particular,
our method yields the spacetime dynamics naturally, 
without making reference to
an ``effective Hubble constant'' (which has been used
in calculations on a fixed background spacetime \cite{boyanovsky:1997c}).

With additional couplings (see \cite{ramsey:1997a}), our
method may also be used to study preheating in ``new'' inflationary scenarios
\cite{linde:1982a}.  In new inflation, the vacuum-dominated expansion of
the Universe is typically driven by the classical potential energy of the
mean field as it rolls towards the symmetry-broken ground
state.  In one of the group 2B studies (Boyanovsky {\em et al.\/} 
\cite{boyanovsky:1997c}),
a quench-induced phase transition is studied with small initial mean-field
amplitude, in which the classical terms in the mean-field equation are 
{\em dominated\/} by spinodal fluctuations.  As a result, the mean field in
their model does not oscillate about the symmetry-broken ground 
state as is generally expected in a new inflation preheating scenario (this
point was emphasized in \cite{kofman:1996b}).
The initial conditions studied in \cite{boyanovsky:1997c} are more
appropriate to a study of defect formation in a 
quench-induced phase transition
than preheating dynamics in new inflation.

In addition, the renormalization
scheme employed in \cite{boyanovsky:1997c}
is not generally covariant (as can be seen by comparing it with 
\cite{bunch:1978a}), and covariant conservation of the renormalized 
energy-momentum tensor is put in by hand.  The regularization scheme
employed here is the well-tested adiabatic regularization 
\cite{hu:1974a,parker:1974a,fulling:1974a,fulling:1974b},
which is simple to use and physically intuitive. It also, in the one-loop
case, ensures both covariant conservation of the regularized energy-momentum
tensor and agreement with manifestly covariant regularization procedures such
as point splitting \cite{bunch:1978a}.  

A related difference between our approach and that of some of the 
group 2B studies is the choice of vacuum state.  
The choice of initial conditions for the quantum mode functions in most
studies of reheating in FRW spacetime
\cite{khlebnikov:1997a,boyanovsky:1994a,khlebnikov:1996b}
has been to instantaneously diagonalize the matter-field Hamiltonian at the
initial-data hypersurface.  However, as has been pointed out long ago
\cite{fulling:1989a}, this method does not correspond
to the vacuum state which registers the least particle flux on 
a comoving detector.  In our work we use the de~Sitter-invariant
(or Bunch-Davies) vacuum, obtained via the adiabatic construction; the
adiabatic vacuum most closely aligns with an intuitive notion of vacuum state 
in a cosmological spacetime \cite{birrell:1982a}.

In many of the group 2B
studies \cite{boyanovsky:1995a,boyanovsky:1996c,boyanovsky:1997c} the
large-$N$ equations for the mean field and variance are derived using a
factorization method which does not readily 
generalize to next-to-leading order in the $1/N$ expansion.  
In all of the group 2B studies of nonperturbative inflaton 
dynamics of which we are aware, the equations for the mean field and variance
are not derived using methods which encompass higher-order correlations
in the Schwinger-Dyson hierarchy. As found in earlier studies of
phase transitions \cite{hu:1987b,calzetta:1989b}, this is necessary in
order to derive the correct infrared behavior of a quantum field
in a study of critical phenomena. In the present work, we use the result of
Chapter~\ref{chap-oncst} in which the CTP
two-particle-irreducible (2PI) formalism is derived.  It has a direct
generalization in terms of the $n$-particle-irreducible ($n$PI) ``master''
effective action \cite{calzetta:1995b}.  The master effective action
can be used to derive a self-consistent truncation of the Schwinger-Dyson 
equations to arbitrary order in the correlation hierarchy 
\cite{calzetta:1995b}.  The techniques
employed here are, therefore, most readily generalized to the study of phase 
transitions in curved spacetime, where higher-order correlation functions 
can become important \cite{calzetta:1997a}.

In summary, our approach to the inflaton dynamics problem
has the following advantages:  
it is nonperturbative and fully covariant; it is
based on rigorous methods of nonequilibrium field theory in curved spacetime;
we use the correct adiabatic vacuum construction; and 
we employ an approximation scheme which can be systematically 
generalized beyond leading order, within a fully covariant and
self-consistent theoretical framework.

Our results are obtained by solving the mean-field
and spacetime dynamics self-consistently using the coupled matter-field
and semiclassical Einstein equations in a FRW spacetime,
including the effect of back reaction
of the variance on the mean field.  Within the leading-order, large $N$
approximation used here, 
we find that (using the conventional value for the self-coupling,
$\lambda = 10^{-14}$) for sufficiently large initial mean-field amplitude,
parametric amplification of quantum fluctuations is not an efficient mechanism
of energy transfer from the mean field to the inhomogeneous field modes.
In this case the energy
density of the inhomogeneous modes remains negligible in comparison to the 
mean-field energy density for all times.  This can be understood from the time
scales for the competing processes of parametric resonance and cosmic
expansion.  When the time scale for parametric amplification of
quantum fluctuations $\tau_1$ is of the same order as (or greater than)
the time scale for cosmic expansion $H^{-1}$, cosmic expansion redshifts 
the energy density of the inhomogeneous modes faster than it increases
due to parametric resonance.  We find that this occurs when 
$\phih \gtrsim \Mpl/300$, for the model and coupling studied here.  

In many chaotic inflation scenarios, the mean-field amplitude at the
end of the slow-roll period can be as large as
$\Mpl/3$ \cite{linde:1990a,linde:1994a}.
In light of our result, in such models, it is
clearly essential to include the effect of spacetime dynamics in order to 
study mean-field dynamics and resonant 
particle production during reheating.  In addition, our result indicates
that for the case of a minimally coupled $\lambda\Phi^4$ inflaton with
unbroken symmetry, parametric amplification of its own quantum fluctuations
is not a viable mechanism for reheating, unless the coupling is significantly
strengthened (see \cite{calzetta:1995a,matacz:1996a}).
Parametric amplification of quantum fluctuations may still
play a dominant role in the reheating of chaotic inflaton models with an
inflaton coupled to other fields, e.g., a $\phi^2 \chi^2$ model
\cite{kofman:1994a,shtanov:1995a,kofman:1997a,khlebnikov:1997a,allahverdi:1997a}.  

For more moderate cosmic expansion, where $H^{-1} \gtrsim 100 \; \tau_1$,
parametric
amplification of quantum fluctuations is an efficient mechanism of energy
transfer to the inhomogeneous modes, and the asymptotic effective equation
of state is found to agree with the prediction of a two-fluid model consisting
of the elliptically oscillating mean field and relativistic
energy density contained in the inhomogeneous mode occupations. In a
collisionless approximation, the mean field eventually decouples from the
mean-squared fluctuations (variance) and at late times undergoes asymptotic
oscillations which are damped solely by cosmic expansion \cite{linde:1994a}.
For the case when cosmic expansion is subdominant, $H^{-1} \gg \tau_1$, the 
mean-field dynamics and the growth of quantum fluctuations are in agreement
with results of studies of preheating in Minkowski space 
\cite{boyanovsky:1996b}.  In particular, the total adiabatically regularized
energy density is found to be constant (to within the limits of 
numerical precision) for the case of $H^{-1} \rightarrow \infty$, in agreement
with the predictions of field theory in Minkowski space.

While there has been a large volume of work on understanding the preheating 
period and parametric particle creation, the thermalization of inflationary 
models has not yet been understood from first principles 
\cite{brandenberger:1997a}.  Because of the absence of a separation
of microscopic and macroscopic time scales at the end of the preheating
stage, the Boltzmann equation is inadequate for studying collisional
thermalization in most inflationary models \cite{boyanovsky:1996b}.
In particular, in the leading-order, large-$N$ approximation employed here
(and in the one-loop approximation which it contains), 
this model also does not thermalize.  However, it still may approach
a radiation-dominated effective equation of state
(as found in \cite{boyanovsky:1996b} for the case of Minkowski space).
Clearly, a first-principles analysis of thermalization is necessary.
Continuing the early work of kinetic field theory \cite{calzetta:1988b},
and the recent work on correlation hierarchy \cite{calzetta:1995b},
we know that such a first-principles analysis should involve at a minimum
the full two-loop, two-particle-irreducible effective action
(or alternatively, next-to-leading order in the large-$N$ approximation).
Since it  represents a rigorous truncation of
the full Schwinger-Dyson hierarchy, in this sense it is a natural
generalization of the collisionless approximations used previously to study 
reheating. However, the equations derived from it for the mean-field and 
two-point function are nonlocal and hence difficult to solve
even numerically \cite{cooper:1994a}.  This Chapter is, therefore, concerned 
only with preheating via parametric resonance particle creation.  A 
possible approach to the thermalization problem is described in 
Chapter~\ref{chap-entropy}.

This Chapter is organized as follows.
In Sec.~\ref{sec-lpffrw} we present the general
theory of nonequilibrium dynamics of a scalar field in curved spacetime,
including a summary discussion of reheating in inflationary cosmology.
In Sec.~\ref{sec-onfrw} we specialize to the case of spatially
flat FRW spacetime, and derive the dynamical equations.
The initial conditions used and the results obtained from numerically
solving the dynamical equations are described in Sec.~\ref{sec-analysis}.
Discussion and conclusions follow in Sec.~\ref{sec-discussion}.

\section{$\lambda \Phi^4$ Inflaton Dynamics in FRW spacetime}
\label{sec-lpffrw}
\subsection{$\lambda \Phi^4$ quantum fields in curved spacetime}
As a simple model of inflation, let us consider a scalar $\lambda \Phi^4$
field in semiclassical gravity, where the matter field is quantized on a
classical, dynamical background spacetime.  The classical action has the form
\begin{equation}
S[\phi,g^{\mu\nu}] = S^{\text{{\tiny G}}}[g^{\mu\nu}] + S^{\text{{\tiny F}}}[
\phi,g^{\mu\nu}],
\label{eq-lpfcsta}
\end{equation}
where $S^{\text{{\tiny F}}}$ is the matter field action defined in
Eq.~(\ref{eq-lpfcstca}) and $S^{\text{{\tiny G}}}$ is the gravity action 
defined in Eq.~(\ref{eq-lpfcstga}).

The inflaton field $\phi$ is then quantized on the classical
background spacetime; we denote the Heisenberg field operator by 
$\Phi_{\text{{\tiny H}}}$, and the quantum state by $|\Omega\rangle$.  
Of particular importance in a study of inflaton dynamics are the mean field
\begin{equation}
\phih(x) \equiv \langle\Omega|\Phi_{\text{{\tiny H}}}(x)|\Omega\rangle,
\label{eq-ifmf}
\end{equation}
the fluctuation field
\begin{equation}
\varphi_{\text{{\tiny H}}}(x) \equiv \Phi_{\text{{\tiny H}}}(x) - \phih(x),
\label{eq-dff2}
\end{equation}
and the mean-squared fluctuations, or variance
\begin{equation}
\langle\Omega| \varphi_{\text{{\tiny H}}}^2 (x) | \Omega \rangle = 
\langle\Omega| \Phi_{\text{{\tiny H}}}^2(x) | \Omega \rangle - 
\langle\Omega| \Phi_{\text{{\tiny H}}}(x) | \Omega \rangle^2.
\label{eq-ifvr}
\end{equation}
In Chapter~\ref{chap-oncst}, a systematic procedure was presented
for deriving real and causal evolution equations for the mean field, 
two-point function, 
and the metric tensor in semiclassical gravity.  Assuming a globally hyperbolic
spacetime, one can evolve the coupled evolution equations forward from initial 
data specified at an initial Cauchy hypersurface.  The evolution equations
follow from functional differentiation (and subsequent field identifications)
of the closed-time-path (CTP) two-particle-irreducible (2PI) effective action,
$\Gamma[\phih,G,g^{\mu\nu}]$.  The CTP-2PI effective
action is a functional of the mean field $\phih$, two-point function $G$, and 
metric tensor $g^{\mu\nu}$, which now carry not only spacetime labels but
also {\em time branch\/} labels, which have an index set $\{+, -\}$.  The 
evolution equations for $\phih$, 
$\langle \varphi_{\text{{\tiny H}}}^2 \rangle$, and $g_{\mu\nu}$
then follow from Eqs.~(\ref{eq-lpfseeb}), (\ref{eq-lpfseec}), and 
(\ref{eq-lpfseea}), respectively.  The variance 
$\langle \vphi_{\text{{\tiny H}}}^2 \rangle$ 
is related to the coincidence limit of any of the four
components of the CTP two-point function through Eq.~(\ref{eq-dvrc}).
The energy-momentum tensor $\langle T_{\mu\nu} \rangle$ is defined by
Eq.~(\ref{eq-lpfcstemt}), and it is this quantum expectation value which
(after renormalization) enters as the source of the semiclassical 
Einstein field equation (\ref{eq-lpfcstsee2}).
Eqs.\ (\ref{eq-lpfseea})--(\ref{eq-lpfseec}) 
constitute a set of coupled, nonlocal, nonlinear
equations for the mean field, two-point function, and metric tensor.
The renormalized versions are what enter into the description of inflaton
dynamics.  The CTP-2PI effective action can be computed using 
diagrammatic methods described in Chapter~\ref{chap-oncst}, where a covariant 
expression for $\Gamma[\phih,G,g^{\mu\nu}]$ was computed in 
a general curved spacetime (truncated at two loops).

\subsection{Inflaton dynamics in FRW spacetime}
\label{sec-idfrw}
We now consider a spatially
flat Friedmann-Robertson-Walker (FRW) spacetime, which
is spatially homogeneous, isotropic, and conformally flat.
Its line element can be written in the form
\begin{equation}
ds^2 = a(\eta)^2 \left[d\eta^2 - \sum_{i=1}^3 (dx^i)^2\right],
\label{eq-frwcrd}
\end{equation}
where $a$ is the scale factor,  $x^i$ ($i \in \{1,2,3\}$) are the physical
position coordinates on the spatial hypersurfaces
of constant conformal time $\eta$ (related to the cosmic time $t$ by
$\eta = \int dt/a$).
The Hubble parameter, which measures the rate of cosmic expansion, is
\begin{equation}
H(\eta) = \frac{\dot{a}}{a},
\end{equation}
where the over-dot denotes differentiation with respect to cosmic time $t$.
Given our choice of sign convention and metric signature, the Ricci tensor 
in the FRW coordinates is given by
\begin{align}
R_{00} &= 3\left[\frac{a''}{a} - \frac{(a')^2}{a^2}\right],  \\
R_{ij} &= -\left[\frac{a''}{a} + \frac{(a')^2}{a^2}\right]\delta_{ij},
\end{align}
where the prime denotes differentiation with respect
to $\eta$, and $R_{00}$ is the component of the Ricci tensor
proportional to $d\eta \otimes d\eta$. The scalar curvature is
\begin{equation}
R  = \frac{6a''}{a^3}, 
\end{equation}
and the Einstein tensor is
\begin{align}
G_{00} &= -\frac{3(a')^2}{a^2}, \\
G_{ij} &= \left[\frac{2 a''}{a} - \frac{(a')^2}{a^2}\right]\delta_{ij}.
\end{align}
Finally, the volume form on $M$ is
\begin{equation}
{\mathbb \epsilon}_{\text{{\tiny M}}} = a^4 (d\eta \wedge dx^1 \wedge 
dx^2 \wedge dx^3).
\end{equation}
The higher-order (e.g., $R^2$)
geometric terms in the geometrodynamical field equation are not
shown because the renormalized constants $b$ and $c$ are set to zero
in Sec.~\ref{sec-rsee}.

In restricting the spacetime to be a spatially flat FRW,
we are reducing the number of degrees of freedom in the metric:
\begin{equation}
g_{\mu\nu} \rightarrow a(\eta)^2 \eta_{\mu\nu}.
\end{equation}
This reduction should not be carried out in the 2PI generating functional
$\Gamma[\phih,G,g^{\mu\nu}]$, but
only in the equations of motion
(\ref{eq-lpfseea})--(\ref{eq-lpfseec}).  
This is because functional differentiation 
of $\Gamma[\phih,G,a^{-2}\eta^{\mu\nu}]$ with respect to the scale factor 
$a$ gives only the {\em trace\/} 
of the energy-momentum tensor,
$a^{-2} \eta^{\mu\nu} \langle T_{\mu\nu} \rangle$,
and not the additional constraint equation which the initial data must 
satisfy.

The spatial homogeneity and isotropy of FRW spacetime permits only
two algebraically independent components of the energy-momentum tensor,
which in the FRW coordinates of Eq.\ (\ref{eq-frwcrd}) are given by
$\langle T_{00} \rangle$ and $\langle T_{ii}\rangle$; all other components
are zero.  These must be functions of $\eta$ only (due to spatial
homogeneity).
For the purpose of numerically solving the semiclassical Einstein equation,
it is convenient to work with the trace 
\begin{equation}
{\mathcal T} = g^{\mu\nu}\langle T_{\mu\nu}\rangle = a^{-2} \eta^{\mu\nu}
\langle T_{\mu\nu} \rangle,
\end{equation}
instead of $\langle T_{ii} \rangle$.  The trace
${\mathcal T}$ enters into the dynamical equation for $a(\eta)$, and
$\langle T_{00} \rangle$ enters into the constraint equation.

Another consequence of the spatial symmetries of FRW spacetime is
the restriction on the generality with which we may specify initial data
for dynamical evolution.  
Let us choose to specify initial data on a Cauchy
hypersurface $\Sigma_{\eta_0}$ of constant conformal time $\eta_0$.
In the Heisenberg picture,\footnote{As discussed in
Sec.~\ref{sec-initcond}, for our purposes it is sufficient to consider only
the case of a pure state.  The analysis can, however, be easily extended
to encompass a mixed state with density matrix ${\mathbb \rho}$.}
for consistency with spatial homogeneity, the quantum state 
$|\Omega\rangle$ must satisfy 
\begin{align}
\langle\Omega|\Phi_{\text{{\tiny H}}}(\eta_0,\vec{x}) |\Omega\rangle &= 
\phih(\eta_0), 
\label{eq-symcdmf} \\
\langle\Omega|\Phi_{\text{{\tiny H}}}'(\eta_0,\vec{x}) |\Omega\rangle &= 
\phih'(\eta_0), 
\label{eq-symcdmfb}
\end{align}
for all $\vec{x} \in {\mathbb R}^3$, where $\Phi_{\text{{\tiny H}}}$ is
the Heisenberg field operator for the scalar field.  The values of 
$\phih(\eta_0)$ and $\phih'(\eta_0)$ constitute initial data for the 
mean field.  In addition, the quantum state must satisfy  
\begin{align}
\langle \Omega| \vphi_{\text{{\tiny H}}}(\eta_0,\vec{x})
\vphi_{\text{{\tiny H}}}(\eta_0,\vec{x}') | \Omega\rangle
&= F(\eta_0,|\vec{x}-\vec{x}'|), 
\label{eq-symcdgf}  \\
\frac{\partial}{\partial \eta}_{|\eta_0}\!\langle\Omega | 
\vphi_{\text{{\tiny H}}}(\eta,\vec{x})
\vphi_{\text{{\tiny H}}}(\eta,\vec{x}')|
\Omega\rangle &= F'(\eta_0,|\vec{x}-\vec{x}'|), 
\label{eq-symcdgfb}
\end{align}
in terms of an equal-time correlation function $F(\eta_0,|\vec{x}-\vec{x}'|)$ 
which is invariant under simultaneous translations and rotations of
$\vec{x}$ and $\vec{x}'$.  As defined in Eq.\ (\ref{eq-dff2}), 
$\vphi_{\text{{\tiny H}}}$ denotes the Heisenberg field operator for the 
fluctuation field.  The spatial Fourier transform of $F$ is 
related to the power spectrum of quantum fluctuations at $\eta_0$ for
the quantum state $|\Omega\rangle$.  Alternatively, we may say that 
$F(\eta_0,r)$ and $F'(\eta_0,r)$ give initial data for the evolution of
the two-point function $G_{++}$ via the gap equation (\ref{eq-lpfseec}).
The symmetry conditions (\ref{eq-symcdmf}), (\ref{eq-symcdmfb}), 
(\ref{eq-symcdgf}), (\ref{eq-symcdgfb}),
along with the spatial symmetries of the classical action
in FRW spacetime, guarantee that the mean field and two-point function
satisfy spatial homogeneity and isotropy for {\em all time,\/} i.e.,
\begin{align}
\langle \Phi_{\text{{\tiny H}}}(x)\rangle &= \phih(\eta),
\label{eq-gfssym} \\
G_{++}(x,x') &= G_{++}(\eta,\eta',|\vec{x}-\vec{x'}|),
\label{eq-gfssymb}
\end{align}
for all $x \in M$.
The conditions (\ref{eq-gfssym}), (\ref{eq-gfssymb})
permit a formal solution of the gap equation (\ref{eq-lpfseec})
for $G_{++}$ in terms of homogeneous mode functions, via a Fourier
transform in comoving momentum $\vec{k}$, as shown in
Sec.~\ref{sec-resfrw}.  By rotational invariance, the Fourier transform
depends only on the magnitude $k \equiv \sqrt{\vec{k}\cdot\vec{k}}$.
Of course, the quantum state $|\Omega\rangle$ is not uniquely defined by
the spatial symmetries; a unique choice of the initial conditions for
$\phih$ and $G_{ab}$ at $\Sigma_{\eta_0}$ is (in the Gaussian wave-functional
approximation) equivalent to choosing $|\Omega\rangle$.
The choice of quantum state depends on the physics of the problem we wish
to study.

As a consequence of covariant conservation of the energy-momentum tensor
\begin{equation}
\nabla^{\mu} \langle T_{\mu\nu} \rangle = 0,
\label{eq-emtcc}
\end{equation}
the functions $\langle T_{00}(\eta) \rangle$ and 
$\langle T_{ii}(\eta) \rangle$ satisfy 
\begin{equation} 
\frac{d}{d\eta} \Bigl( a \langle T_{00} \rangle \Bigr) = 
- \frac{\langle T_{ii} \rangle}{a^2} \frac{d}{d\eta} \left( a^3 \right),
\end{equation}
which comes from taking the $\nu = 0$ component of Eq.\ (\ref{eq-emtcc}).
In analogy with the continuity relation for a classical perfect fluid in FRW
spacetime,
\begin{equation}
\frac{d}{d\eta} \left( a^3 \rho \right) = -p\frac{d}{d\eta}\left( a^3 \right),
\label{eq-fltd}
\end{equation}
we may define the energy density $\rho$ and pressure $p$, by
\begin{align}
\rho(\eta) &= \frac{1}{a^2}\langle T_{00}(\eta) \rangle, \\
p(\eta) &= \frac{1}{a^2}\langle T_{ii}(\eta) \rangle.
\label{eq-frwrho} 
\end{align}
However, the quantity $p$ should not be interpreted
as the true hydrodynamic pressure until a perfect-fluid
condition is shown to exist;
otherwise, bulk viscosity corrections can
enter into Eq.\ (\ref{eq-frwrho}) \cite{weinberg:1972a}.
The effective 
equation of state is defined as a time average (over the time scale
$\tau_1$ for the matter field dynamics, to be discussed in
Sec.~\ref{sec-reheating}) of the ratio $p/\rho$,
\begin{equation}
\bar{\gamma} \equiv \frac{p}{\rho}.
\label{eq-defeos}
\end{equation}
The effective 
equation of state $\bar{\gamma}$ (where the bar denotes a time average)
is an important quantity
in differentiating between the various stages of inflationary cosmology.

Several solutions to the semiclassical Einstein equation (\ref{eq-seesimp})
for idealized equations of state are of particular interest in cosmology.
The effective equation of state $\bar \gamma = -1$ (eternally 
``vacuum dominated'')
leads to a solution  $a(\eta) = -1/(H\eta)$, for $-\infty < \eta < 0$, 
where $H = \sqrt{8 \pi G \rho/3}$ and $\rho$ is a
constant.  This solution
corresponds to the ``steady-state'' coordinatization covering 
one-half of the 
de~Sitter manifold \cite{birrell:1982a}.
The effective 
equation of state $\bar{\gamma} = 0$ 
corresponds to nonrelativistic matter, in which case the scale factor
conformal-time dependence is $a \propto \eta^2$.  The 
effective equation of state
$\bar{\gamma} = 1/3$ corresponds to relativistic matter, and its 
scale factor conformal-time dependence is $a \propto \eta$.

\subsection{Initial conditions for post-inflation dynamics}
\label{sec-initcond}
In most realizations of inflationary cosmology, the Universe evolves through
a period in which a dominant portion of the energy density $\rho$ comes from
a quantum field $\Phi_{\text{{\tiny H}}}$, the {\em inflaton field,\/} whose 
effective equation of state [defined as in Eq.\ (\ref{eq-defeos})] is  
$\bar{\gamma} \simeq -1$.  
In chaotic inflation, this condition is due to the fact that the inflaton 
field is in a quantum state $|\Omega\rangle$
in which the Heisenberg field operator $\Phi_{\text{{\tiny H}}}$ acquires a 
large (approximately spatially homogeneous) expectation value, defined by
Eq.~(\ref{eq-gfssym}).
A requirement for chaotic inflation is that the potential energy $V(\phih)$ 
of the expectation value $\phih$ dominates over both the spatial
gradient energy [coming from $\langle (\nabla \vphi_{\text{{\tiny H}}})^2 
\rangle$] and kinetic energy for the inflaton field, and the energy
density of all other quantum fields coupled to the inflaton.
The potential energy $V(\phih)$ gives a contribution
to the energy-momentum tensor satisfying precisely $\gamma = -1$.
During inflation, the scale factor grows by a factor of approximately
$\exp(H \Delta t)$, where $\Delta t$ is the 
interval of inflation in cosmic time, typically larger than
$60 H^{-1}$. While the Universe is inflating, the
expectation value $\langle \Phi_{\text{{\tiny H}}} \rangle$
is slowly rolling toward the true minimum of the effective 
potential. (In reality, the situation is much more complicated than
this.  The effective potential is an inadequate tool for
studying out-of-equilibrium mean-field dynamics \cite{hu:1986d,mazenko:1985a}.)
Assuming the Universe was in local thermal
equilibrium prior to inflation, the temperature during inflation decreases
in proportion to $1/a$.  The energy density of any relativistic
(nonrelativistic) fields coupled to the inflaton is proportional to
$1/a^4$ ($1/a^3$).
The contribution to the quantum energy density from spatial gradients
of fluctuations about the inflaton field
is proportional to $1/a^4$ (see Sec.~\ref{sec-onfrw} below).
Most importantly, any inhomogeneous modes $\delta \phih_k$ of the 
{\em mean field\/} which might exist at the onset of inflation are
redshifted. The physical momentum of a quantum mode,
$k_{\text{{\tiny phys}}} = k/a,$
decreases as $1/a$ relative to the comoving momentum $k$.  
The quantum state of any field coupled to the inflaton 
at the end of inflation is, therefore, approximately given by
the vacuum state. The inflaton field is well approximated by
a spatially homogeneous mean field, with vacuum fluctuations around the 
mean-field configuration.   The mean field
can be thought of as representing the coherent oscillations of a
condensate of zero-momentum inflaton particles.

Let us consider the case of inflation driven by a 
single self-interacting scalar field $\phi$ (with
unbroken symmetry) in spatially flat FRW spacetime. 
The above arguments imply that one can model post-inflationary physics with a
quantum state $|\Omega\rangle$ which at $\eta_0$ corresponds
to a coherent state for the field operator $\Phi_{\text{{\tiny H}}}$
[in which $\langle \Omega|\Phi_{\text{{\tiny H}}}(x) | \Omega\rangle = 
\phih(\eta)$], and the fluctuation field $\vphi_{\text{{\tiny H}}}$ 
is very nearly in the vacuum state.\footnote{
Though this is a pure quantum state, the methods
employed in this study can be used to treat a quantum field theory in
a mixed state  (for example, a system initially in thermal equilibrium
with a heat bath).} 
Then for $\eta < \eta_0$, $\langle T_{00} \rangle$ is dominated by
the classical energy density of the mean field $\phih$.
The 00 component of the Einstein equation then yields
\begin{equation}
\frac{a'}{a^2} = \sqrt{\frac{8 \pi G \rho_{\text{{\tiny C}}}}{3}},
\end{equation}
where $\rho_{\text{{\tiny C}}}$ is the classical energy density of the mean
field, defined by
\begin{equation}
\rho_{\text{{\tiny C}}} = \frac{1}{2 a^2} (\phih')^2 + V(\phih).
\label{eq-dfrc}
\end{equation}
The mean field $\phih$ satisfies the classical equation
\begin{equation}
\phih'' + \frac{2 a'}{a}\phih' + a^2 V'(\phih) = 0,
\end{equation}
where $V(\phih)$ denotes the classical potential.
For the $\lambda \Phi^4$ theory, the potential is [from the
Minkowski-space limit of Eq.\ (\ref{eq-lpfcsta})]
\begin{equation}
V(\phih) = \frac{1}{2} m^2 \phih^2 + \frac{\lambda}{24}\phih^4.
\end{equation}
The assumption that the Universe is inflating (i.e., $\bar{\gamma} \simeq -1$)
for $\eta < \eta_0$ requires that the energy density $\rho_{\text{{\tiny C}}}$
be potential dominated,
\begin{equation}
V(\phih) \gg \frac{1}{2 a^2}(\phih')^2,
\label{eq-pdas}
\end{equation}
and that the mean field satisfies the slow-roll condition,
\begin{equation}
\phih'' \ll \frac{2 a'}{a} \phih'.
\label{eq-sras}
\end{equation}
Given Eqs.\ (\ref{eq-pdas}) and (\ref{eq-sras}),  
an approximate ``0th adiabatic order''
solution to the Einstein equation can be obtained [normalized
to $a(\eta_0) = 1$],
\begin{equation}
a(\eta) \simeq \frac{1}{1 + H(\eta)(\eta - \eta_0)},
\label{eq-asol}
\end{equation}
where $H$ is a slowly varying function of $\eta$, given by
\begin{equation}
H(\eta) = \sqrt{\frac{8\pi G \rho_{\text{{\tiny C}}}(\eta)}{3}}.
\label{eq-defhinv}
\end{equation}
From Eq.\ (\ref{eq-defhinv}), we can evaluate the expansion rate 
nonadiabaticity
parameter $\bar{\Omega}_H$ \cite{hu:1993d} for 
$\eta < \eta_0$ using Eq.\ (\ref{eq-defhinv}). 
During slow-roll it follows from conditions (\ref{eq-pdas}) and
(\ref{eq-sras}) that
\begin{equation}
\bar{\Omega}_H \equiv \frac{H'}{H^2} 
= \frac{V'(\phih) \phih'}{\sqrt{\frac{32 \pi G}{3} V(\phih)^3}} \ll 1.
\end{equation}
The solution (\ref{eq-asol}) for $a(\eta)$ is exact in the limit of constant
$H$ (corresponding to a constant $\phih$ at the tree level).  For simplicity,
let us assume that $\phih$ goes to a constant value
$\gtrsim \Mpl$ in the asymptotic past, $\eta \rightarrow -\infty$.  
The spacetime is then asymptotically
de Sitter, with the scale factor having an asymptotic cosmic-time 
dependence $a(t) \simeq \exp(H t)$.
Because the enormous cosmic expansion during the slow-roll period redshifts
away all nonvacuum energy in the Universe, it is reasonable to assume
that the quantum state $|\Omega\rangle$ would register no particles for
a comoving detector coupled to the fluctuation field $\vphi$ 
at conformal-past infinity; i.e., that the fluctuation field $\vphi$ is in
the vacuum state at $\eta \rightarrow -\infty$.  This would mean that
$a \simeq 1/(H\eta)$ at $\eta \rightarrow -\infty$.  This spacetime is
{\em not\/} asymptotically static in the past, but it is conformally static
with a conformal factor whose nonadiabaticity parameter vanishes at
conformal-past infinity.  Therefore, the best approximation to a
``no-particle'' state for a comoving detector in the asymptotic past is
given by the adiabatic vacuum \cite{parker:1974a} constructed via matching
at $\eta \rightarrow -\infty$.  All higher-order adiabatic vacua will in this
case agree at conformal past infinity.

To construct the $n$th-order adiabatic vacuum matched at an
equal-time hypersurface $\Sigma_{\eta_m}$, one first exactly solves the 
conformal-mode function equation for the quantum field 
[see Eq.\ (\ref{eq-onhmfe}) below].  Since the mode-function equation is 
second order, each $k$ mode has two independent
solutions, which can be represented as $u_k$ and $u_k^{\star}$.  A particular
solution consists of a linear combination of $u_k$ and $u_k^{\star}$.
The adiabatic vacuum is constructed by choosing (for each $k$) a linear
combination which smoothly matches the $n$th-order positive frequency WKB
mode function at $\Sigma_{\eta_m}$.
The resulting orthonormal basis of mode functions is then used to expand
the Heisenberg field operator $\Phi_{\text{{\tiny H}}}(x)$ in terms of $a_k$
and $a_k^{\dagger}$.
The vacuum state is defined by $a_k |\text{vac}\rangle = 0$ for all $k$,
which can be shown to correspond (in the $\eta_m \rightarrow -\infty$ limit)
to the de~Sitter-invariant, adiabatic (Bunch-Davies) vacuum.

\subsection{Post-inflation preheating}
\label{sec-reheating}
Inflation ends when the mean field has rolled down to the point where 
condition (\ref{eq-sras}) ceases to be valid, which we assume occurs at 
conformal time $\eta_0$.  The inflaton mean field then begins to 
oscillate about the true minimum of the effective potential, leading to a 
change in the effective equation of state.  A harmonically
oscillating scalar mean field ($m^2 \gg \lambda \phih^2/6$) 
has an effective equation of state $\bar{\gamma} = 0$, and
a scalar inflaton undergoing extreme large-amplitude oscillations
($\lambda \phih^2/6 \gg m^2$) has an effective equation of state 
$\bar{\gamma} = 1/3$ \cite{kolb:1990a}.  
In realistic models, the inflaton field is coupled to 
various lighter fields consisting of fermions and/or bosons.  
These couplings, as well as the inflaton's self-coupling,
provide mechanisms for damping of the mean-field oscillations
via back reaction from quantum particle production, and energy transfer
to the lighter fields and the inflaton's inhomogeneous modes.

Let us consider the scalar $\lambda \Phi^4$ 
field theory with unbroken symmetry in Minkowski space 
[with classical action given by the Minkowski-space limit of 
Eq.\ (\ref{eq-lpfcsta})],
and suppose that the mean field $\phih$ oscillates about the stable
equilibrium configuration $\phih = 0$  with initial amplitude $\phih_0$.  
For the moment we are neglecting the effect of spacetime dynamics, 
i.e., assuming $a(\eta)=1$.  The
time scale for oscillations of the mean field is given by 
\cite{boyanovsky:1996b}
\begin{equation}
\tau_0 = \frac{4 K(k)}{m \sqrt{1 + f^2}},
\label{eq-deftau0}
\end{equation}
where $f$ and $k$ are defined by
\begin{align}
f &= \sqrt{\frac{\lambda}{6}} \frac{\phih_0}{m}, 
\label{eq-deff}  \\
k &= \frac{f}{\sqrt{2(1 + f^2)}},
\end{align}
and $K(k)$ is the complete elliptic integral of the first kind 
\cite{gradshteyn:1964a}. For harmonic oscillations where
\begin{equation}
\frac{\lambda}{6} \phih_0^2 \ll m^2,
\label{eq-tdptas1}
\end{equation}
time-dependent perturbation theory was used in the group 1A studies
(see Sec.~\ref{sec-intro}) to compute the damping
rate $\Gamma$ for the mean field in the $\lambda \Phi^4$ model.   At lowest
order in $\lambda$, the damping rate for the mean field $\phih$
corresponds to the rate for four zero-momentum, free-field excitations of the
inflaton to decay into a $\vphi$ (fluctuation field) particle pair, due to the
$\lambda \Phi^4$ self-coupling \cite{shtanov:1995a},
\begin{equation}
\Gamma_{\phi} \simeq O(1) \frac{(\lambda \phih_0)^2}{4\pi m},
\label{eq-tdptdr}
\end{equation}
with vacuum initial state for $\vphi$.
The symbol $O(1)$ denotes a constant of order unity.
In addition to the assumption (\ref{eq-tdptas1}), 
there is another crucial assumption in the derivation
of Eq.\ (\ref{eq-tdptdr}), namely, that the dominant contribution to the decay
rate is given by the lowest order, $|\text{vac}\rangle \rightarrow
|1_{\vec{k}}, \; 1_{-\vec{k}} \rangle$ process, where the occupation numbers  
are for the fluctuation field $\vphi$.
It can be shown  \cite{parker:1969a,hu:1987a} that for this (bosonic) case, the
perturbative decay rate into the $\vec{k}$ momentum shell for the
fluctuation field $\vphi$ is enhanced by $1+2n$ when the occupation of
the $\vec{k}$ shell is $n$.  This is a stimulated 
emission effect due to Bose statistics.\footnote{
In contrast with the case with Bose fields, 
the use of time-dependent perturbation theory to study 
inflaton decay into fermions via a Yukawa coupling does not require the
condition $n_{\vec{k}} \ll 1$, because of the Pauli exclusion principle
\cite{dolgov:1990a,kofman:1994a,shtanov:1995a}.
It is still necessary, however, to assume weak coupling (or small 
mean-field amplitude)
in order to use perturbation theory
\cite{dolgov:1990a,shtanov:1995a}.}
The use of Eq.\ (\ref{eq-tdptdr}) to estimate the damping rate thus implicitly
assumes that for all $\vec{k}$, the fluctuation field occupation numbers
are small, i.e., $n_{\vec{k}} \ll 1$.   This is because
time-dependent perturbation theory in terms of the $\lambda \Phi^4$
interaction corresponds to an expansion of the field theory around
the vacuum configuration.  Equivalently, it corresponds to an amplitude
expansion (in powers of the ``classical field'' $\phihpm$) of the 1PI
closed-time-path effective action $\Gamma[\phih_{+},\phih_{-}]$,
which is defined in Eq.\ (2.11) in Ref.\ \cite{ramsey:1997a}. 
When $\lambda \phih_0^2$ is sufficiently large
at $\eta_0$, or on a time scale for $n_{\vec{k}}$ to grow to order unity,
the perturbative expansion in $\lambda$ breaks down.

In many inflationary scenarios, condition (\ref{eq-tdptas1}) does not hold
at $\eta_0$.
A correct analysis of the dynamics of the inflaton field must, therefore,
be nonperturbative, if the inflaton is self-interacting and/or coupled to
Bose fields.  Again, of interest in ``preheating'' 
is the time scale for damping of the
mean field $\phih$ due to back reaction from particle production
into the inhomogeneous modes of the fluctuation field.  
This quantum particle production is known
to occur by parametric amplification of quantum vacuum fluctuations, for
the zero-temperature, unbroken symmetry system under study here.
Boyanovsky {\em et al.\/} \cite{boyanovsky:1996b} have
obtained an approximate analytic expression (in Minkowski space)
for the time scale $\tau_1$ for the variance 
$\langle \vphi_{\text{{\tiny H}}}^2 \rangle$ to
grow to the point where $\lambda \langle \vphi_{\text{{\tiny H}}}^2 \rangle/2$
is of the same order of magnitude as the tree-level effective mass
$m^2 + \lambda \phih^2/6$,
\begin{equation}
\tau_1 = \frac{m^{-1}}{B(f)} \text{ln} \left( \frac{ (1 + 
f^2/2)}{\lambda \sqrt{B(f)}/(8 \pi^2)} \right).
\label{eq-deftau1}
\end{equation}
The function $B(f)$ is of order unity, and in terms of the asymptotic value
of $f$ at $\eta \rightarrow \infty$, $B[f(\eta\rightarrow \infty)] \simeq
0.285 \, 953$.
Their result is valid in flat space and based on
a solution of the one-loop dynamics which neglects the back reaction
of particle production on the mode functions.
The essential feature of the time scale $\tau_1$ is that it depends on
the $\text{ln}(\lambda^{-1})$.  As a consequence of the analytic solution
to the classical mean-field equation and the estimate for $\tau_1$,
it is possible to estimate (for the case of Minkowski space) the effective
equation of state $\bar{\gamma}_{\text{{\tiny C}}}$
for the mean field \cite{boyanovsky:1996b},
\begin{equation}
\bar{\gamma}_{\text{{\tiny C}}} 
\equiv \left(\overline{\frac{p_{\text{{\tiny C}}}}{
\rho_{\text{{\tiny C}}}}}\right) =
\frac{ -\frac{1}{6} f_0^2 \left[ 1 - \frac{1}{2} f_0^2 \right] + \frac{2}{3}
(1 + f_0^2) \left[ 1 - \frac{E(k)}{K(k)} \right]}{\frac{1}{2} f_0^2 \left[
1+ \frac{1}{2} f_0^2 \right]},
\label{eq-aseos}
\end{equation}
where $E(k)$ is the complete elliptic integral of the second kind 
\cite{gradshteyn:1964a}, $p_{\text{{\tiny C}}}$ is the pressure of
the mean field, and $\rho_{\text{{\tiny C}}}$ is the energy density of
the mean field, defined in Eq.\ (\ref{eq-dfrc}).  The late-time effective equation
of state can
be studied using an idealized two-fluid model consisting of the classical
mean-field oscillations $\bar{\gamma}_{\text{{\tiny C}}}$ and a relativistic
component corresponding to the energy density in the quantum modes
$\rho_{\text{{\tiny Q}}}$ [defined in Eq.\ (\ref{eq-rhoqdef}) below].

The physical processes discussed above neglect collisional scattering
of excitations of the inhomogeneous modes due to the
$\lambda \Phi^4$ self-interaction, for example, binary
scattering.  These scattering processes ultimately lead to thermalization
of the system.  
A quantitative understanding of the time scales for such processes
in the nonperturbative regime studied here within a
rigorous field-theoretic framework is at present lacking.
A perturbative treatment of collisional thermalization of the system using the
Boltzmann equation assumes a separation of time scales for collisionless 
processes ($\tau_1$) and thermalization.  However, due to the 
nonperturbatively large occupation numbers which arise in the resonance
band of the inhomogeneous field modes on the time scale $\tau_1$, such a
naive approach would predict that the time scale for thermalization is on
the order of (or earlier than) the preheating time scale $\tau_1$.  A 
nonperturbative approach to studying the collisional thermalization of the
system is, therefore, required.  However, within the $1/N$ expansion 
(to be discussed in Sec.~\ref{sec-onlna2}), the collisional scattering 
processes are subleading order in $1/N$, and thus the separation of time 
scales is assured within this controlled expansion \cite{boyanovsky:1996b}.
Let us denote the time scale for scattering by $\tau_2$.  

In typical inflationary scenarios, the self-coupling $\lambda$ of the 
inflaton is very weak \cite{kolb:1990a}, in the range $10^{-12}$--$10^{-14}$
(see, however, \cite{calzetta:1995a,matacz:1996a}).  In our numerical
work, $f$ is initially unity, in which case the three time scales
$\tau_0$, $\tau_1$ and $\tau_2$ separate dramatically,
\begin{align}
& \tau_1/\tau_0  \simeq O\left[\text{ln}\left(\frac{1}{\lambda}
\right)\right],
\label{eq-adtau1}\\
& \tau_2/\tau_1  \simeq O(N).
\end{align}
The period leading up to $\tau_1$ is called {\em preheating\/}, because (i)
the energy transfer from the mean field is entirely nonequilibrium in origin,
and (ii) the occupation numbers of the fluctuation field are extremely
nonthermal.  In this regime, since $\tau_2 \gg \tau_1$, collisional effects
can be neglected.  In a collisionless approximation, the damping of the mean
field is due to energy transfer into the inhomogeneous quantum modes, 
a process similar to Landau damping in plasma physics \cite{cooper:1994a}.

So far in this section we have not included 
the effect of spacetime dynamics on the
particle production and back reaction processes.  Cosmic expansion introduces
an additional time scale $H^{-1}$, where $H$ is the Hubble parameter
defined in Eq.\ (\ref{eq-defhinv}).
In typical chaotic inflation scenarios, the initial inflaton amplitude
can be as large as $\Mpl/3$, leading to
\begin{equation}
H^{-1} \simeq \frac{3 \tau_0}{\sqrt{2\pi}}
\end{equation} 
at the onset of reheating.  In this case, $H^{-1} \ll \tau_1$
when $\lambda$ is very small.  Clearly, for sufficiently large initial inflaton
amplitude, it is necessary to include the effect of spacetime dynamics
in a systematic study of preheating dynamics of the inflaton field.

\section{O$(N)$ inflaton dynamics in FRW spacetime}
\label{sec-onfrw}
In this section, we study the nonequilibrium dynamics of a quartically
self-interacting, minimally coupled, O$(N)$ field theory (with unbroken
symmetry) in spatially flat FRW spacetime.  We use the covariant evolution 
equations derived in \cite{ramsey:1997a}, in order to study the dynamics of 
the mean field, variance, and the spacetime, at leading order in the $1/N$
expansion.

\subsection{The O$(N)$ model in the $1/N$ expansion}
\label{sec-onlna2}
The classical action for the unbroken symmetry O$(N)$ model in a general curved
spacetime is
\begin{equation}
S^{\text{{\tiny F}}}[\phi^i, g_{\mu\nu}] = 
-\frac{1}{2} \int_M \! d^{\, 4} x \sqmg
\left[ \vecphi \cdot (\square + m^2 + \xi R ) \vecphi + \frac{\lambda}{4 N} 
( \vecphi \cdot \vecphi )^2 \right], \label{eq-onsm2} 
\end{equation}
where the O$(N)$ inner product is defined by\footnote{
In our index notation, the latin letters
$i,j,k,l,m,n$  are used to designate O$(N)$ indices 
(with index set $\{1,\ldots, N\}$), while the latin letters
$a,b,c,d,e,f$ are used below to designate CTP indices 
(with index set $\{+,-\}$).}
\begin{equation}
\vecphi \cdot \vecphi = \phi^i \phi^j \delta_{ij}.
\end{equation}
As in Eq.\ (\ref{eq-lpfcsta}), $\lambda$ is a coupling constant with dimensions
of $1/\hbar$, and $\xi$ is the dimensionless coupling to gravity (and is 
necessary in order for the quantized theory to be renormalizable).

In \cite{ramsey:1997a}, the covariant mean-field equation, gap equation, 
and geometrodynamical field equation were computed for this model at leading
order in the $1/N$ expansion.  The evolution equations follow from 
Eqs.\ (\ref{eq-lpfseea})--(\ref{eq-lpfseec}), 
with the 2PI, CTP effective action truncated at
leading order in the $1/N$ expansion.  At leading order in the $1/N$ expansion,
we need only keep track of one component of the CTP two-point function
$G_{ab}(x,x')$; we choose $G_{++}(x,x')$, which is the Green function with
Feynman boundary conditions.  The covariant gap equation for $G_{++}$ at 
leading order in the $1/N$ expansion is
\begin{equation}
\left( \square_x + m^2 + \xi R(x) + \frac{\lambda}{2}\phih^2(x) + 
\frac{\hbar \lambda}{2} G(x,x) 
\right) G_{++}(x,x')  = 
\delta^4(x-x') \frac{-i}{\sqmgp},
\label{eq-onige2}
\end{equation}
plus terms of $O(1/N)$.  The covariant $\delta$ function is defined
in Ref.\ \cite{birrell:1982a}.
The mean-field equation is, at leading order in $1/N$,
\begin{equation}
\left(\square + m^2 + \xi R + \frac{\lambda}{2} \phih^2 + 
\frac{\hbar \lambda}{2}
G(x,x) \right) \phih(x) = 0.
\end{equation}
Recall that all four components of $G_{ab}(x,x')$ are the same in the 
coincidence limit, which we are denoting by $G(x,x)$.
The coincidence limit $G(x,x)$ is divergent in four spacetime dimensions,
and the regularization method is described in Sec.~\ref{sec-onrenorm} below.
The geometrodynamical field equation is given by Eq.~(\ref{eq-lpfcstsee2}).
in terms of the (unrenormalized) energy-momentum tensor computed at leading 
order in the $1/N$ expansion, which is shown in 
Eqs.\ (5.37) and (5.38) in Ref.\ \cite{ramsey:1997a}.

\subsection{Restriction to FRW spacetime}
\label{sec-resfrw}
Let us now specialize to the spatially flat FRW universe, with initial
conditions appropriate to post-inflation dynamics of the inflaton field.  
As discussed in Sec.~\ref{sec-initcond}, initial Cauchy data for
$\phih$, $G_{++}$, and $a$ are specified on a spacelike hypersurface
$\Sigma_{\eta_0}$ (at conformal time $\eta_0$).  
The spatial symmetries of $\phih$ and $G_{++}$ for a quantum state 
$|\phi\rangle$ consistent with a spatially homogeneous and isotropic 
cosmology are given in (\ref{eq-gfssym}--b).  As a consequence of these
symmetries, both the mean field $\phih$ and variance
$\langle \vphi_{\text{{\tiny H}}}^2 \rangle$ are spatially homogeneous,
i.e., functions of conformal time only. 

Eq.\ (\ref{eq-onige2}) for $G_{++}$ in spatially flat FRW spacetime
has the formal solution
\begin{multline}
G_{++}(x,x') = a(\eta)^{-1} a(\eta')^{-1} \int \frac{d^{\,3}k}{(2\pi)^3}
e^{i \vec{k} \cdot (\vec{x} - \vec{x}')} \bigl[
\Theta(\eta' - \eta)   \tilde{u}_k(\eta)^{\star} \tilde{u}_k(\eta')  
 \\  
+ \Theta(\eta - \eta') \tilde{u}_k(\eta')\tilde{u}_k(\eta)^{\star} \bigr],
\end{multline}
in terms of conformal-mode functions $\tilde{u}_k$ which satisfy a
harmonic oscillator equation with conformal-time-dependent effective
frequency,
\begin{equation}
\left( \frac{d^2}{d\eta^2} + \Omega_k^2(\eta) \right) \tilde{u}_k = 0.
\label{eq-onhmfe}
\end{equation}
The fact that $\tilde{u}_k(\eta)$ depends only on $\eta$ and $k$ (where $k$
is comoving momentum) implies that $G_{++}$ is invariant under simultaneous
spatial translations and rotations of $\vec{x}$ and $\vec{x}'$.
The effective frequency $\Omega_k(\eta)$ appearing in Eq.\ (\ref{eq-onhmfe}) is
defined by
\begin{align}
\Omega_k^2(\eta) &= k^2 + a^2 {\mathfrak M}^2(\eta), 
\label{eq-oneffrq}  \\
{\mathfrak M}^2(\eta) &= M^2(\eta) + \left( \xi - \frac{1}{6} \right) R(\eta),
\label{eq-oneffrqb}  \\
M^2(\eta) &= m^2 + \frac{\lambda}{2}\phih^2(\eta) + 
\frac{\lambda}{2}\langle \vphi_{\text{{\tiny H}}}^2(\eta) \rangle. 
\label{eq-oneffrqc} 
\end{align}
Initial conditions for the positive frequency conformal mode functions
$\tilde{u}_k(\eta)$ must be specified (for all $k$) at $\eta_0$.  A choice
of initial conditions corresponds to a choice of quantum state 
$|\Omega\rangle$ for the fluctuation field $\vphi_{\text{{\tiny H}}}$; initial
conditions are discussed in Sec.~\ref{sec-onbc} below.
The (bare) variance $\langle \vphi_{\text{{\tiny H}}}^2 \rangle$ 
has a simple representation in terms of the conformal-mode functions:
\begin{equation}
\langle \Omega | \vphi_{\text{{\tiny H}}}(x)^2 
| \Omega \rangle = \hbar G(x,x) 
 = \frac{\hbar}{a^2} \int \frac{d^3k}{(2\pi)^3} | \tilde{u}_k(\eta) |^2.
\label{eq-frwfluc}
\end{equation}
It should be noted that this expression is divergent, in
consequence of our having computed the variance in terms
of the bare (unrenormalized) constants of the theory.  In terms of a
physical upper momentum cutoff $K$, the variance $G(x,x)$ diverges like
$K^2$.  Even after removing the $O(K^2)$ divergence, there remains
a logarithmic dependence on $K$ which must be regularized.
In addition, the mode functions $\tilde{u}_k$ depend on $\langle 
\vphi_{\text{{\tiny H}}}^2 \rangle$ through Eq.\ (\ref{eq-oneffrqc}).
The leading-order, large-$N$, mean-field equation in spatially flat FRW 
spacetime becomes 
\begin{equation}
\phih'' + \frac{2 a'}{a} \phih' + a^2 M^2(\eta)
\phih = 0,
\label{eq-bmfeqfrw}
\end{equation}
where the time-dependent bare effective mass $M(\eta)$ is given by 
Eq.\ (\ref{eq-oneffrqc}).  For simplicity of notation, we will henceforth 
write $M$ instead of $M(\eta)$, and similarly for ${\mathfrak M}(\eta)$.

Finally, we can express the bare energy-momentum tensor in terms of
the conformal-mode functions $\tilde{u}_k(\eta)$.  As discussed in
Sec.~\ref{sec-idfrw}, it is convenient to work with the 00 
component and the trace of the energy-momentum tensor.  
The components of the classical part of the energy-momentum 
tensor are spatially homogeneous, and given by
\begin{align}
T_{00}^{\text{{\tiny C}}}(\eta) &=  \frac{1}{2} (\phih')^2
- \frac{3 \xi}{2}  \left( \phih'' + \frac{2 a'}{a} \phih' \right) \phih 
+ \frac{1}{2} a^2 \left( m^2 + \frac{\lambda}{4} \phih^2 
+ \frac{3 \xi (a')^2}{2 a^4} \right) \phih^2,
\label{eq-onemtcla}  \\
{\mathcal T}^{\text{{\tiny C}}}(\eta) &= 
\frac{1}{a^2} \left\{(6\xi-1)(\phih')^2 + 6\xi\left( \phih'' 
+ \frac{2a'}{a}\phih' \right) \phih \right\} 
+ 2 \left( m^2 + \frac{\lambda}{4}\phih^2 + \frac{\xi}{2}
R \right) \phih^2.
\label{eq-onemtclb} 
\end{align}
The quantum energy-momentum tensor components are also spatially homogeneous.
We find for the 00 component,
\begin{multline}
T^{\text{{\tiny Q}}}_{00}(\eta) = \frac{\hbar}{2 a^2} 
\int \frac{d^{\,3}k}{(2\pi)^3} \Biggl[  | \tilde{u}_k'|^2  +
\Bigl( k^2 + a^2 M^2 + (1  -6\xi) \frac{(a')^2}{a^2} \Bigr) |
\tilde{u}_k|^2  \\ 
 + (6 \xi - 1) \frac{a'}{a}\Bigl[ 
(\tilde{u}_k')^{\star}\tilde{u}_k
+ \tilde{u}_k'\tilde{u}_k^{\star} \Bigr] \Biggr],
\label{eq-onfrwetc}
\end{multline}
and for the trace,
\begin{multline}
{\mathcal T}^{\text{{\tiny Q}}}(\eta) = \frac{\hbar}{a^4} 
\int \frac{d^{\,3}k}{(2\pi)^3} \Biggl[  (6\xi - 1)\biggl\{
| \tilde{u}_k' |^2 - (k^2  + a^2 M^2) |\tilde{u}_k|^2 
- \frac{a'}{a}\Bigl[ (\tilde{u}_k')^{\star}
\tilde{u}_k + \tilde{u}_k'\tilde{u}_k^{\star}\Bigr] \\ 
+ \biggl( \frac{(a')^2}{a^2} - \xi a^2 R \biggr) | \tilde{u}_k|^2 \biggr\}
+ a^2 M^2 | \tilde{u}_k|^2 \Biggr].
\label{eq-onfrwetcp}
\end{multline}
It can be shown by asymptotic analysis that, in terms of
a physical upper momentum cutoff $K$, the bare $T_{00}^{\text{{\tiny Q}}}$
is quartically divergent, i.e., $O(K^4)$, and that (for minimal coupling)
${\mathcal T^{\text{{\tiny Q}}}}$ is quadratically divergent.
In addition, the components of the bare energy-momentum
tensor contain the effective mass $M^2$, which contains the divergent
variance $\langle \vphi_{\text{{\tiny H}}}^2 \rangle$.  The energy density 
$\rho_{\text{{\tiny Q}}}$ of quantum modes of the $\vphi$ field
is defined in terms of $T^{\text{{\tiny Q}}}_{00}$ by
\begin{equation}
\rho_{\text{{\tiny Q}}} = \frac{1}{a^2} T^{\text{{\tiny Q}}}_{00}
- \frac{\lambda}{8}\langle \vphi_{\text{{\tiny H}}}^2 \rangle^2.
\label{eq-rhoqdef}
\end{equation}
We shall also refer to $\rho_{\text{{\tiny Q}}}$ as the energy density of the
``fluctuation field.'' 

\subsection{Renormalization of the dynamical equations}
\label{sec-onrenorm2}
The variance $\langle \vphi_{\text{{\tiny H}}}^2 \rangle$ and
quantum energy-momentum tensor components $T_{00}^{\text{{\tiny Q}}}$
and ${\mathcal T}^{\text{{\tiny Q}}}$ are divergent in four spacetime 
dimensions, and must be regularized within the context of a systematic, 
covariant renormalization procedure.  In the ``in-out'' formulation of 
quantum field theory, renormalization may be carried out via addition
of counterterms to the effective action, which amounts to renormalization 
of the constants in the classical action \cite{ramond:1990a}.
The closed-time-path formulation of the effective dynamics is renormalizable
provided the theory is renormalizable in the ``in-out'' formulation
\cite{hartle:1979a,calzetta:1987a}, as is the case with the O$(N)$ field
theory in curved spacetime 
\cite{ramsey:1997a,toms:1982a,mazzitelli:1989b}.
For our purposes it is convenient (in this model) to carry out
renormalization in the leading-order, large-$N$, evolution equations,
rather than in the CTP effective action \cite{jordan:1986a}.

In this study we employ the adiabatic regularization method of
Parker, Fulling, and Hu \cite{parker:1974a,fulling:1974a}.  The idea
is to define an adiabatic approximation to the conformal mode function, 
and then to construct a regulator for the integrands of the bare 
energy-momentum tensor and
variance from the adiabatic mode functions \cite{fulling:1989a}.   
Renormalization occurs when we define the renormalized variance
and energy-momentum tensor to be the difference between the bare expressions
and the regulators and simultaneously replace the bare quantities
$m,$ $\lambda,$ $G,$ $b,$ $c,$ $\Lambda,$ and $\xi$ by their renormalized
counterparts.  The equivalence of this procedure to other manifestly
covariant methods (such as dimensional continuation) is well established 
\cite{bunch:1980a}.
We implement renormalization as a two-step process: First, we  adiabatically
regularize the variance and renormalize $\xi$, $m$, and
$\lambda$; Second, we adiabatically regularize the energy-momentum tensor
and renormalize the semiclassical geometrodynamical field equation.

We define the adiabatic order of a conformal mode function as follows: 
let $\Omega_k(\eta) \rightarrow \Omega_k(\eta/T)$, where $T$ is introduced
as a time scale which is formally taken to be unity at the end of the
calculation.  Then the adiabatic order of an expression involving
derivatives of $\Omega_k$ is simply the inverse power of $T$, of the
leading-order term in an asymptotic expansion about $T \rightarrow \infty$.
However, in order for the
adiabatically regulated energy-momentum tensor for an interacting scalar
field theory to agree with the renormalized energy-momentum tensor
obtained by manifestly covariant methods (e.g., covariant point splitting 
\cite{bunch:1978a}), it is 
necessary to define the adiabatic order of expressions involving $\lambda$ and
derivatives with respect to $\eta$,
such as $\lambda (\phih^2)''$, as the sum of the exponent of $\phih$ and
the number of conformal time differentiations \cite{paz:1988a}.  Therefore,
$\lambda \langle \vphi_{\text{{\tiny H}}}^2 \rangle''$ is considered fourth
adiabatic order, as is $\lambda (\phih^2)''$.  

Having defined adiabatic order, we now construct the 
adiabatic mode functions. It is well known that the WKB {\em Ansatz}
\begin{equation}
\tilde{u}_k(\eta) = \frac{1}{\sqrt{2 W(\eta)}} \exp \left(
-i \int^{\eta} d\eta' W(\eta') \right)
\end{equation}
turns the harmonic oscillator equation (with time-dependent frequency 
$\Omega_k$) into a nonlinear differential equation for $W$,
\begin{equation}
W(\eta)^2 = \Omega_k^2(\eta) + \frac{3 [W'(\eta)]^2}{4 W^2(\eta)} -
\frac{W''(\eta)}{2 W(\eta)}.
\end{equation}
Starting with the lowest-order {\em Ansatz} $W^{(0)}(\eta) = \Omega_k(\eta)$,
one can iterate this equation; the $n$th-order iteration
yields the $n$th-order WKB approximation for $\tilde{u}_k$.
For the free field theory, the $n$th-order WKB approximation gives
an expression for $\tilde{u}_k$ which is of adiabatic order $2n$.
In the interacting case, the above definition of adiabatic order calls for 
removing terms such as $\lambda (\phih^2)''''$ at 4th adiabatic order.
Thus we have a method of deriving expressions for 
$T^{\text{{\tiny Q}}}_{00}$, ${\mathcal T}^{\text{{\tiny Q}}}$,
and $\langle \vphi_{\text{{\tiny H}}}^2 \rangle$ at fourth, fourth,
and second adiabatic orders, respectively.  One then sets $T=1$ in the
truncated expression.  
We can thus obtain a fourth-order adiabatic approximation to the quantum
energy-momentum tensor $(T_{\mu\nu}^{\text{{\tiny Q}}})_{\text{{\tiny ad4}}}$,
and a second-order adiabatic approximation to the variance 
$\langle \vphi_{\text{{\tiny H}}} \rangle_{\text{{\tiny ad2}}}$.
By subtracting $(T_{\mu\nu}^{\text{{\tiny Q}}})_{
\text{{\tiny ad4}}}$ from the divergent 
$T_{\mu\nu}^{\text{{\tiny Q}}}$ and
$\langle \vphi_{\text{{\tiny H}}}^2 \rangle_{\text{{\tiny ad2}}}$ from
the divergent
$\langle \vphi_{\text{{\tiny H}}}^2 \rangle$, finite expressions for the
renormalized energy-momentum tensor and variance are obtained.

First we regularize the variance $\langle \vphi_{
\text{{\tiny H}}}^2 \rangle$, and carry out a renormalization 
of $\lambda$, $m$, and $\xi$.
In the leading-order, large-$N$ approximation, no terms appear 
in the mode-function equation (\ref{eq-onhmfe}) which would permit
addition of counterterms; therefore, $\Omega_k$ must be finite 
\cite{root:1974a}.  The effective frequency $\Omega_k$ which appears in 
Eq.\ (\ref{eq-oneffrq})
is the ``bare'' effective frequency, which we denote by
$(\Omega_k)_{\text{{\tiny B}}}$.
In conjunction with the adiabatic regularization procedure,
we fix the renormalization scheme by demanding equivalence of the bare and 
renormalized effective mass \cite{cooper:1994a},
\begin{equation}                                             
(\Omega_k^2)_{\text{{\tiny R}}} = (\Omega_k^2)_{\text{{\tiny B}}},
\label{eq-rencst}
\end{equation}
where the ``R'' subscripted quantities are renormalized.  Using
Eqs.\ (\ref{eq-oneffrq}) and (\ref{eq-oneffrqb}), we have
\begin{equation}
\xi_{\text{{\tiny R}}} R + M^2_{\text{{\tiny R}}} =
\xi_{\text{{\tiny B}}} R + M^2_{\text{{\tiny B}}},
\label{eq-rencst2}
\end{equation}
where $M^2_{\text{{\tiny B}}}$ is defined in Eq.\ (\ref{eq-oneffrqc}),
\begin{equation}
M^2_{\text{{\tiny B}}} = m_{\text{{\tiny B}}}^2 +
\frac{\lambda_{\text{{\tiny B}}}}{2} \phih^2 + 
\frac{\lambda_{\text{{\tiny B}}}}{2} \langle \vphi_{\text{{\tiny H}}}^2
\rangle_{\text{{\tiny B}}},
\end{equation}
and $M^2_{\text{{\tiny R}}}$ is defined similarly,
\begin{equation}
M^2_{\text{{\tiny R}}} = m_{\text{{\tiny R}}}^2 +
\frac{\lambda_{\text{{\tiny R}}}}{2} \phih^2 + 
\frac{\lambda_{\text{{\tiny R}}}}{2} \langle \vphi_{\text{{\tiny H}}}^2
\rangle_{\text{{\tiny R}}}.
\label{eq-reneff}
\end{equation}
Now, $\lambda_{\text{{\tiny B}}}$, $m_{\text{{\tiny B}}}$,
and $\xi_{\text{{\tiny B}}}$ are the bare constants of the theory
which appeared (without B's) in the classical action (\ref{eq-onsm2}).
The renormalized quantities in Eq.\ (\ref{eq-oneffrq}) are defined below.
The bare 
$\langle \vphi_{\text{{\tiny H}}}^2\rangle_{\text{{\tiny B}}}$ 
is a conformal-time-dependent function defined by Eq.\ (\ref{eq-frwfluc}),
\begin{equation}
\langle \vphi_{\text{{\tiny H}}}^2(\eta) \rangle_{\text{{\tiny B}}} =
\frac{\hbar}{a^2} \int \frac{d^{\,3}k}{(2\pi)^3} | \tilde{u}_k(\eta) |^2,
\label{eq-renbfle}
\end{equation}
where the conformal-mode functions $\tilde{u}_k(\eta)$ obey 
Eq.\ (\ref{eq-onhmfe}).  Now we demand that the renormalized
theory be minimally coupled, i.e., we set $\xi_{\text{{\tiny R}}} = 0$.
Because of Eq.\ (\ref{eq-rencst}), we can formally use
$(\Omega_k^2)_{\text{{\tiny R}}}$ in computing the adiabatic regulator for
the variance $\langle \vphi_{\text{{\tiny H}}}^2 \rangle_{\text{{\tiny B}}}$.
Computing the asymptotic series (in $1/T$) of the
quantity $|\tilde{u}_k(\eta)|^2$ to $O(1/T^2)$, where
$\Omega_k^2(\eta/T)$ is the effective frequency, we obtain
(after setting $T=1$)
\begin{equation}
\langle \vphi_{\text{{\tiny H}}} \rangle_{\text{{\tiny ad2}}}
= \frac{\hbar}{2 C} \int \frac{d^{\,3}k}{(2\pi)^3}
\left[ \frac{1}{\tilde{\omega}_k}
- \frac{(C')^2 - 2C C''}{8 C^2 \tilde{\omega}_k^3} + \frac{M^2_{ 
\text{{\tiny R}}} C''}{8
\tilde{\omega}_k^5} - \frac{5 M^4_{\text{{\tiny R}}} 
(C')^2}{32 \tilde{\omega}_k^7} \right],
\label{eq-onaregfluc}
\end{equation}
in terms of an auxiliary function
\begin{equation}
C(\eta) = a^2(\eta).
\end{equation}
In Eq.\ (\ref{eq-onaregfluc}) the symbol $\tilde{\omega}_k$ is defined 
as follows
\begin{equation}
\tilde{\omega}_k^2 = k^2 + a^2 M^2_{\text{{\tiny R}}}.
\label{eq-tilomk}
\end{equation}
In the adiabatic prescription, the renormalized variance
$\langle \vphi_{\text{{\tiny H}}}^2 \rangle_{\text{{\tiny R}}}$
appearing in Eq.\ (\ref{eq-reneff}) is defined by
\begin{equation}
\langle \vphi_{\text{{\tiny H}}}^2 \rangle_{\text{{\tiny R}}} =
\langle \vphi_{\text{{\tiny H}}}^2 \rangle_{\text{{\tiny B}}} -
\langle \vphi_{\text{{\tiny H}}}^2 \rangle_{\text{{\tiny ad2}}},
\end{equation}
where the first term on the right-hand side is given by 
Eq.\ (\ref{eq-renbfle}), and the second term on the right-hand 
side is given by Eq.\ (\ref{eq-onaregfluc}).
Everything on the right hand side can be expressed 
in terms of renormalized quantities, so this procedure is well defined.
Written out explicitly, the renormalized variance satisfies the equation
\begin{equation}
\langle \vphi_{\text{{\tiny H}}}^2 \rangle_{\text{{\tiny R}}} =
\frac{\hbar}{C} \int \frac{d^{\,3}k}{(2\pi)^3} \left[
|\tilde{u}_k|^2 - \frac{1}{2\tilde{\omega}_k} - 
\frac{(C')^2 - 2 C C''}{16 C^2 \tilde{\omega}_k^3} +
\frac{M^2_{\text{{\tiny R}}} C''}{16 \tilde{\omega}_k^5} -
\frac{5 M^4_{\text{{\tiny R}}} (C')^2}{64 \tilde{\omega}_k^7}
\right]. 
\label{eq-renrfluc}
\end{equation}
One can use the WKB approximation for $\tilde{u}_k(\eta)$ to compute
the asymptotic series for the integrand in Eq.\ (\ref{eq-renrfluc})
in the limit $k \rightarrow \infty$, and show that the integral is
convergent.
Since $M^2_{\text{{\tiny R}}}$ is contained in the
integrand above, Eq.\ (\ref{eq-renrfluc}) leads to an integral 
equation for the renormalized effective mass
$M_{\text{{\tiny R}}}$,
\begin{multline}
M_{\text{{\tiny R}}}^2 = m_{\text{{\tiny R}}}^2 + \frac{
\lambda_{\text{{\tiny R}}}}{2} \phih^2 + \frac{\hbar \lambda_{
\text{{\tiny R}}}}{2 C} \int \frac{d^{\,3}k}{(2\pi)^3} \biggl[ 
|\tilde{u}_k|^2 - \frac{1}{2\tilde{\omega}_k} - 
\frac{(C')^2 - 2 C C''}{16 C^2 \tilde{\omega}_k^3} \\  +
\frac{M^2_{\text{{\tiny R}}} C''}{16 \tilde{\omega}_k^5} -
\frac{5 M^4_{\text{{\tiny R}}} (C')^2}{64 \tilde{\omega}_k^7}
\biggr].
\label{eq-rem}
\end{multline}
Eqs.\ (\ref{eq-renrfluc}) and (\ref{eq-rencst2})  together define
$\lambda_{\text{{\tiny R}}}$ and $m_{\text{{\tiny R}}}$.  All physical
quantities should now be expressed in terms of the renormalized parameters
$m_{\text{{\tiny R}}}$ and $\lambda_{\text{{\tiny R}}}$ of the theory.
The renormalized mean-field equation now reads
\begin{equation}
\phih'' + \frac{2 a'}{a} \phih' + a^2 M^2_{\text{{\tiny R}}} \phih = 0,
\label{eq-eveqp}
\end{equation}
where $M_{\text{{\tiny R}}}^2$ is given by Eq.\ (\ref{eq-rem}), and the mode
functions in Eq.\ (\ref{eq-rem}) obey the homogeneous equation,
\begin{equation}
\left(\frac{d^2}{d\eta^2} + k^2 + a^2 M_{\text{{\tiny R}}}^2\right)
\tilde{u}_k = 0.
\label{eq-eveqmf}
\end{equation}
The initial conditions for the conformal-mode functions at $\eta_0$ are 
discussed in Sec.~\ref{sec-onbc} below.

Having obtained a renormalized mean-field equation, we now turn our
attention to regularizing the quantum energy-momentum tensor.
As a consequence of Eq.\ (\ref{eq-rencst}), we can substitute 
$M \rightarrow M_{\text{{\tiny R}}}$ and
$\xi \rightarrow \xi_{\text{{\tiny R}}}$ in the equations for the
components of the quantum energy-momentum tensor, 
Eqs.\ (\ref{eq-onfrwetc},
\ref{eq-onfrwetcp}).  Since we wish to study the minimal coupling case,
we set $\xi_{\text{{\tiny R}}}= 0$.  To avoid confusion we denote 
the bare energy-momentum tensor components (\ref{eq-onfrwetc}) and
(\ref{eq-onfrwetcp}) by $(T^{\text{{\tiny Q}}}_{00})_{\text{{\tiny B}}}$
and $({\mathcal T}^{\text{{\tiny Q}}})_{\text{{\tiny B}}}$, respectively.
Let us also relabel the bare constants $b$, $c$, $G$, and $\Lambda$
appearing in the bare semiclassical Einstein equation
(\ref{eq-lpfcstsee2}) as $b_{\text{{\tiny B}}}$,
$c_{\text{{\tiny B}}}$, $G_{\text{{\tiny B}}}$, and
$\Lambda_{\text{{\tiny B}}}$.
Applying the method described above to construct the adiabatic regulator, 
for $T^{\text{{\tiny Q}}}_{00}$ we find
\begin{multline}
\left(T_{00}^{\text{{\tiny Q}}}\right)_{\text{{\tiny ad4}}} = 
\frac{\hbar}{4 C} \int \frac{d^{\,3}k}{(2\pi)^3} 
\Biggl\{   2 \tilde{\omega}_k +
\frac{(C')^2}{4 C^2 \tilde{\omega}_k} + \biggl[ \frac{M_{\text{{\tiny R}}}^2 
(C')^2}{4C}
- \frac{9 (C')^4}{64 C^4}    \\  + \frac{C' (M_{\text{{\tiny R}}}^2)'}{4} +
\frac{(C')^2 C''}{4 C^3} + \frac{(C'')^2}{16 C^2} - 
\frac{C'C'''}{8 C^2}\biggr] \frac{1}{\tilde{\omega}_k^3}    \\  +
\biggl[ \frac{M_{\text{{\tiny R}}}^4 (C')^2}{16} + 
\frac{M_{\text{{\tiny R}}}^2}{32 C^3}
\Bigl( -5 (C')^4 + 4 C^4 (C') (M_{\text{{\tiny R}}}^2)'    \\  
+ 10 C (C')^2 C''
+ 2 C^2 (C'')^2 - 4 C^2 C' C''' \Bigr) \biggr] \frac{1}{\tilde{\omega}_k^5} 
   \\  +
\biggl( \frac{M_{\text{{\tiny R}}}^4}{128 C^2} \biggr)\biggl[ -5(C')^4 +
40 (C')^2 C'' + 2 C^2 (C'')^2    \\  
- 4 C^2 C' C''' \biggr] \frac{1}{\tilde{\omega}_k^7}
+ \frac{7 M_{\text{{\tiny R}}}^6 (C')^2}{128 C} \Bigl( -5 (C')^2 + 2 C C'' 
\Bigr)
\frac{1}{\tilde{\omega}_k^9}    \\ 
- \frac{105 M_{\text{{\tiny R}}}^8 (C')^4}{1024} \frac{1}{
\tilde{\omega}_k^{11}} \Biggr\},
\label{eq-onaregt00}
\end{multline}
where $\tilde{\omega}_k$ is defined in Eq.\ (\ref{eq-tilomk}). For ${\mathcal T}$, we find
\begin{align}
({\mathcal T}^{\text{{\tiny Q}}})_{\text{{\tiny ad4}}} = 
\frac{\hbar}{2 C^2} & \int  \frac{d^{\,3}k}{(2\pi)^3}  
\Biggl\{ \Bigl( M_{\text{{\tiny R}}}^2 C - \frac{(C')^2}{2 C^2} + 
\frac{C''}{2C}\Bigr) \frac{1}{\tilde{\omega}_k}    \notag \\ &  +
\biggl[ \frac{M_{\text{{\tiny R}}}^2}{8C}\Bigl( -3(C')^2 + 4 C C'' \Bigr) +
\frac{1}{16 C^4} \Bigl( 9(C')^4 + 4 C^4 C' (M_{\text{{\tiny R}}}^2)'   \notag \\ &  -
21 C (C')^2 C'' + 6 C^2 (C'')^2 + 4 C^5 (M_{\text{{\tiny R}}}^2)'' + 8 C^2 C' 
C'''    \notag \\ & 
- 2 C^3 C''''\Bigr) \biggr]\frac{1}{\tilde{\omega}_k^3} 
+ \biggl[ \frac{M_{\text{{\tiny R}}}^4}{8}\Bigl( -3 (C')^2 + C C''\Bigr)    \notag \\ &  +
\frac{M_{\text{{\tiny R}}}^2}{128 C^3}\Bigl( 87 (C')^4 - 64 C^4 C' 
(M_{\text{{\tiny R}}}^2)' 
- 208 C (C')^2 C''    \notag \\ &  + 60 C^2 (C'')^2 + 16 C^5 (M_{\text{{\tiny R}}}^2)'' + 
80 C^2 C' C''' - 16 C^3 C'''' \Bigr) \biggr] \frac{1}{\tilde{\omega}_k^5}    \notag \\ & 
+ \biggl[ -\frac{5 C M_{\text{{\tiny R}}}^6}{32} (C')^2 + 
\frac{M_{\text{{\tiny R}}}^4}{32 C^2}
\Bigl( 15 (C')^4 - 10 C^4 C' (M_{\text{{\tiny R}}}^2)'    \notag \\ &  - 40 C (C')^2 C'' 
+ 15 (C')^2 (C'')^2 + 20 C^2 C' C''' - C^3 C'''' \Bigr) \biggr] 
\frac{1}{\tilde{\omega}_k^7}     \notag \\ &  + \biggl[ 
\frac{7 M_{\text{{\tiny R}}}^6}{256 C}
\Bigl( 15 (C')^4 - 80 C (C')^2 C'' + 6 C^2 (C'')^2 + 8 C^2 C' C''' \Bigr)
\Biggr] \frac{1}{\tilde{\omega}_k^9}    \notag \\ &  +
\frac{21 M_{\text{{\tiny R}}}^8 (C')^2}{256}\Bigl( 15 (C')^2 - 11 C C'' \Bigr)
\frac{1}{\tilde{\omega}_k^{11}} + \frac{1155 C M_{\text{{\tiny R}}}^{10}}{2048}
(C')^4 \frac{1}{\tilde{\omega}_k^{13}}  \Biggr\}.
\label{eq-onaregtr}
\end{align}
In the free-field limit ($\lambda_{\text{{\tiny R}}} = 0$), the regulators
(\ref{eq-onaregt00}) and (\ref{eq-onaregtr})
agree with the minimal-coupling, spatially flat limit of the adiabatic
regulators obtained by Bunch \cite{bunch:1980a}.

The renormalization procedure for the semiclassical Einstein equation 
(\ref{eq-lpfcstsee2}) can now be precisely stated.  According to the adiabatic
prescription, we define the quantum energy-momentum tensor by
\begin{equation}
(T_{\mu\nu}^{\text{{\tiny Q}}})_{\text{{\tiny R}}} =
(T_{\mu\nu}^{\text{{\tiny Q}}})_{\text{{\tiny B}}} -
(T_{\mu\nu}^{\text{{\tiny Q}}})_{\text{{\tiny ad4}}}.
\label{eq-tqren}
\end{equation}
It can be checked that the momentum-integral expressions for the two 
independent components of Eq.\ (\ref{eq-tqren}) are convergent.
In terms of $(T^{\text{{\tiny Q}}}_{\mu\nu})_{\text{{\tiny R}}}$,
the total energy-momentum tensor (after renormalization) is
\begin{equation}
\langle T_{\mu\nu} \rangle_{\text{{\tiny R}}} = 
(T^{\text{{\tiny C}}}_{\mu\nu})_{
\text{{\tiny R}}} + (T^{\text{{\tiny Q}}}_{\mu\nu})_{\text{{\tiny R}}}
- \frac{\lambda_{\text{{\tiny R}}}}{8} \left(\langle \vphi_{\text{{\tiny H}}}^2
\rangle_{\text{{\tiny R}}}\right)^2 g_{\mu\nu},
\end{equation}
where $(T^{\text{{\tiny C}}}_{\mu\nu})_{\text{{\tiny R}}}$ stands for
$T^{\text{{\tiny C}}}_{\mu\nu}$, and renormalized quantities are substituted
for bare quantities.  The bare quantities $G_{\text{{\tiny B}}}$,
$\Lambda_{\text{{\tiny B}}}$, $b_{\text{{\tiny B}}}$, and
$c_{\text{{\tiny B}}}$ are now replaced by $G_{\text{{\tiny R}}}$,
$\Lambda_{\text{{\tiny R}}}$, $b_{\text{{\tiny R}}}$, and
$c_{\text{{\tiny R}}}$ in the renormalized semiclassical geometrodynamical
field equation,
\begin{equation}
G_{\mu\nu} + \Lambda_{\text{{\tiny R}}} g_{\mu\nu} +
c_{\text{{\tiny R}}} \, ^{(1)}H_{\mu\nu} +
b_{\text{{\tiny R}}} \, ^{(2)}H_{\mu\nu} = -8\pi 
G_{\text{{\tiny R}}} \langle T_{\mu\nu} \rangle_{\text{{\tiny R}}}.
\label{eq-rscee}
\end{equation}
\subsection{Renormalized semiclassical Einstein equation}
\label{sec-rsee}
Using semiclassical methods to study the dynamics of the inflaton field
in FRW spacetime requires that the Hubble parameter be much less than the
Planck mass, $H \ll \Mpl.$
On dimensional grounds, $c_{\text{{\tiny R}}}$ and $b_{\text{{\tiny R}}}$
are likely to be of order $\hbar^2 \Mpl^{-2}$, in which case
$R \gg c_{\text{{\tiny R}}} R^2,$ 
and $R \gg b_{\text{{\tiny R}}} R^{\alpha\beta}
R_{\alpha\beta}$, provided $R_{\alpha\beta} \neq 0$.
Let us, therefore, set $b_{\text{{\tiny R}}} = 0$ and
$c_{\text{{\tiny R}}} = 0$, and additionally, let us choose
$\Lambda_{\text{{\tiny R}}} = 0$, so that Eq.\ (\ref{eq-rscee}) becomes
the renormalized semiclassical Einstein equation (without cosmological 
constant),
\begin{equation}
G_{\mu\nu} = -8\pi G_{\text{{\tiny R}}}\left[
(T_{\mu\nu}^{\text{{\tiny C}}})_{\text{{\tiny R}}} +
(T_{\mu\nu}^{\text{{\tiny Q}}})_{\text{{\tiny R}}} -
\frac{\lambda_{\text{{\tiny R}}}}{8} (\langle \vphi_{\text{{\tiny H}}}^2
\rangle_{\text{{\tiny R}}})^2 \right].
\label{eq-newsee}
\end{equation}
Taking the trace of Eq.\ (\ref{eq-newsee}) in spatially flat FRW spacetime, 
we find
\begin{equation}
\frac{6a''}{a^3} = 8\pi G_{\text{{\tiny R}}}\left[
({\mathcal T}^{\text{{\tiny C}}})_{\text{{\tiny R}}} +
({\mathcal T}^{\text{{\tiny Q}}})_{\text{{\tiny R}}} -
\frac{\lambda_{\text{{\tiny R}}}}{2} \langle (\vphi_{\text{{\tiny H}}}^2
\rangle_{\text{{\tiny R}}})^2 \right].
\label{eq-eveqa}
\end{equation}
Recalling that $\xi_{\text{{\tiny R}}} = 0$, and using Eq.\ (\ref{eq-onemtclb}),
the classical part of the
trace of the renormalized energy-momentum tensor is given by
\begin{equation}
({\mathcal T}^{\text{{\tiny C}}})_{\text{{\tiny R}}} =
\frac{1}{a^2} \left[ -(\phih')^2 + 2 \left( m_{\text{{\tiny R}}}^2
+ \frac{\lambda_{\text{{\tiny R}}}}{4} \phih^2 \right) \phih^2 \right],
\end{equation}
and the quantum trace of the renormalized energy-momentum tensor is given by
\begin{equation}
\begin{split}
({\mathcal T}^{\text{{\tiny Q}}})_{\text{{\tiny R}}} =
-\frac{\hbar}{a^4} \int \frac{d^{\,3}k}{(2\pi)^3} \Biggl[
|\tilde{u}'_k|^2 - (k^2 - 2 a^2 M_{\text{{\tiny R}}}^2)|\tilde{u}_k|^2
- \frac{a'}{a}\left[ (\tilde{u}_k')^{\star} \tilde{u}_k +
\tilde{u}_k' \tilde{u}_k^{\star}\right]  + \frac{(a')^2}{a^2}
|\tilde{u}_k|^2 \Biggr] \\  - ({\mathcal T}^{\text{{\tiny Q}}})_{
\text{{\tiny ad4}}},
\end{split}
\end{equation}
where $({\mathcal T}^{\text{{\tiny Q}}})_{\text{{\tiny ad4}}}$ is
defined in Eq.\ (\ref{eq-onaregtr}).  As discussed in Sec.~\ref{sec-idfrw},
the 00 component of the semiclassical Einstein equation is a constraint,
and is given by
\begin{equation}
\frac{3 (a')^2}{a^2} = 8\pi G_{\text{{\tiny R}}} \left[
(T^{\text{{\tiny C}}}_{00})_{\text{{\tiny R}}} +
(T^{\text{{\tiny Q}}}_{00})_{\text{{\tiny R}}} -
\frac{\lambda_{\text{{\tiny R}}}}{8} a^2 (\langle \vphi_{\text{{\tiny H}}}^2
\rangle_{\text{{\tiny R}}})^2 \right].
\label{eq-cneqa}
\end{equation}
From Eq.\ (\ref{eq-onemtcla}), the expression for the classical part of the 00
component of the renormalized energy-momentum tensor is given by
\begin{equation}
(T^{\text{{\tiny C}}}_{00})_{\text{{\tiny R}}} =
\frac{1}{2} (\phih')^2 + \frac{1}{2} a^2 \left( m_{\text{{\tiny R}}}^2
+ \frac{\lambda_{\text{{\tiny R}}}}{4} \phih^2 \right) \phih^2,
\end{equation}
and the quantum part of the 00 component of the renormalized energy-momentum
tensor is given by
\begin{equation}
(T^{\text{{\tiny Q}}}_{00})_{\text{{\tiny R}}} = \frac{\hbar}{2a^2}
\int \frac{d^{\,3}k}{(2\pi)^3} \left[ |\tilde{u}'_k|^2 +
\left( k^2 + a^2 M_{\text{{\tiny R}}}^2 \right)|\tilde{u}_k|^2
-\frac{a'}{a}\left[ (\tilde{u}_k')^{\star}\tilde{u}_k +
\tilde{u}_k' \tilde{u}_k^{\star}\right]\right] -
(T^{\text{{\tiny Q}}}_{00})_{\text{{\tiny ad4}}},
\end{equation}
where $(T^{\text{{\tiny Q}}}_{00})_{\text{{\tiny ad4}}}$ is defined in
Eq.\ (\ref{eq-onaregt00}).

Eqs.\ (\ref{eq-eveqa}) and (\ref{eq-eveqp}) are coupled differential
equations for $a$ and $\phih$, involving complex homogeneous
conformal-mode functions $\tilde{u}_k$ which satisfy Eq.\ (\ref{eq-eveqmf}).
The conformal mode functions are related to the variance
$\langle \vphi_{\text{{\tiny H}}}^2 \rangle_{\text{{\tiny R}}}$ by
Eq.\ (\ref{eq-renrfluc}).  This is a closed, time-reversal-invariant system of 
equations.  The initial data at $\eta_0$ must satisfy the constraint equation
(\ref{eq-cneqa}).  We now drop all ``R'' subscripts, because we
will henceforth work only with renormalized quantities.

\subsection{Reduction of derivative orders}
The adiabatic regulators (\ref{eq-onaregfluc}),
(\ref{eq-onaregt00}), (\ref{eq-onaregtr}) 
for the variance and energy-momentum
tensor contain derivatives of up to
fourth order in $a$ and up to second order in $\phih^2$ and 
$\langle \vphi_{\text{{\tiny H}}}^2 \rangle$.  The presence of the former
can be understood as resulting in part from the well-known trace anomaly for
a quantum field in curved spacetime \cite{hu:1979a}, which contains 
higher-derivative local geometric terms, e.g., $\square R$.
In addition, there are nonanomalous finite terms
which result from the renormalization of the energy-momentum tensor and
the choice of minimal coupling.

The effect of higher derivatives in the semiclassical Einstein equation
has been much studied in the literature
\cite{anderson:1985a,suen:1987a,suen:1987b,simon:1990a,parker:1993a}.
The higher-derivative evolution equations for $a$ and $\phih$ have
a much larger solution space than the classical Einstein and mean-field
equations, and in general, the higher-derivative semiclassical Einstein
equation is expected to have many solutions which are unphysical.  In addition,
the semiclassical Einstein equation (which is fourth order in $a$) requires
more initial data than the classical Einstein equation in order to uniquely 
specify a solution.  However, Simon and Parker \cite{simon:1990a,parker:1993a},
following the methods of Ja\'{e}n, Llosa, and Molina \cite{jaen:1986a},
have shown that in one-loop semiclassical gravity,
there exists a procedure for consistently removing the unphysical solutions
within the perturbative ($\hbar$) expansion in which the equations are
derived.   The procedure corresponds to the addition of perturbative 
constraints, thereby
yielding second-order equations which require the same amount of initial data 
as does the classical Einstein equation.
Their method involves reducing the order of the $a'''$ and $a''''$ terms
in the semiclassical Einstein equation using strict perturbation theory
in $\hbar$.

In this study we follow the approach of Simon and Parker 
to reduce the order of the equations for $\phih$, $a$, and 
$\langle \vphi_{\text{{\tiny H}}}^2
\rangle$ to second order.  We replace all expressions involving
$a'''$ and $a''''$ with expressions $a_{\text{{\tiny cl}}}'''$
and $a_{\text{{\tiny cl}}}''''$ obtained from the {\em classical\/}
Einstein equation, i.e., Eq.\ (\ref{eq-eveqa})
with $\hbar = 0$.  This procedure is physically justifiable in this model for
the following reason:  At early times, the dominant contribution to the 
energy-momentum tensor is classical, 
$T^{\text{{\tiny C}}}_{\mu\nu}$.  Therefore, the deviations
$a''' - a'''_{\text{{\tiny cl}}}$ and $a'''' - a''''_{\text{{\tiny cl}}}$,
which are entirely quantum in origin and $\propto \hbar$, are at
early times expected to be very small.  In addition, at late times
the Universe is expected to become asymptotically radiation dominated,
in which case $a''' = a'''' = 0$.  The classical approximations to the
late-time behavior of $a'''$ and $a''''$ should also have this property,
regardless of whether the mean-field oscillations are harmonic or elliptic.
This procedure is, therefore, 
physically justifiable in the system studied here.

\section{Analysis}
\label{sec-analysis}
Having derived coupled dynamical equations (\ref{eq-eveqp}),
(\ref{eq-eveqa}), (\ref{eq-eveqmf}) for the mean field $\phih$, 
scale factor $a$, and conformal-mode functions $\tilde{u}_k$, 
respectively, we now proceed to solve them.
\subsection{Initial conditions}                              
\label{sec-onbc}
At the Cauchy hypersurface at $\eta_0$, 
we specify initial conditions on the conformal-mode functions
$\tilde{u}_k$ which
correspond to a choice of quantum state for the fluctuation field
$\vphi_{\text{{\tiny H}}}$.  
Based on the analysis in Sec.~\ref{sec-initcond},
we choose boundary conditions at $\eta_0$ which correpsond to the
adiabatic vacuum state for $\vphi_{\text{{\tiny H}}}$ at $\eta \rightarrow 
-\infty$.
From the semiclassical Einstein equation (\ref{eq-seesimp}),
the slow-roll condition (\ref{eq-sras}), the
potential-dominated condition (\ref{eq-pdas}), and assuming that 
the variance $\langle \vphi_{\text{{\tiny H}}}^2 \rangle$ satisfies
\begin{equation}
\frac{\lambda}{2}\langle \vphi_{\text{{\tiny H}}}^2 \rangle \ll
m^2 + \frac{\lambda}{2} \phih^2
\end{equation}
for $\eta < \eta_0$, it follows that the spacetime is asymptotically 
de~Sitter at conformal-past infinity.
Using the approximate solution (\ref{eq-asol}) for the scale factor 
for $\eta < \eta_0$, we can solve the mode function equation (\ref{eq-eveqmf})
for $\eta < \eta_0$ at the same (0th) adiabatic order.  The general
solution is of the form
\begin{equation}
\begin{split}
\tilde{u}_k(\eta) \simeq \left( \frac{\pi(\eta - H^{-1} - \eta_0)}{4}\right)^{
\frac{1}{2}} \Bigl[ & 
c^{1}_{k} H_{\nu}^{(1)}\bigl\{k(\eta - H^{-1}(\eta) - \eta_0) \bigr\}\\ 
& + c^{2}_{k} H_{\nu}^{(2)}\bigl\{k(\eta - H^{-1}(\eta) - \eta_0)\bigr\}
\Bigr],
\end{split}
\end{equation}
where $H^{(1)}$ and $H^{(2)}$ are the Hankel functions of first and second
kind, respectively \cite{gradshteyn:1964a}, and $\nu$ is defined by
\begin{equation}
\nu^2 = \frac{9}{4} - \frac{M^2}{H^2}.
\end{equation}
The function $H(\eta)$ is defined as in Eq.\ (\ref{eq-defhinv}),
\begin{equation}
H(\eta) = \sqrt{\frac{8 \pi G \rho_{\text{{\tiny C}}}}{3}},
\end{equation}
where now $\rho_{\text{{\tiny C}}} = a^2 T^{\text{{\tiny C}}}_{00}$.
The Hubble parameter must be slowly varying for this approximation to hold,
i.e., the expansion rate nonadiabaticity parameter \cite{hu:1993d}
\begin{equation}
\bar{\Omega}_H \equiv \frac{H'}{H^2} \ll 1.
\label{eq-areq}
\end{equation}
The Wronskian condition on the mode functions (which comes from
the canonical commutation relations for the fluctuation field operator) 
requires that
\begin{equation}
|c^1_k|^2 + |c^2_k|^2 = 1.
\end{equation}
By choosing $c^1_k$ and $c^2_k$, different vacua are obtained.
The 0th-order adiabatic vacuum (matched at $\eta = -\infty$)
is constructed by choosing $c^1_k$ and $c^2_k$ so that $\tilde{u}_k$
smoothly matches the positive-frequency 0th-order WKB mode
function at $\eta = -\infty$.  This corresponds to $c^2_k = 1$
and $c^1_k = 0$, for all $k$.
Using the asymptotic properties of the Hankel function, the adiabatic limit
$k, |\eta| \rightarrow \infty$, can be derived, and verified to have the
correct form,
\begin{equation}
\lim_{k,|\eta| \rightarrow \infty} 
\tilde{u}_k \simeq \frac{1}{\sqrt{2 k}} e^{-i k \eta}.
\end{equation}
In addition, the high-momentum, flat-space limit
($k, H^{-1} \rightarrow \infty$) gives the
same result.  The initial conditions for the $\tilde{u}_k$ at $\eta_0$
are then defined by demanding that the $\tilde{u}_k$ functions 
smoothly match the 
approximate adiabatic mode function solutions (for $\eta < \eta_0$)
at $\eta = \eta_0$.  This leads to the following initial conditions
for the conformal-mode functions:
\begin{align}
\tilde{u}_k(\eta_0) &= \left( \frac{-\pi}{4H_0} \right)^{1/2}
H_{\nu}^{(2)}(-k H^{-1}_0),  \\
\tilde{u}'_k(\eta_0) &= \frac{d}{d\eta} \left[ \left( \frac{\pi \eta}{4}
\right)^{1/2} H_{\nu}^{(2)}(k\eta)\right]_{|\eta = -H^{-1}_0},
\end{align}
where $H_0 = H(\eta_0)$.  The above conditions are valid only at 0th order
in the above-defined adiabatic approximation, where terms of order $H'/H$
are discarded.  It is straightforward to show that Eq.\ (\ref{eq-areq}) is valid
given the slow-roll (\ref{eq-sras}) and inflation (\ref{eq-pdas}) assumptions.
In addition to the initial conditions for $\tilde{u}_k$ at $\eta_0$,
we may freely choose initial values for $\phih(\eta_0)$ and
$\phih'(\eta_0)$, subject to the constraint that $\phih'$ must be
small enough that conditions (\ref{eq-pdas}) and
(\ref{eq-sras}) are valid.  We are already assuming that $a(\eta_0) = 1$.
The initial value of $a'(\eta_0)$ is fixed by the constraint
equation (\ref{eq-cneqa}).

\subsection{Numerical solution}
\label{sec-numerics}
In this section we describe the methods we used to 
solve the coupled evolution equations for $\phih$ [Eq.\ (\ref{eq-eveqp})],
$a$ [Eq.\ (\ref{eq-eveqa})], and $\tilde{u}_k$ [Eq.\ (\ref{eq-eveqmf})]
numerically.\footnote{Henceforth, we set $\hbar=1$ and work in units of energy
where $m = 1$.}
We evolved a representative sampling of mode functions
$\tilde{u}_k$ for the region of momentum space
$0 \leq k \leq K a$, where $K$ is a physical upper momentum cutoff.\footnote{
A finite momentum cutoff is necessary in any case 
due to the crossing of the Landau point (in the large-$N$ approximation to the 
theory) when the cutoff is taken to infinity 
\cite{collins:1984a,cooper:1994a}.}
Employing a physical \cite{ringwald:1987a}, as opposed to comoving, momentum
cutoff is necessary because a comoving cutoff would require the
use of the renormalization group
equation to track how the renormalized parameters flow as the 
scale factor $a$ increases at each time step.  
For a comoving cutoff the quadratic divergence in the variance would be 
proportional to $1/a^2$, requiring a time-dependent renormalization
(see \cite{boyanovsky:1994a}, for example).
The use of a physical upper
momentum cutoff yields a quadratic divergence which can be removed by
a non-time-dependent mass renormalization \cite{ringwald:1987a}.

We chose a variety of values of $K/m$ between $50$ and $70$.
The sampling of momentum-space is carried out with a uniform binning,
with total number of bins $N_{\text{{\tiny bins}}}$.  Various values of 
$N_{\text{{\tiny bins}}}$ were used, all greater than $10^4$.  Eq.\
(\ref{eq-rem}) was solved by iteration, and the momentum space integrations
were performed numerically using the 
$O(1/N_{\text{{\tiny bins}}}^4)$
extended Simpson rule.  The differential
equations (\ref{eq-eveqa}) and (\ref{eq-eveqp}) were evolved using
4th-order Runge-Kutta with adaptive step-size control; the target precision 
for the time steps varied between $10^{-6}$ and $10^{-8}$.  Cutoff independence
was verified {\em a posteriori\/} by explicitly checking that
the results of the numerical solution were insensitive to a rescaling
of $K/m$.  The solutions were computed to a conformal-time scale of 
$400 \; m^{-1}$.  A typical solution computed according  to the
above methods required on the order of 300 h of CPU time on a modern  
workstation.

\subsection{Results}
\label{sec-results}
\begin{figure}[htb]
\begin{center}
\epsfig{file=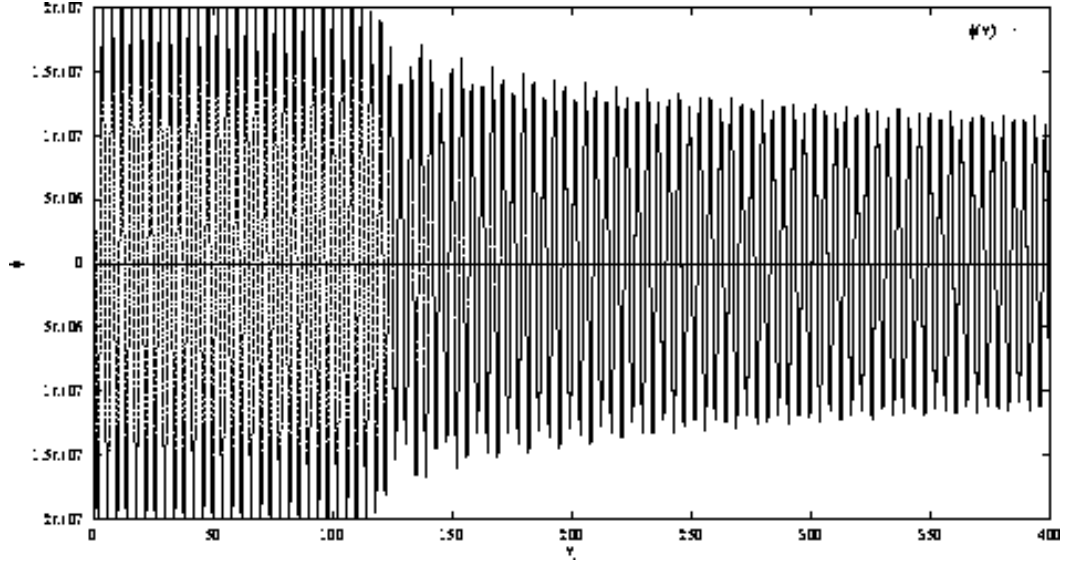,width=5.50in} 
\end{center}
\caption{Plot of $\phi$ vs $\eta$, with $\Mpl/m = 1.0 \times 10^{14}$.}
\label{fig-run19phi}
\end{figure}

\begin{figure}[htb]
\begin{center}
\epsfig{file=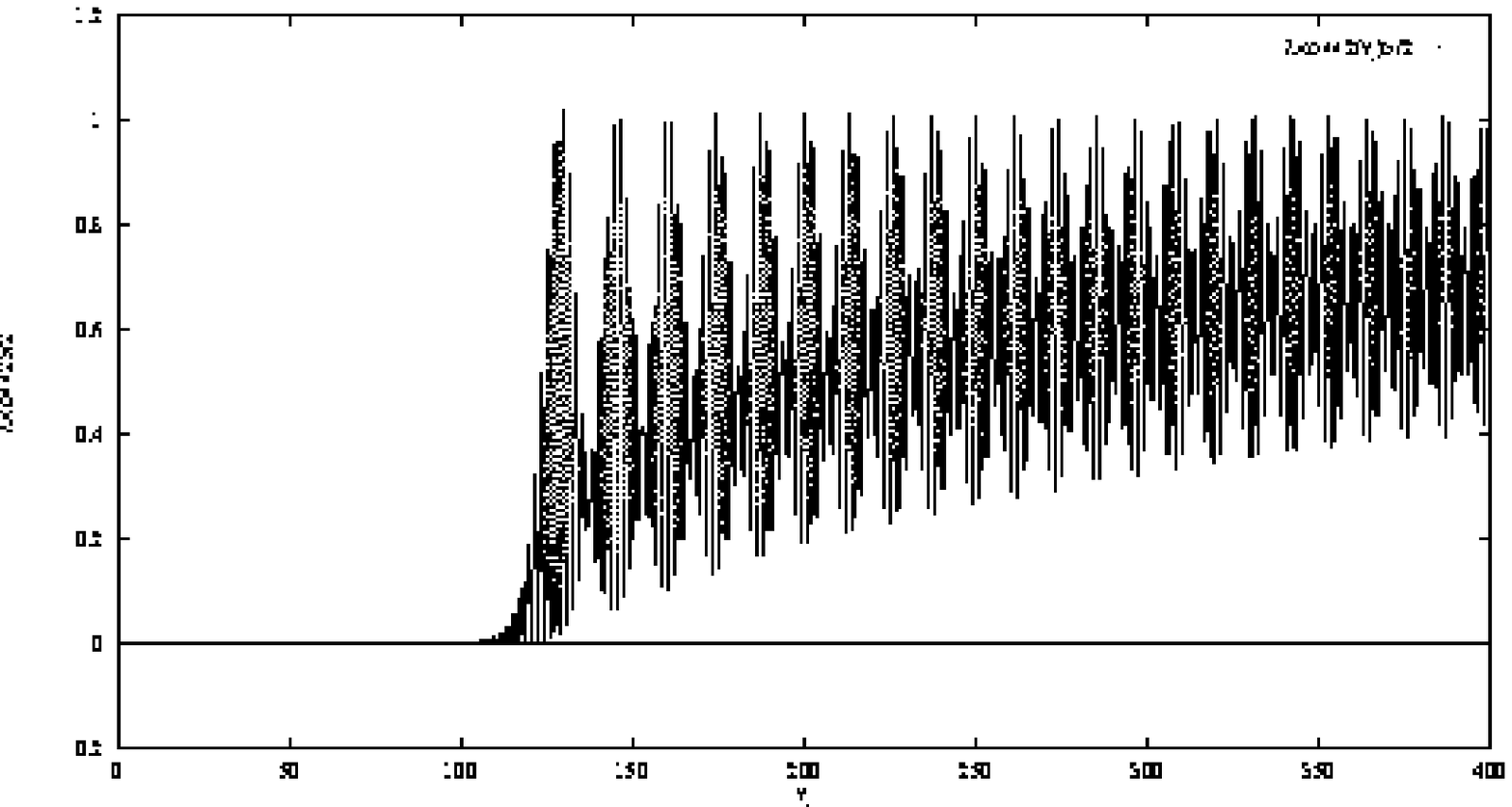,width=5.50in} 
\end{center}
\caption{Plot of $\lambda \langle \vphi^2 \rangle/2$ vs $\eta$, 
with $\Mpl/m = 1.0 \times 10^{14}$.}
\label{fig-run19fluc}
\end{figure}

\begin{figure}[htb]
\begin{center}
\epsfig{file=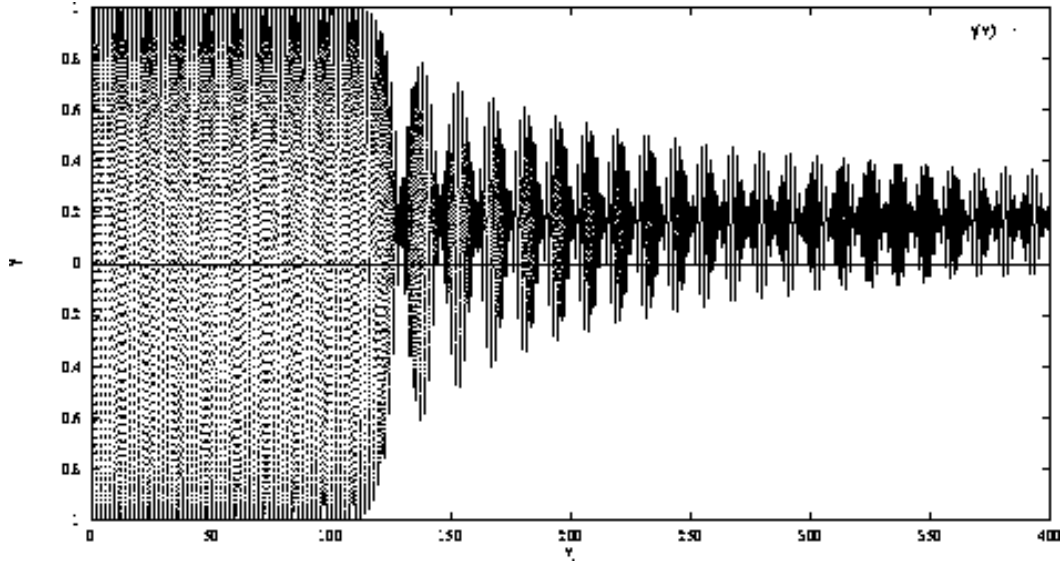,width=5.50in} 
\end{center}
\caption{Plot of $\gamma$ vs $\eta$, 
with $\Mpl/m = 1.0 \times 10^{14}$.}
\label{fig-fun19eos}
\end{figure}

\begin{figure}[htb]
\begin{center}
\epsfig{file=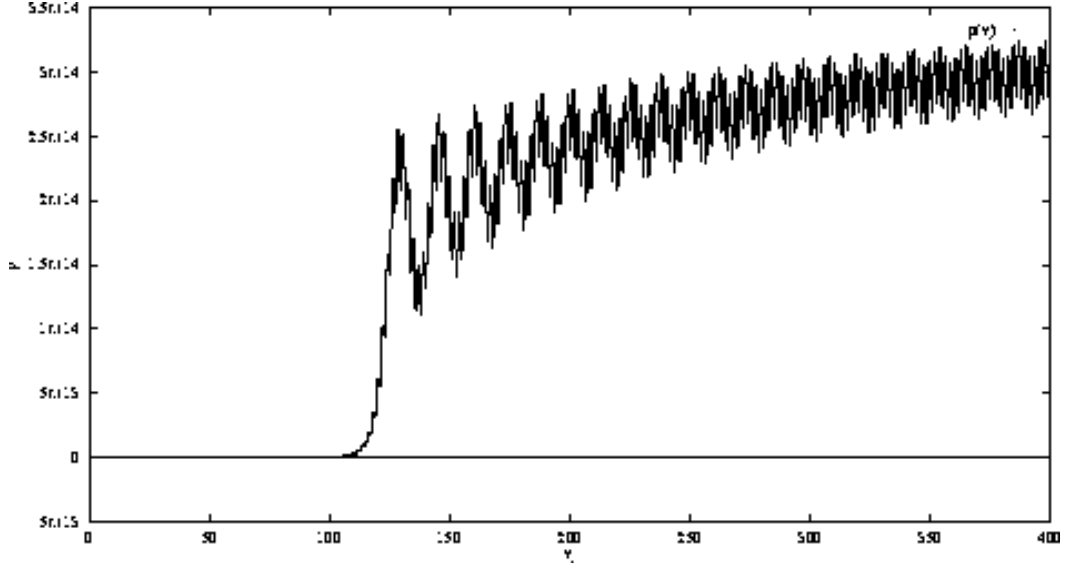,width=5.50in} 
\end{center}
\caption{Plot of $\rho_{\text{{\tiny Q}}}$ vs $\eta$, 
with $\Mpl/m = 1.0 \times 10^{14}$.}
\label{fig-run19rhoq}
\end{figure}

\begin{figure}[htb]
\begin{center}
\epsfig{file=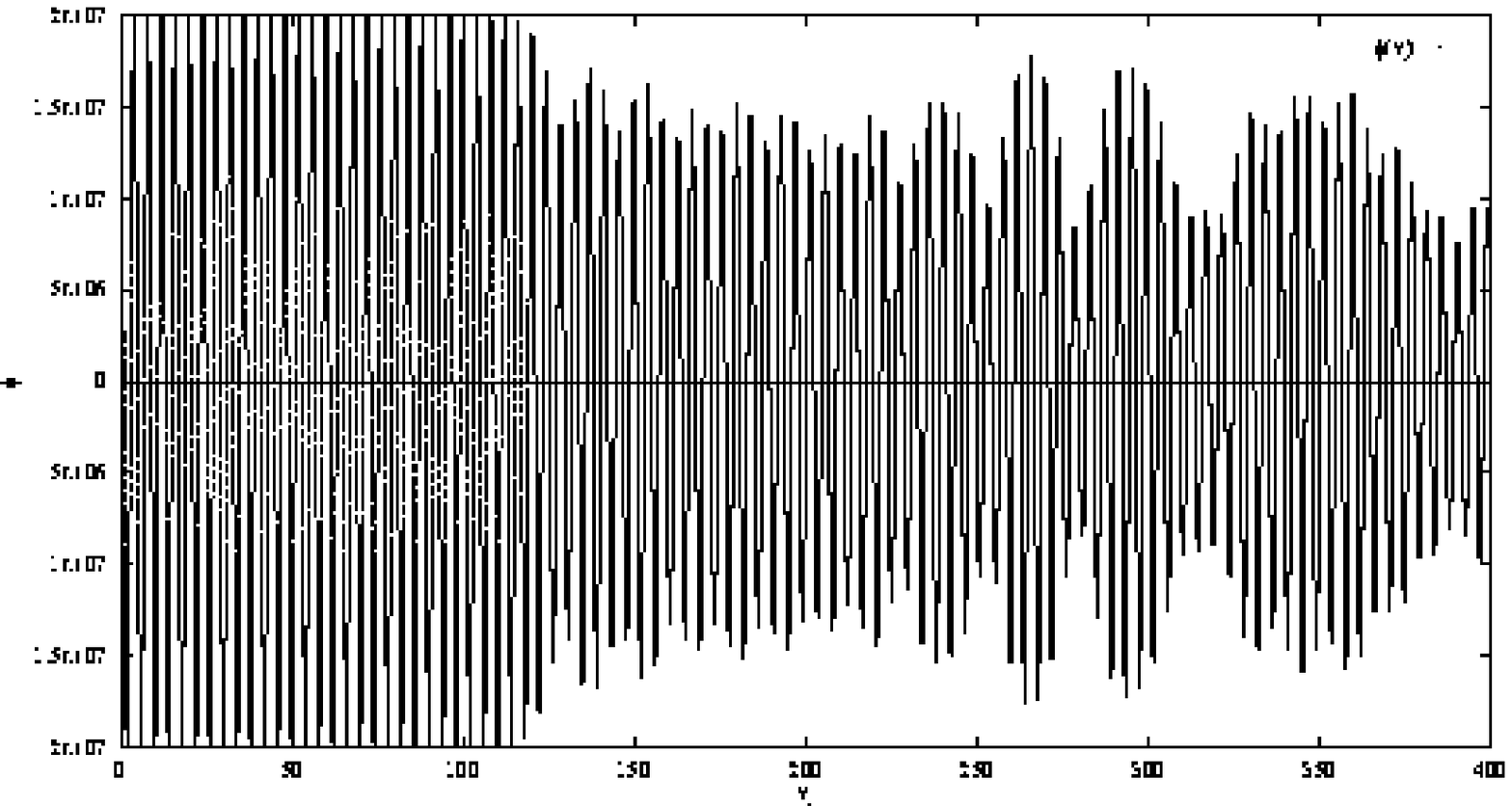,width=5.50in} 
\end{center}
\caption{Plot of $\phi$ vs $\eta$, 
with $\Mpl/m = 1.0 \times 10^{12}$.}
\label{fig-run33phi}
\end{figure}

\begin{figure}[htb]
\begin{center}
\epsfig{file=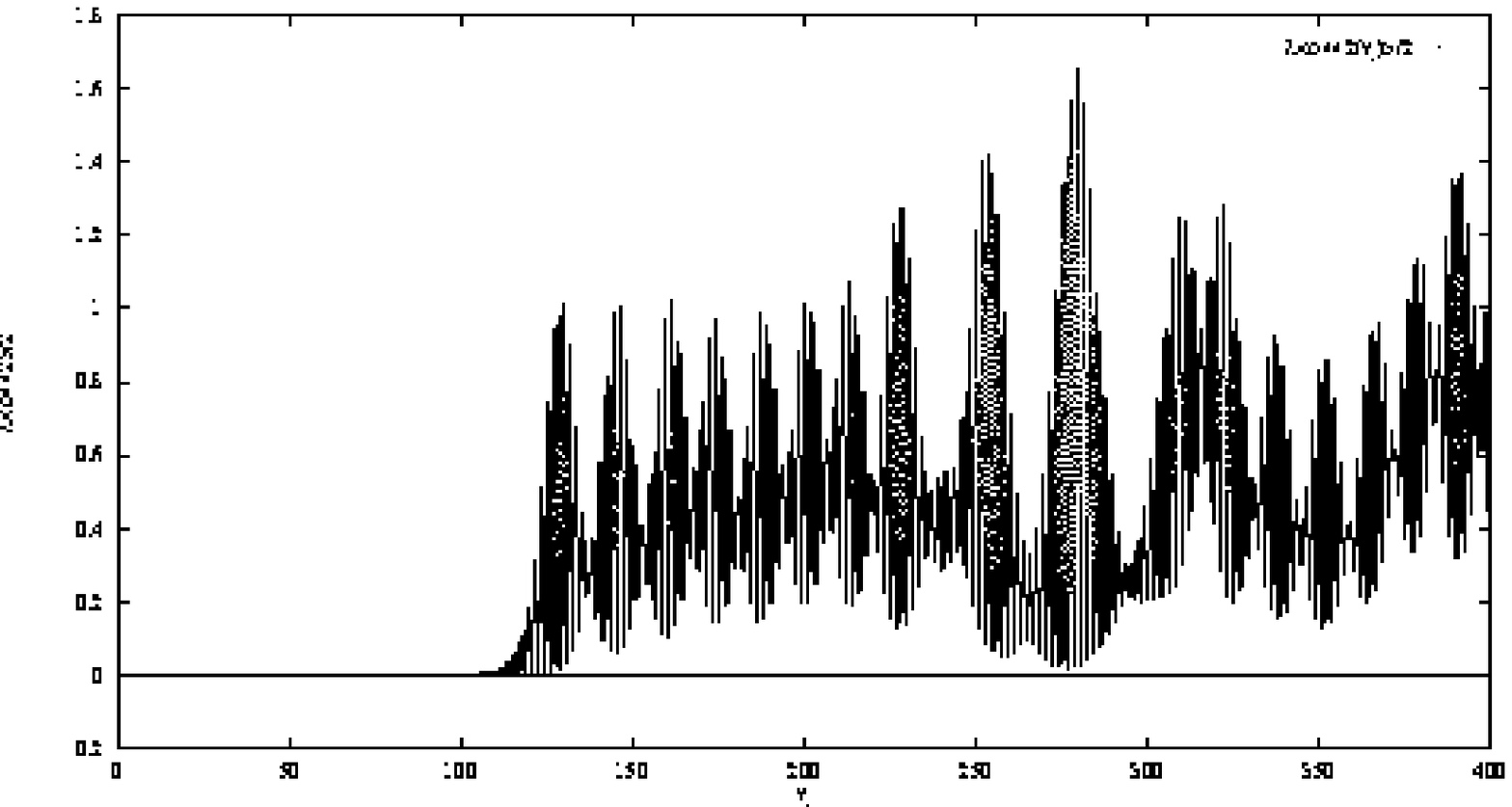,width=5.50in} 
\end{center}
\caption{Plot of $\lambda \langle \vphi^2 \rangle/2$ vs $\eta$,
with $\Mpl/m = 1.0 \times 10^{12}$.}
\label{fig-run33fluc}
\end{figure}

\begin{figure}[htb]
\begin{center}
\epsfig{file=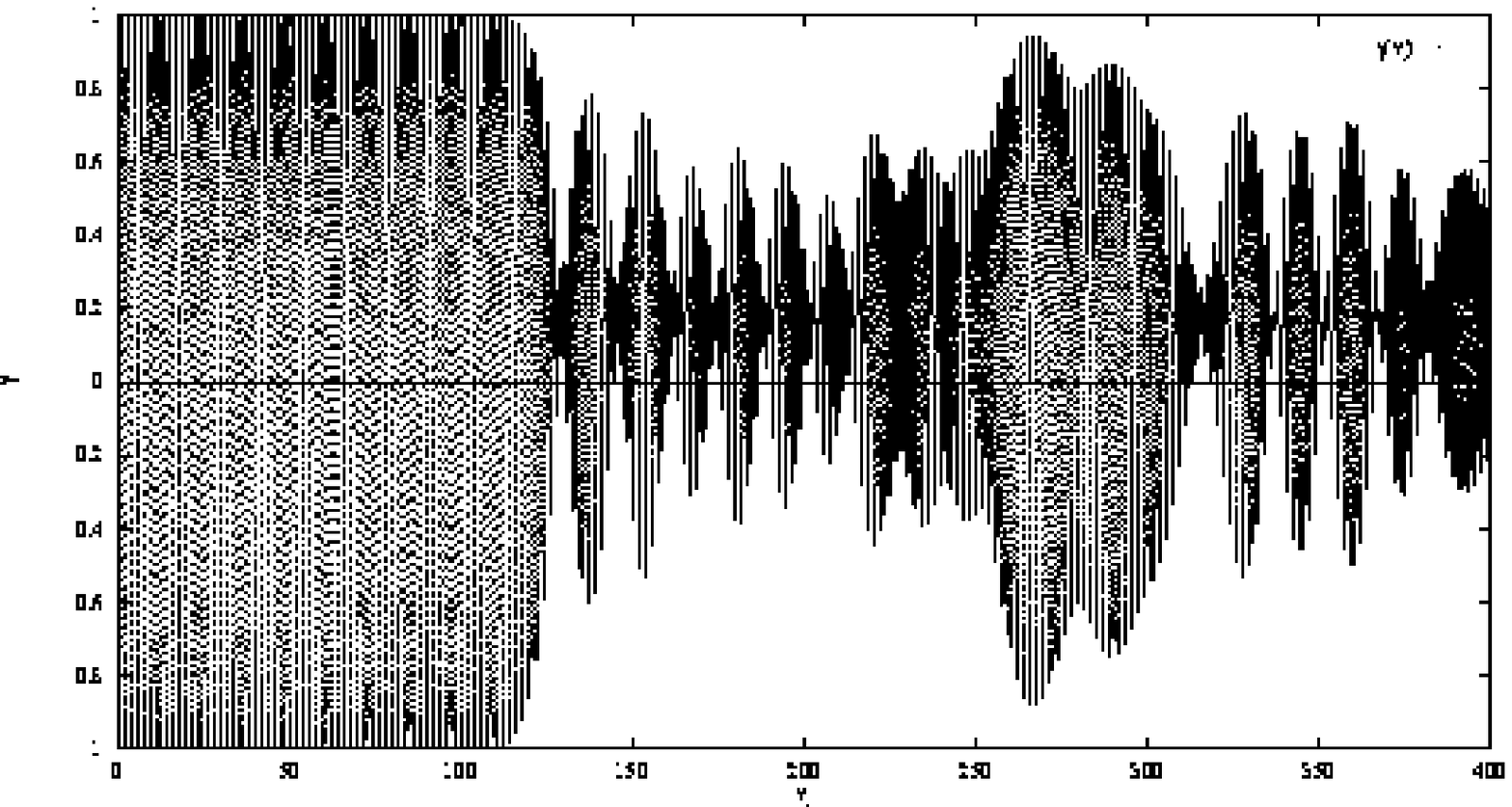,width=5.50in} 
\end{center}
\caption{Plot of $\gamma$ vs $\eta$, 
with $\Mpl/m = 1.0 \times 10^{12}$.}
\label{fig-run33eos}
\end{figure}

\begin{figure}[htb]
\begin{center}
\epsfig{file=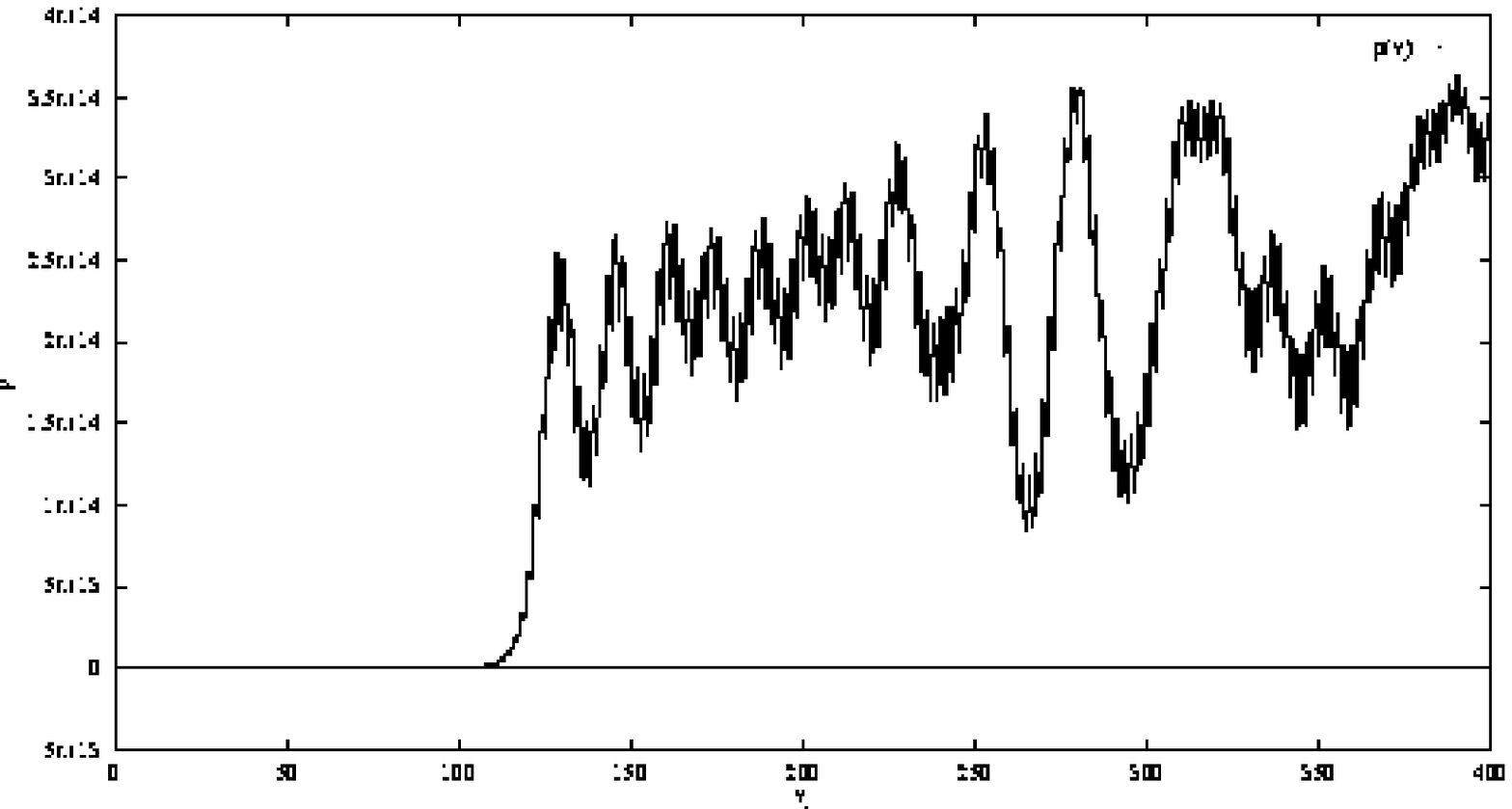,width=5.50in} 
\end{center}
\caption{Plot of $\rho_{\text{{\tiny Q}}}$ vs $\eta$, 
with $\Mpl/m = 1.0 \times 10^{12}$.}
\label{fig-run33rhoq}
\end{figure}

\begin{figure}[htb]
\begin{center}
\epsfig{file=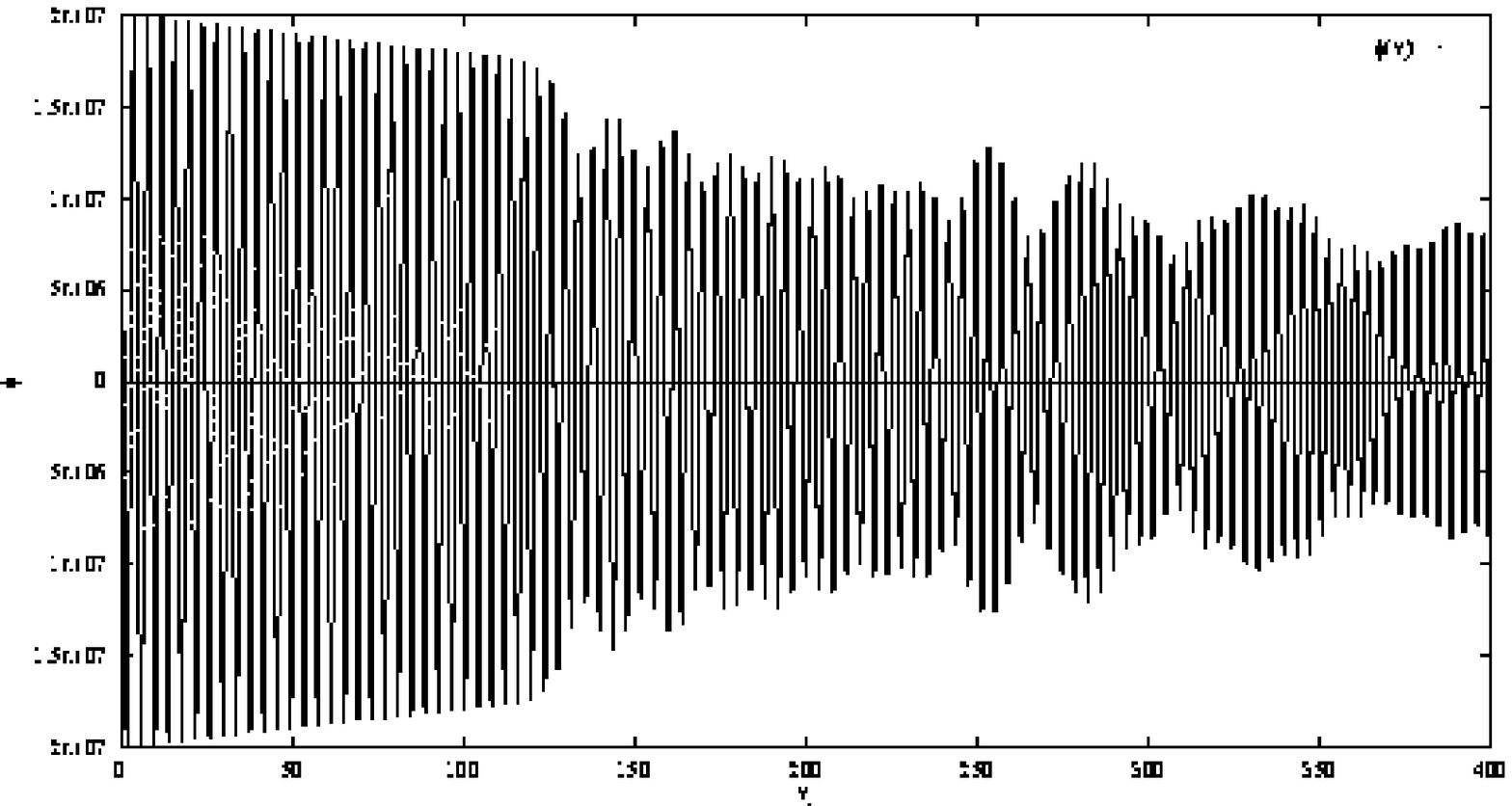,width=5.50in} 
\end{center}
\caption{Plot of $\phi$ vs $\eta$, 
with $\Mpl/m = 6.0 \times 10^{10}$.}
\label{fig-run18phi}
\end{figure}

\begin{figure}[htb]
\begin{center}
\epsfig{file=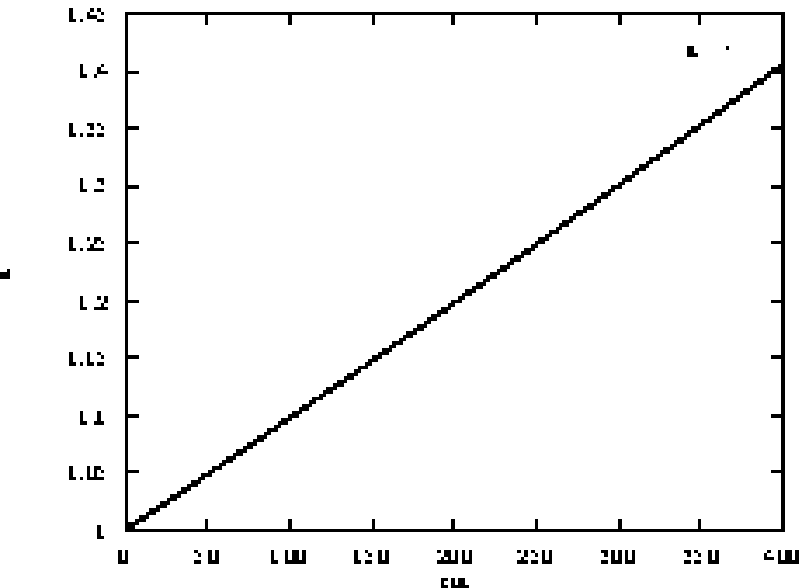,width=3.375in} 
\end{center}
\caption{Plot of $a$ vs $\eta$, 
with $\Mpl/m = 6.0 \times 10^{10}$.}
\label{fig-run18a}
\end{figure}

\begin{figure}[htb]
\begin{center}
\epsfig{file=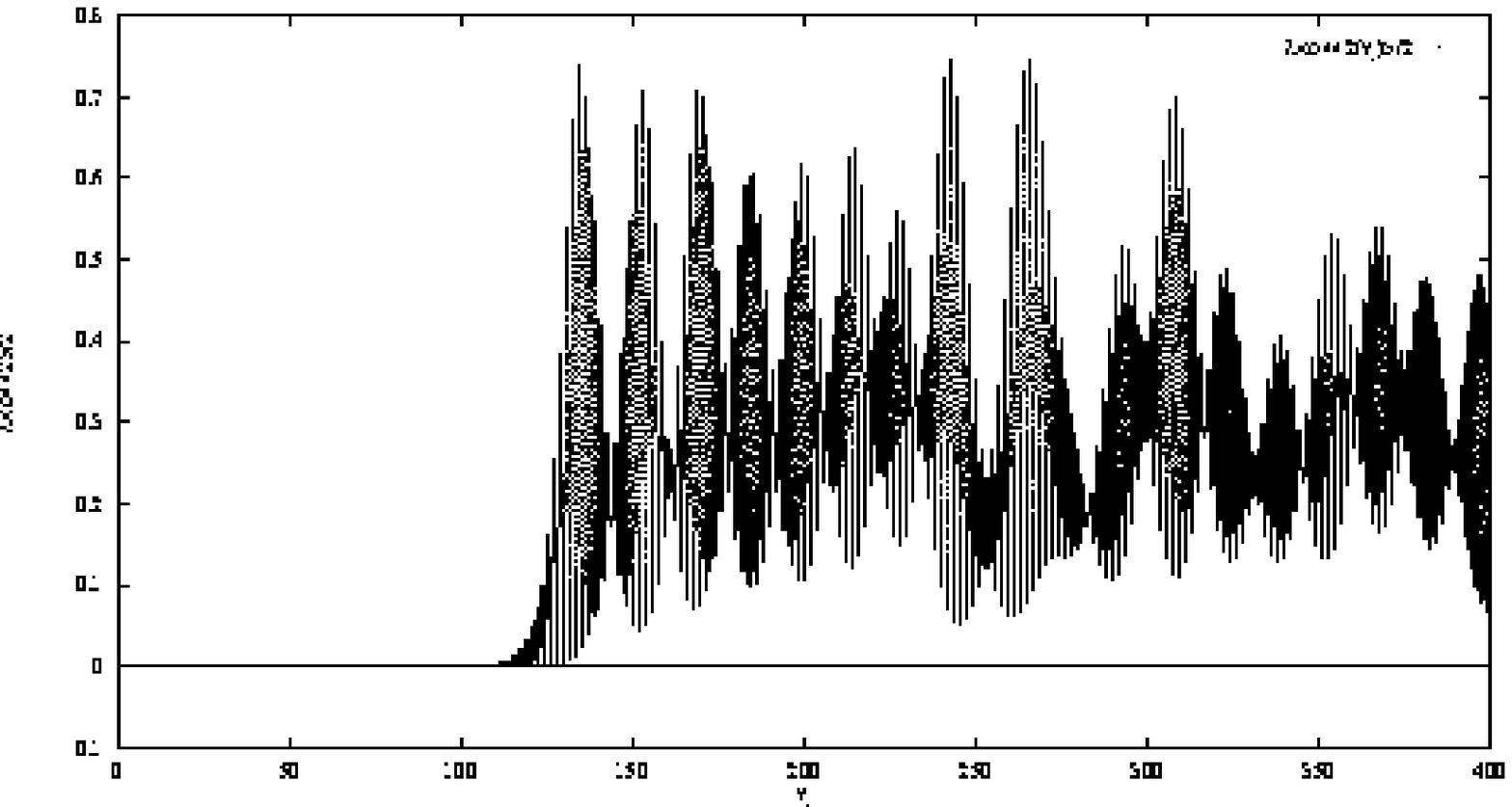,width=5.50in} 
\end{center}
\caption{Plot of $\lambda \langle \vphi^2 \rangle/2$ vs $\eta$,
with $\Mpl/m = 6.0 \times 10^{10}$.}
\label{fig-run18fluc}
\end{figure}

\begin{figure}[htb]
\begin{center}
\epsfig{file=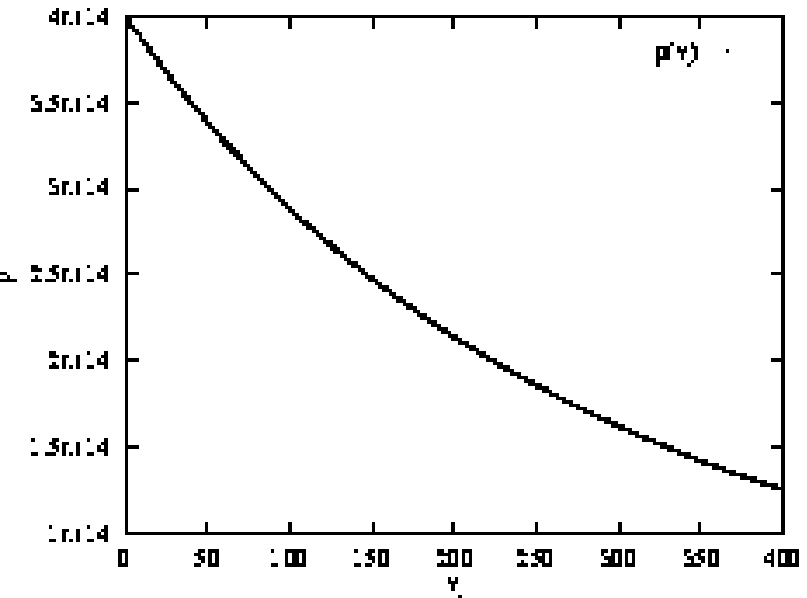,width=3.375in} 
\end{center}
\caption{Plot of $\rho$ vs $\eta$,
with $\Mpl/m = 6.0 \times 10^{10}$.}
\label{fig-run18rho}
\end{figure}

\begin{figure}[htb]
\begin{center}
\epsfig{file=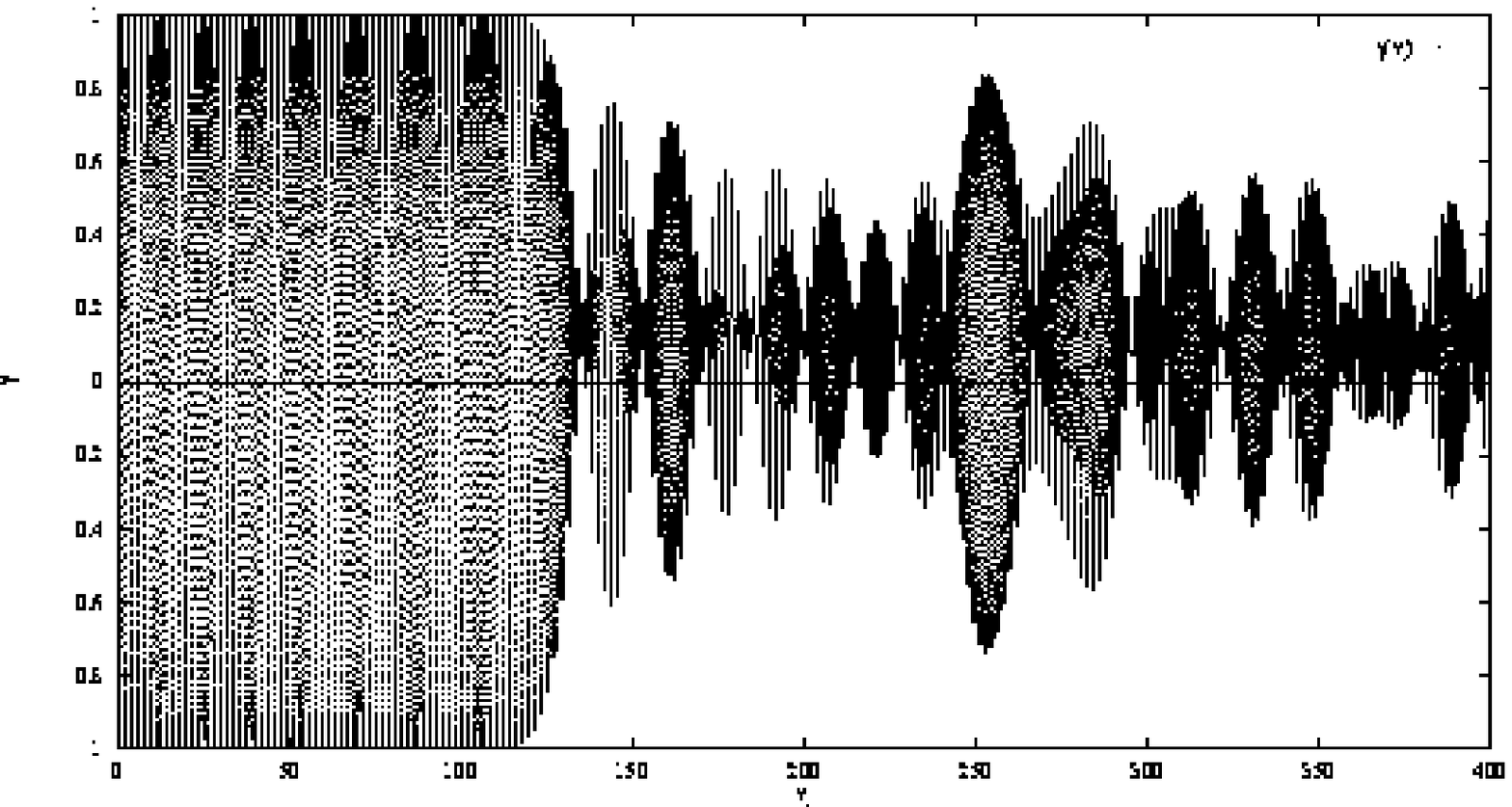,width=5.50in} 
\end{center}
\caption{Plot of $\gamma$ vs $\eta$,
with $\Mpl/m = 6.0 \times 10^{10}$.}
\label{fig-run18eos}
\end{figure}

\begin{figure}[htb]
\begin{center}
\epsfig{file=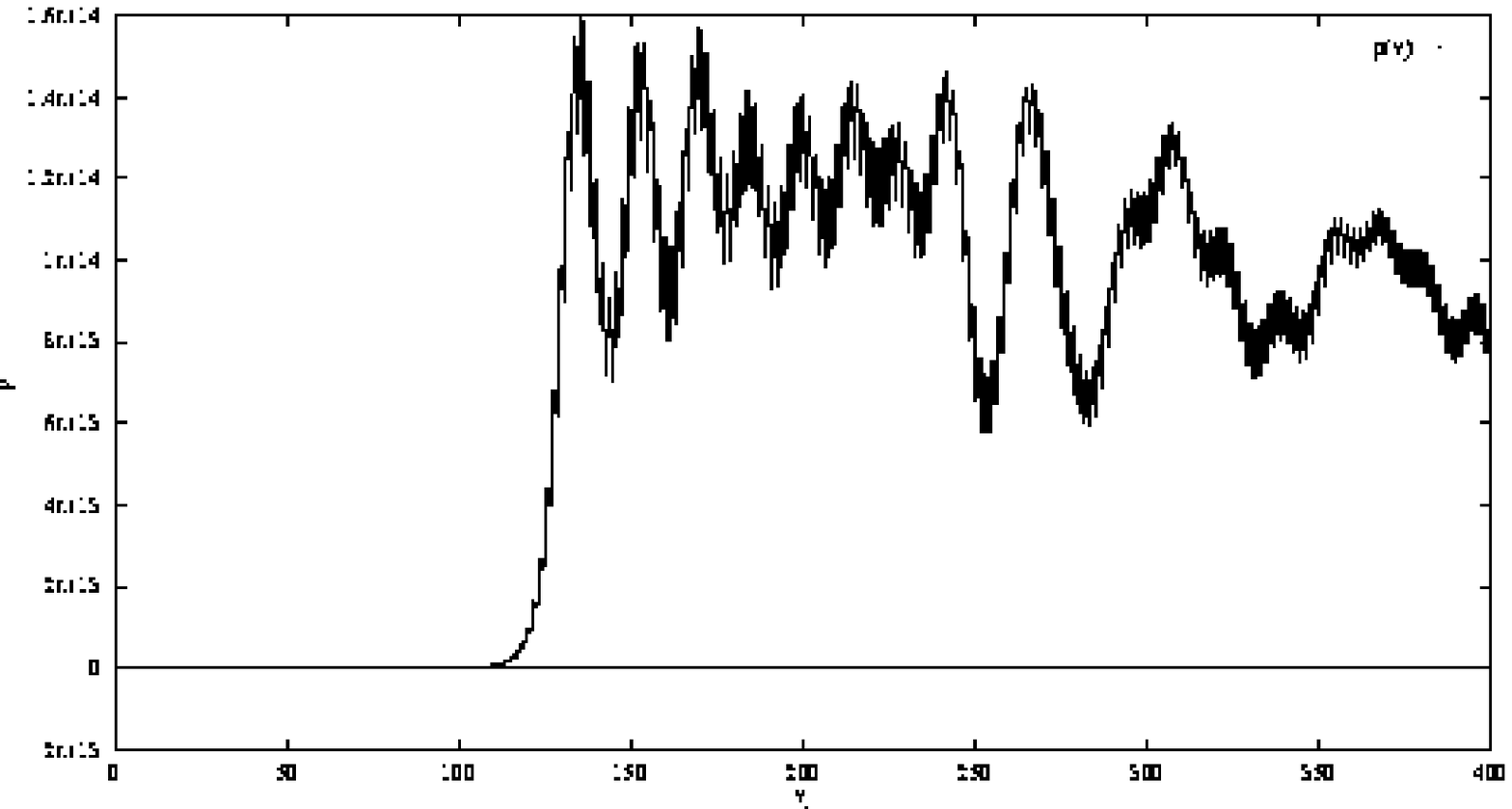,width=5.50in} 
\end{center}
\caption{Plot of $\rho_{\text{{\tiny Q}}}$ vs $\eta$,
with $\Mpl/m = 6.0 \times 10^{10}$.}
\label{fig-run18rhoq}
\end{figure}

\begin{figure}[htb]
\begin{center}
\epsfig{file=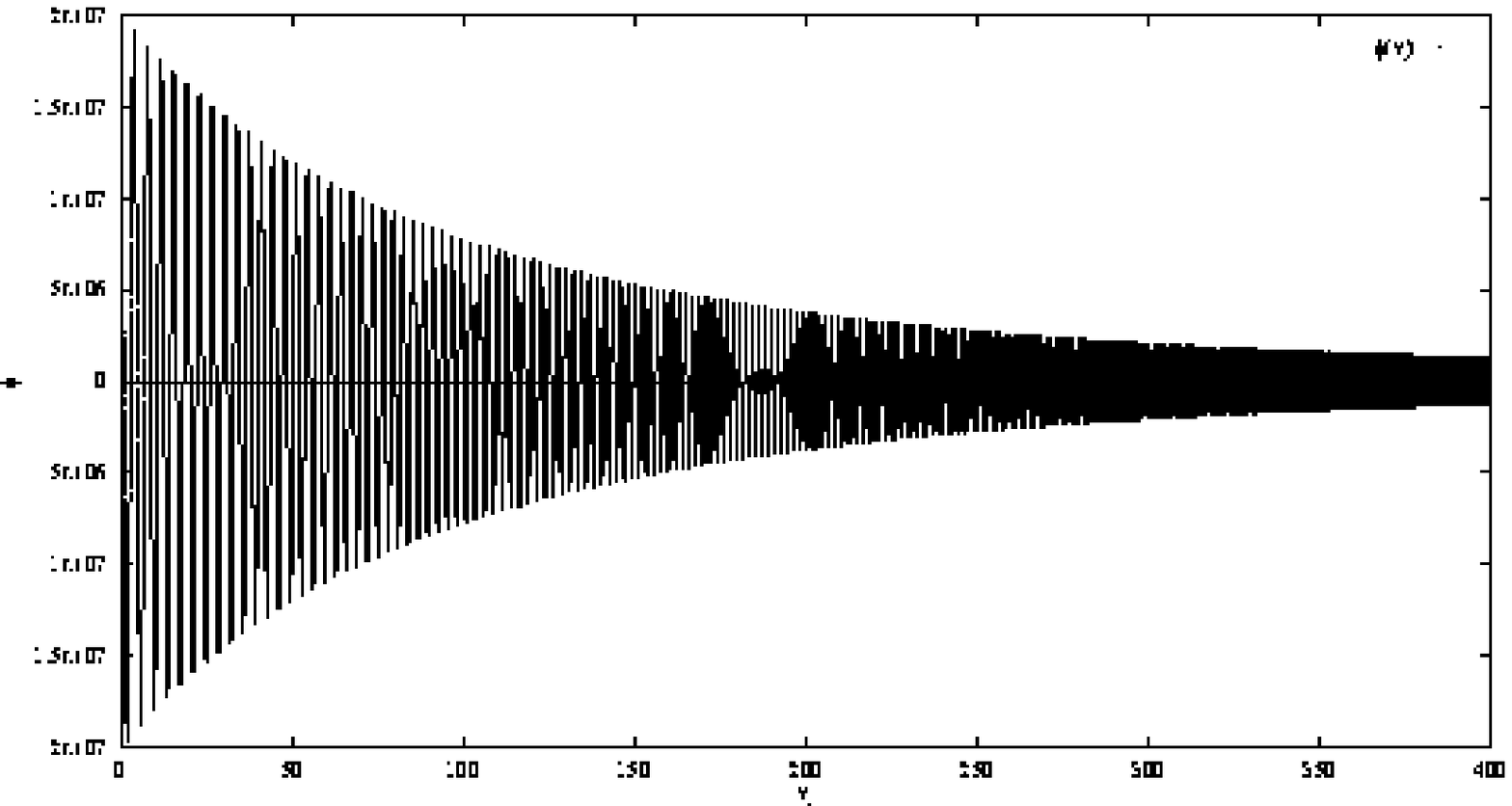,width=5.50in} 
\end{center}
\caption{Plot of $\phi$ vs $\eta$,
with $\Mpl/m = 6.0 \times 10^{9}$.}
\label{fig-run14phi}
\end{figure}

\begin{figure}[htb]
\begin{center}
\epsfig{file=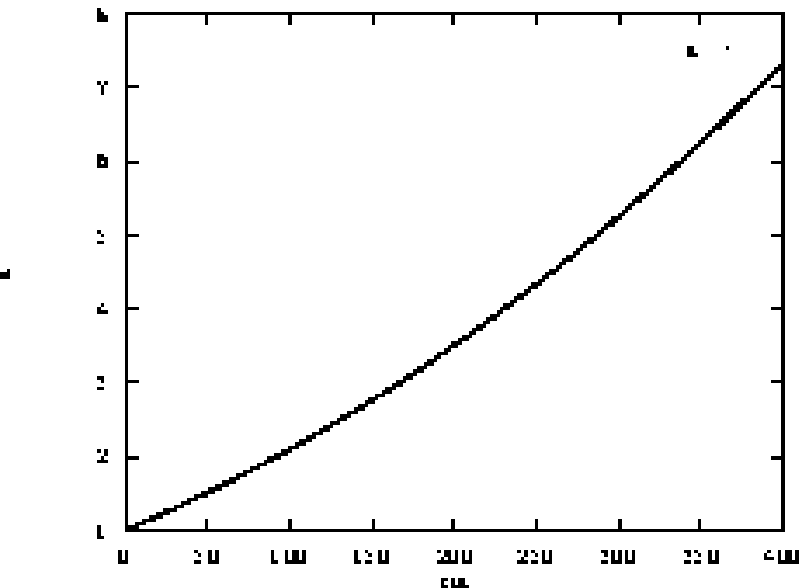,width=3.375in} 
\end{center}
\caption{Plot of $a$ vs $\eta$,
with $\Mpl/m = 6.0 \times 10^{9}$.}
\label{fig-run14a}
\end{figure}

\begin{figure}[htb]
\begin{center}
\epsfig{file=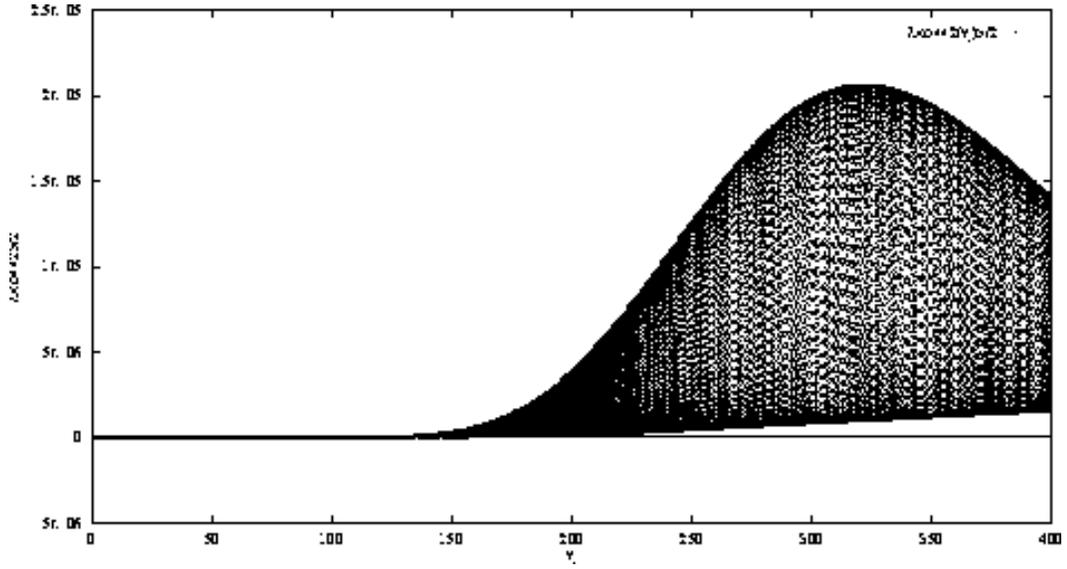,width=5.50in} 
\end{center}
\caption{Plot of $\lambda \langle \vphi^2 \rangle/2$ vs $\eta$,
with $\Mpl/m = 6.0 \times 10^{9}$.}
\label{fig-run14fluc}
\end{figure}

\begin{figure}[htb]
\begin{center}
\epsfig{file=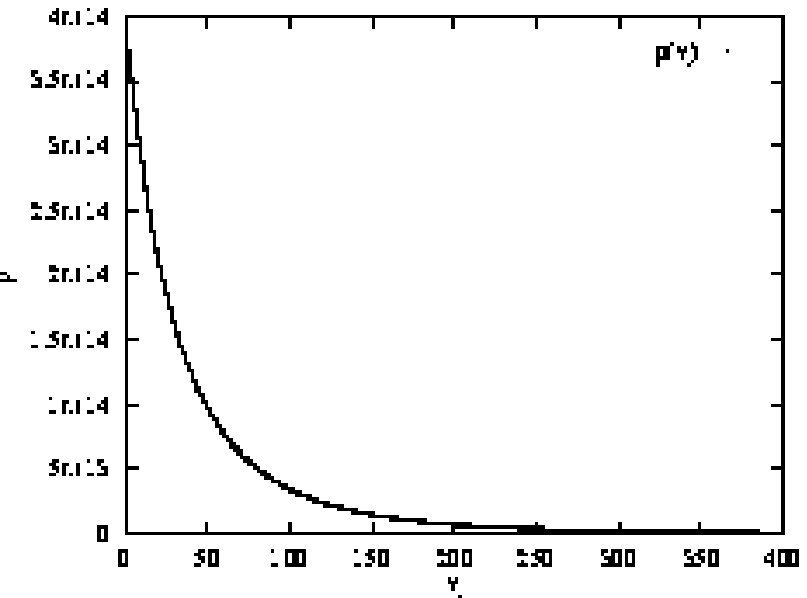,width=3.375in} 
\end{center}
\caption{Plot of $\rho$ vs $\eta$,
with $\Mpl/m = 6.0 \times 10^{9}$.}
\label{fig-run14rho}
\end{figure}

\begin{figure}[htb]
\begin{center}
\epsfig{file=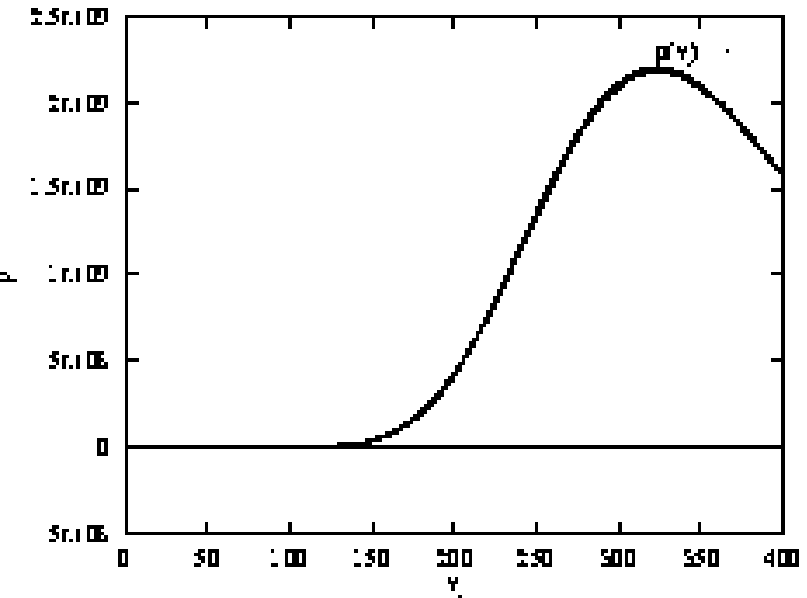,width=3.375in} 
\end{center}
\caption{Plot of $\rho_{\text{{\tiny Q}}}$ vs $\eta$,
with $\Mpl/m = 6.0 \times 10^{9}$.}
\label{fig-run14rhoq}
\end{figure}

A primary goal of this work is the quantitative study of the effect of 
spacetime dynamics on the parametric resonance energy-transfer mechanism
in nonequilibrium zero-mode oscillations of a quantum field.  As discussed in 
Sec.~\ref{sec-reheating}, this energy transfer, and the corresponding 
damping of the mean field due to back reaction, occur on a time scale of order 
$\tau_1$ defined in Eq.\ (\ref{eq-deftau1}).  We numerically evolved the 
evolution equations for $a$, $\phih$, and 
$\langle \vphi_{\text{{\tiny H}}}^2 \rangle$ for various values of $\Mpl/m$, 
ranging from very large values (corresponding to Minkowski space),
to small values (corresponding to a strong-curvature, rapid-expansion regime). 
Figs.~\ref{fig-run19phi}--\ref{fig-run14rhoq} show the resulting time 
dependences for the mean field $\phih$, the scale factor $a$, 
the variance, $\lambda \langle \vphi_{\text{{\tiny H}}}^2 \rangle/2$, the 
energy density $\rho$, the energy density in quantum modes 
$\rho_{\text{{\tiny Q}}}$ [defined in Eq.\ (\ref{eq-rhoqdef})], and the 
pressure-to-energy-density ratio $\gamma$.
The different solutions plotted correspond to different values of $\Mpl/m$,
with $\lambda = 10^{-14}$, $K/m = 50$, and $\phi(\eta_0)/m = 2.0 \times
10^7$.  As discussed in Sec.~\ref{sec-numerics}, a physical momentum
cutoff $K$ was used. The values chosen for $\Mpl/m$ were $10^{14}$,
$10^{12}$, $6 \times 10^{10}$, and $6 \times 10^9$.  
The choice of $\phih(\eta_0)$ and $\lambda$ fixes $\eta_0$ by Eq.\ (\ref{eq-asol})
and $H_0$ by Eq.\ (\ref{eq-defhinv}).  
Table~\ref{table-runparms} shows (in units where $m=1$) 
the values of $\Mpl$, the inverse Hubble constant $H^{-1}(\eta_0)$,
and the figure numbers in which the corresponding solutions are
plotted.   The
$H^{-1}(\eta_0)$ column is the initial inverse Hubble constant, which
gives the initial time scale for cosmic expansion.
Figs.~\ref{fig-run19phi}--\ref{fig-run14rhoq} plot 
the resulting solutions.
\begin{table}[htb]
\begin{center}
\begin{tabular}{|l||l|l|l|l||l|} \hline
Figures & $\phih(\eta_0)$ & $\lambda$ & $\Mpl$ & $K$ & $H^{-1}(\eta_0)$ \\ 
\hline
\ref{fig-run19phi}--\ref{fig-run19rhoq} & $2 \times 10^7$ & $1 \times
10^{-14}$ & $1 \times 10^{14}$ & 50.0 & $1.7275 \times 10^6$ \\
\ref{fig-run33phi}--\ref{fig-run33rhoq} & $2 \times 10^7$ & $1 \times
10^{-14}$ & $1 \times 10^{12}$ & 50.0 & $1.7275 \times 10^4$ \\
\ref{fig-run18phi}--\ref{fig-run18rhoq} & $2 \times 10^7$ & $1 \times
10^{-14}$ & $6 \times 10^{10}$ & 50.0 & $1.0364 \times 10^3$ \\
\ref{fig-run14phi}--\ref{fig-run14rhoq} & $2 \times 10^7$ & $1 \times
10^{-14}$ & $6 \times 10^9$ & 50.0 & 103.65 \\ \hline
\end{tabular}
\end{center}
\caption{Values of parameters for numerical solutions of Eqs.\
(\ref{eq-eveqp}), (\ref{eq-eveqa}), (\ref{eq-eveqmf}).}
\label{table-runparms}
\end{table}

The time scales defined in Sec.~\ref{sec-reheating} can now be explicitly
computed. Using Eqs.\ (\ref{eq-deff}) and (\ref{eq-defhinv}), we have 
$f(\eta_0) = \sqrt{2}$, $\rho_0 = \phih^2_0$, and 
\begin{equation}
H(\eta_0) = \sqrt{\frac{8 \pi \phih_0^2}{3\Mpl^2}} \equiv H_0.
\end{equation}
Using Eq.\ (\ref{eq-deftau0}), we find $\tau_0 \simeq 4.118 \, 32 \; m^{-1}$.  
The value of $\tau_1$ predicted by Eq.\ (\ref{eq-deftau1}) is 132.624 $m^{-1}$,
which is very close to the value predicted by Eq.\ (\ref{eq-adtau1}), 
132.759 $m^{-1}$.  For the cases $\Mpl/m = 10^{14}$ and $10^{12}$, it is clear 
from Table~\ref{table-runparms} that $H^{-1}_0 \gg \tau_1$, so that
the effect of cosmic expansion is expected to be insignificant on the
preheating time scale $\tau_1$.  For the case $\Mpl = 6 \times 10^{10}$,
$1 / ( H_0 \tau_1) \sim 7.8$, so that the effect of cosmic expansion should
be apparent and non-negligible.  For the case $\Mpl = 6 \times 10^9$,
$1 / (H_0 \tau_1) \sim 0.78$, and cosmic expansion should have a 
significant effect on parametric amplification of quantum fluctuations.

Figs.~\ref{fig-run19phi}--\ref{fig-run33rhoq} show the dynamics of the
mean field and variance in the regime of very weak cosmic expansion,
$H^{-1} \ll \tau_1$.  
As expected, under the influence of the elliptically
oscillating mean field, the variance $\langle \vphi_{\text{{\tiny H}}}^2
\rangle$ grows exponentially in time until $\lambda \langle
\vphi_{\text{{\tiny H}}}^2 \rangle/2$ is of the same order as
$m^2 + \lambda \phih^2/2$, at which point back reaction shuts off the
resonant transfer of energy to the inhomogeneous modes.  The time scale
for the variance to become of order unity can be clearly seen to be
$\sim \tau_1$.  As seen previously in studies of preheating dynamics in 
Minkowski
space \cite{boyanovsky:1996b}, on the time scale $\sim \tau_1$, the
mean field decouples from its own fluctuations and oscillates with
an asymptotically finite amplitude, given by \cite{boyanovsky:1996b}
$\lambda \bar{\phih^2}/ (2 m^2) = 0.914$.  In the Minkowski space limit,
corresponding to $\Mpl/m \rightarrow \infty$, covariant conservation
of the energy-momentum tensor implies that $d{\rho}/dt = 0$.  This was 
verified for the case of $\Mpl/m = 10^{14}$, where no change in $\rho$ was 
detected to within the numerical precision of our algorithm,
as expected from dimensional analysis of Eq.\ (\ref{eq-eveqa}).  The increase
in the scale factor for these cases was within a few parts in $10^6$ of
the initial value $a(\eta_0) = 1$.  The asymptotic equation of state
plotted in Fig~\ref{fig-run18eos} is observed to be $\bar{\gamma} \sim
0.18$.  This is exactly what would be predicted for a two-fluid model
consisting of a mean field with equation of state given by
Eq.\ (\ref{eq-aseos}),
$\bar{\gamma}_{\text{{\tiny C}}} \simeq 0.0288$, and a relativistic gas
corresponding to the energy density of the $\vphi$ field, with
$\bar{\gamma}_{\text{{\tiny Q}}} \simeq 0.3333$.  The average 
$\bar{\gamma}_{\text{{\tiny Q}}} + \bar{\gamma}_{\text{{\tiny C}}} = 0.182$.

For the case $\Mpl/m = 6 \times 10^{10}$, the effect of cosmic expansion
is clearly visible in Figs.~\ref{fig-run18phi}--\ref{fig-run18rhoq}.
In Fig.~\ref{fig-run18phi}, the coherent oscillations of the mean field
for the time period $0 < \eta - \eta_0 < \; \sim 27 \tau_0$ are clearly seen
to be redshifted by the usual $1/a$ factor expected from the
Hubble damping term in Eq.\ (\ref{eq-eveqp}).  The expected asymptotic equation
of state (taking into account cosmic expansion) computed from a
simple two-fluid model is $\sim 0.133$, in agreement
with Fig.~\ref{fig-run18eos}.

Figs.~\ref{fig-run14phi}--\ref{fig-run14rhoq} show the solution
for $\Mpl/m = 6 \times 10^9$.  
In this case, $1 /(H_0 \tau_1)
\sim 0.781$.  From Fig.~\ref{fig-run14fluc}, we clearly see that
cosmic expansion renders parametric amplification of quantum fluctuations 
an inefficient mechanism
of energy transfer to the inhomogeneous modes.  The very rapid oscillations
of the mean field at late times are due to the conformal time scale used
here, in which the oscillation period of the mean field decreases inversely
with $a$.  Damping of the mean field due to cosmic expansion is the dominant 
effect in Fig.~\ref{fig-run14phi}.  The power-law decrease in energy density
consistent with matter having an effective equation of state
$\bar{\gamma} \simeq 0.0288$ can be seen in
Fig.~\ref{fig-run14rho}.  At $\eta = 300 \; m^{-1}$, the ratio
$\rho_{\text{{\tiny Q}}} / \rho \sim 0.0002$, so the fraction of
energy density in the inhomogeneous modes is negligible in comparison
to the classical, mean-field contribution.   
Since the variance $\langle \vphi_{\text{{\tiny H}}}^2\rangle$ is never
large enough that it dominates the effective mass $M$, the mode functions
approximately obey the one-loop equation, in which the effective
frequency is $k^2 + a^2 (m^2 + \lambda \phih^2)$, neglecting the $a''/a$ term.
The width of the resonance can then be shown to be approximately
given by $k^2 \leq \lambda \phih^2_0 /2$.  The variance is damped by
$1/a^2$ due to cosmic redshift, so when $H^{-1} \sim \tau_1$, the variance 
never grows to be of order unity.

In addition to varying $\Mpl$, the coupling $\lambda$ was varied, with
results in agreement with Eq.\ (\ref{eq-deftau1}), showing a logarithmic
dependence of $\tau_1$ on $\lambda^{-1}$.  

\section{Summary}
\label{sec-discussion}
In this Chapter we use the minimally coupled,
quartically self-interacting scalar O$(N)$ field theory as a model for the
inflaton field, and study its nonequilibrium dynamics nonperturbatively
in a spatially flat FRW spacetime whose evolution is driven by the quantum
field.  We solve the coupled, self-consistent semiclassical Einstein
equation, mean-field equation, and conformal-mode-function equations 
numerically.  Our goal in this Chapter is to study the effects of spacetime
dynamics on the mean field, and parametric amplification of quantum 
fluctuations. This process of energy transfer from the mean field to the 
inhomogeneous modes is inherently nonperturbative and nonequilibrium.
It requires the use of the closed-time-path formalism
and the two-particle-irreducible effective action.
As our focus in this Chapter is on the parametric amplification of quantum 
fluctuations, we assume unbroken symmetry.  Our analysis is, therefore, most 
relevant to reheating in chaotic inflation scenarios. We use  the two-loop,
covariant equations for the mean field and the two-point function for the 
fluctuation field derived in Chapter~\ref{chap-oncst}
and study the case of leading order in the
$1/N$ expansion, an approximation which is valid on time scales much shorter 
than the mean-free time for multiparticle scattering ($\tau_2$).  
For FRW spacetimes, we use the well-established adiabatic regularization
procedure to obtain finite expressions for the renormalized variance and 
energy-momentum tensor which enter into the mean-field equation, conformal-mode
function equations, and the semiclassical Einstein equation. In our
approach, covariant conservation of the energy-momentum tensor is preserved
at all times, as it should be.  (It should not and
need not be put in by hand, as was done in a recent study of reheating in 
a fixed background FRW spacetime \cite{boyanovsky:1997c}.)
We use the adiabatic vacuum construction (matched at conformal-past infinity)
to define the quantum state for the fluctuation field in FRW spacetime with 
asymptotic de~Sitter initial conditions; this is the most physical vacuum 
construction given the decidedly nonadiabatic conditions which prevail at 
the end of inflation. The instantaneous Hamiltonian diagonalization 
constructions used in earlier studies of reheating in curved spacetime 
\cite{boyanovsky:1996c,khlebnikov:1997a} are known to be problematic
\cite{fulling:1989a}.

We evolved the coupled dynamical equations for the mean field, variance, 
and scale factor using standard numerical methods, for time scales 
of 400 $m^{-1}$, where the initial period of mean-field oscillations is
4.11832 $m^{-1}$.
Several regimes for the parameters of the system were investigated.
From the solutions of the dynamical equations we studied the
behavior of the scale factor $a$, the mean field $\phih$, the energy
density $\rho$, pressure-to-energy-density ratio $\gamma$, and the
inhomogeneous-mode (fluctuation-field) energy density 
$\rho_{\text{{\tiny Q}}}$.  The solutions of the dynamical equations
were analyzed for a variety of values for $\Mpl \tau_0$, the parameter
which controls
the rate of cosmic expansion relative to the time scale for mean-field
oscillations in the model.  In the case of negligible cosmic expansion, 
corresponding to very small initial inflaton amplitude, the dynamics is 
identical to that seen in the group 2B (see Sec.~\ref{sec-backissu})
studies of O$(N)$ preheating in Minkowski space \cite{boyanovsky:1996b}.  
In particular, the conservation of energy and logarithmic dependence of the 
preheating time scale $\tau_1$ on the inverse coupling $\lambda^{-1}$ [as
shown in Eq.\ (\ref{eq-deftau1})] are confirmed.  For the case of 
moderate cosmic expansion, $H(\eta_0) \tau_1 \sim 10$ [where $H(\eta_0)$ is
the Hubble parameter at the initial time $\eta_0$], energy transfer
via parametric amplification of quantum fluctuations
is still efficient, and the dynamics can be understood
using the analytic results of \cite{boyanovsky:1996b} (for Minkowski
space), in terms of the conformally transformed mean-field amplitude 
$\tilde{\phih} = \phih/a$ and oscillation period $\tilde{\tau}_0 = \tau_0/a$. 
The asymptotic effective equation of state is found to be consistent with 
the prediction of a simple two-fluid description of the late-time behavior
of the system.

The most significant physical result concerns the case of rapid cosmic
expansion, where $H^{-1}(\eta_0) \simeq \tau_1$.  In this case
we find that parametric amplification of quantum fluctuations (via parametric 
resonance) is an inefficient mechanism of energy transfer to the inhomogeneous 
modes of the inflaton, because the parametric resonance effect
is inhibited both by redshifting of the mean-field amplitude and by
the redshifting of the physical momenta of the modes out of the resonance 
band.  The energy density of particles produced through parametric resonance is
in this case redshifted so rapidly that, in our model,
the term $\lambda \langle \vphi_{\text{{\tiny H}}}^2\rangle/2$ 
never grows to be of the order of the tree-level effective mass, $m^2 +
\lambda \phih^2/2$.  As the mean-field amplitude is damped ($\propto 1/a$)
due to cosmic expansion,
eventually the resonant particle production ceases, and the mean field
oscillates with a damped envelope at late times.  This leads us to the
following conclusions:  (i) On the physical level, in chaotic inflation 
scenarios with a $\lambda \Phi^4$ inflaton minimally coupled to gravity and 
with a large initial inflaton amplitude at the end of slow roll, parametric 
amplification of the inflaton's {\em own\/} quantum fluctuations is not a 
viable mechanism 
for reheating the Universe, unless the self-coupling is significantly 
increased.\footnote{In recent work on galaxy 
formation from quantum fluctuations, Calzetta, Hu, and Matacz 
\cite{calzetta:1995a,matacz:1996a} report that
$\lambda$ can be as high as $\sim 10^{-5}$.}  
This does not imply that the phenomenon of parametric
amplification of quantum fluctuations does not play a vital role in the 
``preheating'' period of inflationary cosmology, for different models 
and/or couplings.  The interesting case of a $\phi^2 \chi^2$ model 
is currently under investigation.
(ii) On a more methodological level, we conclude that a correct study of the 
reheating period in a chaotic inflation model with large inflaton amplitude 
at the onset of reheating {\em must\/} take into account the effects of 
spacetime dynamics.  This should be carried out {\it self-consistently}
using the coupled semiclassical Einstein equation and matter-field 
equations, so that no {\em ad hoc\/} assumptions need be made about 
the effective equation of state and/or the relevant time scales involved.

A full two-loop treatment of the unbroken symmetry
mean-field dynamics of the O$(N)$ field theory [which involves solving the
nonlocal, integro-differential equations (\ref{eq-on2lge}) 
and (\ref{eq-on2lmfe})] 
includes multiparticle scattering processes, which
provide a mechanism for reheating; but they are of a qualitatively different
nature than the parametric resonance energy-transfer mechanism studied here.
In addition, the nonlocal nature of the gap equation in the full two-loop
analysis makes numerical solution of the coupled Einstein and matter equations
difficult.  In our model, scattering occurs on a time scale 
$\tau_2$ which is significantly longer than the time scale $\tau_1$ 
for parametric amplification of quantum fluctuations.  Therefore, in this model
the leading-order, large-$N$ (collisionless) approximation is sufficient
for a study of parametric amplification of quantum fluctuations.
In addition, realistic models of inflation invariably involve couplings
of the inflaton to other fields, which provides additional mechanisms
of energy transfer away from the inflaton mean field, and into 
its (or other fields') quantum modes. 

The issues involved in a systematic study of the thermalization stage
of post-inflationary physics are more complex.  A quantum kinetic field
theory treatment taking into account multiparticle scattering is required.
The two-loop, 2PI effective action is the simplest and most generally
applicable rigorous formalism which contains the leading-order multiparticle 
scattering processes.  The leading-order, $1/N$ approximation is a 
collisionless subcase of the two-loop, 2PI effective action; it is employed 
in this study in order to obtain local dynamical equations which can be
solved numerically, and is adequate for a study of parametric amplification
of quantum fluctuations.  In addition, the growth of entropy must be 
understood
within the context of a physically meaningful coarse graining of the
full time-reversal-invariant quantum dynamics of the field theory.  This
is discussed in Chapter~\ref{chap-entropy} below.  A first-principles
analysis of the thermalization stage is currently underway 
\cite{ramsey:1997e}.

\chapter{Fermion production, noise, and stochasticity}
\label{chap-fermion}

\section{Introduction}
\subsection{Background and issues}
In Chapter~\ref{chap-preheat} we studied the effect of parametric
particle creation on the inflaton dynamics in the post-inflation,
``preheating'' stage.  It was shown that analysis of the dominant physical 
processes during the early stages of reheating necessitates consideration
of self-consistent back reaction of the inflaton variance on the 
mean field and inhomogeneous modes.  This is a nonperturbative effect,
and its description requires a consistent truncation of the Schwinger-Dyson 
equations \cite{calzetta:1995b,boyanovsky:1996b,ramsey:1997b}.  Assuming 
initial conditions
conducive to efficient parametric particle creation, the end state of the
regime of parametric particle creation consists of a large inflaton variance
(i.e., on the order of the tree level terms in the inflaton's effective mass). 
Thus, both the inflaton mean field and variance should be treated on
an equal footing.  This requires a two-particle-irreducible (2PI)
formulation of the effective action which is a subclass of the master
effective action \cite{calzetta:1995b}.

During the later stages of reheating, the dynamics of the inflaton field is
thought, in the case of unbroken symmetry, to be dominated by 
fermionic particle creation  \cite{kofman:1997a}.  This stage of inflaton 
dynamics is the subject of this chapter.  We consider a model consisting of 
a scalar inflaton field $\phi$ (with $\lambda \phi^4$ self-coupling) 
coupled to a spinor field $\psi$ via a Yukawa interaction and  
we attempt to present as complete and rigorous a treatment as mandated by the 
self-consistency of the formalism and the
actual solvability of the equations. Thus, we adopt a closed-time-path (CTP),
two-particle-irreducible (2PI), coarse-grained effective action (CGEA) to
derive the dynamics of the inflaton field. We have explained the significance
of CTP and 2PI in Chapter~\ref{chap-oncst}
\cite{calzetta:1987a,calzetta:1988b,calzetta:1989a,calzetta:1989b,hu:1989a,calzetta:1995b,ramsey:1997a}, and their relevance to the study of inflaton dynamics
in Chapter~\ref{chap-preheat}.  Here, an added feature of an {\em open 
system\/} is introduced: we wish to include the averaged effect of an 
environment on the system, and a useful method is via the coarse-grained 
effective action 
\cite{hu:1991b,sinha:1991a,lombardo:1996a,dalvit:1996a,greiner:1997a}.
Let us explain the rationale for this.

\subsection{Coarse-Grained Effective Action}
In inflationary cosmology at the onset of the reheating period, the inflaton 
field's zero mode typically has a large expectation value, 
whereas all other fields coupled to the inflaton, as
well as inflaton modes with momenta greater than the Hubble constant, are 
to a good approximation in a vacuum state 
\cite{brandenberger:1985a}. This suggests imposing a physical
coarse graining in which one regards the inflaton field as the system, and the 
various quantum fields coupled to it as the environment.  From the
closed-time-path, coarse-grained effective action (CTP-CGEA) 
\cite{hu:1991b,sinha:1991a,lombardo:1996a,dalvit:1996a,greiner:1997a}
derived in Sec.~\ref{sec-necgidcst} below,
one obtains effective dynamical equations for the inflaton field, taking into 
account its effect on the environment, and back reaction therefrom.
For the present problem, the system consists of the inflaton mean field and 
variance, and the environment consists of the spinor field(s) coupled to the 
system via a Yukawa interaction.

We wish to emphasize here a subtle yet important distinction between the 
system-environment division in nonequilibrium statistical mechanics
and the system-bath 
division assumed in thermal field theory.  In the latter, one assumes that 
the propagators for the bath degrees of freedom are {\em fixed,\/}
finite-temperature equilibrium Green functions, whereas in the case of the
CTP-CGEA, the environmental propagators are {\em slaved\/} (in the sense of
\cite{calzetta:1995b}) to the dynamics of
the system degrees of freedom, and are not fixed {\em a priori\/} to be
equilibrium Green functions for all time.  This distinction is important for
discussions of fermion particle production during reheating, because it is 
only when the inflaton mean field amplitude is small enough for the use of
perturbation theory, that the system-bath split implicit in thermal
field theory can be used.  Otherwise,
one must take into account the effect of the inflaton mean field on the
bath (spinor) {\em propagators.\/}  

\subsection{Earlier work}
The first studies of particle production during reheating in inflationary
cosmology were \cite{abbott:1982a,dolgov:1982a}, where reasonable estimates
of particle production were made, but back reaction effects were not addressed.
The earliest studies of fermionic particle creation during reheating used
time-dependent perturbation theory to compute the imaginary part of
the self-energy for the zero mode of the inflaton field, which was
related to the damping parameter in a friction-type phenomenological 
term in the equation of motion for the inflaton zero mode
\cite{abbott:1982a,albrecht:1982a,morikawa:1985a,dolgov:1990a,shtanov:1995a}.
In these studies, it was
assumed that the effect of fermionic particle production
on the dynamics of the inflaton zero mode is that of a $\Gamma \dot{\phi}$
friction term.  However, it has been shown 
\cite{morikawa:1986a,calzetta:1989a,paz:1990a,stylianopoulos:1991a,kofman:1994a,boyanovsky:1995a,boyanovsky:1995d,kofman:1996a} 
that this assumption is not tenable for a wide variety of field-theoretic
interactions and initial conditions.  Rather, the effect of back reaction from 
particle production must be accounted for by {\em deriving\/} the effective
evolution equation for the quantum expectation value of the inflaton zero mode,
where the dynamics of the degrees of freedom of the produced particles are 
either coarse-grained (as in Sec.~\ref{sec-dsasfo}), or accounted for through 
self-consistent coupled equations (as in Sec.~\ref{sec-necgidcst}).  In 
general, particle creation leads to a {\em nonlocal}
dissipation term in the inflaton mean field equation, and it is only under
rather idealized conditions and specialized cases that one can expect the
dissipation kernel to approach a delta function (i.e., a friction term)
\cite{hu:1992a,hu:1993c,hu:1993a,hu:1994a}.  Therefore, 
\cite{abbott:1982a,dolgov:1990a,shtanov:1995a} missed
the time-nonlocal nature of fermion particle production and its effect
on the dynamics of the inflaton field.  In addition, these
studies computed the self-energy in flat space, thereby neglecting the
effect of curved spacetime on fermion production, and did not examine
the stochastic noise arising from the coarse graining of the fermion
field.  

In addition, most early studies of fermion production during reheating, in
using time-dependent perturbation theory to compute the {\em vacuum\/} particle
production rate \cite{abbott:1982a,dolgov:1990a,shtanov:1995a},
did not include the effect of back reaction of the {\em produced\/} fermion 
particles
on the particle production process itself.  In \cite{boyanovsky:1995d},
the effect of a thermal initial fermion distribution on the particle production
process was investigated (and a Pauli blocking effect was shown), but
their analytic derivation of the Pauli blocking effect involves the same
perturbative expansion (i.e., system-bath split) described above,
and therefore does not incorporate the effect of the {\em produced\/} 
fermion quanta on the particle production process.  In order to
take this effect into account, it is necessary to include the effect of the
time-varying inflaton mean field in the equation of motion for the spinor
propagator, which amounts to a coarse graining of the fermion field, in the
system-environment sense, as described above.  The perturbative amplitude 
expansion of the effective inflaton dynamics, in contrast, amounts to 
a system-bath coarse graining which does not include this back reaction 
effect.

More recent studies of fermion production during reheating 
\cite{boyanovsky:1995b,boyanovsky:1995c,boyanovsky:1995d}
obtained dynamical equations for the inflaton mean field using
a one-loop factorization of the Lagrangian,
and solved them numerically.  However, these studies
were carried out in flat space, and because they studied only the dynamics
of the inflaton mean field (and at one loop), their analysis did not
take into account the back reaction of the inflaton variance on the
fermion quantum modes, nor the back reaction of particle production
on the dynamics of the inflaton two-point function.  The importance
of the curved spacetime effect was addressed in Chapter~\ref{chap-preheat}, 
and we will discuss below the importance of treating the inflaton quantum 
variance on equal footing with the mean field.

\subsection{Present work} In this study, we wish to 
describe the late stages of the reheating period, in which the damping of
the inflaton mean field is dominated by fermionic particle production;
our starting point is the end of the {\em preheating\/} period (in which
the inflaton dynamics was dominated by back reaction from parametric
particle creation, as discussed in Chapter~\ref{chap-preheat}.
Because the inflaton variance $\langle \varphi^2 \rangle$ can (for
sufficiently strong self-coupling) be on the order 
of the square of the amplitude of mean field oscillations at the end of 
preheating (in the case of unbroken symmetry)
\cite{ramsey:1997b}, it is necessary to treat the inflaton
mean field and variance on an equal footing in a study of the subsequent
effective dynamics of the inflaton field.  This requires a two-loop,
two-particle-irreducible formulation of the coarse-grained 
effective action.  At two loops, both the
inflaton mean field {\em and the inflaton variance\/} couple to the spinor 
degrees of freedom, and are damped by back reaction from fermion 
particle production; all the previous studies mentioned 
above, in using the 1PI effective action, missed this possibly important
effect.  
%

In addition to having a large variance, 
the inflaton amplitude at the end of the preheating period 
may be large enough that the usual perturbative expansion 
in powers of the Yukawa coupling constant 
is not valid [see Eq.~(\ref{eq-vpt}) below], 
in which case a nonperturbative derivation of the inflaton
dynamics is required.  
In the construction
of the CTP-2PI, coarse-grained effective action below, the spinor
propagators obey one-loop dynamical equations in which the inflaton mean
field acts as an external source.  
Studies of fermion particle production during reheating which relied 
on the use of perturbation theory in the Yukawa coupling constant
\cite{abbott:1982a,dolgov:1982a,albrecht:1982a,morikawa:1985a,dolgov:1990a,shtanov:1995a,kofman:1994a}
therefore do not apply to the case of fermion particle production
at the end of preheating with unbroken symmetry, 
when the Yukawa coupling is sufficiently large.
The dynamical equations derived
in Sec.~\ref{sec-necgidcst}
below for the inflaton mean field and variance are applicable even when, as
may be the case, the inflaton mean field amplitude is large enough that a 
perturbative expansion in powers of the Yukawa coupling is not justified.  

Although, as discussed above, a proper treatment of the early stage of fermion
production during reheating, starting at the end of preheating with
unbroken symmetry, should in principle employ the CTP-2PI-CGEA to obtain
coupled equations of motion for the inflaton mean field and variance, 
at very late times the inflaton mean field and variance will have been
damped sufficiently (due to the dissipative mechanisms derived below in 
Sec.~\ref{sec-necgidcst}) that the perturbative 1PI effective action
will yield a qualitatively correct description of the inflaton mean-field dynamics.
While curved spacetime effects should in principle be incorporated
self-consistently for a quantitative calculation of the reheating
temperature in a particular inflationary model (as discussed in 
\cite{ramsey:1997b}), for a general discussion of dissipative 
effective dynamics of the inflaton mean field in the case of
weak cosmic expansion (where the Hubble constant $H$ is much smaller
than the frequency of inflaton oscillations), it is reasonable to
neglect curved spacetime effects in computing the spinor propagators.
Therefore in Sec.~\ref{sec-dsasfo}, we derive the perturbative, 
flat-space CTP-1PI-CGEA to 
fourth order in the Yukawa coupling constant, and obtain an
evolution equation for the inflaton mean field with nonlocal dissipation.

Another new feature of our work obtainable only from the stochastic approach
adopted here is the derivation, in Sec.~\ref{sec-nk},
of a Langevin equation for the inflaton mean
field, with clear identification of the dissipation and noise kernels from
the CGEA. We have calculated the energy dissipated and the fluctuations
in the energy. From the latter we obtain the range of parameters where
the conventional ``mean-field'' approach breaks down. We believe the
methodology presented here provides a better theoretical 
framework for the investigation of phase transitions in the early universe, 
as exemplified by our treatment of reheating in inflationary cosmology.

\subsection{Organization}
Our work is organized as follows.  In Sec.~\ref{sec-necgidcst}, we derive
the coupled equations of motion for the inflaton mean field and variance,
in a general curved spacetime, including diagrams in the coarse-grained CTP-2PI
effective action of up to two-loop order.  In Sec.~\ref{sec-dsasfo},
we specialize to Minkowski space and small mean-field amplitude, and obtain
a perturbative mean field equation including dissipative effects up to
fourth order in the Yukawa coupling constant.  In Sec.~\ref{sec-nk}, we 
examine the dissipation and noise kernels obtained in Sec.~\ref{sec-dsasfo},
and show that they obey a fluctuation-dissipation relation.  We then derive a 
Langevin equation which self-consistently includes the effect of
noise on the dynamics of the inflaton zero mode.  We summarize our results in
Sec.~\ref{sec-conc}.

\section{Coarse-grained inflaton dynamics in curved spacetime}
\label{sec-necgidcst}
In this section, we present a first-principles derivation of the 
nonequilibrium, nonperturbative, effective dynamics of a scalar inflaton 
field $\phi$ coupled to a spinor field $\psi$ via a Yukawa interaction, 
in a general, curved classical background spacetime.  The use of the 
Schwinger-Keldysh closed-time-path (CTP) formalism allows formulation
of the nonequilibrium dynamics of the inflaton from an appropriately
defined initial quantum state.
The evolution equations for the inflaton mean field and variance are derived 
from the two-loop, closed-time-path (CTP), two-particle-irreducible (2PI), 
coarse-grained effective action (CGEA).  
As the name suggests, there are
two approximations of a statistical mechanical nature.  One is the 
coarse graining of the environment--- here the inflaton field is the system and
the fermion field is the environment \cite{hu:1991b}.  
The other refers to a 
truncation of the {\em correlation hierarchy\/} for the inflaton field 
\cite{calzetta:1995b}--- the two-particle-irreducible effective action.  This
formulation retains the inflaton mean field and variance as coupled dynamical
degrees of freedom.  
Back Reaction of the scalar and spinor
field dynamics on the spacetime is incorporated using the semiclassical 
Einstein equation, which follows from functional differentiation of the
effective action with respect to the metric.   It is shown that
these dynamical equations are both real and causal, and the
equations are cast in a form suitable for implementation of an explicit
curved-spacetime renormalization procedure.

\subsection{The model}
\label{sec-model}
We study a model consisting of a scalar field $\phi$ (the 
inflaton field) which is Yukawa-coupled to a spinor field $\psi$, in a curved,
dynamical, classical background spacetime.  The total action
\begin{equation}
S[\phi,\bar{\psi},\psi,g^{\mu\nu}] = S^{\text{{\tiny G}}}[g^{\mu\nu}] +
S^{\text{{\tiny F}}}[\phi,\bar{\psi},\psi,g^{\mu\nu}],
\label{eq-ca}
\end{equation}
consists of a part depicting classical gravity, 
$S^{\text{{\tiny G}}}[g^{\mu\nu}]$, and a part for the matter fields,
\begin{equation}
S^{\text{{\tiny F}}}[\phi,\bar{\psi},\psi,g^{\mu\nu}] = 
S^{\phi}[\phi,g^{\mu\nu}] + S^{\psi}[\bar{\psi},\psi,g^{\mu\nu}] +
S^{\text{{\tiny Y}}}[\phi,\bar{\psi},\psi,g^{\mu\nu}], 
\end{equation}
whose scalar (inflaton), spinor (fermion), and Yukawa-interaction 
parts are given by
\begin{align}
\label{eq-sfa}
& S^{\phi}[\phi,g^{\mu\nu}] = - \frac{1}{2} \int d^{\;4}x \sqrt{-g} \left[
\phi ( \square + m^2 + \xi R ) \phi +  \frac{\lambda}{12} \phi^4 \right], \\
& S^{\psi}[\bar{\psi},\psi,g^{\mu\nu}] = \int d^{\;4}x \sqrt{-g} 
\left[ \frac{i}{2} \left(\bar{\psi} \gamma^{\mu} \nabla_{\mu} \psi -
(\nabla_{\mu} \bar{\psi}) \gamma^{\mu}\psi\right)
 - \mu \bar{\psi}\psi \right], \\
& S^{\text{{\tiny Y}}}[\phi,\bar{\psi},\psi,g^{\mu\nu}] =
- \yc \int d^{\;4}x \sqrt{-g} \phi \bar{\psi} \psi.
\label{eq-sfc}
\end{align}
For this theory to be renormalizable in semiclassical gravity,
the bare gravity action $S^{\text{{\tiny G}}}[g^{\mu\nu}]$
of Eq.~(\ref{eq-ca}) should have the form 
given by Eq.~(\ref{eq-lpfcstga}) \cite{dewitt:1975a,birrell:1982a}.
In Eqs.~(\ref{eq-sfa})--(\ref{eq-sfc}), $m$ is the scalar field ``mass'' 
(with dimensions of inverse 
length); $\xi$ is the dimensionless coupling to gravity; $\mu$ is the spinor
field ``mass,'' with dimensions of inverse length; $\square$ is the 
Laplace-Beltrami operator in the curved background spacetime with metric
tensor $g_{\mu\nu}$; $\nabla_{\mu}$ is the covariant derivative compatible with
the metric; $\sqrt{-g}$ is the square root of the absolute value of the
determinant of the metric; $\lambda$ is the self-coupling of the inflaton
field, with dimensions of $1/\sqrt{\hbar}$; 
$\yc$ is the Yukawa coupling constant,
which has dimensions of $1/\sqrt{\hbar}$; and
$R$ is the scalar curvature.
The curved spacetime Dirac matrices $\gamma^{\mu}$ satisfy the anticommutation 
relation
\begin{equation}
\left\{ \gamma^{\mu},\gamma^{\nu} \right\}_{+} = 2 g^{\mu\nu} 1_{\text{{\tiny
sp}}},
\end{equation}
in terms of the contravariant metric tensor $g^{\mu\nu}$.  The symbol 
$1_{\text{{\tiny sp}}}$ denotes the identity element in the Dirac algebra.  

It is assumed that there is a definite separation
of time scales between the stage of ``preheating'' discussed in
Chapter~\ref{chap-preheat}, and fermionic particle production, which is our
primary focus in this work.  However, this does not imply that perturbation
theory in the Yukawa coupling constant
$f$ is necessarily valid, which would require that condition
(\ref{eq-vpt}) (defined in Sec.~\ref{sec-dsasfo} below) be satisfied.
In addition, the fermion field mass $\mu$ is assumed to be 
much lighter than the inflaton field mass $m$, i.e., the renormalized
parameters $m$ and $\mu$ satisfy $m \gg \mu$.

\subsection{Closed-time-path, coarse-grained effective action}
We denote the quantum Heisenberg field operators of the scalar field $\phi$ 
and the spinor field $\psi$ by $\PhiH$ and $\PsiH$, respectively, and the 
quantum state\footnote{Although in this case the particular initial 
conditions constitute a pure state, this formalism can encompass general 
mixed-state initial conditions \cite{calzetta:1988b}.} by $|\Omega\rangle$.  
For consistency with the truncation of the correlation hierarchy at second 
order, we assume $\Phi_{\text{{\tiny H}}}$ to have a Gaussian moment expansion 
in the position basis \cite{mazzitelli:1989b,cooper:1997b}, 
in which case the relevant observables are the scalar mean field 
$\phih$ [defined in Eq.~(\ref{eq-ifmf})], the mean-squared fluctuations,
or variance [defined in Eq.~(\ref{eq-ifvr})].
As discussed above, at the end of the preheating period, 
the inflaton variance can be as large as the square of
the amplitude of mean-field oscillations.
On the basis of our assumption of separation of time scales in 
Sec.~\ref{sec-model},
and the conditions which prevail at the onset of reheating, the initial 
quantum state $|\Omega\rangle$ is assumed to be an appropriately defined
vacuum state for the {\em spinor\/} field.  

The construction of the CTP-2PI-CGEA for the $\phi\psib\psi$ theory
in a general, curved, background spacetime
closely parallels the construction of the CTP-2PI effective action
for the O($N$) model in curved spacetime discussed in Chapter~\ref{chap-oncst}.
Following the approach of Sec.~\ref{sec-lpfcst}, we define a ``CTP manifold''
${\mathcal M}$, and a volume form ${\boldsymbol \epsilon}_{\mathcal M}$
on ${\mathcal M}$, in terms of the discrete set $\{+,-\}$ labeling
the ``time branch,'' and a Cauchy hypersurface $\Sigma_{\star}$ which
is far to the future of the time scales in which we are interested.  The
spacetime manifold $M$ is the past domain of dependence of $\Sigma_{\star}$.
The restrictions of a function $\phi$, defined on ${\mathcal M}$, to the $+$
and $-$ time branches are denoted by $\phi_{+}$ and $\phi_{-}$, respectively.  
We can then define a matter field action on ${\mathcal M}$,
\begin{equation}
{\mathcal S}^{\text{{\tiny F}}}[\phi_{-},\psib_{-},\psi_{-},g^{\mu\nu}_{-};
\phi_{+},\psib_{+},\psi_{+},g^{\mu\nu}_{+}] \equiv
S^{\text{{\tiny F}}}[\phi_{+},\psib_{+},\psi_{+},g^{\mu\nu}_{+}] -
S^{\text{{\tiny F}}}[\phi_{-},\psib_{-},\psi_{-},g^{\mu\nu}_{-}],
\label{eq-sfctp}
\end{equation}
where the spacetime integrations in $S^{\text{{\tiny F}}}$ are now over $M$ 
only.  We use the symbol ${\mathcal S}^{\text{{\tiny F}}}$ to distinguish it 
from the action $S^{\text{{\tiny F}}}$ on $M$.  Let us also simplify
notation by suppressing time branch indices in the argument of functionals
on ${\mathcal M}$, i.e.,
\begin{equation}
{\mathcal S}^{\text{{\tiny F}}}[\phi,\psib,\psi,g^{\mu\nu}] \equiv
{\mathcal S}^{\text{{\tiny F}}}[\phi_{-},\psib_{-},\psi_{-},g^{\mu\nu}_{-};
\phi_{+},\psib_{+},\psi_{+},g^{\mu\nu}_{+}].
\end{equation}
Let us also define the functional 
${\mathcal S}^{\text{{\tiny Y}}}$ on ${\mathcal M}$ by
\begin{equation}
{\mathcal S}^{\text{{\tiny Y}}}[\phi,\psib,\psi,g^{\mu\nu}] \equiv
S^{\text{{\tiny Y}}}[\phi_{+},\psib_{+},\psi_{+},g^{\mu\nu}_{+}] -
S^{\text{{\tiny Y}}}[\phi_{-},\psib_{-},\psi_{-},g^{\mu\nu}_{-}],
\label{eq-syctp}
\end{equation}
in analogy with Eq.~(\ref{eq-sfctp}).  
For a function $\phi$ on ${\mathcal M}$, the restrictions of $\phi$ to the
$+$ and $-$ time branches are subject to the boundary condition
\begin{equation}
(\phi_{+})_{|\Sigma_{\star}} = (\phi_{-})_{|\Sigma_{\star}}
\label{eq-bc}
\end{equation}
at the hypersurface $\Sigma_{\star}$.
The gravity action $S^{\text{{\tiny G}}}$, promoted to
a functional on ${\mathcal M}$, takes the form
\begin{equation}
{\mathcal S}^{\text{{\tiny G}}}[g^{\mu\nu}_{+},g^{\mu\nu}_{-}] = 
S^{\text{{\tiny G}}}[
g^{\mu\nu}_{+}] - S^{\text{{\tiny G}}}[g^{\mu\nu}_{-}],
\label{eq-sgm}
\end{equation}
where the range of spacetime integration in $S^{\text{{\tiny G}}}$ on 
the right-hand side of Eq.~(\ref{eq-sgm}) is understood
to be over $M$.  

To formulate the CTP-2PI-CGEA, our first step is to define a generating 
functional for $n$-point functions of the scalar field, in terms of
the initial quantum state $|\Omega\rangle$ which evolves under the 
influence of a local source $J$, and a non-local source $K$ coupled to 
the scalar field (in the interaction picture with the external sources
being treated as the ``interaction'').  This generating
functional depends on both $J$ and $K$, as well as the
classical background metric $g^{\mu\nu}$.  In the path integral representation,
the generating functional $Z[J,K,g^{\mu\nu}]$ takes the form of a sum over 
scalar field configurations $\phi$ and complex Grassmann-valued configurations 
$\psi$ on the manifold ${\mathcal M}$,
\begin{equation}
\begin{split}
Z[J,K,g^{\mu\nu}] \equiv 
\int_{\text{{\tiny ctp}}} & D\phi_{-} D\bar{\psi}_{-} D\psi_{-} 
D\phi_{+} D\bar{\psi}_{+} D\psi_{+} \exp \Biggl[ 
\frac{i}{\hbar} \biggl(  {\mathcal S}^{\text{{\tiny F}}}[\phi,\bar{\psi},
\psi,g^{\mu\nu}]  \\
& + \int_M d^{\;4}x \sqrt{-g} c^{ab} J_a \phi_b   \\ 
& + \frac{1}{2} \int_M d^{\;4}x \sqrt{-g} \int_M d^{\;4}x' \sqrt{-g'} c^{ab}
c^{cd} K_{ac}(x,x') \phi_b(x) \phi_d(x') \biggr) \Biggr],
\label{eq-zgf}
\end{split}
\end{equation}
where $J_a(x)$ is a local $c$-number source and $K_{ab}(x,x')$ is a nonlocal
$c$-number source.  The subscript CTP on the functional integral denotes a
summation over field configurations $\phi_{\pm}$, $\psib_{\pm}$, and 
$\psi_{\pm}$ which satisfy the boundary condition (\ref{eq-bc}).  The latin 
indices $a$, $b$, $c$, $\ldots\;$, have the discrete index set 
$\{ +,- \}$, and denote the time branch \cite{calzetta:1987a,calzetta:1988b}. 
The boundary conditions on the functional integral of Eq.~(\ref{eq-zgf}) 
at the initial data surface determine the quantum state 
$|\Omega\rangle$. The CTP indices have been dropped from $g^{\mu\nu}$ for ease 
of notation; it will be clear how to reinstate them \cite{ramsey:1997a} 
in the two-loop CTP--2PI effective action shown below in 
Sec.~\ref{sec-dsasfo}.  The two-index symbol $c^{ab}$ is defined by the
$n=2$ case of Eq.~(\ref{eq-cabcdef}).
The generating functional for normalized $n$-point functions is
\begin{equation}
W[J,K,g^{\mu\nu}] = -i\hbar \text{ln} Z[J,K,g^{\mu\nu}],
\label{eq-wgf}
\end{equation}
in terms of which we can define the classical scalar field on ${\mathcal M}$,
\begin{equation}
\label{eq-mf}
\phih_a(x)_{\text{{\tiny $JK$}}} = c_{ab} \frac{1}{\sqrt{-g}} \frac{
\delta W[J,K,g^{\mu\nu}]}{\delta J_b(x)},
\end{equation}
and the scalar two-point function on ${\mathcal M}$ in the presence of
the sources $J_a$ and $K_{ab}$,
\begin{equation}
\label{eq-gf}
\hbar G_{ab}(x,x')_{\text{{\tiny $JK$}}} =
2 c_{ac} c_{bd} \frac{1}{\sqrt{-g}} \frac{1}{\sqrt{-g'}} \frac{\delta W[J,K
g^{\mu\nu}]}{\delta K_{cd}(x,x')} - \phih_a(x) \phih_b(x'),
\end{equation}
where the $JK$ subscripts indicate that $\phih$ and $G$ are functionals
of the $J_a$ and $K_{ab}$ sources.  In the limit $J_a = K_{ab} = 0$, the 
classical field is the same on the two time branches and it is 
equivalent to the mean field $\phih$, as shown in Eq.~(\ref{eq-ljkzmf}).  
In the same limit, $G_{ab}$ becomes the CTP propagator for the 
fluctuation field defined in Eq.~(\ref{eq-dff2}), as shown in
Eqs.~(\ref{eq-ljkzgfa})--(\ref{eq-ljkzgfd}).
In the coincidence limit\footnote{The variance $\langle \varphi(x)^2
\rangle$ is divergent in four spacetime dimensions, and should be regularized
using a covariant procedure \cite{bunch:1979a,birrell:1982a}.} 
$x'=x$, all four components 
(\ref{eq-ljkzgfa})--(\ref{eq-ljkzgfd}) 
are equivalent to the variance $\langle \vphiH^2 \rangle$ defined in 
Eq.~(\ref{eq-ifvr}).  Provided we can invert Eqs.~(\ref{eq-mf}) and 
(\ref{eq-gf}) to obtain $J_a$ and $K_{ab}$ in terms of $\phih_a$ and 
$G_{ab}$, the CTP--2PI effective action can 
be defined as the double Legendre transform (in both $J_a$ and $K_{ab}$) of
$W[J,K,g^{\mu\nu}]$,
\begin{multline}
\Gamma[\phih,G,g^{\mu\nu}]  = W[J,K,g^{\mu\nu}] 
- \int_M d^{\;4}x \sqrt{-g} c^{ab} J_a(x) 
\phih_b(x)  \\
 - \frac{1}{2} \int_M d^{\;4}x \sqrt{-g} \int_M d^{\;4}x' \sqrt{-g'}
c^{ab} c^{cd} K_{ac}(x,x') \left[ \hbar G_{bd}(x,x') + \phih_b(x) \phih_d(x')
\right].
\label{eq-ea}
\end{multline}
The inverses of Eqs.~(\ref{eq-mf}) and (\ref{eq-gf})
can be obtained by functional differentiation of Eq.~(\ref{eq-ea})
with respect to $\phih_a$,
\begin{multline}
\frac{1}{\sqrt{-g}} 
\frac{\delta \Gamma [\phih,G,g^{\mu\nu}]}{\delta \phih_a(x)} \\
= -c^{ab} J_b(x)_{\text{{\tiny $\phih G$}}} 
- \frac{1}{2} c^{ab} c^{cd} \int_M d^{\;4}x' 
\sqrt{-g'} \left[ K_{bd}(x,x')_{\text{{\tiny $\phih G$}}} +
K_{db}(x',x)_{\text{{\tiny $\phih G$}}}\right]\phih_c(x'),
\label{eq-dj} 
\end{multline}
and with respect to $G_{ab}$,
\begin{equation}
\frac{1}{\sqrt{-g}} \frac{1}{\sqrt{-g'}} 
\frac{\delta \Gamma[\phih,G,g^{\mu\nu}]}{\delta G_{ab}(x,x')} 
= -\frac{\hbar}{2}
c^{ac} c^{bd} K_{cd}(x,x')_{\text{{\tiny $\phih G$}}},
\label{eq-dk}
\end{equation}
where the $\phih G$ subscript indicates that $K_{ab}$ and $J_a$ are
functionals of $\phih_a$ and $G_{ab}$.  Inserting Eqs.~(\ref{eq-dj}) and
(\ref{eq-dk}) into Eq.~(\ref{eq-ea}) yields a functional integro-differential
equation for the CTP--2PI effective action in terms of $\phih$ and $G$ only,
so the $JK$ subscripts can be dropped.  It is useful to change the variable
of functional integration to be the fluctuation field about $\phih_a$
defined by Eq.~(\ref{eq-dctpfluc}).
Performing the change-of-variables $D\phi \rightarrow D\varphi$, the
equation for $\Gamma$ is
\begin{equation}
\begin{split}
\Gamma[\phih,G,g^{\mu\nu}] 
& = \int_M d^{\;4}x \int_M d^{\;4}x' \frac{\delta \Gamma
[\phih,G]}{\delta G_{ba}(x',x)}G_{ab}(x,x')   \\
& - i\hbar \text{ln} \Biggl\{ \int_{\text{{\tiny ctp}}} D\varphi_{+}
D\psib_{+} D\psi_{+} D\varphi_{-} D\psib_{-} D\psi_{-} \exp
\biggl[ \frac{i}{\hbar} \Bigl( {\mathcal S}^{\text{{\tiny F}}}[
\varphi + \phih, \psib, \psi, g^{\mu\nu}]  \\
& - \int_M d^{\;4}x \frac{\delta \Gamma[\phih,G,g^{\mu\nu}]}{\delta \phih_a} 
\varphi_a
- \frac{1}{\hbar} \int_M d^{\;4}x \int_M d^{\;4}x' \frac{\delta \Gamma
[\phih,G,g^{\mu\nu}]}{\delta G_{ba}(x',x)} \varphi_a(x) \varphi_b(x') \Bigr) 
\biggr]\Biggr\},
\end{split}
\end{equation}
which has the formal solution
\begin{multline}
\Gamma[\phih,G,g^{\mu\nu}] = 
 {\mathcal S}^{\phi}[\phih] - \frac{i\hbar}{2} \text{ln}\, \text{det} G_{ab} 
- i\hbar \text{ln} \, \text{det} F_{ab} 
+ \Gamma_2[\phih,G]  \\
+ \frac{i\hbar}{2} \int_M d^{\;4}x \sqrt{-g} \int_M d^{\;4}x'
\sqrt{-g'} {\mathcal A}^{ab} (x',x) G_{ab}(x,x'),
\label{eq-fs}
\end{multline}
where ${\mathcal A}^{ab}$ is the second functional derivative of the 
scalar part of the classical action $S^{\phi}$, evaluated at $\phih$,
\begin{equation}
\begin{split}
i {\mathcal A}^{ab}(x,x') &= \frac{1}{\sqrt{-g}}
\left( \frac{\delta^2 {\mathcal S}^{\phi}}{\delta \phi_a(x) \delta \phi_b(x')}
[\phih] \right) \frac{1}{\sqrt{-g'}}  \\ & = 
-\left[ 
c^{ab} ( \square_x + m^2 + \xi R(x) )
+ c^{abcd} \frac{\lambda}{2} \phih_c(x) \phih_d(x) 
\right] \delta(x-x') \frac{1}{\sqrt{-g'}}.
\label{eq-iolscp}
\end{split}
\end{equation}
The symbol $F_{ab}$ denotes the one-loop CTP spinor propagator, 
which is defined by
\begin{equation}
F_{ab}(x,x') \equiv {\mathcal B}_{ab}^{-1}(x,x'),
\label{eq-olspp}
\end{equation}
where we are suppressing spinor indices, and the inverse spinor propagator
${\mathcal B}^{ab}$ is defined by
\begin{equation}
\begin{split}
i {\mathcal B}^{ab}(x,x') &= \frac{1}{\sqrt{-g}} \left[
\frac{\delta^2 ({\mathcal S}^{\psi}[\psib,\psi] + 
{\mathcal S}^{\text{{\tiny Y}}}[\psib,\psi;\phih])}{\delta \psi_a(x) 
\delta \psib_b(x')} \right] \frac{1}{\sqrt{-g'}} \\
&=
\left[ c^{ab} \left( i \gamma^{\mu} \nabla'_{\mu} - \mu \right) - c^{abc} 
\yc \phih_c(x')
\right] \delta(x'-x)
\frac{1}{\sqrt{-g}} {1}_{\text{{\tiny sp}}}.
\label{eq-iolspp}
\end{split}
\end{equation}
It is clear from Eq.~(\ref{eq-iolspp}) that the use of the
one-loop spinor propagators in the construction of the CTP-2PI-CGEA 
represents a nonperturbative resummation in the Yukawa 
coupling constant, which (as discussed above) goes beyond the standard 
time-dependent perturbation theory.
The boundary conditions which define the inverses of Eqs.~(\ref{eq-iolscp})
and (\ref{eq-iolspp}) are the boundary conditions at the initial data surface
in the functional integral in Eq.~(\ref{eq-zgf}), which in turn, define the 
initial quantum state $|\Omega\rangle$.  The one-loop spinor propagator 
$F_{ab}$ is related to the expectation values of the spinor Heisenberg 
field operators for a spinor field in the presence of the c-number background
field $\hat{\phi}$,
\begin{align}
&\hbar \left.F_{++}(x,x')\right|_{\phihp = \phihm = \phih}
 = \langle \Omega | T(\PsiH(x) \PsibH(x'))|\Omega \rangle, \\
&\hbar \left.F_{--}(x,x')\right|_{\phihp = \phihm = \phih}
 = \langle \Omega | \tilde{T} (\PsiH(x) \PsibH(x'))|\Omega
\rangle, \\
&\hbar \left.F_{+-}(x,x')\right|_{\phihp = \phihm = \phih}
 = -\langle \Omega | \PsibH(x') \PsiH(x) | \Omega \rangle, \\
&\hbar \left.F_{-+}(x,x')\right|_{\phihp = \phihm = \phih}
 = \langle \Omega | \PsiH(x) \PsibH(x') | \Omega \rangle,
\end{align}
where the spinor Heisenberg field operators obey the equations
\begin{align}
& (i \gamma^{\mu} \nabla_{\mu} - \mu - f \phih ) \Psi = 0, \\
& (-i \gamma^{\mu} \nabla_{\mu} - \mu - f \phih ) \bar{\Psi} = 0.
\end{align}
The CTP spinor propagator components satisfy the relations 
(valid only when $\phihp = \phihm = \phih$)
\begin{align}
& F_{++}(x,x')^{\dagger} = F_{--}(x',x),\\
& F_{--}(x,x')^{\dagger} = F_{++}(x',x),\\
& F_{-+}(x,x')^{\dagger} = F_{-+}(x',x),\\
& F_{+-}(x,x')^{\dagger} = F_{+-}(x',x).
\end{align}
The functional $\Gamma_2[\phih,G]$ is defined as $-i\hbar$ times the sum of 
all vacuum diagrams drawn according to the following rules:
\begin{enumerate}
\item{Vertices carry spacetime ($x \in M$) and time branch ($a \in \{+,-\}$)
labels.}
\item{Scalar field lines denote $\hbar G_{ab}(x,x')$.}
\item{Spinor lines denote the one-loop CTP spinor propagator 
$\hbar F_{ab}(x,x')$ (spinor indices are suppressed), defined 
in Eq.~(\ref{eq-olspp}).}
\item{There are three interaction vertices, given by $i {\mathcal S}^{
\text{{\tiny I}}} /\hbar$, which is defined by
\begin{align} 
& {\mathcal S}^{\text{{\tiny I}}}[\phih,\varphi,\psib,\psi] =
S^{\text{{\tiny I}}}[\phihp,\vphip,\psibp,\psip] -
S^{\text{{\tiny I}}}[\phihm,\vphim,\psibm,\psim], \\
& S^{\text{{\tiny I}}}
[\phih,\vphi,\psib,\psi] = -\int d^4 x \sqrt{-g} \left[
f \vphi \psib \psi + \frac{\lambda}{24} \vphi^4 +
\frac{\lambda}{6} \phih \vphi^3\right],
\end{align}
where we have followed the notation of Eq.~(\ref{eq-syctp}).}
\item{Only diagrams which are two-particle-irreducible 
with respect to cuts of {\em scalar\/} lines contribute to $\Gamma_2$.}
\end{enumerate}
The distinction between the CTP-2PI, {\em coarse-grained\/}
effective action which arises here, and the fully
two-particle-irreducible effective action (2PI with respect to scalar 
{\em and\/} spinor cuts), is due to the fact that we only Legendre-transformed
sources coupled to $\phi$; i.e., the spinor field is treated as the 
environment.  In Eq.~(\ref{eq-iolscp}), the curved-spacetime
Dirac $\delta$ function is defined as in \cite{birrell:1982a}. Comparison of 
Eq.~(\ref{eq-fs}) above with Eq.~(4.13) of Ref.~\cite{ramsey:1997a}
[which was computed for the $O(N)$
model] shows that the $\text{Tr}\,\text{ln}F_{ab}$ 
in Eq.~(\ref{eq-fs}) differs from the usual one-loop term by a factor of 2,
owing to the difference (in the exponent) between the Gaussian integral 
formulas for real and complex fields \cite{cornwall:1974a}.  

\begin{figure}[htb]
\begin{center}
\epsfig{file=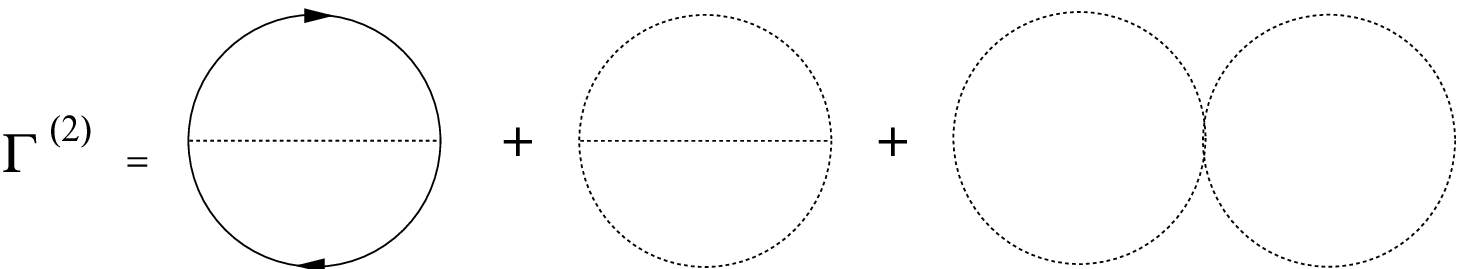,width=4.0in}
\end{center}
\caption{Diagrammatic expansion for $\Gamma^{(2)}$, the two-loop
part of the CTP-2PI coarse-grained effective action.}
\label{fig-gam2}
\end{figure}

The functional $\Gamma_2[\phih,G,g^{\mu\nu}]$ can be evaluated in a loop 
expansion, which corresponds to an expansion in powers of $\hbar$,
\begin{equation}
\Gamma_2[\phih,G,g^{\mu\nu}] = \sum_{l = 2}^{\infty} \hbar^l 
\Gamma^{(l)}[\phih,G,g^{\mu\nu}],
\end{equation}  
starting with the two-loop term, $\Gamma^{(2)}$.  The functional
$\Gamma^{(2)}$ has a diagrammatic expansion shown in Fig.~\ref{fig-gam2},
where solid lines represent the spinor propagator $F$
(as defined in Sec.~\ref{sec-necgidcst}), and dotted lines represent the
scalar propagator $G$.  The vertices terminating three $\phi$ lines
are proportional to the scalar mean field $\phih$.  Each vertex carries
spacetime $(x)$ and CTP $(+,-)$ labels.
The $\lambda \phi^4$ self-interaction leads to two terms in the two-loop
part of the effective action, the second and third graphs
of Fig.~\ref{fig-gam2}.  They are the ``setting sun'' diagram,
which is $O(\lambda^2)$, and the ``double bubble,'' which is
$O(\lambda)$, respectively.   
The Yukawa interaction leads to only one diagram in 
$\Gamma^{(2)}$, the first diagram of Fig.~\ref{fig-gam2},
\begin{equation}
\frac{i f^2}{2} c^{aa'a''} c^{bb'b''} 
\int d^{\,4}x \sqrt{-g} \int d^{\,4}x' \sqrt{-g'} G_{ab}(x,x') 
\text{Tr}_{\text{{\tiny sp}}}\left[ F_{a'b'}(x,x') F_{b''a''}(x',x)\right],
\label{eq-dd2}
\end{equation}
where the trace is understood to be over the spinor indices which are not
shown, and the three-index symbol $c^{abc}$ is defined by the $n=3$ case of
Eq.~(\ref{eq-cabcdef}). Here, we treat the $\lambda$ self-interaction using 
 the time-dependent Hartree-Fock approximation 
\cite{cornwall:1974a}, which is equivalent to retaining the
$O(\lambda)$ (double-bubble) graph and dropping the $O(\lambda^2)$
(setting sun) graph.  We assume for the present study 
that the coupling $\lambda$ is sufficiently
small that the $O(\lambda^2)$ diagram is unimportant on the
time scales of interest in the fermion production regime of the inflaton
dynamics.  The mean-field and gap equations including both the
$O(\lambda)$ and the $O(\lambda^2)$ diagrams were derived for a
general curved spacetime in Chapter~\ref{chap-oncst}.

\subsection{Evolution equations for $\phih$ and $G$ in curved spacetime}
The (bare) semiclassical field equations for the two-point function, 
mean field, and metric can be obtained from the CTP-2PI-CGEA by 
functional differentiation with respect $G_{ab}$, $\phih_a$, and $g^{\mu\nu},$
followed by identifications of $\phih$ and $g^{\mu\nu}$ on the two time
branches, as shown in Eqs.~(\ref{eq-lpfseea})--(\ref{eq-lpfseec}).
These equations constitute the semiclassical
approximation to the full quantum dynamics for the system described by
the classical action (\ref{eq-ca}).  Equation~(\ref{eq-lpfseea}) should be 
understood as following after time branch indices have been reinstated on
the metric tensor in the CTP-2PI-CGEA \cite{ramsey:1997a}.
The field equation of semiclassical gravity (with bare parameters) is obtained
directly from Eq.~(\ref{eq-lpfseec}), and given by Eq.~(\ref{eq-lpfcstsee2}).
The right-hand side of Eq.~(\ref{eq-lpfcstsee2}) is 
the (unrenormalized) quantum energy-momentum tensor defined by 
Eq.~(\ref{eq-lpfcstemt}).
The energy-momentum tensor $\langle T_{\mu\nu}\rangle$ is divergent in
four spacetime dimensions, and must be regularized via a covariant procedure
\cite{birrell:1982a}, as discussed in Sec.~\ref{sec-onrenorm2} above.

Making the two-loop approximation to the CTP-2PI-CGEA,
where we take $\Gamma_2 \simeq \hbar^2 \Gamma^{(2)}$, and dropping the
$O(\lambda^2)$ diagram from $\Gamma_2$, the mean-field equation becomes
\begin{equation}
\left( \square + m^2 + \xi R(x) + \frac{\lambda}{6} \phih^2(x) +
\frac{\lambda\hbar}{2} G(x,x) \right) 
\phih + \hbar f \text{Tr}_{\text{{\tiny sp}}} [F_{ab}(x,x)] - \hbar^2 g^3 
\Sigma(x) = 0,
\label{eq-npmfe}
\end{equation}
where $G(x,x)$ is the coincidence limit of $G_{ab}(x,x')$, and
in terms of a function $\Sigma(y)$ defined by
\begin{equation}
\begin{split}
\Sigma(y) \equiv \int d^{\,4}x \sqrt{-g} \int d^{\,4}x' \sqrt{-g'} 
\biggl\{ & 
G_{++}(x,x') \text{Tr}_{\text{{\tiny sp}}}\left[F_{++}(x,y)
                                           F_{++}(y,x') F_{++}(x',x)\right] 
  \\
- &
G_{-+}(x,x') \text{Tr}_{\text{{\tiny sp}}}\left[F_{-+}(x,y) 
                                           F_{++}(y,x') F_{+-}(x',x)\right] 
  \\
- &
G_{+-}(x,x') \text{Tr}_{\text{{\tiny sp}}}\left[F_{++}(x,y) 
                                           F_{+-}(y,x') F_{-+}(x',x)\right] 
  \\
+ &
G_{--}(x,x') \text{Tr}_{\text{{\tiny sp}}}\left[F_{-+}(x,y)
                                           F_{+-}(y,x') F_{--}(x',x)\right] 
\biggr\}.
\label{eq-sigma}
\end{split}
\end{equation}
Making use of the curved spacetime definitions of the scalar and 
spinor field Hadamard kernels \cite{birrell:1982a}
\begin{align}
G^{(1)}(x,x') &= \langle \Omega | \{ \varphi_{\text{{\tiny H}}}(x), 
\varphi_{\text{{\tiny H}}}(x') \} | \Omega \rangle, \\
F^{(1)}(x,x') &= \langle \Omega | [ \Psi_{\text{{\tiny H}}}(x), 
\bar{\Psi}_{\text{{\tiny H}}}(x')] | \Omega \rangle,
\label{eq-dshf}
\end{align}
and retarded propagators
\begin{align}
G_R(x,x') &= i \theta(x,x') \langle \Omega | [ \varphi_{\text{{\tiny H}}}(x),
\varphi_{\text{{\tiny H}}}(x') ] | \Omega \rangle, \\
F_R(x,x') &= i \theta(x,x') \langle \Omega | \{ \Psi_{\text{{\tiny H}}}(x),
\bar{\Psi}_{\text{{\tiny H}}}(x') \} | \Omega \rangle, 
\label{eq-dsrp}
\end{align}
the function $\Sigma(y)$ can be recast in a manifestly real and causal form,
\begin{multline}
\Sigma(y) = -2 \int d^4 x \sqrt{-g} \int d^4 x' \sqrt{-g'} \, \text{Re} \,
\text{Tr}_{\text{{\tiny sp}}}  \Bigl[\Bigl( \theta(x,x') G^{(1)}(x',x) 
F^{(1)}(x,x')  \\  -
G_R(x,x')^{\dagger} F_R(x,x') \Bigr) F_R(y,x')^{\star} F_R(y,x) \Bigr],
\end{multline}
from which it is clear that the integrand vanishes whenever $x$ or $x'$ is
to the future of $y$.
The ``gap'' equation for $G_{ab}$ is obtained from Eq.~(\ref{eq-lpfseec}), 
\begin{multline}
(G^{-1})^{ba}(x,x') = {\mathcal A}^{ba}(x,x') + 
\frac{i\lambda\hbar}{2} c^{ba} G(x,x)
\delta(x-x')\frac{1}{\sqrt{-g'}}
\\ + \hbar \yc^2 c^{aa'a''}c^{bb'b''}
\text{Tr}_{\text{{\tiny sp}}} \left[ F_{a'b'}(x,x') F_{b''a''}(x',x) \right].
\label{eq-fge}
\end{multline}
Multiplying Eq.~(\ref{eq-fge}) through by $G_{ab}$, performing a spacetime
integration, and taking the $++$ component, we obtain
\begin{multline}
\biggl(( \square + m^2 + \xi R + \frac{\lambda}{2} \phih^2 +  
\frac{\lambda \hbar}{2} G(x,x) \biggr) G_{++}(x,x')   \\
 + \hbar \yc^2 \int dx'' 
\sqrt{-g''} {\mathcal K}(x,x'') G_{++}(x'',x') = -i \delta (x-x')
\frac{1}{\sqrt{-g'}},
\label{eq-dge}
\end{multline}
in terms of a kernel ${\mathcal K}(x,x'')$ defined by
\begin{equation}
{\mathcal K}(x,x') = -i\text{Tr}_{\text{{\tiny sp}}}
\left[ F_{++}(x,x') F_{++}(x',x) - F_{+-}(x,x')F_{-+}(x',x) 
\right].
\end{equation}
Making use of Eqs.~(\ref{eq-dshf}) and (\ref{eq-dsrp}), this kernel takes the
form
\begin{equation}
{\mathcal K}(x,x') = \text{Re} \, \text{Tr}_{\text{{\tiny sp}}} \left[ F_R(x,x') 
F^{(1)}(x',x) \right],
\label{eq-kapker}
\end{equation}
which shows that the gap equation (\ref{eq-dge}) is manifestly real and causal.
As will be shown below in Sec.~\ref{sec-dsasfo} (in a perturbative limit), 
the kernel ${\mathcal K}(x,x')$ is
dissipative, and it reflects back reaction from fermionic particle production
induced by the time-dependence of the inflaton {\em variance}\@.  
The gap equation (\ref{eq-dge}) is therefore damped for modes above 
threshold,\footnote{See \cite{calzetta:1989b} for a similar discussion in the 
context of spinodal decomposition in quantum field theory.} and this damping 
is not accounted for in the 
1PI treatments of inflaton dynamics (where only the inflaton mean field is 
dynamical).  In contrast to previous studies 
\cite{boyanovsky:1995d,boyanovsky:1995b,boyanovsky:1995a} which assumed a 
local equation of motion for the inflaton propagator, the two-loop gap 
equation obtained from the CTP-2PI-CGEA includes a {\em nonlocal\/} kernel, 
which is a generic feature of back reaction from particle
production.  As stressed above, the dissipative dynamics of the inflaton
two-point function can be important when the inflaton variance is on the
order of the square of the inflaton mean-field amplitude; such conditions 
may exist at the end of preheating.

The set of evolution equations (\ref{eq-npmfe}) for $\phih$ and (\ref{eq-dge})
for $G$, is formally complete to two loops.  Dissipation arises due to the
coarse graining of the spinor degrees of freedom.  These dynamical 
equations are useful for general purposes, and are valid in a general 
background spacetime.  However, in order to get explicit results, one needs to
introduce approximations, as we now do.  

\section{Dynamics of small-amplitude inflaton oscillations}
\label{sec-dsasfo}
The effective evolution equations for the inflaton mean field $\phih$
and variance $\langle \varphi^2 \rangle$
derived in the previous section are useful for studying fermion production 
when $\phih_0$, the amplitude of the spatially homogeneous
inflaton mean-field oscillations, is large, and the inflaton variance
is of the same order-of-magnitude as $(\phih_0)^2$.
 As discussed in Sec.~\ref{sec-necgidcst} above, such conditions can 
prevail at the end of the preheating period in chaotic
inflation with unbroken symmetry \cite{boyanovsky:1996b,ramsey:1997b}.
Because of the dissipative kernel ${\mathcal K}(x,x')$ in the gap equation
(\ref{eq-dge}), which damps
the evolution of $G$, and the back reaction terms in the mean-field equation,
which damp the oscillations of $\phih$, eventually the condition,
\begin{equation}
f \phih_0 \ll m
\label{eq-vpt}
\end{equation}
will hold,
at which point it is justifiable to follow the mean-field dynamics
using the perturbative, 1PI, coarse-grained effective action 
\cite{shtanov:1995a}.  Although in principle
one should study this process in a general curved spacetime, 
for simplicity we assume spatial homogeneity, and that the inflaton mass is 
much greater than the Hubble constant, $m \gg H$.  While this condition alone
is in general {\em not\/} sufficient to ensure that curved spacetime 
effects are negligible during reheating
(see, for example, \cite{ramsey:1997b}, where cosmic expansion {\em does\/}
affect preheating dynamics even though $m \gg H$), with the additional 
assumption of condition (\ref{eq-vpt}) it is reasonable to neglect the 
effect of cosmic expansion in the {\em spinor\/} propagators 
\cite{boyanovsky:1995c,boyanovsky:1995d}.  In this and the following section,
we also neglect the self-coupling $\lambda$, because for the case of unbroken
symmetry, the lowest-order $\lambda$-dependent contribution to the 
perturbative inflaton self-energy is $O(\hbar^2)$ \cite{boyanovsky:1995a}, 
and we are only concerned with one-loop dynamics in this section.  

Let us therefore specialize to Minkowski space, and implement a perturbative
expansion of the CTP effective action in powers of the mean field $\phih$.
This formally entails a solution of the gap equation (\ref{eq-dge}) 
for $G$, a back-substitution of the solution into the CTP-2PI 
coarse-grained effective action, and a subsequent expansion of this
expression (now a functional of $\phih$ only) in powers of $\phih$.
The resulting perturbative expansion for the effective action 
contains only free-field  propagators.  For consistency, one should
use an initial density matrix for the spinor degrees of freedom which
corresponds to the end-state particle occupation numbers of the 
nonperturbative dynamics of Sec.~\ref{sec-necgidcst}.  For simplicity,
however, we assume the initial quantum state for the spinor field is the
vacuum state.  Hereafter, $F_{ab}$ denotes the free-field,
Minkowski-space, vacuum spinor CTP propagator, whose components are given by
\cite{chou:1985a,stylianopoulos:1991a,cooper:1994a}
\begin{align}
F_{++}(x,x') &= \int \frac{d^{\,4}p}{(2\pi)^4} e^{-i p (x-x')} \frac{i (\not{p}
+ m)}{p^2 - \mu^2 + i\epsilon}, \\
F_{--}(x,x') &= -\int \frac{d^{\,4}p}{(2\pi)^4} e^{-i p (x-x')} \frac{i (\not{p} 
+ m)}{p^2 - \mu^2 - i\epsilon}, \\
F_{-+}(x,x') &= \int \frac{d^{\,4}p}{(2\pi)^4} e^{-i p (x-x')} 2\pi (\not{p} 
+ m) \delta (p^2 - \mu^2) \theta(p^0), \\
F_{+-}(x,x') &= \int \frac{d^{\,4}p}{(2\pi)^4} e^{-i p(x-x')} 2\pi (\not{p} 
+ m) \delta (p^2 - \mu^2) \theta(-p^0). 
\end{align}
The $++$ and $--$ propagators admit a representation in terms of
a time-ordering function $\theta(x,y) = \theta(x^0 - y^0)$,
\begin{align}
F_{++}(x,x') &= \theta(x,x') F_{-+}(x,x') + \theta(x',x) F_{+-}(x,x'),\\
F_{--}(x,x') &= \theta(x,x') F_{+-}(x,x') + \theta(x',x) F_{-+}(x,x').
\end{align}
The CTP effective action can be expanded in powers of $f^2$, and we find
\begin{equation}
\Gamma[\phih] = {\mathcal S}^{\phi}[\phih] 
- \frac{i\hbar}{2} \text{ln}\,\text{det} ({\mathcal A}^{ab})^{-1}
- i\hbar \text{ln}\,\text{det} F_{ab}
+ \Gamma_1[\phih],
\end{equation}

\begin{figure}[htb]
\begin{center}
\epsfig{file=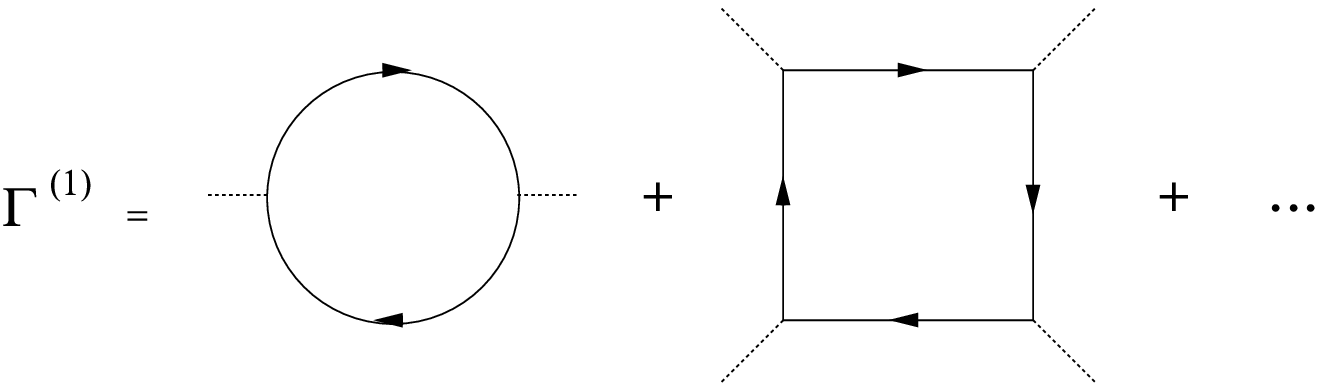,width=3.5in}
\end{center}
\caption{Diagrammatic expansion for $\Gamma^{(1)}$, the one-loop
part of the CTP-1PI coarse-grained effective action.}
\label{fig-gam1}
\end{figure}

The CTP effective action can be expanded in powers of $f^2$, and we find
\begin{equation}
\Gamma[\phih] = {\mathcal S}^{\phi}[\phih] 
- \frac{i\hbar}{2} \text{ln}\,\text{det} (\tilde{{\mathcal A}}^{ab})^{-1}
- i\hbar \text{ln}\,\text{det} F_{ab}
+ \Gamma_1[\phih],
\end{equation}
where the kernel $\tilde{{\mathcal A}}$ is defined by
\begin{equation}
i \tilde{{\mathcal A}}^{ab}(x,x') = -c^{ab}(\square_x + m^2)\delta(x-x'),
\end{equation}
and $\Gamma_1$ is defined as
$-i\hbar$ times the sum of all one-particle-irreducible diagrams constructed
with lines given by $\hbar \tilde{{\mathcal A}}^{-1}$ and 
$\hbar F_{ab}$, and vertices given
by ${\mathcal S}^{\text{{\tiny Y}}}[\phih,\psib,\psi]/\hbar$ and
${\mathcal S}^{\text{{\tiny Y}}}[\varphi,\psib,\psi]/\hbar$.
Because the free-field propagators $\tilde{{\mathcal A}}^{-1}$ and 
$F_{ab}$ do not depend on $\phih$, the $\log(\det)$ 
do not contribute to the variation of $\Gamma[\phih]$
with respect to $\phih$, and therefore, they can be dropped.
The functional $\Gamma_1[\phih]$ can be expanded in powers of $\hbar$,
\begin{equation}
\Gamma_1[\phih] = \sum_{l=1}^{\infty} \hbar^l \Gamma^{(l)}[\phih],
\end{equation}
where the term $\Gamma^{(l)}[\phih]$ is the sum of all 1PI $l$-loop graphs.
Order by order in the loop expansion and the coupling constant, 
the CTP 1PI effective action must satisfy the unitarity condition 
\begin{equation}
\left.\Gamma_1\right|_{\phihp = \phih; \;\; \phihm = \phih} = 0,
\label{eq-uc}
\end{equation}
which has been verified to two-loop order in the case of 
scalar $\lambda \Phi^4$ field theory \cite{jordan:1986a}.
The one-loop term in the loop expansion of the CTP effective action,
$\Gamma^{(1)}[\phih]$, can be further expanded in powers of $f^2$,
\begin{equation}
\Gamma^{(1)}[\phih] = \sum_{n=1}^{\infty} f^{2n} \Gamma^{(1)}_{2n}[\phih],
\end{equation}
which corresponds to the usual amplitude expansion of the CTP
effective action \cite{boyanovsky:1995d}.  Figure~\ref{fig-gam1} shows the
diagrammatic expansion of $\Gamma^{(1)}$,
where solid lines represent the spinor propagator 
$F$ (as defined in Sec.~\ref{sec-dsasfo}), and dotted lines represent 
multiplication by the scalar mean field $\phih$.  Each vertex carries
spacetime $(x)$ and CTP $(+,-)$ labels.
The terms $\Gamma_{2n}^{(1)}[\phih]$
are generally divergent, but since the theory is renormalizable in the 
standard ``in-out'' formulation, it is renormalizable in the closed-time-path,
``in-in'' formulation \cite{jordan:1986a,calzetta:1987a}.

\subsection{One-loop perturbative effective action at $O(f^2)$}
\label{sec-olpefa}
The $O(f^2)$ term in the expansion of the one-loop CTP effective action,
which is the first term in Fig.~\ref{fig-gam1},
takes the form\footnote{Note that there are no nonzero graphs with 
an odd number of vertices in this model.}  
\begin{equation}
\Gamma^{(1)}_2[\phih] = -\frac{i}{2}
c^{abc} c^{a'b'c'} \int d^{\,4}x d^{\,4}x' \phih_a(x)
\phih_{a'}(x') \text{Tr}_{\text{{\tiny sp}}} 
\left[F_{bb'}(x,x') F_{c'c}(x',x)\right].
\end{equation}
Making use of sum and difference variables
\begin{align}
& \Sigma(x) = \frac{1}{2} \left[ \phih_{+}(x) + \phih_{-}(x') \right], 
\label{eq-sdva} \\
& \Delta(x) = \phih_{+}(x) - \phih_{-}(x'),
\label{eq-sdvb}
\end{align}
the functional $\Gamma^{(1)}_2[\phih]$ can be recast in the form
\begin{equation}
\Gamma^{(1)}_2[\phih] = \int d^{\,4}x d^{\,4}x' \Biggl[ \Sigma(x) \Delta(x')
{\mathcal D}_2(x,x') + \frac{i}{2} \Delta(x) \Delta(x')
{\mathcal N}_2(x,x') \Biggr],
\label{eq-d1}
\end{equation}
in terms of manifestly real kernels ${\mathcal D}_2(x,x')$ and 
${\mathcal N}_2(x,x')$ defined by
\begin{align}
&{\mathcal D}_2(x,x') = \,\text{Im} \, \text{Tr}_{\text{{\tiny sp}}}
[F_{++}(x,x')F_{++}(x',x) +
F_{+-}(x,x')F_{-+}(x',x)], \label{eq-dk2} \\
&{\mathcal N}_2(x,x') = -\,\text{Re}\,
\text{Tr}_{\text{{\tiny sp}}}[F_{++}(x,x')F_{++}(x',x)].
\label{eq-dn2}
\end{align}
Only the kernel ${\mathcal D}_2(x,x')$ contributes to the mean-field equation 
of motion.  The kernel ${\mathcal N}_2(x,x')$ constitutes a 
correlator for noise, and will be discussed in Sec.~\ref{sec-nk}.
The unitarity condition (\ref{eq-uc}) requires that the sum of
diagrams proportional to $\Sigma(x)\Sigma(x')$ vanish identically. 
With the definitions of the retarded spinor propagator, Eq.~(\ref{eq-dshf}),
and the spinor Hadamard kernel, Eq.~(\ref{eq-dsrp}),
which in Minkowski space take the form
\begin{align}
& F_R(x,x') = i \theta(x,x') [ F_{-+}(x,x') - F_{+-}(x,x') ], \\
& F^{(1)}(x,x') = F_{-+}(x,x') + F_{+-}(x,x'),
\end{align}
the kernel ${\mathcal D}_2(x,x')$ can be written in a manifestly causal
form,
\begin{equation}
{\mathcal D}_2(x,x') = \frac{1}{2}
\, \text{Re} \, \text{Tr}_{\text{{\tiny sp}}} [F_R(x,x') F^{(1)}(x',x)].
\label{eq-d2rp}
\end{equation}
Using Eq.~(\ref{eq-d2rp}), 
one can compute ${\mathcal D}_2(x,x')$ in an arbitrary curved
background spacetime.  It should be noted that ${\mathcal D}_2(x,x')$ is
just the lowest-order term in the series expansion of ${\mathcal K}(x,x')$ 
[defined in Eq.~(\ref{eq-kapker})] in 
powers of the coupling constant $f$.  The appearance of the retarded 
propagator in Eq.~(\ref{eq-d2rp}) guarantees that the contribution of
$\Gamma^{(1)}_2$ to the mean-field equation of motion is causal.  

Let us now evaluate ${\mathcal D}_2(x,x')$
using dimensional regularization and the modified minimal subtraction 
($\overline{\text{MS}}$) renormalization prescription 
\cite{bardeen:1978a,collins:1984a,peskin:1995a}.  
Dimensional regularization
requires changing the coupling constant so that the interaction
$S^{\text{{\tiny Y}}}$  has the correct dimensions in $n$ spacetime dimensions,
\begin{equation}
f \rightarrow f \Lambda^{\frac{4-n}{2}},
\end{equation}
where we have introduced a parameter $\Lambda$, the renormalization
scale, which has dimensions of mass.   By Lorentz invariance and causality, 
the product of Feynman propagators can be written in terms of an amplitude 
$A_2$ \cite{peskin:1995a},
\begin{equation}
\text{Tr}_{\text{{\tiny sp}}} \, [ F_{++}(x,x') F_{++}(x',x)]
= i \int \frac{d^{\,4}k}{(2\pi)^4} e^{-ik(x-x')} A_2(k^2 + i\epsilon),
\label{eq-spp}
\end{equation}
and with this choice of renormalization prescription, the amplitude
$A_2(k^2)$ takes the form
\begin{equation}
A_2(k^2) = -\frac{3}{4\pi^2} \int_0^1 d\alpha E(\alpha;k^2) \text{log}
\, \Biggl( \frac{E(\alpha;k^2)}{\Lambda^2} \Biggr), \label{eq-dm}
\end{equation}
where $E(\alpha;k^2)$ is defined by\footnote{The notation $E(\alpha;k^2)$ 
used here should not be confused with $E(k)$, the complete elliptic integral of
second kind.}
\begin{equation}
E(\alpha;k^2) = \mu^2 - \alpha(1-\alpha)k^2.
\end{equation}
Note that in Eq.~(\ref{eq-dm}), the $\alpha$ integration shows up via the 
Feynman identity \cite{peskin:1995a}
\begin{equation}
\frac{1}{C_1 \ldots C_N} = (N-1)! \int_0^1 d\alpha_1 \cdots
\int_0^1 d\alpha_N \delta(\alpha_1 + \cdots + \alpha_{N - 1}) 
\bigl[ \alpha_1 C_1 + \cdots + \alpha_N C_N \bigr]^{-N}.
\label{eq-fi}
\end{equation}
The $i\epsilon$ appearing in Eq.~(\ref{eq-spp}) ensures that the amplitude 
$A_2$ is evaluated on the physical sheet \cite{eden:1966a,peskin:1995a}.
The logarithm in Eq.~(\ref{eq-dm}) has a negative real argument when
the two conditions $k^2 > 4 \mu^2$ and $|2\alpha-1| < \sqrt{1-4\mu^2/k^2}$
are both satisfied.  When $|2\alpha -1| < \sqrt{1-4\mu^2/k^2}$,
the amplitude $A_2(k^2)$ has a branch cut (considered as an analytically
continued function of $k^0$) for $(k^0)^2 > \vec{k}^2 + 4\mu^2$.
The discontinuity across the branch cut is related
to the ``cut'' version of the diagram [the second term in 
Eq.~(\ref{eq-dk2})] via the Cutosky rules 
\cite{cutosky:1960a,eden:1966a,thooft:1974a,itzykson:1980a,ramond:1990a,veltman:1994a},
\begin{equation}
\text{Tr}_{\text{{\tiny sp}}} [F_{+-}(x,x') F_{-+}(x',x)] 
= -i \int \frac{d^{\,4}k}{(2\pi)^4} e^{-ik(x-x')}
\text{Disc} [ A_2(k^2) ] \theta(k^0).
\label{eq-spm}
\end{equation}
From Eqs.~(\ref{eq-spp}), (\ref{eq-dm}), and (\ref{eq-spm}), it is 
straightforward to obtain an expression for the dissipation kernel,
\begin{multline}
{\mathcal D}_2(x,x') = \frac{3}{4\pi^2} \int \frac{d^{\,4}k}{(2\pi)^4}
e^{-ik(x-x')} \int_0^1 d\alpha E(\alpha;k^2) 
\Biggl[ \text{log}\,\Biggl(\frac{
|E(\alpha;k^2)|}{\Lambda^2}\Biggr) \\ - 
i \pi \theta[-E(\alpha;k^2)] \text{sgn}(k^0) 
\Biggr],
\label{eq-vf1}
\end{multline}
where we have now taken the limit $\epsilon \rightarrow 0_{+}$.
One can verify by inspection that this kernel is real.  However, the
second term in Eq.~(\ref{eq-vf1}) breaks time-reversal invariance and leads to 
dissipative mean field dynamics.  
The one-loop Fourier-transformed mean-field equation is (dropping the
caret from $\hat{\phi}$)
\begin{equation}
\Biggl[ k^2 - m^2 + i k^0 \tilde{\gamma}_2(k) 
- \frac{3 \hbar f^2}{4\pi^2}
\int^1_0 d\alpha E(\alpha; k^2) 
\text{log}\, \Biggl( \frac{|E(\alpha; k^2)|}{\Lambda^2}
\Biggr) \Biggr]
\tilde{\phi}(k) = -\tilde{J}(k),
\label{eq-mfe}
\end{equation}
where $\tilde{\gamma}_2(k)$ is the dissipation function, defined as 
$-i\hbar f^2/k^0$ times
the Fourier transform of the second term in Eq.~(\ref{eq-vf1}),
\begin{equation}
\tilde{\gamma}_2(k) = \frac{\hbar f^2}{k^0} \text{Im}
\,[\tilde{{\mathcal D}}_2(k)] =
\frac{\hbar f^2}{8\pi} \frac{k^2}{|k^0|} \left(1-\frac{4\mu^2}{k^2}
\right)^{3/2} \theta(k^2 - 4\mu^2).
\label{eq-ddf} 
\end{equation}
The one-loop $O(f^2)$ dissipation kernel agrees with previous calculations
of the probability to produce a fermion particle pair of momentum $k$
\cite{abbott:1982a,dolgov:1990a,itzykson:1980a,berera:1995a}.
In Eq.~(\ref{eq-mfe}), $\tilde{J}(k)$ is an external $c$-number source.
The imaginary term $i k^0 \tilde{\gamma}_2 \tilde{\phi}$ in Eq.~(\ref{eq-mfe}) 
breaks time-reversal invariance and acts as a $k$-dependent dissipative force 
in the mean field equation.
The $\theta$ function enforces the energy threshold for the virtual fermion
pair in the one-loop $O(f^2)$ diagram to go on-shell.  The dissipative
mean-field equation (\ref{eq-mfe}) is essentially the linear-response
approximation to the effective inflaton dynamics.  It should be noted
that the dissipation kernel ${\mathcal D}_2(x,x')$ is {\em nonlocal,\/}
in contrast with the local friction-type dissipation assumed in earlier
studies of post-inflation reheating \cite{kolb:1990a}.  However, in the 
limit $\mu^2 \rightarrow 0$, the dissipation kernel does become time-local,
as there is no longer a length scale in the expression for 
${\mathcal D}_2(x,x')$ which could define a time scale for nonlocal
dissipation \cite{calzetta:1988b}.

\subsection{One-loop perturbative effective action at $O(f^4)$}
The $O(f^4)$ term in the one-loop CTP effective action consists of the
``square'' diagram, which is the second term in Fig.~\ref{fig-gam1},
\begin{multline}
\Gamma^{(1)}_4[\phih] = \frac{i}{4} c^{abc} c^{a'b'c'} 
c^{def} c^{d'e'f'}
\int d^4 x d^4 x' d^4 y' d^4 y \biggl[  \text{Tr}_{\text{{\tiny sp}}} 
\bigl[F_{bb'}(x,x') F_{c'f'}(x',y') F_{e'e}(y',y) F_{fc}(y,x)\bigr] 
 \\ 
\times \phih_a(x) \phih_{a'}(x') \phih_d (y) 
\phih_{d'}(y')
\biggr].
\end{multline}
Expanding out the contracted CTP indices, we obtain
\begin{equation}
\begin{split}
\Gamma^{(1)}_4[\phih] = & \frac{i}{4} \int d^4 x d^4 x' d^4 y' d^4 y 
\biggl[   \\
& \phihp(x) \phihp(x') \phihp(y) \phihp(y') \text{Tr}_{\text{{\tiny sp}}} \bigl\{ 
F_{++}(x,x') F_{++}(x',y') F_{++}(y',y) F_{++}(y,x) \bigr\}  \\
& + \phihm(x) \phihm(x') \phihm(y) \phihm(y') \text{Tr}_{\text{{\tiny sp}}} \bigl\{
F_{--}(x,x') F_{--}(x',y') F_{--}(y',y) F_{--}(y,x) \bigr\}   \\
& - 4 \phihp(x) \phihm(x') \phihm(y') \phihm(y) \text{Tr}_{\text{{\tiny sp}}} \bigl\{
F_{+-}(x,x') F_{--}(x',y') F_{--}(y',y) F_{-+}(y,x) \bigr\}   \\
& - 4 \phihm(x) \phihp(x') \phihp(y') \phihp(y) \text{Tr}_{\text{{\tiny sp}}} \bigl\{
F_{-+}(x,x') F_{++}(x',y') F_{++}(y',y) F_{+-}(y,x) \bigr\}   \\
& + 4 \phihp(x) \phihp(x') \phihm(y') \phihm(y) \text{Tr}_{\text{{\tiny sp}}} \bigl\{
F_{++}(x,x') F_{+-}(x',y') F_{--}(y',y) F_{-+}(y,x) \bigr\}   \\
& + 2 \phihp(x) \phihm(x') \phihp(y') \phihm(y) \text{Tr}_{\text{{\tiny sp}}} \bigl\{
F_{+-}(x,x') F_{-+}(x',y') F_{+-}(y',y) F_{-+}(y,x) \bigr\} 
\biggr].
\end{split}
\end{equation}
When $\Gamma_4^{(1)}$ is expressed in terms of $\Delta$ and $\Sigma$ 
[defined in Eqs.~(\ref{eq-sdva}) and (\ref{eq-sdvb})], only those
terms with one factor of $\Delta$ and three factors of $\Sigma$ contribute
to the mean field equation of motion.  As a consequence of the unitarity
condition (\ref{eq-uc}),
the sum of terms proportional to four factors of $\Sigma$ must vanish.
Keeping only those terms in the effective action which contribute to the mean
field equation or are quadratic in $\Delta$, we find
\begin{equation}
\begin{split}
\Gamma_4^{(1)}[\phih] = \int d^4 x d^4 x' d^4 y' d^4 y \biggl[ &
\Delta(x) \Sigma(x') \Sigma(y') \Sigma(y) {\mathcal D}_4(x,x',y',y) 
 \\ & 
+ \frac{i}{2} \Delta(x) \Delta(x') \Sigma(y') \Sigma(y) {\mathcal N}_4
(x,x',y',y) \biggr],
\end{split}
\end{equation}
in terms of a kernel ${\mathcal D}_4(x,x',y',y)$ defined by
\begin{multline}
{\mathcal D}_4(x,x',y',y) = -\text{Im} \, \text{Tr}_{\text{{\tiny sp}}}
\biggl[  
   F_{++}(x,x') F_{++}(x',y') F_{++}(y',y) F_{++}(y,x)   \\
 + F_{++}(x,x') F_{+-}(x',y') F_{--}(y',y) F_{-+}(y,x)   \\
 + F_{+-}(x,x') F_{--}(x',y') F_{-+}(y',y) F_{++}(y,x)   \\
 + F_{+-}(x,x') F_{-+}(x',y') F_{+-}(y',y) F_{-+}(y,x)   \\
 - F_{+-}(x,x') F_{-+}(x',y') F_{++}(y',y) F_{++}(y,x)   \\
 - F_{++}(x,x') F_{+-}(x',y') F_{-+}(y',y) F_{++}(y,x)   \\
 - F_{++}(x,x') F_{++}(x',y') F_{+-}(y',y) F_{-+}(y,x)   \\
 - F_{+-}(x,x') F_{--}(x',y') F_{--}(y',y) F_{-+}(y,x) \label{eq-fod}
\biggr],
\end{multline}
and a ``noise'' kernel ${\mathcal N}_4(x,x',y',y)$ defined by
\begin{multline}
{\mathcal N}_4(x,x',y',y) = \text{Re} \, \text{Tr}_{\text{{\tiny sp}}}
\biggl[  
   F_{++}(x,x') F_{++}(x',y') F_{++}(y',y) F_{++}(y,x)   \\
 + F_{++}(x,x') F_{+-}(x',y') F_{--}(y',y) F_{-+}(y,x)   \\
 - F_{+-}(x,x') F_{--}(x',y') F_{-+}(y',y) F_{++}(y,x)   \\
 - F_{+-}(x,x') F_{-+}(x',y') F_{+-}(y',y) F_{-+}(y,x)   \\
 + F_{+-}(x,x') F_{-+}(x',y') F_{++}(y',y) F_{++}(y,x)   \\
 - F_{++}(x,x') F_{+-}(x',y') F_{-+}(y',y) F_{++}(y,x)   \\
 - F_{++}(x,x') F_{++}(x',y') F_{+-}(y',y) F_{-+}(y,x)   \\
 + F_{+-}(x,x') F_{--}(x',y') F_{--}(y',y) F_{-+}(y,x)   \\
 + \{F_{++}(x,y') F_{++}(y',x') F_{++}(x',y) F_{++}(y,x) \\
 - F_{+-}(x,y') F_{--}(y',x') F_{-+}(x',y) F_{++}(y,x)   \\
 - F_{++}(x,y') F_{+-}(y',x') F_{--}(x',y) F_{-+}(y,x)   \\
 + F_{+-}(x,y') F_{-+}(y',x') F_{+-}(x',y) F_{-+}(y,x)   \\
 - F_{+-}(x,y') F_{-+}(y',x') F_{++}(x',y) F_{++}(y,x)   \\
 + F_{++}(x,y') F_{+-}(y',x') F_{-+}(x',y) F_{++}(y,x)   \\
 - F_{++}(x,y') F_{++}(y',x') F_{+-}(x',y) F_{-+}(y,x)   \\
 + F_{+-}(x,y') F_{--}(y',x') F_{--}(x',y) F_{-+}(y,x)\}/2
\biggr],
\end{multline}
The noise kernel ${\mathcal N}_4$ does not contribute to the mean field
equation of motion.  There are, of course, terms in $\Gamma^{(1)}_4[\phih]$
which are higher order in $\Delta$, for example, $O(\Delta^4)$, 
but in passing over to a stochastic equation for $\phih$ in Sec.~\ref{sec-nk},
 we will be assuming that $\Delta$ is small, so that
higher-order terms in powers of $\Delta$ can be ignored.  Such terms will 
in general contribute to non-Gaussian noise, 
which will be studied in an upcoming paper \cite{calzetta:1998b}.  

Let us evaluate the first term of Eq.~(\ref{eq-fod}), 
which consists of only Feynman propagators.  The term is logarithmically
divergent, and as in Sec.~\ref{sec-olpefa}, we use dimensional continuation
and the modified minimal subtraction ($\overline{\text{MS}}$) renormalization 
scheme.  Because we are only interested in 
deriving the dissipative terms in the mean-field equation coming from this
diagram, and because we are assuming $m \gg \mu$, 
we include only the one-loop logarithm.  We find
\begin{multline}
\text{Tr}_{\text{{\tiny sp}}} \Bigl[ F_{++}(x,x')  
F_{++}(x',y') F_{++}(y',y) F_{++}(y,x) \Bigr]
_{\text{log only}}
\\  = i \int \frac{d^4 k_1}{(2\pi)^4} \frac{d^4 k_2}{(2\pi)^4} 
\frac{d^4 k_3}{(2\pi)^4}  e^{-i [
-(k_1 + k_2 + k_3) \cdot x + k_1 \cdot x' + k_2 \cdot y' + k_3 \cdot y 
]} A_4(k_1, k_2, k_3),
\end{multline}
where the amplitude $A_4(k_1, k_2, k_3)$ is defined by
\begin{equation}
A_4(k_1, k_2, k_3) = -\frac{3}{2\pi^2} \int 
d\alpha_1 d\alpha_2 d\alpha_3 \log \Biggl[ \frac{ E_4
(\alpha_1,\alpha_2,\alpha_3;
k_1,k_2,k_3)}{\Lambda^2} \Biggr],
\end{equation}
in terms of a function $E_4$ defined by
\begin{equation}
\begin{split}
E_4(\alpha_1,\alpha_2,\alpha_3;k_1,k_2,k_3) = &
\Bigl[ (1-\alpha_1)k_1 + (1-\alpha_1 - \alpha_2)k_2
+ (1-\alpha_1 -\alpha_2-\alpha_3)k_3\Bigr]^2 
\\ & - (1-\alpha_1)k_1^2-
(1-\alpha_1-\alpha_2)(k_2^2 + 2k_1\cdot k_2) 
\\ & 
- (1-\alpha_1-\alpha_2-\alpha_3)(2 k_1 \cdot k_3 + 2 k_2 \cdot k_3 + k_3^2)
+ \mu^2.
\end{split}
\end{equation}
As in Sec.~\ref{sec-olpefa}, the cut diagrams in Eq.~(\ref{eq-fod}) are 
related to the discontinuities in $E_4
(\alpha_1,\alpha_2,\alpha_3;k_1,k_2,k_3)$ 
via the Cutosky rules.  The details are shown in the appendix.
We can then express the dissipation kernel ${\mathcal D}_4(x,x',y',y)$
as a Fourier transform over external momenta,
\begin{equation}
{\mathcal D}_4(x,x',y',y) = 
\int \frac{d^4 k_1}{(2\pi)^4}
\frac{d^4 k_2}{(2\pi)^4} \frac{d^4 k_3}{(2\pi)^4} 
e^{-i[ -(k_1 + k_2 + k_3)\cdot x + k_1 \cdot x' + k_2 \cdot y' + k_3 \cdot y]}
\tilde{{\mathcal D}}_4(k_1,k_2,k_3),
\label{eq-d4ft}
\end{equation}
in terms of a function $\tilde{{\mathcal D}}_4(k_1,k_2,k_3)$ defined by
\begin{equation}
\begin{split}
\tilde{{\mathcal D}}_4(k_1,k_2,k_3) = \frac{3}{2\pi} 
\Biggl[ & \frac{1}{\pi}
\int d\alpha_1 d\alpha_2 d\alpha_3 \log \left( \frac{| E_4(\alpha_1,
\alpha_2,\alpha_3; k_1, k_2,k_3) |}{\Lambda^2} \right)   \\ 
& +  i\text{sgn}(k_2^0 + k_3^0) h[ (k_2 + k_3)^2 ] +
i\text{sgn}(k_1^0 + k_2^0) h[ (k_1 + k_2)^2 ]   \\
& +  i\text{sgn}(k_1^0 + k_2^0 + k_3^0) h[ k_2^2 ] 
+  i\text{sgn}(k_2^0) h[ (k_1 + k_2 + k_3)^2 ]   \\
& +  i\text{sgn}(k_3^0) h[ k_1^2 ] +
i\text{sgn}(k_1^0) h[ k_3^2 ] -  i H(k_1,k_2,k_3) \Biggr],
\label{eq-fpdk}
\end{split}
\end{equation}
and the functions $h(s)$ and $H(k_1,k_2,k_3)$ are defined by
\begin{align}
& \qquad h(s) = \sqrt{1-\frac{4\mu^2}{s}} \theta(s-4\mu^2), \label{eq-dhf} \\
& \begin{split}
H(k_1,k_2,k_3) = \int_{\alpha_1,\alpha_2,\alpha_3 > 0} \bigl\{  
& \theta[ -E_4(\alpha_1,\alpha_2,\alpha_3; k_1,k_2,k_3) ]  \\
& \times [ \theta(k_1^0) \theta(-k_2^0) \theta(k_3^0) -
     \theta(-k_1^0) \theta(k_2^0) \theta(-k_3^0) ] \bigr\}.
\end{split}
\end{align}
Equation~(\ref{eq-fpdk}) leads to the following mean-field equation at
$O(f^4)$,
\begin{equation}
\begin{split}
\Biggl[ & k^2 - m^2 + i k^0 \tilde{\gamma}_2(k) -
\frac{3 \hbar f^2}{4\pi^2}
\int^1_0 d\alpha E(\alpha; k^2) 
\text{log}\, \Biggl( \frac{|E(\alpha; k^2)|}{\Lambda^2}
\Biggr) \Biggr]
\tilde{\phi}(k)   \\
& 
- \frac{3i\hbar f^4}{2\pi} \int \frac{d^4 q}{(2\pi)^4} \frac{d^4 l}{(2\pi)^4}
\tilde{\phi}(k-q-l) \tilde{\phi}(q) \tilde{\phi}(l) \Biggl[ 
\frac{i}{\pi} \int d\alpha_1 d\alpha_2 d\alpha_3 \log \left( \frac{|E_4
(l+q-k,-q,-l)|}{
\Lambda^2} \right)   \\
& \qquad \qquad \qquad\qquad\qquad\qquad \qquad \qquad \qquad \qquad
+ \text{sgn}(q^0 + l^0) h[(q+l)^2] + \text{sgn}(k^0) h(q^2) 
  \\
& \qquad \qquad \qquad\qquad\qquad\qquad \qquad \qquad \qquad \qquad
+ \text{sgn}(k^0 - l^0) h[(k-l)^2] + \text{sgn}(q^0) h(k^2) 
  \\
& \qquad \qquad \qquad\qquad\qquad\qquad \qquad \qquad \qquad \qquad
+ \text{sgn}(k^0 - q^0 - l^0) h(l^2)   \\
& \qquad \qquad \qquad\qquad\qquad\qquad \qquad \qquad \qquad \qquad
+ \text{sgn}(l^0) h[(k-q-l)^2]   \\
& \qquad \qquad \qquad\qquad\qquad\qquad \qquad \qquad \qquad \qquad
+ H(l+q-k,-q,-l) 
\Biggr] = -\tilde{J}(k).
\label{eq-og4dt}
\end{split}
\end{equation}
The presence of terms of the form $i\text{sgn}(p^0)h(p^2)$ in Eq.~(\ref{eq-og4dt})
clearly signifies dissipative dynamics.  The $\theta$-function in Eq.~(\ref{eq-dhf})
enforces the energy threshold for the virtual fermion quanta created at a 
particular vertex to go on-shell.
Comparing Eq.~(\ref{eq-og4dt}) and Eq.~(\ref{eq-mfe}), and assuming
spatial homogeneity, we see that the 
$O(f^4)$ dissipation kernel must be taken into account whenever
the condition (\ref{eq-vpt}) fails to hold for the solution $\phi(t)$ to 
Eq.~(\ref{eq-mfe})\@.

At the end of the regime of parametric resonance in chaotic inflaton, i.e.
the ``preheating'' regime, 
the inflaton mean field may oscillate with an amplitude as large as 
$\sim m/g_{\phi\chi}$, where
$g_{\phi\chi}$ is the coupling to another scalar field $\chi$, typically on
the order of $10^{-4}$ \cite{kofman:1997a}.  Condition~(\ref{eq-vpt}) would then
be violated if $f > g_{\phi\chi}$.   In this case it would be necessary, at a 
minimum, to take into account higher order terms (such as ${\mathcal D}_4$)
in the mean-field equation,  until such time as the amplitude $\phih_0(t)$ 
has decreased to the point where Eq.~(\ref{eq-vpt}) is satisfied.

\section{Noise kernel and stochastic inflaton dynamics}
\label{sec-nk}
Although the kernels ${\mathcal N}_2(x,x')$ and ${\mathcal N}_4(x,x',y',y)$
do not contribute to the mean 
field equation, i.e., the equation of motion for $\phih$, they contain
information about stochasticity in a quasi-classical description of the
effective dynamics of the inflaton field 
\cite{hu:1992a,hu:1993a,hu:1993c,hu:1994a,hu:1995a,calzetta:1995a,calzetta:1994a,gleiser:1994a,lombardo:1996a}.  
In this section, we study the effect of stochasticity
on the dynamics of the inflaton mean field, within the perturbative framework
established above.

\subsection{Langevin equation and fluctuation-dissipation relation at
$O(f^2)$}
In this section we show how to obtain a classical stochastic equation 
for the inflaton field from the $O(f^2)$ perturbative CTP
effective action.  From Eq.~(\ref{eq-d1}), it follows that the $O(f^2)$
one-loop perturbative CTP effective action has the form
\begin{equation}
\Gamma[\phih] = {\mathcal S}^{\phi}[\phih] + \int d^4 x d^4 x' \left[
\Sigma(x) \Delta(x') \mu_2(x,x') + 
\frac{i}{2} \Delta(x) \Delta(x') \nu_2(x,x') \right],
\label{eq-olpea2}
\end{equation}
where for simplicity we have defined
\begin{align}
\nu_2(x,x') &= \hbar f^2 {\mathcal N}_2(x,x'), \\
\mu_2(x,x') &= \hbar f^2 {\mathcal D}_2(x,x').
\end{align}
In order to extract the stochastic noise arising from the
kernel ${\mathcal N}_2(x,x')$, we use the Gaussian identity \cite{ryder:1985a}
\begin{multline}
\exp\biggl[
-\frac{1}{2\hbar} \int d^{\,4}x d^{\,4}x' \Delta(x) \Delta(x') \nu_2
(x,x') \biggr] 
\\ = N \int D\xi_2 \exp \biggl[  -\frac{1}{2\hbar}  \int d^{\,4}x
d^{\,4}x' \xi_2(x)\nu_2^{-1}(x,x') \xi_2(x')  
 + \frac{i}{\hbar}\int 
d^{\,4}x \xi_2(x) \Delta(x)\biggr],
\label{eq-gid}
\end{multline}
where $N$ is a normalization factor which does not depend on $\Delta$,
and $\xi_2$ is a $c$-number functional integration variable.
Following \cite{hu:1995a}, we now define a functional
\begin{equation}
P[\xi_2] = N \exp\biggl[-\frac{1}{\hbar} \int d^{\,4}x d^{\,4}x' \xi_2(x) 
\nu_2^{-1}(x,x') \xi_2(x')\biggr],
\label{eq-dou}
\end{equation}
and it follows from Eq.~(\ref{eq-gid}) that $P[\xi_2]$ is normalized in the 
sense of
\begin{equation}
\int D\xi_2 P[\xi_2] = 1.
\end{equation}
Using Eq.~(\ref{eq-gid}), we can rewrite the $O(f^2)$ one-loop CTP effective 
action, Eq.~(\ref{eq-olpea2}), as
\begin{multline}
\Gamma[\phih] = -i \hbar \log \, \int D\xi_2 P[\xi_2] \exp \biggl[ \frac{i}{\hbar}
\Bigl( {\mathcal S}^{\phi}[\phih] + \int d^{\,4}x d^{\,4}x' \Sigma(x)
\Delta(x') \mu_2(x,x') \\ 
+ \int d^{\,4}x \xi_2(x) \Delta(x) \Bigr) \biggr].
\end{multline}
This suggests defining a new effective action which depends on both
$\xi_2$ and $\phih_{\pm}$ (dropping the carat from $\phih$),
\begin{equation}
\Gamma[\phi,\xi_2] = {\mathcal S}^{\phi}[\phi] + \int d^{\,4}x d^{\,4}x'
\Sigma(x) \Delta(x') \mu_2(x,x') + \int d^{\,4}x \xi_2(x) \Delta(x).
\end{equation}
Let us define a type of ensemble average
\begin{equation}
\langle \langle A \rangle \rangle = \int D\xi_2 P[\xi_2] A(\xi_2),
\label{eq-ntea}
\end{equation}
and note that Eqs.~(\ref{eq-dou}) and (\ref{eq-ntea}) imply that
\begin{align}
& \langle\langle \xi_2(x) \rangle\rangle = 0, \\
& \langle\langle \xi_2(x)\xi_2(x') \rangle\rangle = \hbar \nu_2(x,x').
\label{eq-nc}
\end{align}
Clearly then,
\begin{equation}
\left.\left(\frac{\delta}{\delta \phihp}\langle\langle \Gamma[\phi,\xi_2] 
\rangle\rangle\right)\right|_{\phip = \phim = \phi} =
\left.\left(\frac{\delta}{\delta \phihp}\Gamma[\phi] \right)
\right|_{\phip = \phim = \phi}.
\end{equation}
Taking the variation of $\Gamma[\phi,\xi_2]$ with respect to $\phip$ and
setting $\phip = \phim = \phi$, we obtain (after a Fourier transform)
\begin{equation}
\Biggl[ k^2 - m^2 + i k^0 \tilde{\gamma}_2(k) -
\frac{3 \hbar f^2}{4\pi^2}
\int^1_0 d\alpha E(\alpha;k^2) \text{log}\, 
\Biggl( \frac{|E(\alpha;k^2)|}{\Lambda^2}
\Biggr)\Biggr] \tilde{\phi}(k) + \tilde{\xi}_2(k) = -\tilde{J}(k),
\label{eq-le}
\end{equation}
where $\tilde{\xi}_2(k)$ is defined by
\begin{equation}
\tilde{\xi}_2(k) = \int d^{\,4}x e^{ikx} \xi_2(x).
\end{equation}
We now interpret Eq.~(\ref{eq-le}) as a Langevin equation with 
stochastic force $\xi_2$.  The inflaton Fourier mode $\tilde{\phi}$ 
appearing in Eq.~(\ref{eq-le}) should be viewed as a $c$-number 
stochastic variable, and the presence of the 
stochastic force $\xi_2$ indicates 
spontaneous breaking of spatial translation invariance by a Gaussian
(but not white) noise source $\xi_2$ \cite{hu:1994a}.  
Moreover, this stochastic equation obeys a
zero-temperature fluctuation-dissipation relation, as we now show.
First, let us calculate the one-loop $O(f^2)$ noise kernel, 
${\mathcal N}_2(x,x')$ [defined in Eq.~(\ref{eq-dn2}) above], using dimensional
regularization and modified minimal subtraction,
\begin{equation}
\nu_2(x,x') = \frac{\hbar f^2}{8\pi} \int \frac{d^{\,4}k}{(2\pi)^4}
e^{-ik(x-x')} \tilde{\nu}_2(k),
\end{equation}
in terms of the Fourier-transformed noise kernel
\begin{equation}
\tilde{\nu}_2(k) = \frac{\hbar f^2}{8\pi} k^2 \Bigl( 1-\frac{4\mu^2}{
k^2}\Bigr)^{\frac{3}{2}} \theta(k^2 - 4\mu^2).
\label{eq-dnk}
\end{equation}
The noise kernel $\nu_2(x,x')$ is colored; colored noise has been observed in 
other interacting field theories \cite{hu:1993a,hu:1994a,hu:1995a}.
By inspection of Eqs.~(\ref{eq-ddf}) and (\ref{eq-dnk}), it follows that
\begin{equation}
|k^0| \tilde{\gamma}_2(k) = \tilde{\nu}_2(k),
\label{eq-fdr1}
\end{equation}
which leads to the zero-temperature  
fluctuation-dissipation relation \cite{hu:1995a},
\begin{equation}
\nu_2(t,\vec{k}) = \int_{-\infty}^{\infty} dt' K(t-t') 
\gamma_2(t',\vec{k}),
\end{equation}
in terms of the distribution-valued kernel $K(t)$ defined by
\begin{equation}
K(t) = \int_0^{\infty} \frac{d\omega}{\pi} \omega \cos (\omega t),
\end{equation}
and the spatially Fourier-transformed dissipation function and noise kernel,
\begin{align}
& \nu_2(t,\vec{k}) = \int^{\infty}_{-\infty}
 \frac{d k^0}{2\pi} e^{-ik^0 t} \tilde{\nu}_2(k), \\
& \gamma_2(t,\vec{k}) = \int^{\infty}_{-\infty} 
\frac{d k^0}{2\pi} e^{-ik^0 t} \tilde{\gamma}_2(k).
\end{align}
This shows the physical significance of the noise kernel $\nu_2(x,x')$
in an effective description of the dynamics of the scalar mean field.  

\subsection{Langevin equation and fluctuation-dissipation relation
at $O(f^4)$}
In this section we consider the $O(f^4)$ one-loop noise kernel,
${\mathcal N}_4$.  The non-normal-threshold
singularities in $A_4$ lead to a noise kernel which depends on $\Sigma$, 
which is known to lead to ambiguities in the resulting Langevin equation
\cite{risken:1989a,son:1997a}.  The meaning and interpretation of
the non-normal-threshold parts of ${\mathcal N}_4$ and ${\mathcal D}_4$
will be the subject
of a future study \cite{calzetta:1998b}.  Here, we focus on the effect of the 
{\em normal-threshold\/} singularities of $A_4$, which for the noise kernel, 
${\mathcal N}_4$, contribute a term
\begin{equation}
\frac{i}{2} 
\int d^4x d^4 x' \Delta(x) \Delta(x') \Sigma(x) \Sigma(x') \nu_4(x,x')
\end{equation}
to the CTP effective action, where the kernel $\nu_4(x,x')$ is defined by
\begin{equation}
\nu_4(x,x') = -\frac{3\hbar f^4}{\pi} \int \frac{d^4 q}{(2\pi)^4}
e^{-i q\cdot(x-x')} h(q^2),
\end{equation}
and the function $h(s)$ was defined in Eq.~(\ref{eq-dhf}) above.  The 
normal-threshold singularities of the dissipation kernel, ${\mathcal D}_4$,
[the second and third terms of Eq.~(\ref{eq-fpdk})], lead to the following 
contribution to the CTP effective action,
\begin{equation}
\int d^4 x d^4 x' \Delta(x) \Sigma(x) \left[ \Sigma(x') \right]^2 \mu_4(x,x'),
\end{equation}
where the kernel $\mu_4(x,x')$ is defined by
\begin{equation}
\mu_4(x,x') = -\frac{3 i \hbar f^4}{\pi} \int \frac{d^4 q}{(2\pi)^4} 
e^{-i q\cdot(x'-x)} \text{sgn}(q^0) h(q^2).
\end{equation}
With the definitions
\begin{align}
& \mu_4(x,x') = i \int \frac{d^4 q}{(2\pi)^4} e^{-i q\cdot(x'-x)}
q^0 \tilde{\gamma}_4(q), \\
& \nu_4(x,x') = \int \frac{d^4 q}{(2\pi)^4} e^{-i q\cdot(x'-x)}
\tilde{\nu}_4(q),
\end{align}
it follows immediately that the normal-threshold parts of ${\mathcal D}_4$
and ${\mathcal N}_4$ obey a fluctuation-dissipation
relation identical in form to Eq.~(\ref{eq-fdr1}),
\begin{equation}
|q^0| \tilde{\gamma}_4(q) = \tilde{\nu}_4(q).
\end{equation}
Making use of Eq.~(\ref{eq-gid}), the $O(f^4)$ effective action (including
only normal-threshold contributions) can be written in the form
\begin{multline}
\Gamma[\phi,\xi_2,\xi_4] =  {\mathcal S}^{\phi}[\phi] 
+ \int d^4 x d^4 x' \Delta(x) \Sigma(x') \mu_2(x,x') 
+ \int d^4 x \xi_2(x) \Delta(x)   \\ 
+ \int d^4 x d^4 x' \Delta(x) \Sigma(x) \left[\Sigma(x')\right]^2\mu_4(x,x') 
+ \int d^4 x \xi_4(x) \Delta(x) \Sigma(x),
\label{eq-nea}
\end{multline}
where the stochastic noise source $\xi_4$ satisfies the conditions
\begin{align}
& \langle\langle \xi_4(x) \rangle\rangle = 0 \\ 
& \langle\langle \xi_4(x) \xi_4(x') \rangle\rangle = \hbar \nu_4(x,x').
\end{align}
Taking the functional derivative of Eq.~(\ref{eq-nea}) and making the
usual identification, we obtain a Langevin equation with an additive
noise $\xi_2$ and a multiplicative noise $\xi_4$,
\begin{equation}
\begin{split}
\Biggl[ & q^2 - m^2 + i q^0 \tilde{\gamma}_2(q) - \frac{3 \hbar f^2}{4\pi^2}
\int_0^1 d\alpha E(\alpha;q^2) \log \left( \frac{|E(\alpha;q^2)|}{\Lambda^2}
\right) \Biggr] \tilde{\phi}(q)   \\ & 
+ \int \frac{d^4 k}{(2\pi)^4} 
\frac{d^4 l}{(2\pi)^4} Q(l,q,k) \tilde{\phi}(q-l) \tilde{\phi}(k)
\tilde{\phi}(l-k)  
%
%
\\
& \qquad\qquad\qquad\qquad\qquad\qquad\qquad\qquad
= -\tilde{\xi}_2(q) -\tilde{J}(q) -\int \frac{d^4 k}{(2\pi)^4} \tilde{\xi}_4
(q-k) \tilde{\phi}(k),
\end{split}
\end{equation}
where $d^3 \alpha = d\alpha_1 d\alpha_2 d\alpha_3$,
and $Q(l,q,k)$ is defined as
\begin{equation}
Q(l,q,k) = i l^0 \tilde{\gamma}_4(l) - \frac{3\hbar f^4}{2\pi^2}
\int_0^{1} d^3 \alpha \log\left( \frac{|E_4(
\alpha_1,\alpha_2,\alpha_3; l-q,-k,l-k)|}{\Lambda^2}\right).
\end{equation}
The stochastic force $\xi_4$ is clearly seen to contribute multiplicatively
to the Langevin equation for $\phi$.

\subsection{Homogeneous mean field dynamics at $O(f^2)$}
To make connection with post-inflationary reheating, it is customary to
assume that the mean field $\phih$ is spatially homogeneous 
\cite{kofman:1996a,kofman:1997a,shtanov:1995a}.  In this case,
the Langevin equation~(\ref{eq-le}) takes the form
\begin{equation}
\left[\omega^2 - m^2 + i \omega \beta(\omega) + \eta(\omega) 
\right] \tilde{\phi}(\omega) + \tilde{\xi}_2(\omega) = -\tilde{J}(\omega),
\label{eq-hle}
\end{equation}
where we have defined
\begin{align}
& \beta(\omega) = \frac{\hbar f^2
}{8\pi}\frac{\omega^2}{|\omega|}\left(
1-\frac{4\mu^2}{\omega^2} \right)^{3/2}\theta(\omega^2 - 4\mu^2), \\
& \eta(\omega) = -\frac{3\hbar f^2}{4\pi^2}\int_0^1 d\alpha E(\alpha;\omega^2)
\log \left( \frac{|E(\alpha;\omega^2)|}{\Lambda^2} \right). 
\end{align}
The total energy dissipated to the fermion field over the history of
the dynamical evolution of the mean field is given [at $O(f^2)$] by
\begin{equation}
{\mathcal E} = -\int^{\infty}_{-\infty} dt F_v(t) \frac{d \phi(t)}{dt},
\end{equation}
where the friction force $F_v(t)$ is the Fourier transform of $i\beta
(\omega) \omega \tilde{\phi}(\omega)$.  
After a bit of Fourier algebra, we obtain an expression for the 
ensemble-averaged, total dissipated energy,
\begin{equation}
\langle\langle{\mathcal E}\rangle\rangle = \frac{\hbar f^2
}{8\pi^2} \int_{2\mu}^{\infty} d\omega
\omega^3 \left( 1- \frac{4 \mu^2}{\omega^2} \right)^{3/2} \frac{
|\tilde{J}
(\omega)|^2}{\left[ \omega^2 - m^2 + \eta(\omega) 
\right]^2 + \omega^2 \beta(\omega)^2}.
\end{equation}
It is straightforward to compute the variance in the total dissipated energy.
Making use of Eq.~(\ref{eq-nc}), we find
\begin{equation}
|\langle\langle {\mathcal E}^2 \rangle\rangle - \langle\langle {\mathcal E}
\rangle\rangle^2| = \frac{\hbar^4 f^6}{256 \pi^6}\int_{2\mu}^{\infty}
d\omega \omega^6 I(\omega)
 \left( 1-\frac{4\mu^2}{\omega^2} \right)^{3}
\frac{|\tilde{J}(\omega)|^2}{\left[ (\omega^2 - m^2 + 
\eta(\omega))^2 + \omega^2 \beta(\omega)^2\right]^2},
\end{equation}
where the function $I(\omega)$ is defined by
\begin{equation}
I(\omega) = \int_0^{\sqrt{\omega^2 - 4\mu^2}} dk k^2 (\omega^2 - k^2) \left(
1-\frac{4\mu^2}{\omega^2 - k^2}\right)^{3/2}.
\end{equation}
Following \cite{abbott:1982a}, we assume that the inflaton field
is held fixed via an external, constant
 $c$-number source $J$ for $t < 0$, and that
the source is removed for $t \geq 0$,
\begin{equation}
J(t) = J \theta(-t).
\label{eq-jt}
\end{equation}
Setting $\hbar = 1$, assuming that $m \gg \mu$, and expanding to lowest order 
in $f$, we obtain for the ensemble averaged dissipated energy,
\begin{equation}
\langle\langle {\mathcal E} \rangle\rangle = \frac{f^2 J^2}{16\pi^2 m^2}.
\label{eq-eade}
\end{equation}
Let us now compute the variance in the total dissipated energy.
Performing a regularization via dimensional continuation, we obtain 
\begin{equation}
|\langle\langle {\mathcal E}^2 \rangle\rangle - 
\langle\langle {\mathcal E} \rangle\rangle^2| =
\frac{f^6 J^2 m^2}{960 \pi^6} \delta^2,
\label{eq-vtde}
\end{equation}
where $\delta$ is a constant of order unity defined by
$\delta^2 = | 119/60 - \gamma_{\text{{\tiny EM}}} -
\log (4\pi m^2/\Lambda^2)|$, and 
$\gamma_{\text{{\tiny EM}}}$ is the Euler-Mascheroni
constant, $\approx 0.5772$.
Taking the ratio of the square root of Eq.~(\ref{eq-vtde}) and 
Eq.~(\ref{eq-eade}), we obtain the relative strength of the RMS fluctuations
in the total dissipated energy density, ${\mathcal E}_{\text{{\tiny rms}}}$, 
\begin{equation}
\frac{{\mathcal E}_{\text{{\tiny rms}}}}{
\langle\langle {\mathcal E}\rangle\rangle} \equiv
\frac{\sqrt{|\langle\langle {\mathcal E}^2 \rangle\rangle -
\langle\langle {\mathcal E} \rangle\rangle^2|}}{\langle\langle
{\mathcal E}\rangle\rangle} 
= \frac{2 f m^3 \delta}{\sqrt{15} \pi J}.
\end{equation}
The parameter $J$ is related to the initial inflaton amplitude 
$\phih_0(t_0)$ by
$J = \phih_0(t_0) m^2 /2$, which leads to
\begin{equation}
\frac{{\mathcal E}_{\text{{\tiny rms}}}}{
\langle\langle {\mathcal E} \rangle\rangle} = 
\frac{4 f m \delta}{\sqrt{15} \pi \phih_0(t_0)} \simeq 0.390 
\frac{m f}{\phih_0(t_0)}.
\label{eq-dee}
\end{equation}
The fundamental assumption which justified the perturbative expansion in $f$,
Eq.~(\ref{eq-vpt}), is seen to be independent of 
Eq.~(\ref{eq-dee}).  Therefore, the ratio 
${\mathcal E}_{\text{{\tiny rms}}} /
{\mathcal E}$ is not required to be small by consistency with 
perturbation theory.  As the initial inflaton amplitude $\phi_0$ is
made larger, the relative strength of the rms fluctuations of ${\mathcal E}$
is seen to decrease, in accordance with the correspondence principle.
It has been shown that the fluctuations in the total dissipated energy
density are related to the fluctuations in the occupation numbers of modes
\cite{calzetta:1994a}.
 
Let us now examine whether the rms fluctuations in the total dissipated
energy, as given by Eq.~(\ref{eq-dee}), is significant, given a reasonable
value for the inflaton amplitude at the end of the preheating regime
(the period of parametric resonance-induced particle production).
In chaotic inflaton with a scalar field $\chi$ coupled to the inflaton field
via a coupling constant $g_{\phi\chi}$, the typical inflaton amplitude at
the end of the preheating regime is on the order of $m/g_{\phi\chi}$ 
\cite{kofman:1997a}.  In this case, we would find
${\mathcal E}_{\text{{\tiny rms}}}/\langle\langle {\mathcal E}\rangle\rangle
\simeq f g_{\phi\chi},$
from which it is clear that fluctuations in the {\em total\/} dissipated energy
are not significant relative to the ensemble-averaged total dissipated energy,
and therefore should not appreciably affect the reheating temperature.  
However, in {\em new inflation\/} scenarios where
the inflaton amplitude $\phih_0$ can be on the order of $m$ at the onset of
reheating, the ratio $m/\phih_0$ can be of order unity \cite{kolb:1990a}.  
In this case, the ratio ${\mathcal E}_{\text{{\tiny rms}}}/
\langle\langle{\mathcal E}
\rangle\rangle \simeq f$, which may not be a negligible effect.

Although as shown above, stochasticity does not dramatically affect the
total energy dissipated via fermion production in chaotic inflation, we may
inquire whether the noise term in the Langevin equation for the inflaton 
zero-mode, Eq.~(\ref{eq-hle}), may nonetheless be non-negligible during the 
reheating period.   Let us compute the rms fluctuations in the inflaton 
zero-mode, $\phih_{\text{{\tiny rms}}}$.  
Following methods described above, we find that the rms fluctuations of
the inflaton zero mode are given, to $O(f^2)$, by
\begin{equation}
\phih_{\text{{\tiny rms}}} = \frac{f}{\pi} \frac{m}{\sqrt{60\pi}} \sigma,
\label{eq-frms}
\end{equation}
where $\sigma^2 = |61/30 - \gamma_{\text{{\tiny EM}}} - \log (4\pi m^2/
\Lambda^2)|$.
Equation~(\ref{eq-frms}) is seen to be independent of the inflaton zero-mode 
amplitude $\phih_0$.

In order to determine the relative importance of fluctuations in the 
inflaton zero-mode amplitude $\phih_0$ during and at the end of the
reheating period, we must introduce curved spacetime arguments.  This is
because the end of the reheating period is determined by the time 
$t_{\text{{\tiny end}}}$ at which
the Hubble constant becomes of the order of $3\beta(m)$
for the case of $\lambda=0$ being discussed in this section 
\cite{kolb:1990a,shtanov:1995a}. 
Starting with the semiclassical Einstein equation (\ref{eq-lpfcstsee2}) 
for spatially flat Friedmann-Robertson-Walker (FRW) cosmology, setting 
$b=c=\Lambda_c=0$ (following arguments similar to those of Sec.~III~D of
Ref.~\cite{ramsey:1997b}), retaining only the inflaton zero mode as the
dynamical degree of freedom (for consistency with FRW), and retaining
both the $\psi$ field energy density $\rho_{\psi}$ and the 
{\em classical, stochastic\/} energy density of the inflaton zero mode, we have
\begin{equation}
H^2 = \frac{\dot{a}^2}{a^2} = \frac{8\pi}{3\Mpl^2} \left(
 \rho(\overline{\phih^2}) + \rho_{\psi} \right),
\label{eq-fre}
\end{equation}
where $a$ is the scale factor, the dot denotes a derivative with respect to
cosmic time, and $\rho(\overline{\phih^2})$ is the energy density
as a function of the time-average (over one period of oscillation)
of $\phih^2$, which is given by the virial theorem,
\begin{equation}
\overline{\rho(\phih)} \simeq m^2 \overline{\phih^2} = \frac{1}{2} m^2 
(\phih_0)^2.
\end{equation}
Making use of Eqs.~(\ref{eq-hle}), (\ref{eq-jt}), and (\ref{eq-fre}),
we obtain an approximate expression for the (ensemble-averaged)
energy density of the inflaton zero-mode at the end of the reheating
period\footnote{
We wish to emphasize, however, that this expression does {\em not\/} take
into account the regime of nonperturbative dynamics discussed in
Sec.~\ref{sec-necgidcst}, and therefore should not be expected to yield
a correct reheating temperature in a realistic inflationary scenario.  
However, it suffices for the present discussion of rms fluctuations
of the inflaton amplitude, where we assume an idealized case similar to
Eq.~(106) of Ref.~\cite{shtanov:1995a}.}
 (to lowest order in $f$),
\begin{equation}
\rho(t_{\text{{\tiny end}}}) \simeq \frac{3 f^4 \Mpl^2 m^2}{(8\pi)^3 e},
\label{eq-endrh}
\end{equation}
where $e$ is the base of the natural logarithm.  Note that this expression
is independent of the initial inflaton amplitude \cite{kofman:1997a}.
Equation~(\ref{eq-endrh}) allows us to solve 
for the value of $\phih_0$ at the end of reheating.  We find
\begin{equation}
\phih_0(t_{\text{{\tiny end}}}) \simeq \frac{\sqrt{6/e} \Mpl f^2}{
(8\pi)^{3/2}}.
\label{eq-minp}
\end{equation}
The rms fluctuations in the inflaton zero mode, $\phih_{\text{{\tiny rms}}}$,
can only play a role in the inflaton zero-mode dynamics during reheating
if the ratio $\phih_{\text{{\tiny rms}}} 
/ \phih_0$ is not small relative to higher-order
[e.g., $O(f^4)$] processes which we are neglecting.  In light of the minimum
inflaton zero-mode amplitude attained during reheating, Eq.~(\ref{eq-minp}),
we find that the ratio of fluctuations in the inflaton zero-mode to the
zero-mode amplitude is given by
\begin{equation}
\frac{\phih_{\text{{\tiny rms}}}}{\phih_0(t_{\text{{\tiny end}}})} \simeq
\frac{8 \sigma m}{f \Mpl} \sqrt{\frac{e}{45}} \simeq 2.37 \frac{m}{f \Mpl}.
\label{eq-vizm}
\end{equation}
We estimate the ratio of the mean-squared inflaton amplitude 
fluctuations to the shift in the inflaton mass to be
\begin{equation}
\left|\frac{\hat{\phi}^2_{\mbox{\tiny rms}}}{\eta(m)}\right| \simeq 0.01.
\end{equation}
If, prior to the end of reheating at $t_{\text{{\tiny end}}}$,
 $\phih_{\text{{\tiny rms}}}/\phih_0(t)$ becomes larger than higher-order
terms which are neglected in our perturbative expansion, then
fluctuations in the inflaton zero mode are a non-negligible effect.
This will happen when $\phih_{\text{{\tiny rms}}}/\phih_0(t_{\text{{\tiny 
end}}})$, given by Eq.~(\ref{eq-vizm}), is not $\ll 1$.
This shows that
stochasticity must be taken into account in the dynamics of the
inflaton zero mode, during the late stages of the reheating 
period.

\section{Summary}
\label{sec-conc}
In this chapter, we present the results of a study of (unbroken-symmetry) 
inflaton dynamics during the late stages of reheating, which is dominated by 
fermion particle production to a light spinor field coupled to the inflaton
field via a Yukawa coupling.  We derived coupled nonperturbative 
equations for the inflaton mean field and two-point function, in a 
general curved spacetime, and showed that, in addition to the
dissipative mean-field equation, the gap equation for the two-point
function is also dissipative, due to fermion particle production.
Simultaneous evolution of the inflaton mean-field and two-point function
is necessary for correctly following the inflaton dynamics after the
end of the preheating period, because the large value of the variance
invalidates use of the ordinary perturbative, 1PI effective action.

We also derived the dissipation and noise kernels for the small-amplitude
dynamics of the inflaton field, valid in the late stages of reheating
when the inflaton mean-field amplitude is very small.  The $O(f^2)$ noise
and dissipation kernels, as well as the normal-threshold parts of the
$O(f^4)$ noise and dissipation kernels, are shown to obey a zero-temperature 
fluctuation-dissipation relation.  With the noise and dissipation kernels,
a Langevin equation for the inflaton zero mode is derived, and it is shown
that the noise leads to a variance for the inflaton amplitude which
is non-negligible before the end of reheating.

\chapter{Correlation entropy in effectively open systems}
\label{chap-entropy}

\section{Introduction}
As emphasized in Chapter~\ref{chap-preheat}, an important issue within 
inflationary cosmology is how, and to what temperature, the Universe reheats 
following the period of profuse particle production.  Knowledge of this 
temperature is important for consistency of the inflationary Universe picture 
with the standard big bang cosmology.  Because of the self-coupling
of the inflaton field and its coupling to other quantum fields, it is
expected that the inflaton field and those coupled to it will eventually
come to local thermal equilibrium in the expanding background spacetime,
with the usual equilibrium equation of state depending on the masses
of the particle species relative to the temperature of the plasma 
\cite{kolb:1990a}.  The 
quantitative description of how the system reaches local thermal equilibrium,
and at what temperature, is known as the {\em thermalization problem.}  
Though there have been some previous studies of this problem, most either
employed classical arguments in an essential way \cite{khlebnikov:1997a}, 
or utilized initial conditions which are not appropriate to the end of 
preheating in realistic inflation scenarios \cite{boyanovsky:1996f}.  
In light of the recent, newer understanding of the role that parametric 
resonance effects play during 
reheating, the thermalization problem is at present unsolved 
\cite{brandenberger:1997b}.  
Because of the tremendous variety of different inflationary models, 
and the fact that the details of preheating dynamics of the inflaton field 
is severely model-dependent, any discussion of the thermalization period will 
of necessity be rather general.\footnote{
In fact, in some cases, the preheating and thermalization stages do not 
separate at all, and particle production
and thermalization must be considered simultaneously \cite{linde:1994a},
though in this discussion we will assume that such a separation exists, and
will think of the end state of the preheating stage as constituting the
initial conditions for thermalization. } However, it is useful to explicate 
some of the challenges, both
conceptual and technical, inherent in the thermalization problem.

First and most important, it is not clear that a separation of macroscopic
and microscopic time scales exists at the end of the preheating period
\cite{boyanovsky:1996b}.  In the two-particle-irreducible language 
of Chapter~\ref{chap-preheat}, a separation of macroscopic and microscopic
time scales would require that the initial conditions be such that one can 
assume a {\em quasilocal expansion\/} for the time dependence of the effective 
mass.  In fact, preliminary evidence suggests that for 
many realistic inflation scenarios, such a separation does not exist at
the end of the preheating stage, as can be seen in Chapter~\ref{chap-preheat},
where the time scale for variations of the effective mass is on the order of 
the microscopic time scale for the theory.  Such a separation constitutes the 
basis for passing 
over from the full, closed-system dynamics of quantum fields (given by the 
functional Schr\"{o}dinger or quantum Liouville
equation) to a quantum kinetic theory description in which one obtains  
separate equations for quantum field-theoretic processes on microscopic time 
scales, and relaxation phenomena on macroscopic time scales 
\cite{calzetta:1988b}.  By quantum kinetic
field theory, we are referring to the hierarchy of coupled equations for
the relativistic Wigner function and its higher-correlation analogues, which
are obtained by a Fourier transform of the relative coordinates in the
Schwinger-Dyson equations for correlation functions (or alternatively, in
the master effective action whose variation yields the Schwinger-Dyson
equations).  This is a quantum analogue of the BBGKY hierarchy 
\cite{balescu:1975a}, expressed in a representation convenient for 
distinguishing between microscopic (quantum field-theoretic) and macroscopic 
(transport and relaxation) phenomena.  As such, it does not require 
near-equilibrium conditions, and in fact, is applicable for a rather 
general moment expansion of the initial density matrix \cite{calzetta:1988b}.  
It should be pointed out that in order to {\em identify\/} the relativistic 
Wigner function with a distribution function for quasiparticles, one must 
show that the density matrix has {\em decohered,\/} 
and this is neither guaranteed 
nor required by the existence of a separation of macroscopic and microscopic 
time scales \cite{habib:1990a}.

Let us briefly comment on the important relation between quantum kinetic
field theory in its full generality, and an effective relativistic Boltzmann
description of relaxation phenomena for the one-particle distribution function
of quasiparticles.  In nonequilibrium statistical mechanics, as is well known 
\cite{balescu:1975a,dorfman:1995a}, the act of truncating the BBGKY hierarchy 
does not in itself lead to irreversibility and an $H$-theorem.  One must 
further perform a type of {\em coarse graining\/} of the truncated, coupled 
equations for $n$-particle distribution functions.  For example, if one 
truncates the hierarchy to include only the one-particle and two-particle 
distribution functions, it is the subsequent assumption that the two-particle 
distribution function at some initial time {\em factorizes\/} in terms of a 
product of single-particle distribution functions (which is related to the 
assumption of molecular chaos), and leads to the (irreversible) Boltzmann 
equation \cite{mclennan:1989a}.  
The assumption that the two-particle distribution function factorizes is an 
example of a type of coarse graining called {\em slaving\/} of the 
two-particle distribution function to the single-particle distribution 
function, in the language of Calzetta and Hu \cite{calzetta:1995b}.  
The situation in quantum kinetic field theory is completely analogous.  One 
may choose to work with a truncation of the hierarchy of the Wigner function 
and its higher correlation analogues, or one may instead perform a slaving of, 
for example, the Wigner-transformed four-point function, which leads 
(within the context of perturbation theory) directly to the relativistic 
Boltzmann equation \cite{calzetta:1988b} and the usual $H$-theorem.  Typically
this slaving of the higher correlation function(s) involves imposing causal
boundary conditions to obtain a particular solution for the higher correlation
function(s) in terms of the lower order correlation functions 
\cite{calzetta:1988b,calzetta:1995b}.  The truncation and subsequent slaving 
of the hierarchy within quantum kinetic field theory can be carried out at any 
desired order, as dictated by the initial conditions and relevant interactions.
As with any coarse graining procedure, in implementing the slaving of
a higher correlation/distribution function to lower correlation/distribution
functions, one is going over from a closed system to an {\em effectively 
open system,\/} the hallmarks of which are the emergence of dissipation 
\cite{calzetta:1988b,hu:1989a} and noise/fluctuations 
\cite{hu:1994a,calzetta:1995b}.  
This fact has led some to search for stochastic generalizations of the
Boltzmann equation \cite{kac:1979a,spohn:1983a}, motivated by that fact that
systems in thermal equilibrium always manifest fluctuations, as embodied in the
fluctuation-dissipation relation \cite{callen:1951a,kubo:1991a}.

The essential point about the process of slaving of higher correlation
(or distribution) functions is that it is a step which is wholly 
{\em independent\/} of the assumption of macroscopic and microscopic time 
scales.  In fact, a completely analogous procedure exists at the level
of the Schwinger-Dyson equations (i.e., without Wigner transformation) for 
correlation functions in an interacting quantum field theory 
\cite{calzetta:1995b}.  Recall that the Schwinger-Dyson equations are, in
the context of nonequilibrium field theory formulated using the 
Schwinger-Keldysh closed-time-path, an infinite chain of coupled dynamical
equations for all order correlation functions of the quantum field.  The
importance of the closed-time-path formalism in nonequilibrium situations
is that it ensures that the equations are causal and that the correlation
functions are ``in-in'' expectation values in the appropriate initial 
quantum state or density matrix.  As with the BBGKY hierarchy in 
nonequilibrium statistical mechanics, the general strategy is usually to 
truncate the hierarchy of correlation functions at some finite order. 
A general procedure has been presented for obtaining coupled equations
for correlation functions at any order $l$ in the correlation hierarchy,
which involves a truncation of the {\em master effective action\/} at a
finite order in the loop expansion \cite{calzetta:1995b}.  By working with
an $l$ loop-order truncation of the master effective action, one obtains
a closed, time-reversal invariant set of coupled equations for the first
$l+1$ correlation functions, $\phih$, $G$, $C_3$, \ldots, $C_{l+1}$.
In general, the equation of motion for the  highest order correlation 
function will be linear, and thus can be formally solved using Green's function
methods.  The existence of a unique solution depends on supplying causal
boundary conditions.  When the resulting solution for the highest correlation
function is then back-substituted into the evolution equations for the 
other lower-order correlation functions, the resulting dynamics is 
{\em not\/} time-reversal invariant, and generically dissipative.  
As with the slaving of the higher-order Wigner-transformed
correlation function in quantum kinetic field theory, we have then
gone over from a closed system (the truncated equations for correlation
functions) to an {\em effectively open system.\/}  In addition to dissipation,
one expects that an effectively open system will manifest noise/fluctuations,
as shown by Calzetta and Hu for the case of the slaving of the four-point
function to the two-point function in the symmetry-unbroken $\lambda \Phi^4$
field theory \cite{calzetta:1995b}.  Thus a framework exists for exploring
irreversibility and fluctuations within the context of a unitarily evolving
quantum field theory, using the truncation and slaving of the correlation
hierarchy.  
The effectively open system framework is useful for precisely those situations,
such as thermalization in the post inflationary Universe, where a separation 
of macroscopic and microscopic time scales (which would permit an effective 
kinetic theory description) does {\em not\/} exist.\footnote{At late times 
in the thermalization stage, when the quantum field is near equilibrium, 
an effective kinetic description may be justified, but will likely
require resummation of hard thermal loops \cite{boyanovsky:1996f}.  
Under such circumstances, even the evaluation of transport coefficients
is nontrivial for high temperatures \cite{jeon:1995a}.}

While it is certainly not the only coarse graining scheme which could be
applied to an interacting quantum field,\footnote{
Recently, a paper \cite{boyanovsky:1996f} claimed to study
nonequilibrium relaxation without assuming {\em a priori\/} the existence of 
a separation of microscopic time scales and incorporating hard thermal 
loop resummation.  However, this study assumed a decohered 
initial density matrix {\em and\/} near-equilibrium initial conditions, 
and thus did not encompass the most general initial conditions (and neither 
of the above-stated conditions holds true at the end of preheating in 
realistic inflation scenarios).}
the slaving of higher correlation functions to lower-order correlation 
functions within a particular truncation of the correlation hierarchy, as a 
particular coarse graining method, has several important benefits. 
First, it can be implemented in a truly nonperturbative fashion, which is
essential for post-inflation reheating, where the inflaton variance
can be on the order of the tree-level effective mass at the end of preheating
\cite{boyanovsky:1996b,ramsey:1997b}.
This necessitates a nonperturbative resummation of daisy graphs, which
can be incorporated in the truncation/slaving of the correlation hierarchy
in a natural way.  Second, the truncation of the  correlation hierarchy
accords with our intuition that the degrees of freedom readily accessible
to physical measurements are often limited to the mean field and two-point
function.  For example, the transition rate of a particle detector coupled via 
a $m(\tau) \phi[x^{\mu}(\tau)]$ interaction to a quantum field [where 
$m(\tau)$ is the detector's monopole moment operator] depends on the 
field's positive-frequency Wightman function \cite{birrell:1982a}.

Related to the non-existence of a separation of microscopic and macroscopic
time scales in the conditions which prevail at the end of preheating (where
one cannot treat the time-varying effective mass in the quasilocal 
approximation), is the fact that for any collection of parametric oscillators
with time dependent frequency, the notion of vacuum state, and hence, that
of particle, is ambiguous \cite{birrell:1982a,fulling:1989a}.
While the growth of the
variance of the inhomogeneous modes of the inflaton field during preheating
can be attributed to parametric particle creation \cite{shtanov:1995a},
it is unlikely that a well-defined and unique 
particle concept for the inflaton field
exists at the end of the preheating period.  This is because the inflaton
zero mode amplitude and variance contribute a time dependent term to the
effective mass, which means that modes of the inflaton field (or other
fields coupled to it) perceive a time-dependent effective frequency.  In
addition, the expansion of the background spacetime also contributes a
time-dependence to the effective frequency of quantum modes.  The closest
approximation to a ``no-particle'' state is obtained by defining an 
{\em adiabatic vacuum,\/} \cite{parker:1969a} 
which was discussed in Sec.~\ref{sec-initcond}, and
in terms of this, one can define an adiabatic particle basis
\cite{cooper:1997a}.  In this study, we perform a coarse graining which is 
independent of particle representation, and thereby avoid the 
subtle issues involved in defining particles in time dependent background.

A third challenge presented by the thermalization problem in 
post-inflationary reheating is the issue of how to define the entropy 
of the inflaton field as it proceeds towards thermal equilibrium.  
This is an essential point for post-inflation reheating in cosmology.   
As discussed in Chapter~\ref{chap-preheat}, at the 
end of the slow roll period in inflationary cosmology, the quantum state for 
matter fields (other than the inflaton field, which has a large expectation 
value) is to a good approximation given by an adiabatic vacuum state 
\cite{brandenberger:1985a}.  
This implies an extremely small entropy density per comoving volume element.
At the end of the reheating period, the matter fields have reached
a state of local thermal equilibrium, and the entropy per comoving
volume element should be given very nearly by finite-temperature field
theory calculations.  Therefore, during the reheating period, the
entropy per comoving volume grows by an enormous amount.\footnote{A rough
calculation shows that the entropy per comoving volume element grows by 
a factor of about $10^{130}$ during the reheating period \cite{kolb:1990a}.}
An important criterion of the physicality of a particular coarse graining 
scheme is whether it predicts a monotonically increasing entropy during
the thermalization stage.  It should be pointed out that defining the
entropy of a quantum field is an older problem than inflationary cosmology,
and dates to early studies of particle production and vacuum viscosity
in curved spacetime.  In these studies, hydrodynamic 
transport coefficients such as bulk 
and shear viscosity were computed in finite temperature field theory,
and were related to the rate of entropy growth 
through the first law of thermodynamics 
\cite{weinberg:1972a,hu:1981a,hu:1982a,hu:1984b,morikawa:1985a}.  
This approach required the assumption of  near-equilibrium conditions,
an ``imperfect fluid'' (hydrodynamic) form for the energy-momentum tensor,
and a background temperature $T$.  

Because entropy is of such fundamental importance, it is useful to 
discuss how entropy can be defined for the various ways (outlined above)
of approximating the dynamics of a quantum field.  First, for a unitarily 
evolving quantum field theory whose dynamics is a closed system and 
governed by the quantum Liouville equation, it is well known that the
von~Neumann entropy of the density matrix,
\begin{equation}
S_{\text{{\tiny VN}}} = -\text{Tr}[\dens(t) \ln \dens(t)],
\end{equation}
is exactly conserved.  If one can assume a separation of macroscopic and
microscopic time scales, one can go over to the quantum kinetic field
theory framework.  However, in merely carrying out the Wigner transform,
one has not sacrificed any information, and therefore, one should not expect
any increase in entropy.  Of course, if one additionally makes the 
assumption of factorization (equivalently, slaving of the Wigner-transformed
four-point function), one indeed obtains the relativistic Boltzmann equation
in the binary collision approximation.  The Boltzmann entropy
$S_{\text{{\tiny B}}}$ defined
in terms of the phase space distribution $f(k,X)$ for quasiparticles
can in this case be shown to satisfy a relativistic $H$-theorem 
\cite{degroot:1980a,calzetta:1988b}.  
However, in the case where there does {\em not\/} exist such a separation of 
time scales, how does one define the entropy of a quantum field? For 
nonperturbative truncations of the
dynamics of interacting quantum fields, this is a nontrivial question
\cite{hu:1987a}.  Intuitively, one expects that any 
coarse graining which leads to an effectively open system with irreversible 
dynamics will also lead to the growth of entropy.  This intuition is based
on nonequilibrium quantum statistical mechanics, 
where if one has a specific projection operator $P$ projects out the 
{\em irrelevant\/} degrees of freedom from the density operator 
and retains only the {\em relevant\/} degrees of freedom (thus going over to
an open system),
\begin{equation}
\dens_{\text{{\tiny R}}}(t) = P \dens(t),
\end{equation}
there exists a formalism for deriving the equation of motion of the reduced
density matrix $\dens_{\text{{\tiny R}}}$, and in terms of it, the
coarse-grained entropy
\begin{equation}
S_{\text{{\tiny CG}}} = -\text{Tr}[\dens_{\text{{\tiny R}}}(t) \ln
\dens_{\text{{\tiny R}}}(t)],
\label{eq-cge4}
\end{equation}
which will in general not be conserved \cite{zwanzig:1961a,balescu:1975a}. 
Another equally
powerful method adept to field theory is the Feynman-Vernon 
influence functional formalism \cite{feynman:1963a} \nocite{hu:1994b}
which has been used to treat open systems \cite{koks:1997a}.

Let us give a brief depiction of entropy generated associated with
parametric particle creation for a free quantum field in an expanding
Universe \cite{parker:1969a}, or for an interacting field such as the
$\lambda \Phi^4$ theory in the Hartree-Fock approximation or the
O$(N)$ field theory at leading order in the large-$N$ expansion 
\cite{cooper:1994a,cooper:1997a}, where
the dynamics of the quantum field reduces to a collection of parametric
oscillators, each with a time-dependent frequency.  Since the underlying
dynamics is clearly unitary and time-reversal invariant in this case,
a suitable coarse graining leading to entropy growth is not trivially
evident.  Hu and Pavon \cite{hu:1986e} first
made the observation that a coarse graining is implicitly incorporated when 
one chooses to depict particle numbers in the $n$-particle Fock
(or ``N'') representation or to depict phase coherence in the coherent state 
(or ``P'') representation.
Various proposals for coarse graining the dynamics of parametric oscillators
have followed
\cite{hu:1986e,kandrup:1988a,grischuk:1990a,gasperini:1993a,brandenberger:1993a,keskivakkuri:1994a,habib:1996a,cooper:1997b,koks:1997a}.
The language of squeezed states is particularly useful for describing entropy
growth due to parametric particle creation 
\cite{grischuk:1990a,hu:1994b,matacz:1994a,koks:1997a}.  For our purposes, the 
essential features of entropy growth due to parametric particle creation which
distinguish it from correlational entropy growth to be discussed below,
are that parametric particle creation involves a choice of representation
for the state space of the parametric oscillators, and also usually involves
an explicit coarse graining which can be expressed in terms of a projection
operator acting on the density matrix.

In contrast to entropy growth resulting from parametric particle creation, the
coarse graining implicit in the slaving of the correlation hierarchy 
represents a choice of relevant {\em correlations\/} versus irrelevant
correlations.  In this sense, it accords with our intuitive notion that only
the lower correlation functions are readily accessible to physical 
measurement.  However, it is not clear in what sense the nonperturbative 
 slaving of the correlation hierarchy corresponds to the projection of 
irrelevant variables
from the full density matrix of the quantum field.\footnote{
Recently there has been an interesting attempt to express the slaving of higher
correlation functions in the correlation hierarchy in terms of the Zwanzig
projection operator formalism \cite{anastopoulos:1997a}, though so far it 
has only been implemented within the framework of perturbation theory.}
Nevertheless, it is clear that once one truncates the Schwinger-Dyson
hierarchy, and carries out a subsequent slaving of a higher correlation
function to the lower correlation functions, then the resulting effectively
open system will manifest irreversibility and dissipation.  As such, this
coarse graining scheme should result in entropy growth.  Because the 
equal-time correlation functions determine the moment expansion of the 
(Schr\"{o}dinger-picture) density matrix, one may attempt to associate
a reduced density matrix with the dynamical equations for the equal-time
correlation functions in an effectively open system.  Using Eq.~\ref{eq-cge4},
one may then define a coarse-grained entropy for the effectively open system,
which we call the (Calzetta-Hu)
{\em correlation entropy\/} to emphasize its origin in the
slaving of the correlation hierarchy.  This is the entropy which we endeavor
to compute in this study.

It is useful to compare the coarse graining scheme in the 
Calzetta-Hu correlation entropy with the coarse graining scheme
proposed by Hu and Kandrup in their study of the entropy growth due to 
particle interactions in quantum field theory \cite{hu:1987a}.  
In the language of a collection of coupled parametric oscillators, the
Hu-Kandrup proposal is to define a reduced density matrix by projecting the 
full density operator onto each oscillator's single-oscillator Hilbert space 
in turn, 
\begin{equation}
{\boldsymbol g}(\vk) \equiv \text{Tr}_{\vec{k}' \neq \vk} \dens,
\end{equation}
and defining the reduced density operator as the tensor product of the
projected single-oscillator density operators ${\boldsymbol g}(\vk)$,
\begin{equation}
\dens_{\text{{\tiny R}}} \equiv \bigotimes_{\vk} {\boldsymbol g}(\vk).
\end{equation}
The coarse-grained (Hu-Kandrup) entropy
is then just given by Eq.~(\ref{eq-cge4}), from which we obtain
\begin{equation}
S_{\text{{\tiny CG}}} = -\sum_{\vk} \text{Tr}[{\boldsymbol g}(\vk)
\ln {\boldsymbol g}(\vk)].
\end{equation}
It is interesting to observe that for a spatially translation-invariant
density matrix for a quantum field theory which is Gaussian in the
position basis, the Hu-Kandrup entropy is just the von~Neumann entropy
of the full density matrix, because the spatially translation-invariant,
Gaussian density matrix separates into a product over density 
submatrices for each $\vk$ oscillator.  Like the Calzetta-Hu 
correlation-hierarchy coarse graining scheme, the Hu-Kandrup coarse
graining does not choose or depend on a particular representation for
the single oscillator Hilbert space.  In this sense, it is not 
sensitive to parametric particle creation, but instead, it is
sensitive to the establishment of correlations through the explicit
couplings\footnote{ 
The Hu-Kandrup proposal
can also be applied to both free fields in anisotropic spacetimes, where
the mode-mode interaction is called an {\em intrinsic interaction,\/} 
and a self-interacting quantum field, where the interaction
is an {\em extrinsic interaction\/} \cite{hu:1987a}.}
between the oscillators \cite{hu:1987a}.   
The Hu-Kandrup coarse graining also has a direct interpretation in terms of 
projection operator language, which is very convenient from the standpoint of 
defining a coarse-grained entropy.  

In this study, we are interested in the growth of entropy due to the
coarse graining of the {\em correlation hierarchy\/} by slaving of
a higher correlation function.  The simplest nonperturbative truncation
of the Schwinger-Dyson equations for the $\lambda \Phi^4$ field theory
which contains the time-dependent Hartree-Fock approximation (necessary
for taking into account the large variance of the inhomogeneous modes of
the inflaton field at the end of preheating) is the two-loop truncation
of the master effective action, in which only the mean field $\phih$, the
two-point function $G$, the three-point function $C_3$ are dynamical.  All
higher order correlation functions obey algebraic constraints, and can
thus be expressed in terms of the three dynamical correlation functions.
While this truncation of the Schwinger-Dyson equations is well-defined
and could in principle be solved, it is disadvantageous for two reasons.
First, as stated above, without some coarse graining, the
system will not manifest irreversibility and will not equilibrate.  Secondly,
and on a more pragmatic level, it is much easier to work with a Gaussian
density matrix than a non-Gaussian density matrix.  Therefore we slave
the three-point function to the mean field and two-point function, and
thus arrive at an effectively open system.  In principle, a systematic
analysis of the coarse-grained dynamics of the mean field and two-point 
function should include stochasticity \cite{calzetta:1995b}, but we shall
defer a study of stochasticity to a future investigation.

%
%

This chapter is organized as follows.  In Sec.~\ref{sec-tchei}, we
show how the two-loop truncation of the correlation hierarchy leads
to coupled, time-reversal invariant equations for the mean field $\phih$, the
two-point function $G$, and the three-point function $C_3$.  We then show 
that slaving of $C_3$ to $G$ and $\phih$ leads to an effective open
system in which the dynamics for $G$ and $\phih$ is irreversible and
dissipative.  In Sec.~\ref{sec-detcf}, we show how the two-loop truncation 
of the correlation hierarchy can be reformulated in local equations for
equal-time correlation functions.  In Sec.~\ref{sec-cent}, we argue that
slaving of the correlation hierarchy leads to the growth of correlation 
entropy. In Sec.~\ref{sec-4dis} we summarize our results and discuss their
physical significance, as well as possible extensions of this work.

\section{Correlation hierarchy and effectively open systems}
\label{sec-tchei}
We seek a consistent truncation of the Schwinger-Dyson equations for the 
Minkowski-space $\lambda \Phi^4$ field theory  at third order in the 
correlation hierarchy, which means that the dynamical variables are the mean 
field $\phih$, the two-point function $G$, and the three-point function
$C_3$.  As mentioned above, this is the simplest (nonperturbative)
truncation of the correlation hierarchy which contains explicit
mode-mode interactions.\footnote{It should also be noted that a 
systematic study of
the three-point function is useful in calculating the polarization tensor
for the $\phi^3$ theory in six dimensions, which is a toy model for the
three gluon interaction in QCD \cite{carrington:1996a}.}
In this truncation of the dynamics, higher correlation functions 
obey algebraic constraints which relate them to the dynamical correlations.  
Generalizing the CTP-2PI (closed-time-path, two-particle-irreducible)
effective action to incorporate arbitrary-order nonlocal sources 
(e.g., three-point sources $J_{abc}(x,y,z)$, four-point sources 
$J_{abcd}(w,x,y,z)$, etc.) within the Schwinger-Keldysh framework for
nonequilibrium quantum fields, one obtains the {\em master effective action,\/}
which is a functional which, when properly truncated, yields an arbitrary-order
truncation of the Schwinger-Dyson equations for correlation functions
\cite{calzetta:1988b,hu:1989a,calzetta:1993a,calzetta:1995b}.  
In general, truncating the master effective action
at a finite number of loops $l$ yields a truncation of the Schwinger-Dyson
equations in which only the first $l+1$ correlation functions, $\phih$, $G$, 
$C_3$, \ldots, $C_{l+1}$ are dynamical, and all higher correlation functions
are constrained.  Therefore the coupled equations for $\phih$, $G$, and $C_3$
which we seek correspond to the {\em two-loop\/} 
truncation of the master effective action.  For the $\lambda \Phi^4$ field
theory, the two-loop truncation of the master effective action (denoted 
by $\Gamma_{l=2}[\phih,G,C_3]$ in Chapter~\ref{chap-oncst}) has the 
form\footnote{In this chapter, we set $\hbar = 1$ for notational simplicity.}
\begin{multline}
\Gamma_{l=2}[\phih,G,C_3] = S[\phih] - \frac{i}{2} \text{Tr} \text{ln} G
+ \frac{i}{2} {\mathcal A}^{AB} G_{BA} - \frac{\lambda}{8} \sigma^{ABCD}
G_{AB} G_{CD} \\+ \frac{i}{12} C_{ABC} (G^{-1})^{AA'} (G^{-1})^{BB'}
(G^{-1})^{CC'} C_{A'B'C'} - \frac{\lambda}{6} \sigma^{ABCD} C_{ABC} \phih_D,
\label{eq-2lea}
\end{multline}
where we have (following \cite{calzetta:1988b})
introduced a compact notation using capital letters as indices
to denote both spacetime and CTP labels \cite{calzetta:1988b}, i.e.,
$A = (a,x)$, $B = (b,x')$, etc.  In terms of the CTP notation of the previous
chapters,
\begin{equation}
\phih_A \equiv \phih_a(x),
\end{equation}
and
\begin{equation}
G_{AB} = G_{ab}(x,x').
\end{equation}
In this new notation, the four-index symbol $\sigma^{ABCD}$ denotes
\begin{equation}
\sigma^{ABCD} = c^{abcd} \delta(x-x')\delta(x-x'')\delta(x-x'''),
\end{equation}
and the two-index symbol $\sigma^{AB}$ denotes
\begin{equation}
\sigma^{AB} = c^{ab} \delta(x-x').
\end{equation}
As in Chapter~\ref{chap-oncst}, 
${\mathcal A}^{AB}$ denotes the second functional derivative
of the classical action with respect to $\phi$, evaluated at $\phih$,
\begin{equation}
i{\mathcal A}^{AB} = \frac{\delta^2 S}{\delta \phi_A \delta \phi_B}[\phih] 
= -(\square + m^2)\sigma^{AB} - \frac{\lambda}{2} \sigma^{ABA'B'}\phih_{A'}
\phih_{B'}.
\label{eq-4olp}
\end{equation}
In the two-loop truncation of the master effective action, all
$n$-point correlation functions for $n > 3$ are constrained, and can 
be expressed in terms of $\phih$, $G$, and $C_3$ \cite{calzetta:1995b}.
The evolution equations for $C_3$, $G$, and $\phih$ are obtained by
functional differentiation of the effective action with respect to 
$C_3$, $G$, and $\phih$, respectively:
\begin{align}
& \frac{i}{6} (G^{-1})^{AE}(G^{-1})^{
BF}(G^{-1})^{CG} C_{ABC} - \frac{\lambda}{6} \sigma^{EFGD}\phih_D = 0, 
\label{eq-dc3}
\\ &
\frac{i}{2} (G^{-1})^{FE} - \frac{i}{2}
{\mathcal A}^{FE} + \frac{\lambda}{4} \sigma^{ABEF} G_{AB} \notag \\ & 
\qquad\qquad + \frac{i}{4}
C_{ABC} (G^{-1})^{AE} (G^{-1})^{FA'} (G^{-1})^{BB'} 
(G^{-1})^{CC'} C_{A'B'C'} = 0,
\label{eq-dginv}
\\&
\sigma^{EA}(\square + m^2) \phih_A
+ \frac{\lambda}{6} \sigma^{EABC} \phih_A \phih_B \phih_C + \frac{\lambda}{2}
\sigma^{ABED} G_{BA} \phih_D + \frac{\lambda}{6}\sigma^{ABCE} C_{ABC} = 0.
\end{align}
With a little bit of algebra, this reduces to
\begin{align}
& i (G^{-1})^{AE} C_{ABC} - \lambda \sigma^{EFGD} \phih_D 
G_{FB} G_{GC} = 0, \label{eq-3pf}
\\ &
\sigma^{EA}(\square + m^2) \phih_A + \frac{\lambda}{2} \sigma^{ABED}\left(
\frac{1}{3}\phih_A \phih_B + G_{AB}\right)\phih_D + \frac{\lambda}{6}
\sigma^{ABCE} C_{ABC} = 0,
\label{eq-eemf}
\\&
i\left[ (\square + m^2) \sigma^{EA} + \frac{\lambda}{2} \sigma^{EACD}
(\phih_C \phih_D + G_{CD})\right]G_{AF} + \frac{i \lambda}{2} \sigma^{EBCD}
\phih_D C_{FBC} = \delta^E_{\;\;F}.
\label{eq-eeg}
\end{align}
The above three equations are the coupled, time-reversal invariant,
dynamical equations for the mean field, the two-point function, and
the three-point function.  This truncation of the Schwinger-Dyson
hierarchy does not lead to irreversibility or dissipation.  However,
let us now {\em slave\/} the three-point function to the mean-field 
and two-point function, which means solving Eq.~(\ref{eq-3pf}) for the
history of the three-point function with appropriate initial data for
$C_3$ \cite{calzetta:1988b,hu:1989a,calzetta:1995b}.  
Formally, the particular solution is
\begin{equation}
C_{ABC} = -i \lambda \sigma^{EFGD} G_{EA} G_{FB} G_{GC} \phih_D,
\label{eq-solc3}
\end{equation}
where boundary conditions have to be imposed to get the homogeneous
part of the solution.  It is the causal boundary conditions introduced
to solve Eq.~(\ref{eq-3pf}) which introduces an ``arrow of time'' into
the problem.  If we view Eq.~(\ref{eq-solc3}) as representing
an approximation to the dynamics for $C_3$, we may insert it into the
two-loop effective action (\ref{eq-2lea}), obtaining
\begin{multline}
\Gamma[\phih,G] = S[\phih] - \frac{i}{2} \text{Tr} \text{ln} G
+ \frac{i}{2} {\mathcal A}^{AB} G_{BA} - \frac{\lambda}{8} \sigma^{ABCD}
G_{AB} G_{CD} \\
+\frac{i}{12} \lambda^2 \sigma^{ABCD}\sigma^{A'B'C'D'} \phih_D \phih_{D'}
G_{AA'} G_{BB'} G_{CC'},
\end{multline}
which is precisely the two-loop, 2PI effective action for the
$\lambda \Phi^4$ theory (as discussed in Chapter~\ref{chap-oncst}).  From
previous studies \cite{calzetta:1988b,calzetta:1995b}, it is known that the
order $\lambda^2$ term in this action leads to non-time-reversal-invariant 
dynamics for the mean field and two-point function.  In slaving the 
three-point function to the mean field and two-point function, we have 
introduced a coarse graining which turns the closed system of 
(\ref{eq-3pf})--(\ref{eq-eeg}) into an {\em effective open system.}  To
see this, let us compute the equation of motion for the mean-field $\phih$
by taking the functional derivative with respect to $\phihp$ and then
identifying $\phihp = \phihm = \phih$, as described in 
Chapter~\ref{chap-oncst}.  We obtain a real and causal equation of the form
\begin{equation}
\left( \square + m^2 + \frac{\lambda}{6} \phih^2 + \frac{\lambda}{2} G 
\right) \phih(x) + \frac{\lambda^2}{3} \int d^{\,4}x' \text{Im}\left(
G_{++}(x,x')^3\right) \theta(x,x') = 0.
\label{eq-mfntr}
\end{equation}
The theta function $\theta(x,x')$ appearing in the integrand
dictates that the integral is over the past history of 
$\phih$; this term is clearly not time-reversal invariant.  Furthermore, if one
substitutes free-field propagators for the $G_{ab}$ in Eq.~(\ref{eq-mfntr}),
[an approximation which corresponds to a quadratic expansion of 
$\Gamma[\phih,G]$ in powers of $\phih$], it follows that the equation for 
$\phih$ is dissipative for momentum modes of $\phih$ above the three-particle 
threshold $9m^2$ \cite{bjorken:1965a,calzetta:1988b,calzetta:1995b},
\begin{equation}
\text{Im}(G_{++}(x,x')^3) \theta(x,x') = \int \frac{d^{\,4}k}{(2\pi)^4}
e^{ik\cdot(x-x')} \int^{\infty}_{9m^2} ds \frac{h(s)}{s - (k+i\epsilon)^2},
\end{equation}
where 
\begin{equation}
(k+i\epsilon)^2 = (k_0 + i\epsilon)^2 - \vec{k}^2,
\end{equation} 
and the function $h(s)$ is given by \cite{calzetta:1995b}
\begin{equation}
h(s) = \frac{8}{s} \int^{(\sqrt{s}-m)^2}_{4m^2} dt \sqrt{(s+m^2-t)^2 -4sm^2}
\sqrt{1 - \frac{4m^2}{t}} \theta(s-9m^2).
\end{equation}
Having illustrated how the slaving of higher correlation function(s) to
the lower correlation functions leads to irreversible dynamics, let us
momentarily return to the full two-loop truncation of the Schwinger-Dyson
equations for $\phih$, $G$, and $C_3$. 
Although Eq.~(\ref{eq-3pf}) readily leads to the time-oriented 
{\em Ansatz\/} (\ref{eq-solc3}) for slaving the three-point function to 
$\phih$ and $G$, it is not convenient if, instead, we wish to simultaneously 
solve the coupled Eqs.~(\ref{eq-3pf})--(\ref{eq-eeg}).
This is because one must have a closed expression for $G^{-1}$ in order 
understand (\ref{eq-3pf}) as a differential equation.  Making use of
Eqs.~(\ref{eq-dc3}) and (\ref{eq-dginv}), we have
\begin{multline}
(G^{-1})^{FE} = i(\square + m^2) \sigma^{FE} + \frac{i \lambda}{2}
\sigma^{ABEF} (\phih_A \phih_B + G_{AB})\\ - \frac{i\lambda}{2}
\sigma^{FBCD} (G^{-1})^{AE} \phih_D C_{ABC},
\end{multline}
from which we obtain a formal expression for $G^{-1}$,
\begin{equation}
(G^{-1})^{DE} = i \left[ (\square + m^2)\sigma^{FE} + \frac{\lambda}{2}
\sigma^{ABEF} (\phih_A \phih_B + G_{AB})\right] (M^{-1})^D_{\;\;F},
\end{equation}
in terms of the inverse of the ``matrix'' $M_A^{\;\;B}$ defined by
\begin{equation}
M_A^{\;\;B} = \delta_A^{\;\;B} + 
\frac{i\lambda}{2}\sigma^{BECD}\phih_D C_{AEC}.
\end{equation}
Naturally, the inverse $(M^{-1})^A_{\;\;B}$ has an infinite power series
expansion in $\lambda$, but for our purposes, it is sufficient to work
at lowest order in $\lambda$ in the equation for $C_{ABC}$, so we
will take as a first approximation $M_A^{\;\;B} 
\simeq \delta_A^{\;\;B} + O(\lambda)$,
whereupon the equation for $C_{ABC}$ becomes linear,
\begin{equation}
i {\mathcal A}^{AE} C_{ABC} = \lambda \sigma^{EFGD} \phih_D G_{BF} G_{GC},
\label{eq-eec3}
\end{equation}
where ${\mathcal A}^{AB}$ is defined above in Eq.~(\ref{eq-4olp}).
From Eqs.~(\ref{eq-eemf}), (\ref{eq-eeg}), and (\ref{eq-eec3}), we see 
immediately that this approximation of the two-loop Schwinger-Dyson equations
results in coupled dynamical equations for $\phih$, $G$, and $C_3$ which are 
manifestly time-reversal invariant.  The essential features of 
Eq.~(\ref{eq-eec3}) are that it is linear in $C_3$ and that,
for a spatially translation-invariant quantum state, it does
not couple the various spatial momentum modes of the spatial Fourier
representation of $C_3$ to one another.

\section{Dynamics of equal-time correlation functions}
\label{sec-detcf}
While the previous exposition of the functional integral approach to deriving 
the (truncated and coarse-grained)
correlation dynamics is useful for a study of the origin of dissipation 
in an effective open system \cite{calzetta:1988b,hu:1989a},
it appears to be less convenient for computing the entropy of the quantum
field, and in particular, making a connection between the slaving of $C_3$
and the growth of entropy.  This is because in the functional integral
approach, the dynamical equations derived are time-nonlocal.  Since in this
chapter we are concerned with the problem with defining the entropy of an
effectively open system, it would be useful to have a formulation of the
dynamics which makes reference only to the equal-time expectation values
of various products of the Heisenberg field operators $\Phi_{\text{{\tiny H}}}$
and $\dot{\Phi}_{\text{{\tiny H}}}$, which are the quantities which should 
actually appear in a position-basis moment expansion 
of the Schr\"{o}dinger-picture density operator 
\cite{eboli:1988a,cooper:1997b}.  Let us therefore
reconsider the $\lambda \Phi^4$ field theory in the Heisenberg picture, and 
derive the evolution equations for {\em equal-time correlation functions\/}
and their time derivatives.\footnote{This approach was also applied in 
\cite{mazenko:1985b} to the large-$N$ limit of the O$(N)$ model.}
As in the previous section, we will first work with a two-loop
truncation of the Schwinger-Dyson equations
in the same approximation as Eqs.~(\ref{eq-eec3}), (\ref{eq-eeg}), and
(\ref{eq-eemf}).  As in Chapter~\ref{chap-oncst}, the Heisenberg field 
operator is $\PhiH$ and its conjugate momentum operator, 
$\dot{\Phi}_{\text{{\tiny H}}}$.  
The Heisenberg equation of motion for the field operator has the form
\begin{equation}
\ddot{\Phi}_{\text{{\tiny H}}} + m^2 \PhiH + \frac{\lambda}{6} 
\Phi^3_{\text{{\tiny H}}} = 0.
\label{eq-hem}
\end{equation}
Let us suppose that the quantum state of the system consists of a
density operator $\dens_{\text{{\tiny H}}}$ which is invariant under
spatial translations and rotations, and is Hermitian and has unit trace,
$\text{Tr}(\dens_{\text{{\tiny H}}}) = 1$.  In the Heisenberg picture, the density
operator is time independent.
The expectation value of an observable ${\mathcal O}_{\text{{\tiny H}}}$ 
is given by
\begin{equation}
\langle {\mathcal O} \rangle = \text{Tr}(\dens_{\text{{\tiny H}}} 
{\mathcal O}_{\text{{\tiny H}}}).
\end{equation}
The mean field, defined in Eq.~(\ref{eq-defmf3}) above, is then given by
\begin{equation}
\phih(x^0) = \langle \Phi \rangle = \text{Tr}(\dens_{\text{{\tiny H}}} \PhiH(x)),
\end{equation}
which is spatially homogeneous due to the spatial translation and
rotation invariance of the density matrix and the action for the 
$\lambda \Phi^4$ theory.  Following Eq.~(\ref{eq-deffluc1pi}), we define
the fluctuation field $\vphiH$ by
\begin{equation}
\vphiH(x) = \PhiH(x) - \phih(x^0).
\label{eq-deffluc4}
\end{equation}
The expectation value of $\vphiH$ clearly vanishes as a consequence of
Eqs.~(\ref{eq-defmf3}) and (\ref{eq-deffluc1pi}).  Inserting 
Eq.~(\ref{eq-deffluc4}) into Eq.~(\ref{eq-hem}), and taking the expectation
value, we find
\begin{equation}
\ddot{\phih} + m^2 \phih + \frac{\lambda}{6} (\phih^3 + [A(t)] + 
3 \phih [G(t)]) = 0,
\label{eq-mfe4}
\end{equation}
where we have defined
\begin{align}
&[A(x^0)] = \langle \PhiH(x)^3 \rangle = \text{Tr}(\dens_{\text{{\tiny H}}} \PhiH(x)^3 )\\
&[G(x^0)] = \langle \PhiH(x)^2 \rangle = \text{Tr}(\dens_{\text{{\tiny H}}} \PhiH(x)^2 ),
\end{align}
which are spatially homogeneous because of the translation and rotation
invariance of the density matrix.\footnote{Naturally, $[G]$ and $[A]$ are
divergent and must be regularized within a consistent renormalization 
procedure.  The divergence structure for $[G]$ for a spatially 
translation-invariant quantum state is well known (see, e.g., 
Chapter~\ref{chap-preheat} above).  We will not discuss the divergence
structure of $[A]$.}  Making use of Eqs.~(\ref{eq-hem}) and (\ref{eq-mfe4}), 
we can derive an equation of motion for $\vphiH$,
\begin{equation}
\ddot{\vphi}_{\text{{\tiny H}}}(x) + m^2 \vphiH(x) + \frac{\lambda}{6}
(\vphi_{\text{{\tiny H}}}^3 + 3 \phih^2 \vphiH + 3 \phih \vphi^2_{\text{{\tiny
H}}} - [A] - 3 \phih [G] ) = 0.
\label{eq-hemv}
\end{equation}
Using Eqs.~(\ref{eq-hem}) and (\ref{eq-hemv}), we would like to derive
coupled equations for the equal-time correlation functions and their
derivatives.  To simplify notation, we will use the latin indices
$i,j,k,l,m,n$ to denote spatial coordinates, so that $G_{ij}(t)$ stands for
$G(\vec{x}_i,t;\vec{x}_j,t)$.  We will further simplify notation by dropping
explicit notation of the time $t$, so that $G_{ij}(t)$ will be abbreviated as 
$G_{ij}$.  Finally, we drop the $H$ subscript on Heisenberg field operators.
In this notation, we seek coupled equations for
\begin{align}
& A_{ijk} = \langle \vphi_i \vphi_j \vphi_k \rangle, \label{eq-daijk4} \\
& B_{ijk} = \frac{1}{3} \langle \dvphi_i \vphi_j \vphi_k +
\vphi_i \dvphi_j \vphi_k + \vphi_i \vphi_j \dvphi_k\rangle,\\
& C_{ijk} = \frac{1}{3} \langle \vphi_i \dvphi_j \dvphi_k
+ \dvphi_i \vphi_j \dvphi_k + \dvphi_i \dvphi_j
\vphi_k \rangle, \\
& D_{ijk} = \langle \dvphi_i \dvphi_j \dvphi_k \rangle, \\
& G_{ij} = \langle \vphi_i \vphi_j \rangle,\\
& F_{ij} = \frac{1}{2} \langle \dvphi_i \vphi_j + \vphi_i \dvphi_j \rangle,\\
& E_{ij} = \langle \dvphi_i \dvphi_j \rangle,\label{eq-deij4}
\end{align}
along with $\phih$ and $\dot{\phih} \equiv \pih$.  It should be emphasized
that in the above definitions a limit process is understood which would
avoid the appearance of Schwinger terms \cite{pokorski:1987a,boyanovsky:1996f}.
Clearly, $A_{ijk}$ as defined above
is just the equal-time limit of $C_3$ from the previous section, and
$G_{ij}$ is just the equal-time limit of $G$ from the previous section.
In the two-loop truncation of the correlation hierarchy, the correlation
functions $C_4$ and $C_5$ are constrained \cite{calzetta:1995b}.  The
constraint equations for $C_4$ and $C_5$ are
\begin{equation}
\langle \vphi_i \vphi_j \vphi_k \vphi_l \rangle = G_{ij} G_{kl} +
G_{ik} G_{jl} + G_{il} G_{jk} 
\label{eq-cec44}
\end{equation}
and 
\begin{multline}
\langle \vphi_i \vphi_j \vphi_k \vphi_l \vphi_m \rangle = 
G_{ij} A_{klm} + G_{ik} A_{jlm} + G_{il} A_{jkm} + G_{im} A_{jkl} +
G_{jk} A_{ilm} \\ + G_{jl} A_{ikm} + G_{jm} A_{ikl} + G_{kl} A_{ijm} +
G_{km} A_{ijl} + G_{lm} A_{ijk}.
\label{eq-cec54}
\end{multline}
We now differentiate each of Eqs.~(\ref{eq-daijk4})--(\ref{eq-deij4}) 
(as well as $\pih$ and $\phih$) with respect to time, and apply 
Eqs.~(\ref{eq-hemv}), (\ref{eq-cec44}), and (\ref{eq-cec54}), and we
find that within the approximation where the equations of motion for 
spatial Fourier modes of $C_3$ do not couple to one another 
[i.e., where we require agreement with Eq.~(\ref{eq-eec3})], we obtain a closed
set of dynamical equations for the equal-time correlation functions
and their time derivatives, $A_{ijk}$, $B_{ijk}$, $C_{ijk}$, $D_{ijk}$,
$G_{ij}$, $F_{ij}$, $E_{ij}$, $\phih$, and $\pih$,
\begin{align}
& \dot{A}_{ijk} = 3 B_{ijk} \label{eq-be1} \\
& \dot{B}_{ijk} = 2 C_{ijk} - {\mathfrak M}^2 A_{ijk} - 
\frac{\lambda}{3} \phih ( G_{ij} G_{ik} + G_{jk} G_{ji} + G_{ki} G_{kj} ) \\
& \dot{C}_{ijk} = D_{ijk} - 2 {\mathfrak M}^2 B_{ijk} - \frac{\lambda}{3}
\phih ( G_{ik} F_{ij} + G_{ij} F_{ik} + G_{ij} F_{jk} ) \\
& \dot{D}_{ijk} = - 3 {\mathfrak M}^2 - \lambda \phih ( F_{ij} F_{ik} +
F_{jk} F_{ji} + F_{ki} F_{kj}) \\
& \dot{G}_{ij} = 2 F_{ij} \label{eq-sleq1} \\
& \dot{F}_{ij} = E_{ij} - {\mathfrak M}^2 G_{ij} - \frac{\lambda}{4} \phih
(A_{iij} + A_{ijj}) \label{eq-sleq2} \\
& \dot{E}_{ij} = -2 {\mathfrak M}^2 F_{ij} \label{eq-sleq3} \\
& \dot{\phih} = \pih \label{eq-sleq4} \\
& \dot{\pih} = -\left( m^2 + \frac{\lambda}{6} \phih^2 + \frac{\lambda}{2}
[G] \right) \phih - \frac{\lambda}{6} [A], \label{eq-be2}
\end{align}
where ${\mathfrak M}^2$ is the time-dependent Hartree-Fock effective mass
defined by
\begin{equation}
{\mathfrak M}^2 = m^2 + \frac{\lambda}{2}\phih^2 + \frac{\lambda}{2}[G].
\label{eq-dems4}
\end{equation}
The appearance of $\lambda \phih/6$ in Eq.~(\ref{eq-be2}), as opposed to
the $\lambda \phih/2$ in Eq.~(\ref{eq-onmflm}), is because in this
Chapter we are working in the time-dependent Hartree-Fock approximation
for the $\lambda \Phi^4$ theory instead of the leading-order large-$N$
approximation in the O$(N)$ theory studied in Chapter~\ref{chap-oncst}.
We remind the reader that $[A] = A_{iii}$ and $[G] = G_{ii}$.
It can be verified that Eqs.~(\ref{eq-be1})--(\ref{eq-be2}) are 
time-reversal invariant, where under the time-reversal operator $\Theta$
acts as follows
\begin{align}
& \Theta(A_{ijk}) = A_{ijk}, \\
& \Theta(B_{ijk}) = -B_{ijk}, \\
& \Theta(C_{ijk}) = C_{ijk}, \\
& \Theta(D_{ijk}) = -D_{ijk}, \\
& \Theta(G_{ij}) = G_{ij},\\
& \Theta(F_{ij}) = -F_{ij},\\
& \Theta(E_{ij}) = E_{ij},\\
& \Theta(\phih) = \phih, \\
& \Theta(\pih) = -\pih.
\end{align}
Eqs.~(\ref{eq-be1})--(\ref{eq-be2}) represent a generalization of the
equations derived in \cite{mazenko:1985b} to third order in the correlation
hierarchy, for the case of the $\lambda \Phi^4$ model.\footnote{In 
\cite{mazenko:1985b}, the quartic O$(N)$ model was studied in the large-$N$ 
limit, which at leading order is structurally analogous to the time-dependent
Hartree-Fock approximation.}  Because of the time-reversal invariance of
the dynamical equations, it is expected that the density matrix
whose moment expansion is given by the equal-time correlation functions
(\ref{eq-daijk4})--(\ref{eq-deij4}) would have conserved von Neumann
entropy.  However, this appears to be difficult to prove.

It is interesting to note that the equations of motion 
for the time-dependent Hartree-Fock approximation are contained in the 
equations of motion (\ref{eq-be1})--(\ref{eq-be2}), which is to be
expected since this approximation is known to be a subcase of the
two-loop truncation of the 2PI effective action, as pointed out
in Chapter~\ref{chap-oncst}.  In the present context,
the time-dependent Hartree-Fock approximation consists of setting $A_{ijk}=0$ 
for all time, and in that case the resulting equations of motion for 
$G_{ij}$, $F_{ij}$ and $E_{ij}$ take the simple form
\begin{align}
& \dot{G}_{ij} = 2 F_{ij},\label{eq-tdhf1}\\
& \dot{F}_{ij} = E_{ij} - {\mathfrak M}^2 G_{ij},\\
& \dot{E}_{ij} = -2 {\mathfrak M}^2 F_{ij},\label{eq-tdhf3}
\end{align}
which is seen to agree with the equations of motion in \cite{mazenko:1985b}.
We note that in the time-dependent Hartree-Fock approximation, there is
a first integral which can be obtained from 
Eqs.~(\ref{eq-tdhf1})--(\ref{eq-tdhf3}),
\begin{equation}
\frac{d}{dt} \left( E_{ij} G_{ij} - F_{ij}^2 \right) = 0.
\end{equation}
Therefore it will be useful to define
\begin{equation} 
\sigma_{ij}^2 \equiv 4( E_{ij} G_{ij} - F_{ij}^2),
\label{eq-dsig}
\end{equation}
where the $\sigma_{ij}$ function is a constant of the motion in the
time-dependent Hartree-Fock approximation.  It is not obvious from the
definition (\ref{eq-dsig}) whether $\sigma_{ij}$ is necessarily real.  
However, it was shown in \cite{cooper:1997b} that the spatial Fourier 
transform of $\sigma_{ij}$, denoted by $\sigma_{\vec{k}}$, is real and 
bounded from below, as a consequence of the uncertainty principle.
that its spatial Fourier transform is indeed real, and bounded from below.
We will also show in the next section that by including 
the ``setting-sun'' diagram which goes beyond the Hartree-Fock approximation 
(i.e., in the effectively open system discussed in Sec.~\ref{sec-tchei}), the 
$\sigma_{ijk}$ function defined by Eq.~(\ref{eq-dsig}) is no longer constant.

Let us therefore {\em slave\/} the three-point function to the mean field and 
two-point function as was done in Sec.~\ref{sec-tchei}, in which case 
the equal-time correlation function $A_{ijk}$ takes the form
\begin{equation}
A_{ijk}(t) = 2\lambda \int d^{\,4} y \theta(t,y^0) \text{Im} \left[
G_{++}(y,x_i) G_{++}(y,x_j) G_{++}(y,x_k)\right] \phih(y),
\label{eq-slc34}
\end{equation}
where $x_i$, $x_j$ and $x_k$ are shorthand for $(t,\vec{x}_i)$, 
$(t,\vec{x}_j)$, etc., and the $i,j,k$ label the spatial coordinate vectors
but are not vector indices.  The functions $G_{++}(x,x')$ are the time-ordered
Green functions satisfying the equation
\begin{equation}
\left( \square + {\mathfrak M}^2 \right) G_{++}(x,x') = -i\delta(x-x'),
\label{eq-eqfp4}
\end{equation}
with appropriate ``in-in'' boundary conditions.  The parameter 
${\mathfrak M}^2$ is the time-dependent effective mass defined in
Eq.~(\ref{eq-dems4}). In the effectively open system where the three-point
function is slaved to the mean field and the two-point function, it is 
evidently not possible to represent the coupled equations for 
$\phih$ and $G$ in terms of completely time-local quantities, since
the time-ordered propagators appear.  This accords with intuition
gained from the projection operator formalism in nonequilibrium statistical
mechanics, where  the resulting equation for the ``relevant'' density operator
is time nonlocal \cite{zwanzig:1961a}.  However, the dynamical equations
for $\phih$ and $G$ are still well-defined, provided boundary conditions are 
supplied to get a unique solution to Eq.~(\ref{eq-eqfp4}).  In this case, 
however, the dynamics becomes time-nonlocal and irreversible.

\section{Correlation entropy}
\label{sec-cent}
In the preceding sections we derived coupled equations for the equal-time 
correlation functions within a two-loop truncation of the correlation 
hierarchy for the $\lambda \Phi^4$ field theory, and showed how the slaving 
of the three-point function to the mean field and the two-point function 
leads to irreversibility and dissipation.  In this section we attempt to 
define the entropy associated with this effectively open system, which we call
the (Calzetta-Hu) 
{\em correlation entropy.\/}  Recall that the slaving of the three-point
function leads to coupled equations (\ref{eq-sleq1}), (\ref{eq-sleq2}), 
(\ref{eq-sleq3}), (\ref{eq-sleq4}), and (\ref{eq-be2}) for $G_{ij}$,
$F_{ij}$, $E_{ij}$, $\phih$, and $\pih$, where $A_{ijk}$ is expressed in terms
of the two-point function and mean field by Eq.~(\ref{eq-slc34}).  
Because we have slaved the three-point
function, the reduced density matrix should have a Gaussian moment expansion 
in the position basis.  In this section we investigate the consequences
of this fact for the entropy of the effectively open system.

For a calculation of the correlation entropy, we 
now go over to the Schr\"{o}dinger picture, where $\Phi_{\vk}$ and
$\Pi_{\vk}$ are the spatially Fourier transformed field operator and 
conjugate momentum, respectively.
The most general Gaussian density operator satisfying spatial translation
and rotation invariance has the position-basis matrix element 
\cite{eboli:1988a,cooper:1997b}
\begin{multline}
\langle \phi' | \dens | \phi \rangle = 
\prod_{\veck} (2\pi \xi_{\veck}^2)^{-1/2} \exp \biggl[
i \pih (\phi'_0 - \phi_0) - \frac{\sigma_{\veck}^2 + 1}{8\xi_{\veck}^2}
\left[ (\phi'_{\veck} - \phih \delta_{\veck 0})^2 + (\phi_{\veck} - \phih \delta_{
\veck 0})^2 \right] \\
- \frac{i \eta_{\veck}}{2 \xi_{\veck}}\left[(\phi'_{\veck} - \phih\delta_{\veck 0})^2
- (\phi_{\veck} - \phih \delta_{\veck 0})^2\right] + 
\frac{\sigma_{\veck}^2 -1}{4\xi_{\veck}^2}(\phi'_{\veck} - \phih\delta_{\veck 0})(
\phi_{\veck} - \phih \delta_{\veck 0})\biggr],
\label{eq-gdm}
\end{multline}
where $\xi_{\vk}$, $\sigma_{\vk}$, and $\eta_{\vk}$ are all functions of time.
The $\sigma_{\vk}$ parameter controls the extent to which the density operator
represents a {\em mixed state.\/} It has been called the ``phase
mixing'' parameter by some authors \cite{eboli:1988a}; we eschew this 
nomenclature because of the alternative meanings of ``phase mixing'' in
statistical mechanics.
The equal-time correlation functions can be computed directly from 
Eq.~(\ref{eq-gdm}),
\begin{align}
& \text{Tr}\left[\dens(t)(\Phi_{\vk} - \phih\delta_{\vk 0})^2\right] = 
\xi^2_{\vk}, \\
& \text{Tr}\left[\dens(t)(\Pi_{\vk} - \pih\delta_{\vk 0})^2\right] = 
\eta_{\vk}^2 + \frac{\sigma_{\vk}^2}{4\xi_{\vk}^2},\\
& \text{Tr}\left[\dens(t)(\Phi_{\vk}\Pi_{\vk} + \Pi_{\vk}\Phi_{\vk} - 
2 \phih\pih \delta_{\vk 0})\right] = 2\xi_{\vk} \eta_{\vk}.
\end{align}
Then it is straightforward to determine the relations between the
time-dependent functions in the moment expansion of the density operator
and the equal-time correlation functions,
\begin{align}
& G_{\vk}(t) = \xi_{\vk}^2, \\
& E_{\vk}(t) = \eta_{\vk}^2 + \frac{\sigma_{\vk}^2}{4\xi_{\vk}^2},
\label{eq-dsk4} \\
& F_{\vk}(t) = \xi_{\vk} \eta_{\vk},
\end{align}
where, due to spatial translation invariance,
 the spatially Fourier-transformed two-point functions are defined as
\begin{equation}
G_{ij}(t) = \int \frac{d^{\,3}k}{(2\pi)^3} e^{i\vk\cdot(\vec{x}_i-\vec{x}_j)}
G_{\vk}(t),
\end{equation}
and similarly for $F_{ij}$ and $E_{ij}$.
In the approximation where the three-point function has been slaved to the
mean field and two-point function, we can use Eqs.~(\ref{eq-sleq1}),
(\ref{eq-sleq2}), and (\ref{eq-sleq3}) to obtain
\begin{equation}
\frac{d}{dt}\left( E_{ij} G_{ij} - F_{ij}^2 \right) = 
\frac{d}{dt}\left( \frac{ \sigma_{ij}^2 }{4} \right) =
\frac{\lambda}{4} \phih (A_{iij} + A_{ijj}), \label{eq-s41}
\end{equation}
so that $\sigma_{ij}$ is no longer a constant of the motion.  Note that
the $\sigma_{\vk}$ defined in Eq.~(\ref{eq-dsk4}) above is just the
spatial Fourier transform of $\sigma_{ij}$,
\begin{equation}
\sigma_{ij}(t) = \int \frac{d^{\,3}k}{(2\pi)^3} e^{i \vk \cdot (\vec{x}_i -
\vec{x}_j)} \sigma_{\vk}(t). \label{eq-s42}
\end{equation}
As stated above, due to the uncertainty principle, $\sigma_{\vk}$
is bounded from below \cite{cooper:1997b}, $\sigma_{\vk} \geq 1$.
We are now in a position to compute the coarse-grained entropy of the reduced
density matrix in the effectively open system where the three-point function
has been slaved, the {\em correlation entropy.\/}  
Making use of the calculation of \cite{eboli:1988a,cooper:1997b}, we find
\begin{align}
S_{\text{{\tiny CG}}} = & -\text{Tr}[\dens(t) \ln \dens(t)] \\ & =
\sum_{\vk} \left[ \left( \frac{\sigma_{\vk}(t) + 1}{2} \right) \ln
\left( \frac{\sigma_{\vk}(t) + 1}{2} \right) - 
\left( \frac{\sigma_{\vk}(t) - 1}{2} \right) \ln
\left( \frac{\sigma_{\vk}(t) - 1}{2} \right)\right],
\end{align}
where $\sigma_{\vk}$ is given by Eqs.~(\ref{eq-dsig}) and (\ref{eq-s42}).
From Eq.~(\ref{eq-s41}), we see that $\dot{S}$ is given by
\begin{equation}
\dot{S}_{\text{{\tiny CG}}} = \sum_{\vk} \frac{\dot{\sigma}_{\vk}}{2}
\ln \left( \frac{\sigma_{\vk}+1}{\sigma_{\vk}-1} \right),
\label{eq-dscgt4}
\end{equation}
where $\dot{\sigma}_{\vk}$ is given by Eqs.~(\ref{eq-s41}) and (\ref{eq-s42}).
It is not clear from Eq.~(\ref{eq-dscgt4}) whether 
$\dot{S}_{\text{{\tiny CG}}}$ is positive definite.

\section{Summary}
\label{sec-4dis}
In this chapter we have shown how the truncation of the master effective 
action in the Schwinger-Keldysh formalism leads to coupled,
nonperturbative, and causal dynamical equations for
correlation functions in the $\lambda \Phi^4$ field theory, and that
the slaving of higher correlation functions to the lower correlation
functions leads to irreversibility and dissipation.  We then
showed that a coupled set of equations can be derived for equal-time
correlation functions which appears to be equivalent to the 
two-loop truncation of the master effective action.  Finally, we showed
that the slaving of the three-point function to the mean field and
two-point function leads to nonconserved correlation entropy.
The coupled equations derived in Sec.~\ref{sec-detcf} may be useful in
a further study of the thermalization stage in post-inflation reheating.

There are several directions in which this study might be extended. First, we
intend to investigate whether the two-loop coupled equations for equal-time 
correlation functions, which are manifestly time-reversal invariant, are 
also Hamiltonian.  This would imply a conserved von Neumann entropy for the
density operator in this closed truncation of the correlation hierarchy.
Secondly, we would like to compute the Hu-Kandrup entropy for this truncation
of the dynamics, and to compare it with the correlation entropy computed above
in the case of the effectively open system (where the three-point function
has been slaved to the mean field and the two-point function).  Finally,
it would be useful to investigate under what conditions the correlation
entropy is strictly increasing, as this would help to clarify the role that
the slaving of the correlation hierarchy plays in describing the 
equilibration of nonequilibrium quantum fields.  The dynamical equations
derived for equal-time correlation functions in Sec.~\ref{sec-detcf} may
prove useful in a systematic study of the thermalization stage of a realistic
inflation scenario.

\chapter{Conclusion}
\label{sec-bigconc}

In this dissertation, we have studied the nonequilibrium dynamics of
quantum fields in both curved spacetime and Minkowski space, with particular 
emphasis on the reheating problem in inflationary cosmology.  The primary 
theoretical tools utilized (in various combinations) in this dissertation 
are the Schwinger-Keldysh closed-time-path formalism, the 
two-particle-irreducible and $n$-particle irreducible effective actions, 
and the coarse grained effective action.  

We first derived the coupled dynamical equations for the mean field
and two-point function of a minimally coupled, quartically self-interacting 
O$(N)$-invariant quantum field in a general curved, classical spacetime 
including diagrams up to two loops in the CTP-2PI effective action.
The equations obtained are useful for the study of the dynamics of
the inflaton field during the reheating period of inflationary cosmology,
and, with a changeover to a tachyonic mass, would be useful for a study of
the dynamics of a symmetry-breaking phase transition.  Various
subcases of the two-loop equations were discussed, including the
leading order large-$N$ expansion, which is of particular use in a 
study of parametric particle creation including back reaction effects.

Next we studied the dynamics of the O$(N)$ model in spatially flat
FRW spacetime at leading order in the large-$N$ expansion, where 
the dynamics of the scale factor is determined self-consistently
using the semiclassical Einstein equation.  Initial conditions appropriate
to the end-state of the slow-roll period in chaotic inflation were assumed.
The coupled dynamical equations for the mean field, scale factor, and
inhomogeneous modes of the inflaton field were solved numerically for
different values of the ratio of the inflaton mass $m$  to the Planck mass
$\Mpl$. Nearly relativistic oscillations were assumed, where the
initial inflaton amplitude $\phih_0$ satisfies 
$m^2 \simeq \lambda \phih_0^2/2$.  It was shown that for the case
where $\phih_0 \gtrsim \Mpl/300$, parametric resonance effects are not
an efficient mechanism of energy transfer from the background field
to the inhomogeneous modes because of cosmic expansion.  This shows that
cosmic expansion should be taken into account in a study of preheating
dynamics in inflationary scenarios, such as chaotic inflation, where
the inflaton amplitude is $\Mpl/300$ at the end of slow roll, and more 
generally, when the time scale for growth of the inflaton variance due
to parametric resonance is on the order of the Hubble time.

We then studied fermion particle production in a model consisting of
a scalar $\lambda\Phi^4$ inflaton field coupled via a Yukawa coupling $f$ to
a fermion field $\Psi$.  Fermion particle production is expected to be
important at late stages during reheating in models with unbroken symmetry,
after back reaction has caused parametric resonance effects to cease.
We derived nonperturbative equations for the inflaton mean field and
two-point function which are dissipative due to fermion particle production.
In the small-amplitude limit where perturbation theory is valid, we showed 
that the effective dynamics of the inflaton zero mode, at order $f^2$, can 
be described by a stochastic equation.  The dissipation and noise kernels were
shown to satisfy a zero-temperature fluctuation-dissipation relation (FDR)\@.  
Furthermore, the normal threshold parts of the perturbative coarse-grained 
effective action at $O(f^4)$ were shown to satisfy an FDR, and the noise
kernel contributes multiplicatively to the effective stochastic equation
for the zero mode.  The stochastic variance of the inflaton zero mode was
computed for the late stages of reheating, and it was shown that the
rms fluctuations of the inflaton amplitude can be on the order of the 
inflaton amplitude before the end of reheating, and under such circumstances,
the effect of stochasticity on the zero mode evolution should be taken into 
account.

Finally, we discussed various proposals for defining the entropy of an
interacting quantum field, and in particular, the correlation entropy
which arises when one performs a slaving of a higher correlation function 
within the context of a systematic truncation of the correlation hierarchy.
We showed how the slaving of a higher correlation function leads to
an effectively open system where dissipation necessarily arises.  
We then presented a framework for deriving time-local, coupled equations
for equal-time correlation functions within the context of a two-loop
truncation of the Schwinger-Dyson hierarchy (in which the mean field,
the two-point function, and the three-point function are dynamical).  
We then computed the correlation entropy arising from the slaving of the
three-point function to the mean field and two-point function.

One avenue of ongoing research is the application of the above-described
methods to a study of the dynamics of a nonequilibrium phase transition 
in a field theory with a spontaneously broken symmetry \cite{calzetta:1997a}. 
While phase transitions have long been treated using phenomenological methods 
such as the time-dependent Landau-Ginzberg equation 
\cite{bray:1994a,laguna:1996a}, a complete, first-principles picture of a 
nonequilibrium phase transition from the viewpoint of correlation dynamics, 
including domain growth and/or spinodal decomposition 
(the groundwork for which was set forth in \cite{calzetta:1989b}), has yet to 
emerge.  The master effective action \cite{calzetta:1995b}, which allows 
systematic improvement over mean-field and Hartree-Fock calculations, is the 
preferred tool for this purpose. 

One of the more important cosmological applications of the study of 
nonequilibrium phase transitions is the problem of computing the density of 
topological defects produced in a GUT-scale phase transition in the 
early Universe.  In particular, GUT models with a vacuum manifold which has
nontrivial first homotopy group may give rise to cosmic strings, which are
one of the prevailing candidates for seeding large-scale structure
\cite{brandenberger:1997b}.  Incorporating stochasticity arising from the 
slaving of the four-point function within the three-loop truncation of the 
Schwinger-Dyson equations (with unbroken symmetry) can in principle be used 
to compute the {\em variance\/} in the defect density, which would give an 
indication of the validity of most previous calculations [which focus on only
on the ensemble-averaged defect density \cite{gill:1995a}] 
to observational limits on primordial fluctuations.  In addition, there has 
recently been a great deal in interest 
in (and controversy regarding) the possibility of nonthermal symmetry 
restoration during the preheating dynamics of a quantum field with 
symmetry-breaking potential
\cite{tkachev:1996a,riotto:1996a,boyanovsky:1996h,kofman:1996b,kofman:1996c,boyanovsky:1996c,boyanovsky:1996b}.  The 2PI effective action has already been
applied to this problem \cite{riotto:1996a}.  The CTP-2PI effective action has
also recently proven useful in the study of the dynamics of disoriented 
chiral condensates (DCCs) which might be produced in relativistic heavy-ion 
collisions of sufficient energy to locally restore chiral symmetry
\cite{cooper:1995a,cooper:1997a,boyanovsky:1994b,boyanovsky:1995e,boyanovsky:1996g,boyanovsky:1997a}.  

The use of the coarse grained effective action, in conjunction with the
Schwinger-Keldysh closed-time-path formalism, as in Chapter~\ref{chap-fermion}
where we studied fermion particle production during reheating, may prove useful
in several problems in cosmology and particle physics.  In particular, the 
coarse grained effective action may be used to study the nonperturbative, 
effective dynamics of soft modes of the gauge field in the high temperature, 
symmetry-unbroken phase of the electroweak theory, which is necessary in order
to correctly determine the hot electroweak baryon violation rate 
\cite{son:1997a,arnold:1997a}.  The coarse grained effective action has also
been applied to a study of relaxation, transport, and thermalization 
phenomena \cite{boyanovsky:1996f,greiner:1997a} in scalar field theory with a
quartic interaction, and should be applicable to a study of effective 
dynamics of soft gluon modes in the quark-gluon plasma, where the microscopic
theory is finite-temperature QCD.  Of most direct importance for 
inflationary cosmology is the study of the coarse grained dynamics of 
super-horizon modes of the inflaton field during the slow roll period, where 
decoherence and stochasticity are directly related to the emergence of a 
classical picture of primordial density perturbations \cite{hu:1993a}.  
Work on this problem is ongoing \cite{calzetta:1995a,hu:1996a}.

The problem of thermalization in post-inflation reheating is also
the subject of continued investigation \cite{ramsey:1997e}.  As described
in Chapter~\ref{chap-entropy}, a systematic study of the thermalization stage
will necessarily require inclusion of diagrams beyond the Hartree-Fock
approximation (or equivalently, leading order in the large-$N$ expansion for
an O$(N)$ model).
At this point, temporal non-localities enter into the effective dynamics of 
the two-point function and mean field, as derived from the 2PI effective 
action.  Therefore, the method discussed in Sec.~\ref{sec-detcf} of deriving
coupled time-local equations for equal-time correlation functions should 
prove useful.  More generally, the theoretical techniques involved in 
the study of thermalization in post-inflationary reheating should prove 
useful for any problem which involves thermalization of quantum fields in a 
dynamical background, such as GUT phase transitions in the early Universe.

\appendix

\chapter{Discontinuities of the square diagram}
\label{sec-cutdiag}
In this appendix, the seven terms of Eq.~(\ref{eq-fod}) involving cut 
propagators are explicitly evaluated using the Cutosky rules.
The second and third terms of Eq.~(\ref{eq-fod}) correspond to normal-threshold
singularities in the $t$ and $s$ channels, and are given by
\begin{multline} 
\int \frac{d^4 q}{(2\pi)^4}
\text{Tr}_{\text{{\tiny sp}}}\Bigl[ F_{++}(q) F_{+-}(q+k_1)  
F_{--}(q+k_1+k_2) F_{-+}(q+k_1+k_2+k_3) \Bigr] 
\\ =
 -i \text{Disc}[A_4(k_1,k_2,k_3)_{|\alpha_1 = \alpha_3 = 0}
] \theta(k_2^0 + k_3^0) 
\end{multline}
and
\begin{multline} 
\int \frac{d^4 q}{(2\pi)^4}
\text{Tr}_{\text{{\tiny sp}}}\Bigl[ F_{+-}(q) F_{--}(q+k_1) 
F_{-+}(q+k_1+k_2) F_{++}(q+k_1+k_2+k_3) \Bigr] 
\\ =
 -i \text{Disc}[A_4(k_1,k_2,k_3)_{|\alpha_2=0; \;\; 
\alpha_1 + \alpha_3 = 1}] \theta(k_1^0 + k_2^0),
\end{multline}
respectively.  The third term in Eq.~(\ref{eq-fod}) corresponds to the 
leading-order singularity of the square diagram \cite{eden:1966a} (i.e.,
the solution of the Landau equations in which $\alpha_1, \alpha_2, \alpha_3$
are all nonzero),
\begin{multline}
\int \frac{d^4 q}{(2\pi)^4}
\text{Tr}_{\text{{\tiny sp}}}\Bigl[ F_{+-}(q) F_{-+}(q+k_1)  
F_{+-}(q+k_1+k_2) F_{-+}(q+k_1+k_2+k_3) \Bigr] 
\\ =
 i \text{Disc}[A_4(k_1,k_2,k_3)_{|\alpha_1,\alpha_2,\alpha_3 > 0}
] \theta(k_1^0) \theta(-k_2^0) \theta(k_3^0).
\end{multline}
The last four terms in Eq.~(\ref{eq-fod}) correspond to the four remaining
twice-contracted singularities, and are given by
\begin{multline}
\int \frac{d^4 q}{(2\pi)^4}
\text{Tr}_{\text{{\tiny sp}}}\Bigl[ F_{++}(q) F_{+-}(q+k_1)  
F_{-+}(q+k_1+k_2) F_{++}(q+k_1+k_2+k_3) \Bigr] 
\\ =
 i \text{Disc}[A_4(k_1,k_2,k_3)_{|\alpha_2 = \alpha_3 = 0}
] \theta(k_2^0),
\end{multline}
\begin{multline}
\int \frac{d^4 q}{(2\pi)^4}
\text{Tr}_{\text{{\tiny sp}}}\Bigl[ F_{++}(q) F_{++}(q+k_1)  
F_{+-}(q+k_1+k_2) F_{-+}(q+k_1+k_2+k_3) \Bigr] 
\\ =
 i \text{Disc}[A_4(k_1,k_2,k_3)_{|\alpha_3 = 0; 
\;\; \alpha_1 + \alpha_2 = 1}] \theta(k_3^0),
\end{multline}
\begin{multline}
\int \frac{d^4 q}{(2\pi)^4}
\text{Tr}_{\text{{\tiny sp}}}\Bigl[ F_{+-}(q) F_{-+}(q+k_1)  
F_{++}(q+k_1+k_2) F_{++}(q+k_1+k_2+k_3) \Bigr] 
\\ =
 i \text{Disc}[A_4(k_1,k_2,k_3)_{|\alpha_1 = \alpha_2 = 0}
] \theta(k_1^0),
\end{multline}
\begin{multline}
\int \frac{d^4 q}{(2\pi)^4}
\text{Tr}_{\text{{\tiny sp}}}\Bigl[ F_{+-}(q) F_{--}(q+k_1)  
F_{--}(q+k_1+k_2) F_{-+}(q+k_1+k_2+k_3) \Bigr] 
\\ =
 i \text{Disc}[A_4(k_1,k_2,k_3)_{|\alpha_1 = 0; \;\; 
\alpha_2 + \alpha_3 = 1}] \theta(k_1^0 + k_2^0 + k_3^0).
\end{multline}


\end{document}